\newif\iffigs\figstrue
\documentclass[a4paper,11pt]{article}
\usepackage{latexsym,amssymb,lscape,graphics}
\usepackage{graphicx}        
\usepackage{longtable}
\usepackage{youngtab}
\usepackage{multirow}
\usepackage{color}
\usepackage{slashed,epsfig}
\usepackage{amsfonts}

\newtheorem{definizione}{Definition}[section]

\newtheorem{lemma}{Lemma}[section]

\newcommand{\bd}{\begin{definizione}}
\newcommand{\ed}{\end{definizione}}

\def\IC{\relax\,\hbox{$\inbar\kern-.3em{\rm C}$}}
\def\IG{\relax\,\hbox{$\inbar\kern-.3em{\rm G}$}}
\def\IB{\relax{\rm I\kern-.18em B}}
\def\ID{\relax{\rm I\kern-.18em D}}
\def\IL{\relax{\rm I\kern-.18em L}}
\def\IF{\relax{\rm I\kern-.18em F}}
\def\IH{\relax{\rm I\kern-.18em H}}
\def\II{\relax{\rm I\kern-.17em I}}
\def\IN{\relax{\rm I\kern-.18em N}}
\def\IP{\relax{\rm I\kern-.18em P}}
\def\IQ{\relax\,\hbox{$\inbar\kern-.3em{\rm Q}$}}
\def\bfzero{\relax\,\hbox{$\inbar\kern-.3em{\rm 0}$}}
\def\IK{\relax{\rm I\kern-.18em K}}
\def\IG{\relax\,\hbox{$\inbar\kern-.3em{\rm G}$}}
 \font\cmss=cmss10 \font\cmsss=cmss10 at 7pt
\def\IR{\relax{\rm I\kern-.18em R}}
\def\ZZ{\relax\ifmmode\mathchoice
{\hbox{\cmss Z\kern-.4em Z}}{\hbox{\cmss Z\kern-.4em Z}}
{\lower.9pt\hbox{\cmsss Z\kern-.4em Z}} {\lower1.2pt\hbox{\cmsss
Z\kern-.4em Z}}\else{\cmss Z\kern-.4em Z}\fi}
\def\bfone{\relax{\rm 1\kern-.35em 1}}

\def\Solv{\mathop{\rm Solv}\nolimits}

\def\inbar{\vrule height1.5ex width.4pt depth0pt}
\def\bfzero{\relax{\rm I\kern-.18em 0}}
\def\bfone{\relax{\rm 1\kern-.35em 1}}
\def\twomat#1#2#3#4{\left(\begin{array}{cc}
 {#1}&{#2}\nonumber \\ {#3}&{#4}\nonumber \\
\end{array}
\right)}
\def\twovec#1#2{\left(\begin{array}{c}
{#1}\nonumber \\ {#2}\nonumber \\
\end{array}
\right)}
\def\o#1#2{{{#1}\over{#2}}}
\DeclareFontFamily{U}{rsf}{} \DeclareFontShape{U}{rsf}{m}{n}{
  <5> <6> rsfs5 <7> <8> <9> rsfs7 <10-> rsfs10}{}
\DeclareMathAlphabet\Scr{U}{rsf}{m}{n}

\def\T{T}

\newcommand{\Sp}{\mathop{\rm {}Sp}}

\newcommand{\ft}[2]{{\textstyle\frac{#1}{#2}}}
\def\tilde{\widetilde}

\def\1bar{1\hskip -.275cm -}
\def\2bar{2\hskip -.275cm -}
\def\3bar{3\hskip -.275cm -}

\newsavebox{\uuunit}
\sbox{\uuunit}
                 {\setlength{\unitlength}{0.825em}
                      \begin{picture}(0.6,0.7)
                                      \thinlines
                                      \put(0,0){\line(1,0){0.5}}
                                      \put(0.15,0){\line(0,1){0.7}}
                                      \put(0.35,0){\line(0,1){0.8}}
                                     \multiput(0.3,0.8)(-0.04,-0.02){10}{\rule{0.5pt}{0.5pt}}
                      \end {picture}}

\makeatletter \@addtoreset{equation}{section} \makeatother

\def\bfone{\relax{\rm 1\kern-.35em 1}}

\def\bfone{\relax{\rm 1\kern-.35em 1}}
\font\cmss=cmss10 \font\cmsss=cmss10 at 7pt

\newcommand{\so}{\mathfrak{so}}
\newcommand{\su}{\mathfrak{su}}
\newcommand{\usp}{\mathfrak{usp}}

\newcommand{\sym}{\mathfrak{sp}}
\newcommand{\slal}{\mathfrak{sl}}

\begin{document}
\begin{titlepage}
\begin{center}
\vskip 0.2cm
\vskip 0.2cm
{\Large\sc The $c$-map, Tits Satake subalgebras\\  and \\ the search  for $\mathcal{N}=2$ inflaton potentials  }\\[1cm]
{\sc P.~Fr\'e${}^{\; a}$\footnote{Prof. Fr\'e is presently fulfilling the duties of Scientific Counselor of the Italian Embassy in the Russian Federation, Denezhnij pereulok, 5, 121002 Moscow, Russia.},  A.S.~Sorin$^{\; b}$ and M. Trigiante$^{\; c}$}\\[10pt]
{${}^a$\sl\small Dipartimento di Fisica, Universit\'a di Torino\\INFN -- Sezione di Torino \\
via P. Giuria 1, \ 10125 Torino \ ITALY \\}\emph{e-mail:} \quad {\small {\tt
fre@to.infn.it}}\\
\vspace{5pt}
{{\em $^{b}$\sl\small Bogoliubov Laboratory of Theoretical Physics \& Laboratory of High Energy Physics,}}\\
{{\em Joint Institute for Nuclear Research,}}\\
{\em 141980 Dubna, Moscow Region, Russia}~\quad\\
\emph{e-mail:}\quad {\small {\tt sorin@theor.jinr.ru}}
\quad \\
\vspace{5pt}
{{\em $^c$\sl\small  Dipartimento di Fisica Politecnico di Torino,}}\\
{\em C.so Duca degli Abruzzi, 24, I-10129 Torino, Italy}~\quad\\
\emph{e-mail:}\quad {\small {\tt mario.trigiante@gmail.com}}
\quad \vspace{8pt}
\vspace{15pt}
\begin{abstract}
In this paper
we address the general problem of including inflationary models exhibiting Starobinsky-like potentials into (symmetric) $\mathcal{N}=2$ supergravities.
This is done by gauging suitable abelian isometries of the hypermultiplet sector and then truncating the resulting theory to a single scalar field.
By using the characteristic properties of the global symmetry groups of the $\mathcal{N}=2$ supergravities we are able to make a general statement on the possible $\alpha$-attractor models which can obtained upon truncation. We find that in symmetric $\mathcal{N}=2$ models  group theoretical constraints
restrict the allowed values of the parameter $\alpha$ to be $\alpha=1,\,\frac{2}{3},\, \frac{1}{3}$. This confirms and generalizes results recently obtained in the literature.
Our analysis heavily relies on the mathematical  structure of symmetric $\mathcal{N}=2$ supergravities, in particular on the so called $c$-map connection between  Quaternionic K\"ahler manifolds starting from Special K\"ahler ones.  A general statement on the possible consistent truncations of the gauged models, leading to Starobinsky-like potentials, requires the essential help of Tits Satake universality classes. The paper is mathematically self-contained and aims at presenting the involved mathematical structures to a public not only of physicists but also of mathematicians. To this end the main mathematical structures and the general gauging procedure of $\mathcal{N}=2$ supergravities is reviewed in some detail.
\end{abstract}
\end{center}
\end{titlepage}
\tableofcontents
\noindent {}
\newpage
\part{\sc Introducing the subject  and motivations}
The present one is a research paper and it contains some new original results. These latter are mostly of mathematical-geometrical character and in our opinion they might  be of some interest both for the mathematical scientific community, as well as for  that of the theoretical physicists working in the field of supergravity/superstrings. The motivations for the present study is that of analyzing within a  general geometric framework some of the recent results \cite{thesearch} relative to the inclusion of candidate inflaton potentials into extended supergravity, the aim being that of singling out general mathematical patterns that  eventually can be exported to other examples that include more fields and more multiplets.
\par
From the physicist's viewpoint the obvious ultimate goal is that of singling out possible chains of inclusions of the inflationary models into hierarchically larger and more (super)-symmetric  theories finally sheading  light on the appropriate  place of the inflaton-dynamics, which is revealed by observational cosmology, within a duely unified theory of all interactions.
\par
Notwithstanding the above mentioned specific motivations, the geometrical results presented in this paper, have an intrinsic mathematical interest and moreover admit different applications than those in cosmology within the very framework of supergravities and their gaugings.
For this reason we have chosen to make this paper self-consistent and readable by a wider audience of both mathematicians and physicists working  in  fields different from that of supergravity cosmological models, by presenting in a systematic way all the  mathematical definitions and structures that we utilize in the original part of our work. In the same spirit, in order to provide our reader with orientation, the paper is subdivided into Parts and sections.
\par
The present first part provides a conceptual introduction and the physical motivations in relation to current research.
\par The second part provides the definitions of K\"ahler-Hodge, Special K\"ahler and Quaternionic K\"ahler manifolds. Then introduces the $c$-map \cite{sabarwhal} and illustrates how in Quaternionic K\"ahler manifolds lying in the image of $c$-map, all the quaternionic structures, in particular the HyperK\"ahler two-forms, the $\su(2)$-connection and the tri-holomorphic moment maps of isometries can all be constructed purely in terms of Special geometry data.
\par
The third part discusses abelian gaugings of hypermultiplet isometries in $\mathcal{N}=2$ supergravity. Using the general mathematical formulae derived in Part Two we discuss generic properties and features of the ensuing scalar potentials.
\par
The fourth part presents concrete examples. In the particular case of the $c$-map of the $S^3$ model we retrieve the results obtained in \cite{thesearch}. Another relevant Special K\"ahler Geometry that we consider corresponds to that of the symmetric space $\mathrm{Sp(6,\mathbb{R})}/\mathrm{SU(3)\times U(1)}$. For this  model we provide an in depth, full fledged construction. Analyzing its Quaternionic K\"ahler extension by means of the $c$-map we are able to generalize the results of \cite{thesearch} showing that they fall into a general pattern. Our detailed  construction may have applications both in the cosmological perspective of the present paper and in the classification of Black-Hole solutions, possibly also in other contexts.
\par
Part five  summarizes the results obtained in the two considered examples and shows that they unveil a deep and universal structure. We briefly summarize the organization of  $\mathcal{N}=2$ scalar manifolds that are symmetric spaces into Tits--Satake universality classes
\cite{titsusataku}. Such a concept already proved to be of high value in relation to the construction and classification of supergravity black hole solutions \cite{noinilpotenti}. It is equally effective and precious in relation to the $c$-map and cosmological models. Indeed relying on these structures we are able to show that  the inclusion of Starobinsky-like models into extended supergravity theories has a universal character being associated with the gauging of the universal sub-Tits-Satake subalgebra $\mathbb{G}_{\mathrm{subTS}} \, = \, \slal(2) \times \slal(2) \times \slal(2)$. This also leads to the prediction that available values of $\alpha$ for the so named $\alpha$-attractors \cite{alfatrattori} are just $\alpha \, = \, 1,\, \frac{2}{3},\, \frac{1}{3}$. After this conceptual elaboration, part five contains our conclusions and remarks on further perspectives.
\section{\sc Introduction}
The recent observational results on the power-spectrum of the Cosmological Microwave Background radiation \cite{Ade:2013uln},\cite{Ade:2013zuv},\cite{Hinshaw:2012aka},\cite{biceppo} have stirred renewed interest in one-field inflationary cosmological models \cite{Starobinsky:1980te},\cite{lindefund},\cite{guthfund},\cite{steinhardfund}. Indeed the type of cosmology \cite{cosmology},\cite{pietrocosmobook} that seems to be consistent with observations is that based on the simplest scenario of just one scalar field $\phi$ (\textit{the inflaton}) minimally coupled to Einstein Gravity and endowed with a suitable scalar potential $V(\phi)$.
\par
In view of this, several studies have been devoted to the problem of including into supergravity potentials $V(\phi)$ that produce an early inflationary phase and have good cosmological properties, \cite{johndimitri},\cite{Ketov:2010qz},\cite{Ketov:2012jt},\cite{Kallosh:2013hoa},\cite{Kallosh:2013lkr},\cite{Farakos:2013cqa},
\cite{Kallosh:2013maa},\cite{Kallosh:2013daa},\cite{Kallosh:2013tua},\cite{alfatrattori},\cite{Ferrara:2014cca}. A separate investigation was also devoted to determine a list of one field integrable potentials \cite{augustopietrosasha} and to discuss their possible inclusion into $\mathcal{N}=1$ supergravity by means of suitable superpotentials \cite{noiGaugings}. It was shown that such type of inclusion is quite difficult and can be realized only in very few cases \cite{noiGaugings}.
\par
Notwithstanding the difficulties with the inclusion of inflaton potentials by means of a superpotential (F-terms), approximately one year ago, in the seminal paper \cite{minimalsergioKLP} it was instead pointed out that every positive-definite potential $V(\phi)$ can be minimally included into $\mathcal{N}=1$ supergravity as a D-term (see \cite{cfgv}, \cite{castdauriafre} and \cite{primosashapietro} for the complete structure of matter coupled $\mathcal{N}=1$ supergravity). It suffices to introduce a K\"ahler one-fold $\Sigma$ with an abelian group of isometries $\mathrm{G}_{\mathrm{iso}}$ and a K\"ahler potential $\mathcal{K}$ related to the potential $V$ by an appropriate differential relation which allows to interpret this latter as the square of the moment-map of the holomorphic Killing vector generating $\mathrm{G}_{\mathrm{iso}}$.  From the mathematical point of view this new vision stimulated the formulation of the concept of $D$-map and its extensive study in \cite{primosashapietro} and \cite{piesashatwo}. The essential mechanism behind the $D$-map is the just the Brout-Englert-Higgs mechanism as it is realized in supergravity. Gauging an isometry of the K\"ahlerian scalar manifold $\mathcal{M}_{K}$ one generates both mass terms for the fermions and a scalar potential that is the square of the moment map $\mathcal{P}_k$ of the corresponding Killing vector. The gauge vector field $A_\mu$ utilized to gauge the considered isometry becomes massive by eating up one of the two scalar fields composing the Wess-Zumino multiplet, while its partner remains in the Lagrangian as a degree of freedom of spin zero, self interacting by means of the $D$-term potential \cite{VanProeyen:1979ks},\cite{Freedman:1976uk}.
\par
Among the positive definite scalar potentials that can be included in supergravity in this way there are the Starobinsky-like potentials \cite{Starobinsky:1980te}:
\begin{equation}\label{starobombolo}
   V_{Starobinsky-like} \, = \, \mbox{const} \, \times \, \left( 1\, - \, \exp\left [ - \, \sqrt{\frac{2}{3 \, \alpha}} \, \phi\right]\right)^2\,.
 \end{equation}
 When $\alpha \,  = \, 1$, the potential (\ref{starobombolo}) emerges in the  second derivative supergravity dual of a higher derivative  $R+R^2$ supergravity model  \cite{Whitt:1984pd},
\cite{Cecotti:1987qe},\cite{Cecotti:1987sa},\cite{Ferrara:2013wka},\cite{Ferrara:2013kca}. Due to the high relevance of these potentials in phenomenology, several studies were devoted to the mechanisms for their inclusion in supergravity\footnote{The inclusion of the original Starobinsky $R+R^2$ model in $\mathcal{N}=2$ supergravity was discussed in \cite{thesearch} (see \cite{Ketov:2014qoa} for an earlier discussion) where it is shown to be dual to an $\mathcal{N}=2$  model with two long massive vector multiplets
on a $\mathcal{N}=2$ Minkowski vacuum. The  scalaron is subject to a scalar potential of the form (\ref{starobombolo}) with $\alpha=1$. The models considered here are not dual to the  $R+R^2$ Starobinsky model and share with it only the scalaron potential, which we shall refer to as the \emph{Starobinsky potential}. If $\alpha\neq 1$ the potential (\ref{starobombolo}) will be called \emph{Starobinsky-like}.}. In their minimal $D$-term realization, the Starobinsky-like potentials were shown to be generated by the gauging of the parabolic subgroups of $\mathrm{SL(2,\mathbb{R})}$ in a theory where the K\"ahler one-fold $\Sigma$ is the homogeneous space $\frac{\mathrm{SL(2,\mathbb{R})}}{\mathrm{SO(2)}}$, with a value of the curvature directly related to $\alpha$ by:
\begin{equation}\label{kreptosio}
    R_{\alpha} \, = \, - \, \frac{2}{3\,\alpha}
\end{equation}
The case $\alpha \, = \, 1$ which is the proper Starobinsky potential corresponds to a realization of the $\frac{\mathrm{SL(2,\mathbb{R})}}{\mathrm{SO(2)}}$ geometry which is not only K\"ahlerian but actually Special K\"ahlerian. Indeed it is the case of the $S^3$ model.
\par
In more general terms the minimal supergravity approach and the $D$-map from potentials to K\"ahler geometry posed the question of relating the type of generated potential with the global topology of the isometry whose gauging generates it. This issue was thoroughly studied in \cite{pietrosergiosasha1},\cite{pietrosergiosasha2}.  Relying on notions developed by Gromov et al \cite{Gromov1985},\cite{Gromov1987}, in \cite{pietrosergiosasha1},\cite{pietrosergiosasha2} it was shown that the global topology of the isometry group can be characterized as, \textit{elliptic, parabolic or hyperbolic} on general K\"ahler manifolds $\Sigma$ of non-positive curvature and that the three cases lead to different distinctive properties of the moment-maps, that either have a fixed-point at finite distance (elliptic case) or a fixed point on the boundary (parabolic) or two fixed point on the boundary (hyperbolic). In the case of constant negative curvature K\"ahler one-folds, \textit{i.e.} of $\Sigma_\alpha \, = \, \frac{\mathrm{SL(2,\mathbb{R})}}{\mathrm{SO(2)}}$ manifolds, it was shown in \cite{pietrosergiosasha1} and \cite{pietrosergiosasha2} that one generates the following three potentials:
\begin{equation}\label{ginolullobrigo}
    V(\phi) \, = \, \left \{ \begin{array}{ccccc}
                              V_{elliptic}& = & \left( \cosh \left[\sqrt{\frac{2}{3 \, \alpha}} \, \phi\right] \, - \,  \kappa\right)^2 & \Leftrightarrow & \mbox{from gauging of an elliptic subgroup} \\
                             V_{hyperbolic}& = & \left( \sinh \left[\sqrt{\frac{2}{3 \, \alpha}} \, \phi\right] \, - \,  \kappa\right)^2 & \Leftrightarrow & \mbox{from gauging of a hyperbolic subgroup}\\
                            V_{parabolic}& = &\left( \exp\left [ - \, \sqrt{\frac{2}{3 \, \alpha}} \, \phi\right]\, - \,  \kappa\right)^2 & \Leftrightarrow & \mbox{from gauging of a parabolic subgroup}\\
                             \end{array}
    \right.
\end{equation}
where $\alpha$ parameterizes  the curvature of the one-fold $\Sigma_\alpha$ according to eq. (\ref{kreptosio}) and the parameter $\kappa\, = \, \pm1,0$ is interpreted as the coupling constant of a Fayet Iliopoulos term \cite{Fayet:1974jb}. Utilizing instead a flat K\"ahler one-fold $\Sigma_{flat}\, = \, \mathbb{C}$, it was shown in \cite{pietrosergiosasha1} and \cite{pietrosergiosasha2} that the gauging of an elliptic isometry yields a mexican hat Brout-Englert-Higgs potential, while the quadratic potential of chaotic inflation is generated by a translation parabolic gauging.
\par
In view of the above conclusions an obvious and very much relevant question concerns the possible inclusion of the potentials (\ref{ginolullobrigo}) into extended supergravity, in particular $\mathcal{N}=2$. Such a question amounts to asking whether there are consistent one-field truncations of  appropriate  gauged extended supergravities that produce the considered potentials. Such a question includes two  issues:
\begin{enumerate}
  \item the choice of a gauging,
  \item the existence of an appropriate consistent truncation.
\end{enumerate}
The first of these two issues is relevant for another important problem, that of constructing and classifying de Sitter or Minkowskian, stable or metastable, vacua in supergravity and it was extensively addressed in that context, \cite{mapietoine},\cite{gkp},\cite{uplift},\cite{kklt},
\cite{scrucca},\cite{noscale}. Let us spend some words on the second issue.
\par
The general form of the scalar field equations in supergravity is that of a $\sigma$-model with a potential, namely the  following one:
\begin{equation}\label{guberator}
    0 \, = \, \square \, \phi^I \, + \, \Gamma^I_{JK} (\phi)\,  \partial_\mu \phi^J \, \partial^\mu \, \phi^K \, + \, \frac{\partial}{\partial \phi^I} \, V(\phi)
\end{equation}
where the scalars $\phi^I$ are coordinates of the target Riemaniann  manifold $\mathcal{M}_{scalar}$, by $\Gamma^I_{JK} (\phi)$ we have denoted the Levi-Civita connection on the former and $V$ is the potential. The problem of consistent truncation boils down to the following. We consider the embedding of a $\Sigma_\alpha$ K\"ahlerian one-fold into the scalar manifold:
\begin{equation}\label{pullabacca}
    \pi \, : \, \Sigma_\alpha \, \mapsto \, \mathcal{M}_{scalar}
\end{equation}
and we require that the pull-back of the field equations (\ref{guberator}) onto the surface $\Sigma_\alpha $:
 \begin{equation}\label{guberator2}
    0 \, = \, \phi^\star\left[\square \, \phi^I \, + \, \Gamma^I_{JK} (\phi)\,  \partial_\mu \phi^J \, \partial^\mu \, \phi^K \, + \, \frac{\partial}{\partial \phi^I} \, V(\phi)\right]
\end{equation}
 be consistent, namely that it reproduces always the same equations for all values of $I$. What are the restrictions on the embedding $\pi$ that guarantee such a consistency? This question can be answered in a general form when $\mathcal{M}_{scalar}$ is a symmetric homogeneous space $\mathrm{G/H}$, as it happens most frequently in supergravity models. Considering as usual the symmetric decomposition of the Lie algebra $\mathbb{G}$:
 \begin{eqnarray}
    \mathbb{G} & =& \mathbb{H} \, \oplus \, \mathbb{K} \nonumber\\
     \left [\mathbb{H} \, , \, \mathbb{H}\right] & \subset & \mathbb{H} \nonumber\\
      \left [\mathbb{H} \, , \, \mathbb{K}\right] & \subset & \mathbb{K} \nonumber\\
      \left [\mathbb{K} \, , \, \mathbb{K}\right] & \subset & \mathbb{H} \label{simmetricoGH}
 \end{eqnarray}
 the embedding (\ref{pullabacca}) induces a homomorphism:
 \begin{equation}\label{girolamo}
    \pi \, : \, \slal(2,\mathbb{R}) \, \mapsto \, \mathbb{G} \quad ; \quad \pi \, : \, \mathbb{O}(2) \, \mapsto \, \mathbb{H}\quad ; \quad \pi \, : \, \mathbb{K}_2 \, \mapsto \, \mathbb{K}
 \end{equation}
 where:
 \begin{equation}\label{birillo2}
    \slal(2,\mathbb{R}) \, =\, \mathbb{O}(2)\, \oplus \, \mathbb{K}_2
 \end{equation}
Let us introduce the centralizer algebra of the image of $\mathbb{O}(2)$ in $\mathbb{H}$, namely
\begin{equation}\label{gunther}
 \forall \,\mathfrak{g} \, \in \,  \mathbb{H} \quad  : \quad \mathfrak{g}\, \in \,  \mathbb{N}_\pi \quad \mbox{iff} \quad \left[\mathfrak{g}\, , \, \pi\left(\mathbb{O}(2)\right)\right] \, = \,0
\end{equation}
The subspace $\mathbb{K}$ decomposes into irreducible representations of $\pi\left(\mathbb{O}(2)\right)\, \oplus \, \mathbb{N}_\pi$. The embedding $\pi$ leads to a consistent truncation if such a decomposition has the following structure:
\begin{equation}\label{contrallus}
    \mathbb{K} \, = \, \underbrace{\left(\mathbf{2}_1\,|\, \mathbf{1}\right)}_{\pi(\mathbb{K}_2 )} \, \oplus_{i=1}^m \, \left(\mathbf{2}_{q_i}\,|\, \mathbf{D}_i\right) \, \oplus \, \left(\mathbf{1}\,|\, \mathbf{D}_0\right)
\end{equation}
where $\mathbf{2}_{q}$ denotes a doublet representation of  $\pi\left(\mathbb{O}(2)\right)$ with charge $q$ and all $\mathbf{D}_i$ and $\mathbf{D}_0$ are transitive representations, namely carry a \textit{color} of $\mathbb{N}_\pi$, the only
 $\mathbb{N}_\pi$-singlet being the subspace ${\pi(\mathbb{K}_2 )}$ tangent to the embedded one-fold $\Sigma_\alpha$.
 In this case setting to zero all the colored fields (those not in ${\pi(\mathbb{K}_2 )}$ )  is consistent because the equation of a colored field cannot receive contribution from two colorless singlets.
\par
In the case of $\mathcal{N}=2$ supergravity the scalar manifold is actually the direct product of two manifolds: a special K\"ahler manifold $\mathcal{M}_{SK}$ that contains the vector multiplet scalars and a Quaternionic K\"ahler $\mathcal{M}_{Q}$ manifold describing the hypermultiplet scalars. Hence one comes to the question whether the one-fold $\Sigma_\alpha$ associated with the inflaton is to be embedded in $\mathcal{M}_{SK}$ or in $\mathcal{M}_{Q}$. The significant advance introduced by \cite{thesearch} is the scenario in which $\Sigma_\alpha$ goes into $\mathcal{M}_{Q}$ and its abelian isometry is gauged by means of vector muliplets assigned to the so named Minimal Coupling Special Geometry. This choice allows for a generic stabilization of the vector multiplet scalars and allows to focus only on the properties of the Quaternionic K\"ahler manifold $\mathcal{M}_{Q}$. In \cite{thesearch} the authors  considered the cases where
\begin{equation}\label{g22su2su2}
    \mathcal{M}_{Q} \, = \, \frac{\mathrm{G_{(2,2)}}}{\mathrm{SU(2)} \times \mathrm{SU(2)}}\,\,\,\mbox{and}\,\,\,\,\,\,  \mathcal{M}_{Q} \, =\,\frac{{\rm SO}(1,4)}{{\rm SO}(4)}\,,
\end{equation}
and it was shown how all the potentials (\ref{ginolullobrigo}) can be included with possible values of $\alpha\, = \, 1,\,\frac{2}{3},\, \frac{1}{3}$ (the value $\alpha=2/3$ could be obtained only for the quaternionic projective space). In the present paper we address the same question for generic homogeneous symmetric Quaternionic K\"ahler manifolds and we eventually arrive at the same conclusion. The three potentials (\ref{ginolullobrigo}) can always be embedded with the same possible values of $\alpha$. Our main weapons in reaching such a conclusion and proving its generality are two:
\begin{description}
  \item[a)] The $c$-map from Special K\"ahler manifolds $\mathcal{M}_{\mathcal{SK}}$ to Quaternionic K\"ahler manifolds $\mathcal{M}_{\mathcal{Q}}$
  \item[b)] The Tits-Satake projection and the Tits-Satake universality classes.
\end{description}
Combining these two mathematical instruments we are able to look inside the hypermultiplet manifold and to single out its inner core which is the Special K\"ahler STU-model. Working in this framework we were able to derive simple general formulae for the triholomorphic moment maps which besides their present use in cosmology might admit several other interesting and useful applications. Similarly the Tits Satake structure allows for a deep understanding of the universal character of the Starobinsky like potentials and opens the way to their further uplift to higher $\mathcal{N}$ theories in the perspective of finding \textit{microscopic interpretations} of the gaugings  that generates them.
\newpage
\part{\sc Mathematical Theory of the $c$-map}
As announced in the introductory part, the aim of this section of the paper is two-fold. On the one hand we want to
review the general geometric structure of homogeneous symmetric Quaternionic K\"ahler manifolds which are in the image of the c-map and to present
in a unified fashion general analytic formulae for the complex structures, $\su(2)$-connection and tri-holomorphic moment maps.
The ultimate goal is the use of such mathematical instruments in the quest of retrieving \textit{inflaton potentials} inside properly gauged $\mathcal{N}=2$ supergravity theories. On the other hand we aim at a concise, yet comprehensive presentation of Special K\"ahler geometry, Quaternionic K\"ahler geometry and the $c$-map which might be readable by mathematicians, in particular differential geometers and Lie algebrists. Indeed this  mathematical subject was mostly developed by theoretical physicists and it is not widely known in the mathematical community. This is rather unfortunate, especially in the light of its relevance to the issue of the classification of nilpotent orbits, which, instead, is largely explored by mathematicians, and is quite relevant both to the classification of  black hole solutions  and emerges now as quite important in cosmological issues. Unfortunately the mathematical results on nilpotent orbits are obtained within frameworks that make no reference to the very particular structures of special geometry, quaternionic geometry and the magic relations of the $c$-map. Spreading awareness of this sophisticated and beautiful mathematics among mathematicians might be beneficial to both communities and it is the second aim of the following sections. A further, humbler, yet quite important aim, pursued here, is that of establishing  some unified and systematic notations that might be utilized in subsequent publications devoted to a systematic exploration of the  reach field of \textit{gaugings}, \textit{consistent truncations}, \textit{de Sitter vacua} and \textit{inflaton potentials}.
\section{\sc  Special K\"ahler Geometry}
Special K\"ahler geometry in special coordinates was introduced
in 1984--85 by B. de Wit et al.
and E. Cremmer et al. (see pioneering papers in \cite{SKG}), where the coupling of
$\mathcal{N}=2$ vector multiplets to $\mathcal{N}=2$ supergravity was fully determined. The
more intrinsic definition of special K\"ahler geometry in terms of
symplectic bundles is due to Strominger (1990), who
obtained it in connection with the moduli spaces of
Calabi--Yau compactifications, (see  \cite{defiskg}). The coordinate-independent description
and derivation of special K\"ahler geometry in the context of $\mathcal{N}=2$
supergravity is due to Castellani, D'Auria, Ferrara \cite{skgintrinsic}
and to D'Auria, Ferrara, Fre' (1991)\cite{D'Auria:1991fj}. Homogenous symmetric special K\"ahler manifolds were classified before by Cremmer and Van Proyen in \cite{ToineCremmerOld}. An early review in modern mathematical language is provided by \cite{mylecture}.
The structure of isometry group for both Special K\"ahler and Quaternionic K\"ahler manifolds was extensively studied in \cite{specHomgeo},\cite{vanderseypen}
\par
Let us summarize the relevant concepts and definitions
\subsection{\sc Hodge--K\"ahler manifolds}
\def\mom{{M(k, \IC)}}
Consider a {\sl line bundle} ${\cal L}
{\stackrel{\pi}{\longrightarrow}} {\cal M}$ over a K\"ahler
manifold ${\cal M}$. By definition this is a holomorphic vector
bundle of rank $r=1$. For such bundles the only available Chern
class is the first:
\begin{equation}
c_1 ( {\cal L} ) \, =\, \o{i}{2}
\, {\bar \partial} \,
\left ( \, h^{-1} \, \partial \, h \, \right )\, =
\, \o{i}{2} \,
{\bar \partial} \,\partial \, \mbox{log} \,  h
\label{chernclass23}
\end{equation}
where the 1-component real function $h(z,{\bar z})$ is some
hermitian fibre metric on ${\cal L}$. Let $\xi (z)$ be a
holomorphic section of the line bundle ${\cal L}$: noting that
under the action of the operator ${\bar
\partial} \,\partial \, $ the term $\mbox{log} \left ({\bar \xi}({\bar z})
\, \xi (z) \right )$ yields a vanishing contribution, we conclude that
the formula in eq.(\ref{chernclass23})  for the first Chern class can be
re-expressed as follows:
\begin{equation}
c_1 ( {\cal L} ) ~=~\o{i}{2} \,
{\bar \partial} \,\partial \, \mbox{log} \,\parallel \, \xi(z) \, \parallel^2
\label{chernclass24}
\end{equation}
where $\parallel \, \xi(z) \, \parallel^2 ~=~h(z,{\bar z}) \,
{\bar \xi}({\bar z}) \,
\xi (z) $ denotes
the norm of the holomorphic section $\xi (z) $.
\par
Eq.(\ref{chernclass24}) is the starting point for the definition
of Hodge--K\"ahler manifolds. A K\"ahler manifold ${\cal M}$ is a
Hodge manifold if and only if there exists a line bundle ${\cal L}
{\stackrel{\pi}{\longrightarrow}} {\cal M}$ such that its first
Chern class equals the cohomology class of the K\"ahler two-form
$\mathrm{K}$:
\begin{equation}
c_1({\cal L} )~=~\left [ \, \mathrm{K} \, \right ]
\label{chernclass25}
\end{equation}
\par
In local terms this means that there is a holomorphic section $\xi
(z)$ such that we can write
\begin{equation}
\mathrm{K}\, =\, \o{i}{2} \, g_{ij^{\star}} \, dz^{i} \, \wedge
\, d{\bar z}^{j^{\star}} \, = \, \o{i}{2} \, {\bar \partial}
\,\partial \, \mbox{log} \,\parallel \, \xi (z) \,
\parallel^2
\label{chernclass26}
\end{equation}
Recalling the local expression of the K\"ahler metric
in terms of the K\"ahler potential
$ g_{ij^{\star}}\, =\, {\partial}_i \, {\partial}_{j^{\star}}
{\mathcal{K}} (z,{\bar z})$,
it follows from eq.(\ref{chernclass26}) that if the
manifold ${\cal M}$ is a Hodge manifold,
then the exponential of the K\"ahler potential
can be interpreted as the metric
$h(z,{\bar z}) \, = \, \exp \left ( {\cal K} (z,{\bar z})\right )$
on an appropriate line bundle ${\cal L}$.
\par
\subsection{\sc Connection on the line bundle}
On any complex line bundle ${\cal L}$ there is a canonical hermitian connection defined as :
\begin{equation}
\begin{array}{ccccccc}
{\theta}& \equiv & h^{-1} \, \partial  \, h = {\o{1}{h}}\, \partial_i h \,
dz^{i} &; &
{\bar \theta}& \equiv & h^{-1} \, {\bar \partial}  \, h = {\o{1}{h}} \,
\partial_{i^\star} h  \,
d{\bar z}^{i^\star} \cr
\end{array}
\label{canconline}
\end{equation}
For the line-bundle advocated by the Hodge-K\"ahler structure we have
\begin{equation}
\left  [ \, {\bar \partial}\,\theta \,  \right ] \, = \,
c_1({\cal L}) \, = \, [\mathrm{K}]
\label{curvc1}
\end{equation}
and since the fibre metric $h$ can be identified with the
exponential of the K\"ahler potential we obtain:
\begin{equation}
\begin{array}{ccccccc}
{\theta}& = &  \partial  \,{\cal K} =  \partial_i {\cal K}
dz^{i} & ; &
{\bar \theta}& = &   {\bar \partial}  \, {\cal K} =
\partial_{i^\star} {\cal K}
d{\bar z}^{i^\star}\cr
\end{array}
\label{curvconline}
\end{equation}
To define special K\"ahler geometry,  in addition to the afore-mentioned line--bundle
${\cal L}$ we need a flat holomorphic vector bundle ${\cal SV}
\, \longrightarrow \, {\cal M}$ whose sections play an important role in the construction of the supergravity Lagrangians. For reasons
intrinsic to such constructions the rank of the vector bundle ${\cal SV}$ must be $2\, n_V$ where $n_V$ is the total number of vector fields in the theory. If we have $n$-vector multiplets the total number of vectors is $n_V = n+1$ since, in addition to the vectors of the vector multiplets, we always have the graviphoton sitting in the graviton multiplet. On the other hand the total number of scalars is $2 n$. Suitably paired into $n$-complex fields $z^i$, these scalars span the $n$ complex dimensions of the base manifold ${\cal M}$ to the rank $2n+2$ bundle ${\cal SV}
\, \longrightarrow \, {\cal M}$.
\par
In the sequel we make extensive use of covariant derivatives with
respect to the canonical connection of the line--bundle ${\cal
L}$. Let us review its normalization. As it is well known there
exists a correspondence between line--bundles and
$\mathrm{U(1)}$--bundles. If $\mbox{exp}[f_{\alpha\beta}(z)]$ is
the transition function between two local trivializations of the
line--bundle ${\cal L} {\stackrel{\pi}{\longrightarrow}} {\cal
M}$, the transition function in the corresponding principal
$\mathrm{U(1)}$--bundle ${\cal U} \, \longrightarrow {\cal M}$ is
just $\mbox{exp}[{\rm i}{\rm Im}f_{\alpha\beta}(z)]$ and the
K\"ahler potentials in two different charts are related by: ${\cal
K}_\beta = {\cal K}_\alpha + f_{\alpha\beta}   + {\bar
{f}}_{\alpha\beta}$. At the level of connections this
correspondence is formulated by setting: $\mbox{
$\mathrm{U(1)}$--connection}   \equiv   {\cal Q} \,  = \,
\mbox{Im} \theta = -{\o{\rm i}{2}}   \left ( \theta - {\bar
\theta} \right)$. If we apply this formula to the case of the
$\mathrm{U(1)}$--bundle ${\cal U} \, \longrightarrow \, {\cal M}$
associated with the line--bundle ${\cal L}$ whose first Chern
class equals the K\"ahler class, we get:
\begin{equation}
{\cal Q}  =    {\o{\rm i}{2}} \left ( \partial_i {\cal K}
dz^{i} -
\partial_{i^\star} {\cal K}
d{\bar z}^{i^\star} \right )
\label{u1conect}
\end{equation}
 Let now
 $\Phi (z, \bar z)$ be a section of ${\cal U}^p$.  By definition its
covariant derivative is $ \nabla \Phi = (d - i p {\cal Q}) \Phi $
or, in components,
\begin{equation}
\begin{array}{ccccccc}
\nabla_i \Phi &=&
 (\partial_i + {1\over 2} p \partial_i {\cal K}) \Phi &; &
\nabla_{i^*}\Phi &=&(\partial_{i^*}-{1\over 2} p \partial_{i^*} {\cal K})
\Phi \cr
\end{array}
\label{scrivo2}
\end{equation}
A covariantly holomorphic section of ${\cal U}$ is defined by the equation:
$ \nabla_{i^*} \Phi = 0  $.
We can easily map each  section $\Phi (z, \bar z)$
of ${\cal U}^p$
into a  section of the line--bundle ${\cal L}$ by setting:
\begin{equation}
\tilde{\Phi} = e^{-p {\cal K}/2} \Phi  \,   .
\label{mappuccia}
\end{equation}
  With this position we obtain:
\begin{equation}
\begin{array}{ccccccc}
\nabla_i    \tilde{\Phi}&    =&
(\partial_i   +   p   \partial_i  {\cal K})
\tilde{\Phi}& ; &
\nabla_{i^*}\tilde{\Phi}&=& \partial_{i^*} \tilde{\Phi}\cr
\end{array}
\end{equation}
Under the map of eq.(\ref{mappuccia}) covariantly holomorphic sections
of ${\cal U}$ flow into holomorphic sections of ${\cal L}$
and viceversa.
\subsection{\sc Special K\"ahler Manifolds}
We are now ready to give the first of two equivalent definitions of special K\"ahler
manifolds:
\bd
A Hodge K\"ahler manifold is {\bf Special K\"ahler (of the local type)}
if there exists a completely symmetric holomorphic 3-index section $W_{i
j k}$ of $(T^\star{\cal M})^3 \otimes {\cal L}^2$ (and its
antiholomorphic conjugate $W_{i^* j^* k^*}$) such that the following
identity is satisfied by the Riemann tensor of the Levi--Civita
connection:
\begin{eqnarray}
\partial_{m^*}   W_{ijk}& =& 0   \quad   \partial_m  W_{i^*  j^*  k^*}
=0 \nonumber \\
\nabla_{[m}      W_{i]jk}& =&  0
\quad \nabla_{[m}W_{i^*]j^*k^*}= 0 \nonumber \\
{\cal R}_{i^*j\ell^*k}& =&  g_{\ell^*j}g_{ki^*}
+g_{\ell^*k}g_{j i^*} - e^{2 {\cal K}}
W_{i^* \ell^* s^*} W_{t k j} g^{s^*t}
\label{specialone}
\end{eqnarray}
\label{defspecial}
\ed
In the above equations $\nabla$ denotes the covariant derivative with
respect to both the Levi--Civita and the $\mathrm{U(1)}$ holomorphic connection
of eq.(\ref{u1conect}).
In the case of $W_{ijk}$, the $\mathrm{U(1)}$ weight is $p = 2$.
\par
Out of the $W_{ijk}$ we can construct covariantly holomorphic
sections of weight 2 and - 2 by setting:
\begin{equation}
C_{ijk}\,=\,W_{ijk}\,e^{  {\cal K}}  \quad ; \quad
C_{i^\star j^\star k^\star}\,=\,W_{i^\star j^\star k^\star}\,e^{  {\cal K}}
\label{specialissimo}
\end{equation}
The flat bundle mentioned in the previous subsection apparently does not appear in this definition of special geometry.
Yet it is there. It is indeed the essential ingredient in the second definition whose equivalence to the first we shall
shortly provide.
\par
Let ${\cal L} {\stackrel{\pi}{\longrightarrow}} {\cal M}$ denote
the complex line bundle whose first Chern class equals the
cohomology class of the K\"ahler form $\mathrm{K}$ of an
$n$-dimensional Hodge--K\"ahler manifold ${\cal M}$. Let ${\cal
SV} \, \longrightarrow \,{\cal M}$ denote a holomorphic flat
vector bundle of rank $2n+2$ with structural group
$\mathrm{Sp(2n+2,\mathbb{R})}$. Consider   tensor bundles of the
type ${\cal H}\,=\,{\cal SV} \otimes {\cal L}$. A typical
holomorphic section of such a bundle will be denoted by ${\Omega}$
and will have the following structure:
\begin{equation}
{\Omega} \, = \, {\twovec{{X}^\Lambda}{{F}_ \Sigma} } \quad
\Lambda,\Sigma =0,1,\dots,n
\label{ololo}
\end{equation}
By definition
the transition functions between two local trivializations
$U_i \subset {\cal M}$ and $U_j \subset {\cal M}$
of the bundle ${\cal H}$ have the following form:
\begin{equation}
{\twovec{X}{ F}}_i=e^{f_{ij}} M_{ij}{\twovec{X}{F}}_j
\end{equation}
where   $f_{ij}$ are holomorphic maps $U_i \cap U_j \, \rightarrow
\,\IC $
while $M_{ij}$ is a constant $\mathrm{Sp(2n+2,\mathbb{R})}$ matrix. For a consistent
definition of the bundle the transition functions are obviously
subject to the cocycle condition on a triple overlap:
$e^{f_{ij}+f_{jk}+f_{ki}} = 1 $ and $ M_{ij} M_{jk} M_{ki} = 1 $.
\par
Let ${\rm i}\langle\ \vert\ \rangle$ be the compatible
hermitian metric on $\cal H$
\begin{equation}
{\rm i}\langle \Omega \, \vert \, \bar \Omega \rangle \, \equiv \,-
{\rm i} \Omega^\T \twomat {0} {\bfone} {-\bfone}{0} {\bar \Omega}
\label{compati}
\end{equation}
\bd
We say that a Hodge--K\"ahler manifold ${\cal M}$
is {\bf special K\"ahler} if there exists
a bundle ${\cal H}$ of the type described above such that
for some section $\Omega \, \in \, \Gamma({\cal H},{\cal M})$
the K\"ahler two form is given by:
\begin{equation}
\mathrm{K}= \o{\rm i}{2}
 \partial \bar \partial \, \mbox{\rm log} \, \left ({\rm i}\langle \Omega \,
 \vert \, \bar \Omega
\rangle \right )=\frac{i}{2}\,g_{i, j^*}\,dz^i\wedge d\bar{z}^{j^*}\,.
\label{compati1} .
\end{equation}
\ed
From the point of view of local properties, eq.(\ref{compati1})
implies that we have an expression for the K\"ahler potential
in terms of the holomorphic section $\Omega$:
\begin{equation}
{\cal K}\,  = \,  -\mbox{log}\left ({\rm i}\langle \Omega \,
 \vert \, \bar \Omega
\rangle \right )\,
=\, -\mbox{log}\left [ {\rm i} \left ({\bar X}^\Lambda F_\Lambda -
{\bar F}_\Sigma X^\Sigma \right ) \right ]
\label{specpot}
\end{equation}
The relation between the two definitions of special manifolds is
obtained by introducing a non--holomorphic section of the bundle
${\cal H}$ according to:
\begin{equation}
V \, = \, \twovec{L^{\Lambda}}{M_\Sigma} \, \equiv \, e^{{\cal K}/2}\Omega
\,= \, e^{{\cal K}/2} \twovec{X^{\Lambda}}{F_\Sigma}
\label{covholsec}
\end{equation}
so that eq.(\ref{specpot}) becomes:
\begin{equation}
1 \, = \,  {\rm i}\langle V  \,
 \vert \, \bar V
\rangle  \,
= \,   {\rm i} \left ({\bar L}^\Lambda M_\Lambda -
{\bar M}_\Sigma L^\Sigma \right )
\label{specpotuno}
\end{equation}
Since $V$ is related to a holomorphic section by eq.(\ref{covholsec})
it immediately follows that:
\begin{equation}
\nabla_{i^\star} V \, = \, \left ( \partial_{i^\star} - {\o{1}{2}}
\partial_{i^\star}{\cal K} \right ) \, V \, = \, 0
\label{nonsabeo}
\end{equation}
On the other hand, from eq.(\ref{specpotuno}), defining:
\begin{eqnarray}
U_i  & = &  \nabla_i V  =   \left ( \partial_{i} + {\o{1}{2}}
\partial_{i}{\cal K} \right ) \, V   \equiv
\twovec{f^{\Lambda}_{i} }{h_{\Sigma\vert i}}\nonumber\\
{\bar U}_{i^\star}  & = &  \nabla_{i^\star}{\bar V}  =   \left ( \partial_{i^\star} + {\o{1}{2}}
\partial_{i^\star}{\cal K} \right ) \, {\bar V}   \equiv
\twovec{{\bar f}^{\Lambda}_{i^\star} }{{\bar h}_{\Sigma\vert i^\star}}
\label{uvector}
\end{eqnarray}
it follows that:
\begin{equation}
\label{ctensor}
\nabla_i U_j  = {\rm i} C_{ijk} \, g^{k\ell^\star} \, {\bar U}_{\ell^\star}
\end{equation}
where $\nabla_i$ denotes the covariant derivative containing both
the Levi--Civita connection on the bundle ${\cal TM}$ and the
canonical connection $\theta$ on the line bundle ${\cal L}$.
In eq.(\ref{ctensor}) the symbol $C_{ijk}$ denotes a covariantly
holomorphic (
$\nabla_{\ell^\star}C_{ijk}=0$) section of the bundle
${\cal TM}^3\otimes{\cal L}^2$ that is totally symmetric in its indices.
This tensor can be identified with the tensor of eq.(\ref{specialissimo})
appearing in eq.(\ref{specialone}).
Alternatively, the set of differential equations:
\begin {eqnarray}
&&\nabla _i V  = U_i\\
 && \nabla _i U_j = {\rm i} C_{ijk} g^{k \ell^\star} U_{\ell^\star}\\
 && \nabla _{i^\star} U_j = g_{{i^\star}j} V\\
 &&\nabla _{i^\star} V =0 \label{defaltern}
\end{eqnarray}
with V satisfying eq.s (\ref{covholsec}, \ref {specpotuno}) give yet
another definition of special geometry.
In particular it is easy to find eq.(\ref{specialone})
as integrability conditions of(\ref{defaltern})
\subsection{\sc The vector kinetic matrix $\mathcal{N}_{\Lambda\Sigma}$ in special geometry}
\label{scrittaN}
In the construction of supergravity actions another essential item is the complex symmetric matrix $\mathcal{N}_{\Lambda\Sigma}$ whose real and imaginary parts are necessary in order to write the kinetic terms
of the vector fields. The matrix $\mathcal{N}_{\Lambda\Sigma}$ constitutes an integral part of the Special Geometry set up and we provide its general definition in the following lines.
Explicitly $\mathcal{N}_{\Lambda\Sigma}$ which, in relation to its interpretation in the case of Calabi-Yau threefolds, is named
the {\it period matrix}, is defined by means of the following relations:
\begin{equation}
{\bar M}_\Lambda = {{\cal N}}_{\Lambda\Sigma}{\bar L}^\Sigma \quad ;
\quad
h_{\Sigma\vert i} = { {\cal N}}_{\Lambda\Sigma} f^\Sigma_i
\label{etamedia}
\end{equation}
which can be solved introducing the two $(n+1)\times (n+1)$ vectors
\begin{equation}
f^\Lambda_I = \twovec{f^\Lambda_i}{{\bar L}^\Lambda} \quad ; \quad
h_{\Lambda \vert I} =  \twovec{h_{\Lambda \vert i}}{{\bar M}_\Lambda}
\label{nuovivec}
\end{equation}
and setting:
\begin{equation}
{{\cal N}}_{\Lambda\Sigma}= h_{\Lambda \vert I} \circ \left (
f^{-1} \right )^I_{\phantom{I} \Sigma}
\label{intriscripen}
\end{equation}
\par
Let us now consider the case where the Special K\"ahler manifold $\mathcal{SK}_n$ of complex dimension $n$  has some isometry group
$\mathrm{U}_{\mathcal{SK}}$. Compatibility with the Special Geometry structure requires the existence of a $2n+2$-dimensional symplectic representation of such a group that we name the  $\mathbf{W}$ representation.
In other words that  there necessarily exists  a symplectic embedding of the isometry group $\mathcal{SK}_n$
\begin{equation}
  \mathrm{U}_{\mathcal{SK}} \mapsto \mathrm{Sp(2n+2, \mathbb{R})}
\label{sympembed}
\end{equation}
such that for each element $\xi \in \mathrm{U}_{\mathcal{SK}}$ we have its
representation by means of a  suitable real symplectic matrix:
\begin{equation}
  \xi \mapsto \Lambda_\xi \equiv \left( \begin{array}{cc}
     A_\xi & B_\xi \\
     C_\xi & D_\xi \
  \end{array} \right)
\label{embeddusmatra}
\end{equation}
satisfying the defining relation (in terms of the symplectic antisymmetric metric $\mathbb{C}$):
\begin{equation}
  \Lambda_\xi ^T \, \underbrace{\left( \begin{array}{cc}
     \mathbf{0}_{n \times n}  & { \mathbf{1}}_{n \times n} \\
     -{ \mathbf{1}}_{n \times n}  & \mathbf{0}_{n \times n}  \
  \end{array} \right)}_{ \equiv \, \mathbb{C}} \, \Lambda_\xi = \underbrace{\left( \begin{array}{cc}
     \mathbf{0}_{n \times n}  & { \mathbf{1}}_{n \times n} \\
     -{ \mathbf{1}}_{n \times n}  & \mathbf{0}_{n \times n}  \
  \end{array} \right)}_{\mathbb{C}}
\label{definingsympe}
\end{equation}
which implies the following relations on the $n \times n$ blocks:
\begin{eqnarray}
A^T_\xi \, C_\xi - C^T_\xi \, A_\xi & = & 0 \nonumber\\
A^T_\xi \, D_\xi - C^T_\xi \, B_\xi& = & \mathbf{1}\nonumber\\
B^T_\xi \, C_\xi - D^T_\xi \, A_\xi& = & - \mathbf{1}\nonumber\\
B^T_\xi \, D_\xi - D^T_\xi \, B_\xi & =  & 0 \label{symplerele}
\end{eqnarray}
Under an element of the isometry group the symplectic section $\Omega$ of Special Geometry transforms
as follows:
\begin{equation}
\Omega\left( \xi \, \cdot \, z\right) \, = \, \Lambda_\xi \, \Omega\left ( z \right )
\end{equation}
As a consequence of its definition, under the same isometry the matrix ${\cal N}$ transforms  by means of a generalized linear fractional
transformation:
\begin{equation}
  \mathcal{N}\left(\xi \cdot z,\xi \cdot \bar{z}\right) = \left(  C_\xi + D_\xi \, \mathcal{N}(z,\bar{z})\right) \, \left( A_\xi + B_\xi \,\mathcal{N}(z,\bar{z})\right) ^{-1}
\label{Ntransfa}
\end{equation}
\subsection{\sc The holomorphic moment map on K\"ahler manifolds}
The concept of holomorphic moment map applies to all K\"ahler manifolds, not necessarily special. Indeed it can be constructed just in terms of the K\"ahler potential without advocating any further structure. In this subsection we review its properties and definition, as usual in order to fix conventions, normalizations and notations.
\par
Let  $g_{i {j^\star}}$ be the K\"ahler metric of a K\"ahler
manifold ${\cal M}$ and let us assume that  $g_{i {j^\star}}$ admits
a non trivial group of continuous isometries ${\cal G}$
generated by Killing vectors $k_\mathbf{I}^i$ ($\mathbf{I}=1, \ldots, {\rm dim}
\,{\cal G} )$ that define the infinitesimal variation of the complex
coordinates $z^i$ under the group action:
\begin{equation}
\label{urca1}
z^i \to z^i + \epsilon^\mathbf{I} k_\mathbf{I}^i (z)
\end{equation}
Let $k^i_{\mathbf{I}} (z)$ be a basis of holomorphic Killing vectors for
the metric $g_{i{j^\star}}$.  Holomorphicity means the following
differential constraint:
\begin{equation}
\partial_{j^*} k^i_{\mathbf{I}} (z)=0
\leftrightarrow \partial_j k^{i^*}_{\mathbf{I}} (\bar z)=0 \label{holly}
\end{equation}
while the generic Killing equation (suppressing the
gauge index $\mathbf{I}$):
\begin{equation}
\nabla_\mu k_\nu +\nabla_\mu k_\nu=0
\end{equation}
in holomorphic indices reads as follows:
\begin{equation}
\begin{array}{ccccccc}
\nabla_i k_{j} + \nabla_j k_{i} &=&0 & ; &
\nabla_{i^*} k_{j} + \nabla_j k_{i^*} &=& 0
\label{killo}
\end{array}
\end{equation}
where the covariant components are defined as
$k_{j }=g_{j i^*} k^{i^*}$ (and similarly for
$k_{i^*}$).
\par
The vectors $k_{\mathbf{I}}^i$ are generators of infinitesimal
holomorphic coordinate transformations $\delta z^i = \epsilon^\mathbf{I} k^i_{\mathbf{I}} (z)$
which leave the metric invariant. In the same way as the metric is
the derivative of a more fundamental
object, the Killing vectors in a K\"ahler manifold are the
derivatives of suitable prepotentials. Indeed the first of
eq.s (\ref{killo})  is automatically satisfied by holomorphic vectors
and the second equation reduces to the following one:
\begin{equation}
k^i_{\mathbf{I}}=i g^{i j^*} \partial_{j^*} {\cal P}_{\mathbf{I}},
\quad {\cal P}^*_{\mathbf{I}} = {\cal P}_{\mathbf{I}}\label{killo1}
\end{equation}
In other words if we can find a real function ${\cal P}^\mathbf{I}$ such
that the expression $i g^{i j^*} \partial_{j^*}
{\cal P}_{(\mathbf{I})}$ is holomorphic, then eq.(\ref{killo1}) defines a
Killing vector.
\par
The construction of the Killing prepotential can be stated in a more
precise geometrical fashion through the notion of {\it moment map}.
Let us review this construction.
\par
Consider a K\"ahlerian manifold ${\cal M}$ of real dimension $2n$.
Consider an isometry group ${\cal G}$ acting on
 ${\cal M}$  by means of Killing vector
fields $\overrightarrow{X}$ which are holomorphic
with respect to the  complex structure
${ J}$ of ${\cal M}$; then these vector
fields preserve also the K\"ahler 2-form
\begin{equation}
\begin{array}{ccc}
\matrix{
{\cal L}_{\scriptscriptstyle\overrightarrow{X}}g = 0 & \leftrightarrow &
\nabla_{(\mu}X_{\nu)}=0 \cr
{\cal L}_{\scriptscriptstyle\overrightarrow{X}}{  J}= 0 &\null &\null \cr }
  \Biggr \} & \Rightarrow &  0={\cal L}_{\scriptscriptstyle\overrightarrow{X}}
K = i_{\scriptscriptstyle\overrightarrow{X}}
dK+d(i_{\scriptscriptstyle\overrightarrow{X}}
K) = d(i_{\scriptscriptstyle\overrightarrow{X}}K) \cr
\end{array}
\label{holkillingvectors}
\end{equation}
Here ${\cal L}_{\scriptscriptstyle\overrightarrow{X}}$ and
$i_{\scriptscriptstyle\overrightarrow{X}}$
denote respectively the Lie derivative along
the vector field $\overrightarrow{X}$ and the contraction
(of forms) with it.
\par
If ${\cal M}$ is simply connected,
$d(i_{\overrightarrow{X}}K)=0$ implies the existence
of a function ${\cal P}_{\overrightarrow{X}}$ such
that
\begin{equation}
-\frac{1}{2}d{\cal P}_{\overrightarrow{X}}=
i_{\scriptscriptstyle\overrightarrow{X}}K
\label{mmap}
\end{equation}
The function ${\cal P}_{\overrightarrow{X}}$ is defined up to a constant,
which can be arranged so as to make it equivariant:
\begin{equation}
\overrightarrow{X} {\cal P}_{\overrightarrow{Y}} =
{\cal P}_{[\overrightarrow{X},\overrightarrow{Y}]}
\label{equivarianza}
\end{equation}
${\cal P}_{\overrightarrow{X}}$ constitutes then a {\it moment map}.
This can be regarded as a map
\begin{equation}
{\cal P}: {\cal M} \, \longrightarrow \,
\mathbb{R} \otimes
{\mathbb{G} }^*
\end{equation}
where ${\mathbb{G}}^*$ denotes the dual of the Lie algebra
${\mathbb{G} }$ of the group ${\cal G}$.
Indeed let $x\in {\mathbb{G} }$ be the Lie algebra element
corresponding to the Killing vector $\overrightarrow{X}$; then, for a given
$m\in {\cal M}$
\begin{equation}
\mu (m)\,  : \, x \, \longrightarrow \,  {\cal P}_{\overrightarrow{X}}(m) \,
\in  \, \mathbb{R}
\end{equation}
is a linear functional on  ${\mathbb{G}}$.
If we expand
$\overrightarrow{X} = a^\mathbf{I} k_\mathbf{I}$ in a basis of Killing vectors
$k_\mathbf{I}$ such that
\begin{equation}
[k_\mathbf{I}, k_\mathbf{L}]= f_{\mathbf{I} \mathbf{L}}^{\ \ \mathbf{K}} k_\mathbf{K}
\label{blio}
\end{equation}
we have also
\begin{equation}
{\cal P}_{\overrightarrow{X}}\, = \, a^\mathbf{I} {\cal P}_\mathbf{I}
\end{equation}
In the following we  use the
shorthand notation ${\cal L}_\mathbf{I}, i_\mathbf{I}$ for the Lie derivative
and the contraction along the chosen basis of Killing vectors $ k_\mathbf{I}$.
\par
From a geometrical point of view the prepotential,
or moment map, ${\cal P}_\mathbf{I}$ is the Hamiltonian function providing the Poissonian
realization  of the Lie algebra on the K\"ahler manifold. This
is just another way of stating the already mentioned
{\it  equivariance}.
Indeed  the  very  existence  of the closed 2-form $K$ guarantees that
every K\"ahler space is a symplectic manifold and that we can define  a
Poisson bracket.
\par
Consider eqs.(\ref{killo1}). To every generator of the abstract  Lie algebra
${\mathbb{G}}$ we have associated a function  ${\cal P}_\mathbf{I}$ on
${\cal M}$; the Poisson bracket of
${\cal P}_\mathbf{I}$ with ${\cal P}_\mathbf{J}$ is defined as follows:
\begin{equation}
\{{\cal P}_\mathbf{I} , {\cal P}_\mathbf{J}\} \equiv 4\pi K
(\mathbf{I}, \mathbf{J})
\end{equation}
where $K(\mathbf{I}, \mathbf{J})
\equiv K (\vec k_\mathbf{I}, \vec k_\mathbf{J})$ is
the value of $K$ along the pair of Killing vectors.
\par
In reference \cite{D'Auria:1991fj}  the following
lemma was proved:
\begin{lemma}
{\it{The following identity is true}}:
\begin{equation}
\{{\cal P}_\mathbf{I}, {\cal
P}_\mathbf{J}\}=f_{\mathbf{I}\mathbf{J}}^{\ \ \mathbf{L}}{\cal
P}_\mathbf{L} + C_{\mathbf{I} \mathbf{J}} \label{brack}
\end{equation}
{\it{where $C_{\mathbf{I} \mathbf{J}}$ is a constant fulfilling the
cocycle condition}}
\begin{equation}
f^{\ \ \mathbf{L}}_{\mathbf{I}\mathbf{M}} C_{\mathbf{L} \mathbf{J}} +
f^{\ \ \mathbf{L}}_{\mathbf{M}\mathbf{J}} C_{\mathbf{L} \mathbf{I}}+
f_{\mathbf{J}\mathbf{I}}^{\ \  \mathbf{L}} C_{\mathbf{L} \mathbf{M}}=0
\label{cocy}
\end{equation}
\end{lemma}
If the Lie algebra ${\mathbb{G}}$ has a trivial second cohomology group
$H^2({\mathbb{G}})=0$, then the cocycle $C_{\mathbf{I} \mathbf{J}}$ is a
coboundary; namely we have
\begin{equation}
C_{\mathbf{I} \mathbf{J}} = f^{\ \ \mathbf{L}}_{\mathbf{I} \mathbf{J}} C_\mathbf{L}
\end{equation}
where $C_\mathbf{L}$ are suitable constants. Hence, assuming
$H^2 (\mathbb{G})= 0$
we can reabsorb $C_\mathbf{L}$ in  the definition of ${\cal
P}_\mathbf{I}$:
\begin{equation}
{\cal P}_\mathbf{I} \rightarrow {\cal P}_\mathbf{I}+ C_\mathbf{I}
\end{equation}
and we obtain the stronger equation
\begin{equation}
\{{\cal P}_\mathbf{I}, {\cal P}_\mathbf{J}\} =
f_{\mathbf{I}\mathbf{J}}^{\ \  \mathbf{L}} {\cal P}_\mathbf{L}
\label{2.39}
\end{equation}
Note that $H^2({\mathbb{G}}) = 0$ is true for all semi-simple Lie
algebras.
Using eq.(\ref{brack}), eq.(\ref{2.39})
can be rewritten in components as follows:
\begin{equation}
{i\over 2} g_{ij^*}(k^i_\mathbf{I} k^{j^*}_\mathbf{J} -
k^i_\mathbf{J} k^{j^*}_\mathbf{I})=
{1\over 2} f_{\mathbf{I} \mathbf{J}}^{\  \  \mathbf{L}} {\cal
P}_\mathbf{L}
\label{2.40}
\end{equation}
Equation (\ref{2.40}) is identical with the equivariance condition
in eq.(\ref{equivarianza}).
\par
Finally let us recall the explicit general way of solving eq.(\ref{mmap}) obtaining the real valued function ${\cal P}_\mathbf{I}$ which satisfies eq.(\ref{killo1}). In terms of the K\"ahler potential $\mathcal{K}$ we have:
\begin{equation}\label{sisalvichipuo}
    \mathcal{P}_{{\bf I}}{}^x=-\frac{i}{2}\left(k_{{\bf I}}^i\partial_i \mathcal{K}-k_{{\bf I}}^{\bar{\imath}}\partial_{\bar{\imath}} \mathcal{K}\right)+{\rm Im}(f_{{\bf I}})\,,
\end{equation}
where $f_{{\bf I}}=f_{{\bf I}}(z)$ is a holomorphic transformation on the line-bundle, defining a compensating K\"ahler transformation:
\begin{equation}
k_{{\bf I}}^i\partial_i \mathcal{K}+k_{{\bf I}}^{\bar{\imath}}\partial_{\bar{\imath}} \mathcal{K}=-f_{{\bf I}}(z)-\bar{f}_{{\bf I}}(\bar{z})\,.\label{sisalvichipuo2}
\end{equation}
 We also have:
\begin{eqnarray}
\mathfrak{T}_{{\bf I}}\cdot \Omega &=&\mathfrak{T}_{{\bf I}}\cdot \Omega+f_{{\bf I}}\,\Omega\,,\label{sisalvichipuo30}\\
\mathfrak{T}_{{\bf I}}\cdot V+i\,{\rm Im}(f_{{\bf I}})\,V
&=& k_{{\bf I}}^i\partial_i V+k_{{\bf I}}^{\bar{\imath}}\partial_{\bar{\imath}} V\,,\label{sisalvichipuo3}
\end{eqnarray}
where $\mathfrak{T}_{{\bf I}}\cdot \Omega$ denotes the symplectic action of the isometry on the section $V$. If $\mathfrak{T}_{{\bf I}}$ is represented by the symplectic matrix $(\mathfrak{T}_{{\bf I}})_\alpha{}^\beta=-(\mathfrak{T}_{{\bf I}})^\beta{}_\alpha$, $\alpha,\,\beta=1,\dots,\,2n+2$:
 \begin{equation}\label{cosasimplettica}
    \mathfrak{T}_{\mathbf{I}}^T \, \mathbb{C} \, + \, \mathbb{C} \, \mathfrak{T}_{\mathbf{I}} \, = \, 0
\end{equation}
we have $(\mathfrak{T}_{{\bf I}}\cdot V)^\alpha=-\mathfrak{T}_{{\bf I}\,\beta}{}^\alpha\, V^\beta=\mathfrak{T}_{{\bf I}}^\alpha{}_\beta\, V^\beta$. From (\ref{sisalvichipuo3}) and (\ref{sisalvichipuo}) we derive the following useful symplectic-invariant expression for the moment maps:
\begin{equation}
  \mathcal{P}_{{\bf I}}{}^x=-\bar{V}^\alpha\,\mathfrak{T}_{{\bf I}\,\alpha}{}^\beta\mathbb{C}_{\beta\gamma}\,V^\gamma\,.
\end{equation}
Eq.s (\ref{sisalvichipuo}), (\ref{sisalvichipuo2}), (\ref{sisalvichipuo3}) generalize the corresponding formulae given in sections 7.1 and 7.2 of \cite{Andrianopoli:1996cm}, where the condition $f_{{\bf I}}=0$ was imposed, to gaugings of non-compact isometries which are associated with non-trivial compensating K\"ahler transformations and/or to gauged (non-compact) isometries whose symplectic action is not diagonal.

\section{\sc Quaternionic geometry}
\label{hypgeosec}
Next we turn our attention to the geometry that pertains to the hypermultiplet sector of an
$\mathcal{N}=2$ supersymmetric theory. Each hypermultiplet contains $4$ real scalar fields
and, at least locally, they can be regarded as the
four components of a quaternion. The locality caveat is, in this
case, very substantial because global quaternionic coordinates can be
constructed only occasionally even on those manifolds that are
denominated quaternionic in the mathematical literature
\cite{hklr}, \cite{gal}. Anyhow, what is important  is that, in
the hypermultiplet sector, the scalar manifold $\mathcal{QM}$ has
dimension multiple of four:
\begin{equation}
\mbox{dim}_{\bf R} \, \mathcal{QM} \, = \, 4 \, m \,\equiv \, 4 \, \# \,
\mbox{of hypermultiplets}
\label{quatdim}
\end{equation}
and, in some appropriate sense, it has a quaternionic structure.
\par
We name {\it Hypergeometry} that pertaining to the
hypermultiplet sector, irrespectively whether we deal with global or
local $\mathcal{N}$=2 theories. Yet there are two kinds of hypergeometries.
Supersymmetry requires the existence
of a principal $\mathrm{SU}(2)$--bundle
\begin{equation}
{\cal SU} \, \longrightarrow \, \mathcal{QM} \label{su2bundle}
\end{equation}
The bundle ${\cal SU}$ is
{\bf flat} in the {\it rigid supersymmetry case} while its curvature is
proportional to the K\"ahler forms in the {\it local case}.
\par
These two versions of hypergeometry were already known in mathematics prior to
their use \cite{D'Auria:1991fj},\cite{specHomgeo}, \cite{vanderseypen}, \cite{Andrianopoli:1996cm}, \cite{hklr}, \cite{gal}  in the context of $\mathcal{N}=2$
supersymmetry and are identified as:
\begin{eqnarray}
\mbox{rigid hypergeometry} & \equiv & \mbox{HyperK\"ahler geometry.}
\nonumber\\ \mbox{local hypergeometry} & \equiv & \mbox{Quaternionic K\"ahler
geometry} \label{picchio}
\end{eqnarray}
\subsection{\sc Quaternionic K\"ahler, versus HyperK\"ahler manifolds}
Both a Quaternionic K\"ahler or a HyperK\"ahler manifold $\mathcal{QM}$
is a $4 m$-dimensional real manifold endowed with a metric $h$:
\begin{equation}
d s^2 = h_{u v} (q) d q^u \otimes d q^v   \quad ; \quad u,v=1,\dots,
4  m \label{qmetrica}
\end{equation}
and three complex structures
\begin{equation}
(J^x) \,:~~ T(\mathcal{QM}) \, \longrightarrow \, T(\mathcal{QM}) \qquad
\quad (x=1,2,3)
\end{equation}
that satisfy the quaternionic algebra
\begin{equation}
J^x J^y = - \delta^{xy} \, \bfone \,  +  \, \epsilon^{xyz} J^z
\label{quaternionetta}
\label{quatalgebra}
\end{equation}
and respect to which the metric is hermitian:
\begin{equation}
\forall   \mbox{\bf X} ,\mbox{\bf Y}  \in   T\mathcal{QM}   \,: \quad h
\left( J^x \mbox{\bf X}, J^x \mbox{\bf Y} \right )   = h \left(
\mbox{\bf X}, \mbox{\bf Y} \right ) \quad \quad
  (x=1,2,3)
\label{hermit}
\end{equation}
From eq. (\ref{hermit}) it follows that one can introduce a triplet
of 2-forms
\begin{equation}
\begin{array}{ccccccc}
K^x& = &K^x_{u v} d q^u \wedge d q^v & ; & K^x_{uv} &=&   h_{uw}
(J^x)^w_v \cr
\end{array}
\label{iperforme}
\end{equation}
that provides the generalization of the concept of K\"ahler form
occurring in  the complex case. The triplet $K^x$ is named the {\it
HyperK\"ahler} form. It is an $\mathrm{SU}(2)$ Lie--algebra valued
2--form  in the same way as the K\"ahler form is a $\mathrm{U(1)}$
Lie--algebra valued 2--form. In the complex case the definition of
K\"ahler manifold involves the statement that the K\"ahler 2--form is
closed. At the same time in Hodge--K\"ahler manifolds  the K\"ahler 2--form can be
identified with the curvature of a line--bundle which in the case of
rigid supersymmetry is flat. Similar steps can be taken also here and
lead to two possibilities: either HyperK\"ahler or Quaternionic K\"ahler manifolds.
\par
Let us  introduce a principal $\mathrm{SU}(2)$--bundle ${\cal SU}$ as
defined in eq. (\ref{su2bundle}). Let $\omega^x$ denote a connection
on such a bundle. To obtain either a HyperK\"ahler or a Quaternionic K\"ahler manifold we must impose the condition that the HyperK\"ahler 2--form
is covariantly closed with respect to the connection $\omega^x$:
\begin{equation}
\nabla K^x \equiv d K^x + \epsilon^{x y z} \omega^y \wedge K^z    \,
= \, 0 \label{closkform}
\end{equation}
The only difference between the two kinds of geometries resides in
the structure of the ${\cal SU}$--bundle.
\begin{definizione} A
HyperK\"ahler manifold is a $4 m$--dimensional manifold with the
structure described above and such that the ${\cal SU}$--bundle is
{\bf flat}
\end{definizione}
 Defining the ${\cal SU}$--curvature by:
\begin{equation}
\Omega^x \, \equiv \, d \omega^x + {1\over 2} \epsilon^{x y z}
\omega^y \wedge \omega^z \label{su2curv}
\end{equation}
in the HyperK\"ahler case we have:
\begin{equation}
\Omega^x \, = \, 0 \label{piattello}
\end{equation}
Viceversa \begin{definizione} A Quaternionic K\"ahler manifold is a $4
m$--dimensional manifold with the structure described above and such
that the curvature of the ${\cal SU}$--bundle is proportional to the
HyperK\"ahler 2--form \end{definizione} Hence, in the quaternionic
case we can write:
\begin{equation}
\Omega^x \, = \, { {\lambda}}\, K^x \label{piegatello}
\end{equation}
where $\lambda$ is a non vanishing real number.
\par
As a consequence of the above structure the manifold $\mathcal{QM}$ has
a holonomy group of the following type:
\begin{eqnarray}
{\rm Hol}(\mathcal{QM})&=& \mathrm{SU}(2)\otimes \mathrm{H} \quad
(\mbox{Quaternionic K\"ahler}) \nonumber \\ {\rm Hol}(\mathcal{QM})&=& \bfone
\otimes \mathrm{H} \quad (\mbox{HyperK\"ahler}) \nonumber \\ \mathrm{H} &
\subset & \mathrm{Sp (2m,\mathbb{R}) }\label{olonomia}
\end{eqnarray}
In both cases, introducing flat indices $\{A,B,C= 1,2\}
\{\alpha,\beta,\gamma = 1,.., 2m\}$  that run, respectively, in the
fundamental representation of $\mathrm{SU}(2)$ and of
$\mathrm{Sp}(2m,\mathbb{R})$, we can find a vielbein 1-form
\begin{equation}
{\cal U}^{A\alpha} = {\cal U}^{A\alpha}_u (q) d q^u
\label{quatvielbein}
\end{equation}
such that
\begin{equation}
h_{uv} = {\cal U}^{A\alpha}_u {\cal U}^{B\beta}_v
\mathbb{C}_{\alpha\beta}\epsilon_{AB} \label{quatmet}
\end{equation}
where $\mathbb{C}_{\alpha \beta} = - \mathbb{C}_{\beta \alpha}$ and $\epsilon_{AB}
= - \epsilon_{BA}$ are, respectively, the flat $\mathrm{Sp}(2m)$ and
$\mathrm{Sp}(2) \sim \mathrm{SU}(2)$ invariant metrics. The vielbein
${\cal U}^{A\alpha}$ is covariantly closed with respect to the
$\mathrm{SU}(2)$-connection $\omega^z$ and to some
$\mathrm{Sp}(2m,\mathbb{R})$-Lie Algebra valued connection
$\Delta^{\alpha\beta} = \Delta^{\beta \alpha}$:
\begin{eqnarray}
\nabla {\cal U}^{A\alpha}& \equiv & d{\cal U}^{A\alpha} +{i\over 2}
\omega^x (\epsilon \sigma_x\epsilon^{-1})^A_{\phantom{A}B}
\wedge{\cal U}^{B\alpha} \nonumber\\ &+& \Delta^{\alpha\beta} \wedge
{\cal U}^{A\gamma} \mathbb{C}_{\beta\gamma} =0 \label{quattorsion}
\end{eqnarray}
\noindent where $(\sigma^x)_A^{\phantom{A}B}$ are the standard Pauli
matrices. Furthermore ${ \cal U}^{A\alpha}$ satisfies  the reality
condition:
\begin{equation}
{\cal U}_{A\alpha} \equiv ({\cal U}^{A\alpha})^* = \epsilon_{AB}
\mathbb{C}_{\alpha\beta} {\cal U}^{B\beta} \label{quatreality}
\end{equation}
Eq.(\ref{quatreality})  defines  the  rule to lower the symplectic
indices by means   of  the  flat  symplectic   metrics
$\epsilon_{AB}$   and $\mathbb{C}_{\alpha \beta}$. More specifically we can
write a stronger version of eq. (\ref{quatmet}) \cite{sugkgeom_3}:
\begin{eqnarray}
({\cal U}^{A\alpha}_u {\cal U}^{B\beta}_v + {\cal U}^{A\alpha}_v
{\cal
 U}^{B\beta}_u)\mathbb{C}_{\alpha\beta}&=& h_{uv} \epsilon^{AB}\nonumber\\
 \label{piuforte}
\end{eqnarray}
\noindent We have also the inverse vielbein ${\cal U}^u_{A\alpha}$
defined by the equation
\begin{equation}
{\cal U}^u_{A\alpha} {\cal U}^{A\alpha}_v = \delta^u_v \label{2.64}
\end{equation}
Flattening a pair of indices of the Riemann tensor ${\cal
R}^{uv}_{\phantom{uv}{ts}}$ we obtain
\begin{equation}
{\cal R}^{uv}_{\phantom{uv}{ts}} {\cal U}^{\alpha A}_u {\cal
U}^{\beta B}_v = -\,{{\rm i}\over 2} \Omega^x_{ts} \epsilon^{AC}
 (\sigma_x)_C^{\phantom {C}B} \mathbb{C}^{\alpha \beta}+
 \mathbb{R}^{\alpha\beta}_{ts}\epsilon^{AB}
\label{2.65}
\end{equation}
\noindent where $\mathbb{R}^{\alpha\beta}_{ts}$ is the field strength
of the $\mathrm{Sp}(2m) $ connection:
\begin{equation}
d \Delta^{\alpha\beta} + \Delta^{\alpha \gamma} \wedge \Delta^{\delta
\beta} \mathbb{C}_{\gamma \delta} \equiv \mathbb{R}^{\alpha\beta} =
\mathbb{R}^{\alpha \beta}_{ts} dq^t \wedge dq^s \label{2.66}
\end{equation}
Eq. (\ref{2.65}) is the explicit statement that the Levi Civita
connection associated with the metric $h$ has a holonomy group
contained in $\mathrm{SU}(2) \otimes \mathrm{Sp}(2m)$. Consider now
eq.s (\ref{quatalgebra}), (\ref{iperforme}) and (\ref{piegatello}).
We easily deduce the following relation:
\begin{equation}
h^{st} K^x_{us} K^y_{tw} = -   \delta^{xy} h_{uw} +
  \epsilon^{xyz} K^z_{uw}
\label{universala}
\end{equation}
that holds true both in the HyperK\"ahler and in the quaternionic
case. In the latter case, using eq. (\ref{piegatello}), eq.
(\ref{universala}) can be rewritten as follows:
\begin{equation}
h^{st} \Omega^x_{us} \Omega^y_{tw} = - \lambda^2 \delta^{xy} h_{uw} +
\lambda \epsilon^{xyz} \Omega^z_{uw} \label{2.67}
\end{equation}
Eq.(\ref{2.67}) implies that the intrinsic components of the curvature
 2-form $\Omega^x$ yield a representation of the quaternion algebra.
In the HyperK\"ahler case such a representation is provided only by
the HyperK\"ahler form. In the quaternionic case we can write:
\begin{equation}
\Omega^x_{A\alpha, B \beta} \equiv \Omega^x_{uv} {\cal U}^u_{A\alpha}
{\cal U}^v_{B\beta} = - i \lambda \mathbb{C}_{\alpha\beta}
(\sigma_x)_A^{\phantom {A}C}\epsilon _{CB} \label{2.68}
\end{equation}
\noindent Alternatively eq.(\ref{2.68}) can be rewritten in an
intrinsic form as
\begin{equation}
\Omega^x =\,-{\rm i}\, \lambda \mathbb{C}_{\alpha\beta} (\sigma
_x)_A^{\phantom {A}C}\epsilon _{CB} {\cal U}^{\alpha A} \wedge {\cal
U}^{\beta B} \label{2.69}
\end{equation}
\noindent whence we also get:
\begin{equation}
{i\over 2} \Omega^x (\sigma_x)_A^{\phantom{A}B} = \lambda{\cal
U}_{A\alpha} \wedge {\cal U}^{B\alpha} \label{2.70}
\end{equation}
\subsection{\sc The triholomorphic moment map on quaternionic manifolds}
Next, following closely the original derivation of \cite{D'Auria:1991fj,mylecture}
let us turn to a discussion of the triholomorphic isometries of the manifold $\mathcal{QM}$
associated with hypermultiplets.
In $D=4$  supergravity the manifold of hypermultiplet scalars $\mathcal{QM}$ is  a Quaternionic
K\"ahler manifold and we can gauge only those of its isometries
that are  triholomorphic  and that  either generate an abelian group $\mathcal{G}$
or are \emph{suitably realized}  as isometries also on the special manifold $\widehat{\mathcal{SK}}_n$.
This means  that on $\mathcal{QM}$ we have Killing vectors:
\begin{equation}
\vec k_\mathbf{I} = k^u_\mathbf{I} {\vec \partial\over \partial q^u}
\label{2.71}
\end{equation}
\noindent
satisfying the same Lie algebra  as the corresponding Killing
vectors on $\widehat{\mathcal{SK}}_n$. In other words
\begin{equation}
{\vec{\mathfrak{K}}}_\mathbf{I} =
\hat{k}^i_\mathbf{I} \vec \partial_i + \hat{k}^{i^*}_\mathbf{I}
\vec\partial_{i^*} + k_\mathbf{I}^u \vec\partial_u
\label{2.72}
\end{equation}
\noindent
is a Killing vector of the block diagonal metric:
\begin{equation}
 \mathfrak{g} = \left (
\matrix { \widehat{g}_{ij^\star} & \quad 0 \quad \cr \quad 0
\quad & h_{uv} \cr } \right )
\label{2.73}
\end{equation}
defined on the product manifold\footnote{Following the notations described in the introduction, the Special K\"ahler manifold which describes the interaction of vector multiplets is denoted $\widehat{\mathcal{SK} }$ and all the Special Geometry Structures are endowed with a hat in order to distinguish this Special K\"ahler manifold from the other one which is incapsulated into the Quaternionic K\"ahler manifold $\mathcal{QM}$ describing the hypermultiplets when this latter happens to be in the image of the $c$-map.} $\widehat{\mathcal{SK} }\otimes\mathcal{QM}$.
\par
Let us first focus on the manifold $\mathcal{QM}$.
Triholomorphicity means that the Killing vector fields leave
the HyperK\"ahler structure invariant up to $\mathrm{SU(2)}$
rotations in the $\mathrm{SU(2)}$--bundle defined by eq.(\ref{su2bundle}).
Namely:
\begin{equation}
\begin{array}{ccccccc}
{\cal L}_\mathbf{I} K^x & = &\epsilon^{xyz}K^y
W^z_\mathbf{I} & ; &
{\cal L}_\mathbf{I}\omega^x&=& \nabla W^x_\mathbf{I}
\end{array}
\label{cambicchio}
\end{equation}
where $W^x_\mathbf{I}$ is an $\mathrm{SU(2)}$ compensator associated with the
Killing vector $k^u_\mathbf{I}$. The compensator $W^x_\mathbf{I}$ necessarily
fulfills  the cocycle condition:
\begin{equation}
{\cal L}_\mathbf{I} W^{x}_\mathbf{J} - {\cal L}_\mathbf{J} W^x_\mathbf{I} + \epsilon^{xyz}
W^y_\mathbf{I} W^z_\mathbf{J} = f_{\mathbf{I} \mathbf{J}}^{\cdot \cdot \mathbf{L}}
W^x_\mathbf{L}
\label{2.75}
\end{equation}
In the HyperK\"ahler case the $\mathrm{SU(2)}$--bundle is flat and the
compensator can be reabsorbed into the definition of the
HyperK\"ahler forms. In other words we can always find a
map
\begin{equation}
\mathcal{QM} \, \longrightarrow \, L^x_{\phantom{x}y} (q)
\, \in \, \mathrm{SO(3)}
\end{equation}
that trivializes the ${\cal SU}$--bundle globally. Redefining:
\begin{equation}
K^{x\prime} \, = \, L^x_{\phantom{x}y} (q) \, K^y
\label{enfantduparadis}
\end{equation}
the new HyperK\"ahler form  obeys the stronger equation:
\begin{equation}
{\cal L}_\mathbf{I} K^{x\prime} \, = \, 0
\label{noncambio}
\end{equation}
On the other hand, in the quaternionic case, the non--triviality of the
${\cal SU}$--bundle forbids to eliminate the $W$--compensator
completely. Due to the identification between HyperK\"ahler
forms and $\mathrm{SU(2)}$ curvatures eq.(\ref{cambicchio}) is rewritten
as:
\begin{equation}
\begin{array}{ccccccc}
{\cal L}_\mathbf{I} \Omega^x& = &\epsilon^{xyz}\Omega^y
W^z_\mathbf{I} & ; &
{\cal L}_\mathbf{I}\omega^x&=& \nabla W^x_\mathbf{I}
\end{array}
\label{cambiacchio}
\end{equation}
In both cases, anyhow, and in full analogy with the case of
K\"ahler manifolds, to each Killing vector
we can associate a triplet ${\cal
P}^x_\mathbf{I} (q)$ of 0-form prepotentials.
Indeed we can set:
\begin{equation}
{\bf i}_\mathbf{I}  K^x =
- \nabla {\cal P}^x_\mathbf{I} \equiv -(d {\cal
P}^x_\mathbf{I} + \epsilon^{xyz} \omega^y {\cal P}^z_\mathbf{I})
\label{2.76}
\end{equation}
where $\nabla$ denotes the $\mathrm{SU(2)}$ covariant exterior derivative.
\par
As in the K\"ahler case eq.(\ref{2.76}) defines a moment map:
\begin{equation}
{\cal P}: {\cal M} \, \longrightarrow \,
\mathbb{R}^3 \otimes
{\mathcal{G} }^*
\end{equation}
where ${\mathcal{G}}^*$ denotes the dual of the Lie algebra
${\mathcal{G} }$ of the group ${\cal G}$.
Indeed let $x\in {\mathcal{G} }$ be the Lie algebra element
corresponding to the Killing
vector $\overrightarrow{X}$; then, for a given
$m\in {\cal M}$
\begin{equation}
\mu (m)\,  : \, x \, \longrightarrow \,  {\cal P}_{\overrightarrow{X}}(m) \,
\in  \, \mathbb{R}^3
\end{equation}
is a linear functional on  ${\mathcal{G}}$. If we expand
$\overrightarrow{X} = a^\mathbf{I} k_\mathbf{I}$ on a basis of Killing vectors
$k_\mathbf{I}$ such that
\begin{equation}
[k_\mathbf{I}, k_\mathbf{L}]= f_{\mathbf{I} \mathbf{L}}^{\ \ \mathbf{K}} k_\mathbf{K}
\label{blioprime}
\end{equation}
and we also choose a basis ${\bf i}_x \, (x=1,2,3)$ for $\mathbb{R}^3$
we get:
\begin{equation}
{\cal P}_{\overrightarrow{X}}\, = \, a^\mathbf{I} {\cal P}_\mathbf{I}^x \, {\bf i}_x
\end{equation}
Furthermore we need a generalization of the equivariance defined
by eq.(\ref{equivarianza})
\begin{equation}
\overrightarrow{X} \circ {\cal P}_{\overrightarrow{Y}} \,=  \,
{\cal P}_{[\overrightarrow{X},\overrightarrow{Y}]}
\label{equivarianzina}
\end{equation}
In the HyperK\"ahler case, the left--hand side of
eq.(\ref{equivarianzina})
is defined as the usual action of a vector field on a $0$--form:
\begin{equation}
\overrightarrow{X} \circ {\cal P}_{\overrightarrow{Y}}\, =  \, {\bf i}_{\overrightarrow{X}} \, d
{\cal P}_{\overrightarrow{Y}}\, = \,
X^u \, {\frac{\partial}{\partial q^u}} \, {\cal P}_{\overrightarrow{Y}}\,
\end{equation}
The equivariance condition   implies
that we can introduce a triholomorphic Poisson bracket defined
as follows:
\begin{equation}
\{{\cal P}_\mathbf{I}, {\cal P}_\mathbf{J}\}^x \equiv 2 K^x (\mathbf{I},
\mathbf{J})
\label{hykapesce}
\end{equation}
leading to the triholomorphic Poissonian realization of the Lie
algebra:
\begin{equation}
\left \{ {\cal P}_\mathbf{I}, {\cal P}_\mathbf{J} \right \}^x \, = \,
f^{\mathbf{K}}_{\phantom{\mathbf{K}}\mathbf{I}\mathbf{J}} \, {\cal P}_\mathbf{K}^{x}
\label{hykapescespada}
\end{equation}
which in components reads:
\begin{equation}
K^x_{uv} \, k^u_\mathbf{I} \, k^v_\mathbf{J} \, = \, {\frac{1}{2}} \,
f^{\mathbf{K}}_{\phantom{\mathbf{K}}\mathbf{I}\mathbf{J}}\, {\cal P}_\mathbf{K}^{x}
\label{hykaide}
\end{equation}
In the quaternionic case, instead, the left--hand side of
eq.(\ref{equivarianzina})
is interpreted as follows:
\begin{equation}
\overrightarrow{X} \circ {\cal P}_{\overrightarrow{Y}}\, =  \, {\bf i}_{\overrightarrow{X}}\,  \nabla
{\cal P}_{\overrightarrow{Y}}\, = \,
X^u \, {\nabla_u} \, {\cal P}_{\overrightarrow{Y}}\,
\end{equation}
where $\nabla$ is the $\mathrm{SU(2)}$--covariant differential.
Correspondingly, the triholomorphic Poisson bracket is defined
as follows:
\begin{equation}
\{{\cal P}_\mathbf{I}, {\cal P}_\mathbf{J}\}^x \equiv 2 K^x (\mathbf{I},
\mathbf{J})  - { {\lambda}} \, \varepsilon^{xyz} \,
{\cal P}_\mathbf{I}^y  \, {\cal P}_\mathbf{J}^z
\label{quatpesce}
\end{equation}
and leads to the Poissonian realization of the Lie algebra
\begin{equation}
\left \{ {\cal P}_\mathbf{I}, {\cal P}_\mathbf{J} \right \}^x \, = \,
f^{\mathbf{K}}_{\phantom{\mathbf{K}}\mathbf{I}\mathbf{J}} \, {\cal P}_\mathbf{K}^{x}
\label{quatpescespada}
\end{equation}
which in components reads:
\begin{equation}
K^x_{uv} \, k^u_\mathbf{I} \, k^v_\mathbf{J} \, - \,
{ \frac{\lambda}{2}} \, \varepsilon^{xyz} \,
{\cal P}_\mathbf{I}^y  \, {\cal P}_\mathbf{J}^z\,= \,  {\frac{1}{2}} \,
f^{\mathbf{K}}_{\phantom{\mathbf{K}}\mathbf{I}\mathbf{J}}\, {\cal P}_\mathbf{K}^{x}
\label{quatide}
\end{equation}
Eq.(\ref{quatide}), which is the most convenient way of
expressing equivariance in a coordinate basis was originally written
in \cite{D'Auria:1991fj} and has played a fundamental
role in the construction of  supersymmetric actions  for gauged $\mathcal{N}=2$ supergravity
both in $D=4$ \cite{D'Auria:1991fj,Andrianopoli:1996cm} and in $D=5$ \cite{ceregatta}.
\section{\sc The Quaternionic K\"ahler Geometry in the image of the $c$-map}
The main object of study in the present paper are those Quaternionic K\"ahler manifolds that are in the image of the $c$-map.\footnote{Not all non-compact,  homogeneous Quaternionic K\"ahler manifolds which are relevant to supergravity (which are \emph{normal}, i.e. exhibiting a solvable group of isometries having a free and transitive action on it) are in the image of the c-map, the only exception being the quaternionic  projective spaces \cite{alek,Cecotti:1988ad,vanderseypen}.} This latter
\begin{equation}\label{cimappo}
    \mbox{c-map} \, \, : \,\, \mathcal{SK}_n \, \Longrightarrow \, \mathcal{QM}_{4n+4}
\end{equation}
is a universal construction that starting from an arbitrary  Special K\"ahler manifold $\mathcal{SK}_n$ of complex dimension $n$, irrespectively whether it is homogenoeus or not, leads to a unique Quaternionic K\"ahler manifold $\mathcal{QM}_{4n+4}$ of real dimension $4n+4$ which contains $\mathcal{SK}_n$ as a submanifold. The precise modern definition of the $c$-map, originally introduced in \cite{sabarwhal}, is provided below.
\begin{definizione}
Let $\mathcal{SK}_n$ be a special K\"ahler manifold whose complex coordinates we denote by $z^i$ and whose K\"ahler metric we denote by $g_{ij^\star}$. Let moreover $\mathcal{N}_{\Lambda\Sigma}(z,{\bar z})$ be the symmetric period matrix defined by eq.(\ref{intriscripen}), introduce the following set of $4n+4$ coordinates:
\begin{equation}\label{finnico}
    \left\{q^u \right\} \, \equiv \, \underbrace{\{U,a\}}_{\mbox{2 real}}\, \bigcup \,\underbrace{\underbrace{\{ z^i\}}_{\mbox{n complex}}}_{\mbox{2n real}} \, \bigcup\, \underbrace{\mathbf{Z} \, = \, \{ Z^\Lambda \, , \, Z_\Sigma \}}_{\mbox{(2n+2) real}}
\end{equation}
Let us further introduce the following $(\mathrm{2n+2})\times(\mathrm{2n+2}) $ matrix  ${\cal M}_4^{-1}$:
\begin{eqnarray}
\mathcal{M}_4^{-1} & = &
\left(\begin{array}{c|c}
{\mathrm{Im}}\mathcal{N}\,
+\, {\mathrm{Re}}\mathcal{N} \, { \mathrm{Im}}\mathcal{N}^{-1}\, {\mathrm{Re}}\mathcal{N} & \, -{\mathrm{Re}}\mathcal{N}\,{ \mathrm{Im}}\,\mathcal{N}^{-1}\\
\hline
-\, { \mathrm{Im}}\mathcal{N}^{-1}\,{\mathrm{Re}}\mathcal{N}  & { \mathrm{Im}}\mathcal{N}^{-1} \
\end{array}\right) \label{inversem4}
\end{eqnarray}
which depends only on the coordinate of the Special K\"ahler manifold.
The $c$-map image of $\mathcal{SK}_n$ is the unique Quaternionic K\"ahler manifold $\mathcal{QM}_{4n+4}$ whose coordinates are the $q^u$ defined in (\ref{finnico}) and whose metric is given by the following universal formula
\begin{eqnarray}
ds^2_{\mathcal{QM}} &=&\frac{1}{4} \left(  d{U}^2+\, 4 g_{ij^\star} \,d{z}^j\, d{{\bar z}}^{j^\star}
+ e^{-2\,U}\,(d{a}+{\bf Z}^T\mathbb{C}d{{\bf
Z}})^2\,-\,2 \, e^{-U}\,d{{\bf
Z}}^T\,\mathcal{M}_4^{-1}\,d{{\bf Z}}\right)
\label{geodaction}
\end{eqnarray}
\end{definizione}
The metric (\ref{geodaction}) has the following positive definite signature
\begin{equation}
\mbox{sign}\left[ds^2_{\mathcal{QM}}\right] \, = \, \left(\underbrace{+,\dots,+}_{4+4\mathrm{n}}\right)
\end{equation}
since the matrix $\mathcal{M}_4^{-1} $ is negative definite.
\par
It is wort mentioning that if we utilize the same $4n+4$ coordinates (\ref{finnico}) and instead of (\ref{geodaction}) we introduce the alternative Lorentzian metric:
 \begin{eqnarray}
ds^2_{\mathcal{QM^\star}} &=&\frac{1}{4}\left( d{U}^2+\, 4 g_{ij^\star} \,d{z}^j\, d{{\bar z}}^{j^\star}
+ e^{-2\,U}\,(d{a}+{\bf Z}^T\mathbb{C}d{{\bf
Z}})^2\,+\,2 \, e^{-U}\,d{{\bf
Z}}^T\,\mathcal{M}_4\,d{{\bf Z}}\right)
\label{BHaction}
\end{eqnarray}
that has signature:
\begin{equation}
\mbox{sign}\left[ds^2_{\mathcal{QM}^\star}\right] \, = \, \left(\underbrace{+,\dots,+}_{2\mathrm{n}+2},\underbrace{-,\dots ,-}_{2\mathrm{n}+2}\right)
\end{equation}
we obtain the pseudo-quaternionic manifold $\mathcal{QM}$ which constitutes the target manifold in the $3$-dimensional $\sigma$-model description of $D=4$ supergravity Black-Hole solutions \cite{PietroSashaMarioBH1},\cite{noinilpotenti},\cite{miosasha}. In the case the Special K\"ahler pre-image is a symmetric space $\mathrm{U}_{\mathcal{SK}}/\mathrm{H}_{\mathcal{SK}}$, both $\mathcal{QM}$ and $\mathcal{QM}^\star$ turn out to be  symmetric spaces as well, $\mathrm{U}_{Q}/\mathrm{H}_{Q}$ and $\mathrm{U}_{Q}/\mathrm{H}_{Q}^\star$, the numerator group being the same. We will come back to the issue of symmetric homogeneous Quaternionic K\"ahler manifolds in section \ref{omosymmetro}
\subsection{\sc The HyperK\"ahler two-forms and the $\su(2)$-connection}
The reason why we state that $\mathcal{QM}_{4n+4}$ is Quaternionic K\"ahler is that, by utilizing only the identities of Special K\"ahler Geometry we can construct the three complex structures $J_u^{x|v}$ satisfying the quaternionic algebra (\ref{quaternionetta}) the corresponding HyperK\"ahler two-forms $K^x$  and the $\su(2)$ connection $\omega^x$ with respect to which they are covariantly constant.
\par
The construction is extremely beautiful and it is the following one.
\par
Consider the K\"ahler connection $\mathcal{Q}$ defined by eq. (\ref{u1conect}) and furthermore introduce the following differential form:
\begin{equation}
\label{Phidiffe}
    \Phi \, = \, da + \mathbf{Z}^T \, \mathbb{C}\, \mathrm{d}\mathbf{Z}
\end{equation}
Next define the two dimensional representation of both the $\su(2)$ connection and of the HyperK\"ahler $2$-forms as it follows:
\begin{eqnarray}
  \omega  &=& \frac{\rm i}{\sqrt{2}}\,\sum_{x=1}^3 \, \omega^x \, \gamma_x \label{cunnettasu2}\\
  \mathbf{K} &=& \frac{\rm i}{\sqrt{2}}\,\sum_{x=1}^3 \, K^x \, \sigma_x  \label{HypKalmatra}
\end{eqnarray}
where $\gamma_x$ denotes a basis of $2\times 2$ euclidian $\gamma$-matrices for which we utilize the following basis
which is  convenient in the explicit calculations  we perform in Part Four\footnote{The chosen $\gamma$-matrices are a permutation of the standard pauli matrices divided by $\sqrt{2}$ and multiplied by $\frac{\rm i}{2}$ can be used as a basis of anti-hermitian generators for the $\su(2)$ algebra in the fundamental defining representation.}:
\begin{eqnarray}
  \gamma_1&=& \left(
\begin{array}{ll}
 \frac{1}{\sqrt{2}} & 0 \\
 0 & -\frac{1}{\sqrt{2}}
\end{array}
\right) \nonumber\\
   \gamma_2&=& \left(
\begin{array}{ll}
 0 & -\frac{i}{\sqrt{2}} \\
 \frac{i}{\sqrt{2}} & 0
\end{array}
\right)\nonumber\\
 \gamma_3 &=& \left(
\begin{array}{ll}
 0 & \frac{1}{\sqrt{2}} \\
 \frac{1}{\sqrt{2}} & 0
\end{array}
\right) \label{gammini}
\end{eqnarray}
These $\gamma$-matrices satisfy the following Clifford algebra:
\begin{equation}\label{cliffordus}
    \left\{ \gamma_x \, , \, \gamma_y \right \} \, = \, \delta^{xy} \, \mathbf{1}_{2 \times 2}
\end{equation}
and $\frac{\rm i}{2} \, \gamma_x$ provide a basis of generators of the $\su(2)$ algebra.
\par
Having fixed these conventions the expression of the quaternionic $\su(2)$-connection in terms of Special Geometry structures is encoded in the following expression for the $2\times 2$-matrix valued $1$-form $\omega$. Explicitly we have:
\begin{equation}\label{omegaSu2}
    \omega \, = \, \left( \begin{array}{cc}
                            -\frac{\rm i}{2} \, \mathcal{Q}  \, - \, \frac{\rm i}{4} \,e^{-U} \, \Phi  & e^{-\frac{U}{2}} \, V^T \, \mathbb{C} \, \mathrm{d}\mathbf{Z}\\
                            - \, e^{-\frac{U}{2}} \, \overline{V}^T \, \mathbb{C} \, \mathrm{d}\mathbf{Z} & \frac{\rm i}{2} \, \mathcal{Q}  \, + \, \frac{\rm i}{4} \,e^{-U} \, \Phi
                          \end{array}
    \right)
\end{equation}
where $V$ and $\overline{V}$ denote the covariantly holomorphic sections of Special geometry defined in eq.s (\ref{covholsec}).
The curvature of this connection is obtained from a straight-forward calculation:
\begin{eqnarray}
\label{K2per2}
  \mathbf{K} &\equiv& d\omega \, + \, \omega \, \wedge \, \omega \nonumber \\
  \null &=& \left(\begin{array}{cc}
                 \mathfrak{u}   & \mathfrak{v} \\
                    - \,\overline{\mathfrak{v}}& -\,\mathfrak{u}
                  \end{array}
   \right)
\end{eqnarray}
the independent $2$-form matrix elements being given by the following explicit formulae:
\begin{eqnarray}
\label{uvvb}
 \mathfrak{u} &=& -{\rm i} \frac{1}{2} \, K \, -\frac{1}{8} dS \,\wedge \, d\bar{S}\, - \, e^{-U} \, V^T \, \mathbb{C} \,\mathrm{d}\mathbf{Z}\, \wedge\, \bar{V}^T \, \mathbb{C} \,\mathrm{d}\mathbf{Z} \, - \, \frac{1}{4} \, e^{-U} \,\mathrm{d}\mathbf{Z}^T \, \wedge \, \mathbb{C} \, \mathrm{d}\mathbf{Z} \nonumber\\
  \mathfrak{v }&=& e^{-\frac{U}{2}} \left( \, DV^T \, \wedge \, \mathbb{C} \, \mathrm{d}\mathbf{Z}\, - \, \frac{1}{2} dS \, \wedge \,
  V^T \, \mathbb{C} \, \mathrm{d}\mathbf{Z}\right)\nonumber\\
  \overline{\mathfrak{v }} &=& e^{-\frac{U}{2}} \left( \, D\overline{V}^T \, \wedge \, \mathbb{C} \, \mathrm{d}\mathbf{Z}\, - \, \frac{1}{2} d\overline{S} \, \wedge \,
  \overline{V}^T \, \mathbb{C} \, \mathrm{d}\mathbf{Z}\right)
\end{eqnarray}
where
\begin{equation}\label{kalleforma}
    K \, = \, \frac{ {\rm i}}{2} \, g_{ij^\star} \, dz^i \, \wedge \, d\bar{z}^{j^\star}
\end{equation}
is the K\"ahler $2$-form of the Special K\"ahler submanifold and where we have used the following short hand notations:
\begin{eqnarray}
  dS &=& dU \, + \, {\rm i} \, e^{-U}\, \left(da \, + \, \mathbf{Z}^T \, \mathbb{C} \, \mathrm{d}\mathbf{Z}\right) \label{firbone1} \\
  d\overline{S} &=& dU \, - \, {\rm i} \, e^{-U}\, \left(da \, + \, \mathbf{Z}^T \, \mathbb{C} \, \mathrm{d}\mathbf{Z}\right) \label{firbone2} \\
  DV &=& dz^i \, \nabla_i V \label{firbone3} \\
  D\overline{V} &=& d\bar{z}^{i^\star} \, \nabla_{i^\star} V \label{firbone4}
\end{eqnarray}
The three HyperK\"ahler forms $K^x$ are easily extracted from eq.s (\ref{K2per2}-\ref{uvvb}) by collecting the coefficients of the $\gamma$-matrix expansion and we need not to write their form which is immediately deduced. The relevant thing is that the components of $K^x$ with an index raised through multiplication with  the inverse of the quaternionic metric $h^{uv}$ exactly satisfy the algebra of quaternionic complex structures (\ref{quatalgebra}). Explicitly we have:
\begin{eqnarray}
  K^x &=& - \, {\rm i} \, 4 \sqrt{2} \, \mbox{Tr} \, \left( \gamma^x \, \mathbf{K}\right) \, \equiv \, K^x_{uv} \, dq^u \,\wedge \, dq^v \nonumber\\
  J^{x|s}_u &=& K^x_{uv}\, h^{vs} \nonumber \\
  J^{x|s}_u \, J^{y|v}_s &=& - \delta^{ xy} \, \delta^v_u \, + \, \epsilon^{xyz} \, J^{z|v}_u \label{quatKvera}
\end{eqnarray}
The above formulae are not only the general proof that the Riemaniann manifold $\mathcal{QM}$ defined by the metric (\ref{geodaction}) is indeed a Quaternionic K\"ahler  manifold, but, what is most relevant, they  also provide an algorithm to write in terms of Special Geometry structures  the tri-holomorphic moment map of the principal isometries possessed by $\mathcal{QM}$.
\subsection{\sc Isometries of $\mathcal{QM}$ in the image of the $c$-map and their tri-holomorphic moment maps}
\label{triholoformul}
Let us now consider the isometries of the metric (\ref{geodaction}). There are three type of  isometries:
\begin{description}
  \item[a)] The isometries of the $(2n+3)$--dimensional Heisenberg algebra $\mathrm{\mathbb{H}eis}$ which is always present and is universal  for any $(4n+4)$--dimensional Quaternionic K\"ahler manifold in the image of the $c$-map. We describe it below.
  \item[b)] All the isometries of the pre-image Special K\"ahler manifold $\mathcal{SK}_n$ that are promoted to isometries of the image manifold in a way described below.
  \item[c)] The additional $2n+4$ isometries that occur only when $\mathcal{SK}_n$ is a symmetric space and such, as a consequence, is also the $c$-map image $\mathcal{QM}_{4n+4}$. We will discuss these isometries
       in section (\ref{omosymmetro}).
\end{description}
For the first two types of isometries a) and b) we are able to write general expressions for the tri-holomorphic moment maps that utilize only the structures of Special Geometry. In the case that the additional isometries c) do exist we have another universal formula which can be used for all generators of the isometry algebra $\mathbb{U}_\mathcal{Q}$ and which relies on the identification of the generators of the $\su(2) \subset \mathbb{H}$ subalgebra with the three complex structures. We will illustrate the details of such an identification while discussing the example of the $S^3$-model.
\par
First of all let us fix the notation writing the general form of a Killing vector. This a tangent vector:
\begin{eqnarray}\label{killingus}
    \vec{\mathbf{k}}& = & k^u (q) \, {\partial}_u \nonumber\\
    &=& k^\diamond \, \frac{\partial}{\partial U} \, + \, k^i \, \frac{\partial}{\partial z^i}\, + \, k^{i^\star} \, \frac{\partial}{\partial \bar{z}^{i^\star}}\, + \, k^\bullet \, \frac{\partial}{\partial a}\, + \, k^\alpha \, \frac{\partial}{\partial \mathbf{Z}^\alpha}\nonumber\\
    & \equiv & k^\diamond \, \partial_\diamond + \, k^i \, \partial_i \, + \, k^{i^\star} \,\partial_{i^\star} + \, k^\bullet \, \partial_\bullet \, + \, k^\alpha \, \partial_\alpha
\end{eqnarray}
with respect to which the Lie derivative of the metric element (\ref{geodaction}) vanishes:
\begin{equation}\label{isoverissima}
    \ell_{\vec{\mathbf{k}}}\, ds^2_{\mathcal{QM}} \, = \,0
\end{equation}
\subsubsection{\sc Tri-holomorphic moment maps for the Heisenberg algebra translations}
First let us consider the isometries associated with the Heisenberg algebra. The  transformation:
\begin{equation}\label{infinoLam}
    Z^\alpha \, \mapsto \, Z^\alpha \, + \, \Lambda^\alpha \quad ; \quad a \, \mapsto \, a \, - \, \Lambda^T \, \mathbb{C} \, \mathbf{Z}
\end{equation}
where $\Lambda^\alpha$ is an arbitrary set of $2n+2$ real infinitesimal parameters is an infinitesimal isometry for the metric $ ds^2_{\mathcal{QM}}$ in (\ref{geodaction}). It corresponds to the following Killing vector:
\begin{eqnarray}\label{KillusW}
    \overrightarrow{\mathbf{k}}_{[\Lambda]} & = & \Lambda^\alpha \,  \overrightarrow{\mathbf{k}}_\alpha \nonumber\\
    & = & \Lambda^\alpha \, \partial_\alpha \, - \, \Lambda^T \, \mathbb{C} \, \mathbf{Z} \, \partial_\bullet
\end{eqnarray}
whose components are immediately deduced by comparison of eq.(\ref{KillusW}) with eq.(\ref{killingus}).
\par
We are interested in determining the expression of the tri-holomorphic moment map  $\mathfrak{P}_{[\Lambda]}$ which satisfies the defining equation:
\begin{eqnarray}
\mathbf{i}_{[\Lambda]}\, \mathbf{K} \, \equiv \,
  \left(\begin{array}{cc}
                 \mathbf{i}_{[\Lambda]}\,\mathfrak{u}   & \mathbf{i}_{[\Lambda]}\,\mathfrak{v} \\
                    - \,\mathbf{i}_{[\Lambda]}\,\overline{\mathfrak{v}}& -\,\mathbf{i}_{[\Lambda]}\,\mathfrak{u}
                  \end{array}
   \right) &=& \mathrm{d} \mathfrak{P}_{[\Lambda]} \, + \, \left[ \omega \, , \, \mathfrak{P}_{[\Lambda]}\right ]
\label{pullusLam}
\end{eqnarray}
The general solution to this problem is
\begin{eqnarray}
\label{triholoHeis}
  \mathfrak{P}_{[\Lambda]}&=&\left(\begin{array}{cc}
                 - \, \frac{\rm i}{4}\, e^{-U}\, \Lambda^T \, \mathbb{C} \, \mathbf{Z}   &\frac{1}{2} \, e^{-\frac{U}{2}} \, \Lambda^T \, C \, V \\
                    - \,\frac{1}{2} \, e^{-\frac{U}{2}} \, \Lambda^T \, C \, \overline{V} & \, \frac{\rm i}{4}\, e^{-U}\, \Lambda^T \, \mathbb{C} \, \mathbf{Z}
                  \end{array}
   \right) \nonumber\\
\end{eqnarray}

\subsubsection{\sc Tri-holomorphic moment map for the Heisenberg algebra central charge}
Consider next the isometry associated with the Heisenberg algebra central charge. The  transformation:
\begin{equation}\label{infinoZeta}
     a \, \mapsto \, a \, + \, \varepsilon
\end{equation}
where $\varepsilon$ is an arbitrary  real small parameter is an infinitesimal isometry for the metric $ ds^2_{\mathcal{QM}}$ in (\ref{geodaction}). It corresponds to the following Killing vector:
\begin{eqnarray}\label{KillusZ}
  \varepsilon \,  \overrightarrow{\mathbf{k}}_{[\bullet]} & = & \varepsilon \, \partial_\bullet
\end{eqnarray}
whose components are immediately deduced by comparison of eq.(\ref{KillusZ}) with eq.(\ref{killingus}).
\par
We are interested in determining the expression of the tri-holomorphic moment map  $\mathfrak{P}_{[\bullet]}$ which satisfies the defining equation analogous to eq.(\ref{pullusLam}):
\begin{eqnarray}
\mathbf{i}_{[\bullet]}\, \mathbf{K}  &=& \mathrm{d} \mathfrak{P}_{[\bullet]} \, + \, \left[ \omega \, , \, \mathfrak{P}_{[\bullet]}\right ]
\label{pullusBull}
\end{eqnarray}
The solution of this problem is even simpler than in the previous case. Explicitly we obtain:
\begin{eqnarray}
\label{triholoHeisZ}
  \mathfrak{P}_{[\bullet]}&=&\left(\begin{array}{cc}
                 - \, \frac{\rm i}{8}\, e^{-U} &0 \\
                   0 & \, \frac{\rm i}{8}\, e^{-U}
                  \end{array}
   \right) \nonumber\\
\end{eqnarray}
The explicit expression of the moment maps and Killing vectors  associated with the Heisenberg isometries was used in the  gauging of abelian subalgebras of the Heisenberg algebra,  which is relevant to the description of  compactifications of
Type II superstring on a generalized Calabi-Yau manifold \cite{heisenberg}.
\subsubsection{\sc Tri-holomorphic moment map for the extension of $\mathcal{SK}_n$ holomorphic isometries}
Next we consider the question how to write the moment map associated with those isometries that where already present in the original
Special K\"ahler manifold $\mathcal{SK}_n$ which we $c$-mapped to a Quaternionic K\"ahler manifold.
\par
Suppose that ${\mathcal{SK}}_n$ has a certain number of holomorphic Killing vectors $k_{\mathbf{I}}^i(z)$ satisfying eq.s (\ref{killo},\ref{killo1},\ref{holkillingvectors}) necessarily closing some Lie algebra $\mathfrak{g}_{\mathcal{SK}}$ among themselves. Their  holomorphic momentum-map is provided by eq.(\ref{sisalvichipuo}).  Necessarily every isometry of a special K\"ahler manifold has a linear symplectic $(2n+2)$-dimensional realization on the holomorphic section $\Omega(z)$ up to an overall holomorphic factor. This means that for each holomorphic Killing vector we have (see Eq. (\ref{sisalvichipuo30})):
\begin{equation}\label{bellacosa}
   k_{\mathbf{I}}^i(z)\, \partial_i \, \Omega(z) \, = \, \exp\left[f_{\mathbf{I}}(z)\right] \, \mathfrak{T}_{\mathbf{I}} \, \Omega(z) \,.
\end{equation}
where $f_{\mathbf{I}}(z)$ the  holomorphic K\"ahler compensator.
Then it can be easily checked that the transformation:
\begin{equation}\label{kQupKil}
    z^i \, \mapsto \, z^i \, + \, k^i_{\mathbf{I}}(z) \quad ; \quad \mathbf{Z} \, \mapsto \, \mathbf{Z} \, + \, \mathfrak{T}_{\mathbf{I}} \, \mathbf{Z}
\end{equation}
is an infinitesimal isometry of the metric (\ref{geodaction}) corresponding to the Killing vector:
\begin{equation}\label{promossoKil}
    \vec{\mathbf{k}}_{\mathbf{I}} \, = \, k^i_{\mathbf{I}}(z) \, \partial_i \, + \, k^{i^\star}_{\mathbf{I}}(\bar{z}) \,\partial_{i^\star} \, + \, \left(\mathfrak{T}_{\mathbf{I}}\right)^\alpha_{\phantom{\alpha}\beta} \, \mathbf{Z}^\beta \, \partial_\alpha
\end{equation}
Also in this case we are interested in determining the expression of the tri-holomorphic moment map  $\mathfrak{P}_{[\mathbf{I}]}$ satisfying the defining equation:
\begin{eqnarray}
\mathbf{i}_{\vec{\mathbf{k}}_{\mathbf{I}}}\, \mathbf{K}  &=& \mathrm{d} \mathfrak{P}_{[\mathbf{I}]} \, + \, \left[ \omega \, , \,
\mathfrak{P}_{[\mathbf{I}]}\right ]
\label{pullusBull2}
\end{eqnarray}
The solution  is given by the expression below:
\begin{eqnarray}
\label{triholoSK}
  \mathfrak{P}_{[\mathbf{I}]}&=&\left(\begin{array}{cc}
             \frac{\rm i}{4} \left(\mathcal{P}_{\mathbf{I}} \, + \, \frac{1}{2} \, e^{-U} \, \mathbf{Z}^T \, \mathbb{C} \, \mathfrak{T}_{\mathbf{I}} \, \mathbf{Z} \right)& - \, \frac{1}{2} \, e^{-U/2} \, V^T \, \mathbb{C} \, \mathfrak{T}_{\mathbf{I}} \, \mathbf{Z} \\
                    \, \frac{1}{2} \,  \, e^{-U/2} \,\overline{V}^T \, \mathbb{C} \, \mathfrak{T}_{\mathbf{I}} \, \mathbf{Z} & \, -  \frac{\rm i}{4} \left(\mathcal{P}_{\mathbf{I}} \, + \, \frac{1}{2} \, e^{-U} \, \mathbf{Z}^T \, \mathbb{C} \, \mathfrak{T}_{\mathbf{I}} \, \mathbf{Z} \right)
                  \end{array}
   \right) \nonumber\\
\end{eqnarray}
where $\mathcal{P}_{\mathbf{I}}$ is the moment map of the same Killing vector in pure Special Geometry.
\subsection{\sc Homogeneous Symmetric Special Quaternionic K\"ahler manifolds}
\label{omosymmetro}
When the Special K\"ahler manifold $\mathcal{SK}_n$ is a symmetric coset space, it turns out that the metric (\ref{geodaction}) is actually the symmetric metric on an enlarged symmetric coset manifold
\begin{equation}\label{qcosetto}
    \mathcal{QM}_{4n+4} \, = \, \frac{\mathrm{U}_Q}{\mathrm{H}_Q} \, \supset \, \frac{\mathrm{U}_{\mathcal{SK}}}{\mathrm{H}_{\mathcal{SK}}}
 \end{equation}
\par
Naming $\Lambda[\mathfrak{g}]$ the $\mathbf{W}$-representation of any finite element of the $\mathfrak{g}\in\mathrm{U}_{\mathcal{SK}}$ group, we have that the matrix $\mathcal{M}_4(z,\bar{z})$ transforms as follows:
\begin{equation}\label{traduco}
    \mathcal{M}_4\left( \mathfrak{g}\cdot z,\mathfrak{g}\cdot \bar{z}  \right)\, = \, \Lambda[\mathfrak{g}] \, \mathcal{M}_4\left(z,\bar{z}\right ) ]\, \Lambda^T[\mathfrak{g}]
\end{equation}
where $\mathfrak{g}\cdot z$ denotes the non linear action of $\mathrm{U}_{\mathcal{SK}}$ on the scalar fields. Since the space $\frac{\mathrm{U}_{\mathcal{SK}}}{\mathrm{H}_{\mathcal{SK}}}$ is homogeneous, choosing any reference point $z_0$ all the others can be reached by a suitable group element $\mathfrak{g}_z$ such that $\mathfrak{g}_z\cdot z_0 \, = \, z$ and we can write:
\begin{equation}\label{turnaconto}
 \mathcal{M}_4^{-1}(z,\bar{z}) \,  = \,  \Lambda^T[\mathfrak{g}_z^{-1}] \, \mathcal{M}_4^{-1}(z_0,\bar{z}_0) ]\, \Lambda[\mathfrak{g}^{-1}_z]
\end{equation}
This allows to introduce a set of $4n+4$ vielbein defined in the following way:
\begin{equation}\label{filibaine}
    E^I_{\mathcal{QM}} \, = \, \frac{1}{2} \, \left\{ dU \, , \, \underbrace{e^i(z)}_{2\,n} \, , \, e^{-U} \,\left(d{a}+{\bf Z}^T\mathbb{C}d{{\bf
Z}}\right) \, , \, \underbrace{e^{-\frac {U} {2}}\, \Lambda[\mathfrak{g}_z^{-1}] \, \mathrm{d}\mathbf{Z}}_{2n+2} \right\}
\end{equation}
and rewrite the metric (\ref{geodaction}) as it follows:
\begin{equation}\label{cornish}
    ds^2_{\mathcal{QM}} \, = \, E^I_{\mathcal{QM}} \, \mathfrak{q}_{IJ} \, E^J_{\mathcal{QM}}
\end{equation}
where the quadratic symmetric constant tensor  $\mathfrak{q}_{IJ}$ has the following form:
\begin{equation}\label{quadrotta}
    \mathfrak{q}_{IJ} \, = \, \left( \begin{array}{c|c|c|c}
                                       1 & 0 & 0 & 0 \\
                                       \hline
                                       0 & \delta_{ij} & 0 & 0 \\
                                       \hline
                                       0 & 0 & 1 & 0 \\
                                       \hline
                                       0 & 0 & 0 & -\, 2\, \mathcal{M}_4^{-1}(z_0,\bar{z}_0)
                                     \end{array}
    \right)
\end{equation}
The above defined vielbein are endowed with a very special property namely they identically satisfy a set of  Maurer Cartan equations:
\begin{equation}\label{MCSolv}
    dE^I_{\mathcal{QM}} \, - \, \frac{1}{2} f^I_{\phantom{I}JK} \, E^J_{\mathcal{QM}} \, \wedge \,  E^K_{\mathcal{QM}} \, = \, 0
\end{equation}
where $f^I_{\phantom{I}JK}$ are the structure constants of a solvable Lie algebra $\mathfrak{A}$ which can be identified as follows:
\begin{equation}\label{solvableGHalg}
    \mathfrak{A} \, = \, Solv\left( \frac{\mathrm{U}_\mathcal{Q}}{\mathrm{H}_\mathcal{Q}} \right)
\end{equation}
In the above equation $Solv\left( \frac{\mathrm{U}_\mathcal{Q}}{\mathrm{H}_\mathcal{\mathcal{Q}}} \right)$ denotes the Lie algebra of the solvable group manifold metrically equivalent to the non-comapact coset manifold $\frac{\mathrm{U}_\mathcal{Q}}{\mathrm{H}_\mathcal{Q}}$ according to a well developed mathematical theory extensively used in supergravity theories\cite{solvableparam}. In the case ${\mathrm{U}_{\mathcal{SK}}}$ is a \textit{maximally split} real form of a complex Lie algebra, then  also  ${\mathrm{U}_{\mathcal{Q}}}$ is maximally split and  we have:
\begin{equation}\label{solvableGHalg2}
    Solv\left( \frac{\mathrm{U}_\mathcal{Q}}{\mathrm{H}_\mathcal{Q}} \right) \, =\, \mbox{Bor}\left ( \mathbb{U}_\mathcal{Q} \right)
\end{equation}
where $\mbox{Bor}\left ( \mathbb{U}_\mathcal{Q} \right)$ denotes the \textit{Borel subalgebra} of the semi-simple Lie algebra $\mathbb{G}$, generated by its Cartan generators and by the step operators associated with all positive roots.
\par
According to the general mathematical theory mentioned above, the very fact that the vielbein (\ref{filibaine}) satisfies the Maurer-Cartan equations of $Solv\left( \frac{\mathrm{U}_\mathcal{Q}}{\mathrm{H}_\mathcal{Q}} \right)$ implies that the metric (\ref{cornish}) is the symmetric metric on the coset manifold $\frac{\mathrm{U}_\mathcal{Q}}{\mathrm{H}_\mathcal{Q}}$ which therefore admits continuous isometries associated with all the generators of the Lie algebra $\mathbb{U}_\mathcal{Q}$. This latter admits the following general decomposition:
\begin{equation}
\mbox{adj}(\mathbb{U}_{\mathcal{Q}}) =
\mbox{adj}(\mathbb{U}_{\mathcal{SK}})\oplus\mbox{adj}(\mathrm{SL(2,\mathbb{R})_E})\oplus
\mathbf{W}_{(2,\mathbf{W})}
\label{gendecompo}
\end{equation}
where $\mathbf{W}$ is the {\bf symplectic} representation of
$\mathbb{U}_{\mathcal{SK}}$ in  which the symplectic section of Special Geometry transforms and which was used to construct the vielbein (\ref{cornish}).  Denoting the generators of
$\mathbb{U}_{\mathcal{SK}}$ by $T^a$, the generators of
$\mathrm{SL(2,\mathbb{R})_E}$ by $\mathrm{L^x}$ and denoting by
$\mathbf{W}^{i\alpha}$ the generators in $\mathbf{W}_{(2,\mathbf{W})}$, the commutation
relations that correspond to the decomposition (\ref{gendecompo})
have the following general form \cite{MarioPietroKsenyaKM}:
\begin{eqnarray}
\nonumber && [T^a,T^b] = f^{ab}_{\phantom{ab}c} \, T^c  \\
\nonumber && [L^x_E,L^y_E] = f^{xy}_{\phantom{xy}z} \, L^z , \\
&&\nonumber [T^a,\mathbf{W}^{i\alpha}] = (\Lambda^a)^\alpha_{\,\,\,\beta} \, \mathbf{W}^{i\beta},
\\ \nonumber && [L^x_E, \mathbf{W}^{i\alpha}] = (\lambda^x)^i_{\,\, j}\, \mathbf{W}^{j\alpha}, \\
&&[\mathbf{W}^{i\alpha},\mathbf{W}^{j\beta}] = \epsilon^{ij}\, (K_a)^{\alpha\beta}\, T^a + \,
\mathbb{C}^{\alpha\beta}\, k_x^{ij}\, L^x_E \label{genGD3pre}
\end{eqnarray}
where the $2 \times 2$ matrices $(\lambda^x)^i_j$, are the canonical generators of $\mathrm{SL(2,\mathbb{R})}$
in the fundamental, defining representation:
\begin{equation}
  \lambda^3 = \left(\begin{array}{cc}
     \ft 12 & 0 \\
     0 & -\ft 12 \
  \end{array} \right) \quad ; \quad \lambda^1 = \left(\begin{array}{cc}
     0 & \ft 12  \\
     \ft 12 & 0\
  \end{array} \right) \quad ; \quad \lambda^2 = \left(\begin{array}{cc}
     0 & \ft 12  \\
     -\ft 12 & 0\
  \end{array} \right)
\label{lambdax}
\end{equation}
while $\Lambda^a$ are the generators
of $\mathbb{U}_{\mathcal{SK}}$ in the symplectic representation $\mathbf{W}$. By
\begin{equation}
  \mathbb{C}^{\alpha\beta} \equiv \left( \begin{array}{c|c}
     \mathbf{0}_{(n+1)\times (n+1)} & \mathbf{1}_{(n+1)\times (n+1)} \\
     \hline
     -\mathbf{1}_{(n+1)\times (n+1)} & \mathbf{0}_{(n+1)\times (n+1)} \
  \end{array}\right)
\label{omegamatra}
\end{equation}
we denote the antisymmetric symplectic metric in $2n+2$ dimensions, $n$
being the complex dimension of the Special K\"ahler manifold $\frac{\mathbb{U}_{\mathcal{SK}}}{\mathbb{H}_{\mathcal{SK}}}$. The symplectic character
of the representation $\mathbf{W}$ is asserted by the identity:
\begin{equation}
  \Lambda^a\, \mathbb{C} + \mathbb{C}\, \left( \Lambda^a \right )^T = 0
\label{Lamsymp}
\end{equation}
The fundamental doublet representation of $\mathrm{SL(2,\mathbb{R})_E}$
is also symplectic and we have denoted by $\epsilon^{ij}= \left( \begin{array}{cc}
  0 & 1 \\
  -1 & 0
\end{array}\right) $ the
$2$-dimensional symplectic metric, so that:
\begin{equation}
    \lambda^x\, \epsilon + \epsilon\, \left( \lambda^x \right )^T = 0,
\label{lamsymp}
\end{equation}
The matrices
$\left(K_a\right)^{\alpha\beta}=\left(K_a\right)^{\beta\alpha}$ and
$\left(k_x\right)^{ij}=\left(k_y\right)^{ji}$ are just symmetric matrices
in one-to-one correspondence with the generators of $\mathbb{U}_{\mathcal{Q}}$ and
$\mathrm{SL(2,\mathbb{R})}$, respectively. Implementing Jacobi
identities we find the following relations:
\begin{eqnarray}
  && \nonumber K_a\Lambda^c +
\Lambda^c K_a = f^{bc}_{\phantom{bc}a}K_b, \quad k_x\lambda^y + \lambda^y k_x
= f^{yz}_{\phantom{yz}x}k_z,
\label{jacobrele}
\end{eqnarray}
which admit the unique solution:
\begin{equation}
    K_a = c_1 \, \mathbf{g}_{ab} \,\Lambda^b\mathbb{C}, \quad ; \quad k_x
= c_2 \, \mathbf{g}_{xy} \, \lambda^y \epsilon
\label{uniquesolutK&k}
\end{equation}
where $\mathbf{g}_{ab}$, $\mathbf{g}_{xy}$ are the Cartan-Killing metrics
on the algebras $\mathbb{U}_{\mathcal{SK}}$ and $\mathrm{SL(2,\mathbb{R})}$, respectively
and  $c_1$ and $c_2$ are two arbitrary constants. These latter
can always be reabsorbed into the normalization of the generators
$\mathbf{W}^{i\alpha}$ and correspondingly set to one. Hence the algebra
(\ref{genGD3pre}) can always be put into the following elegant form:
\begin{eqnarray}
 && [T^a,T^b] = f^{ab}_{\phantom{ab}c} \, T^c  \nonumber\\
&& [L^x,L^y] = f^{xy}_{\phantom{xy}z} \, L^z , \nonumber\\
&&
[T^a,\mathbf{W}^{i\alpha}] = (\Lambda^a)^\alpha_{\,\,\,\beta} \, \mathbf{W}^{i\beta},
\nonumber\\
&& [L^x, \mathbf{W}^{i\alpha}] = (\lambda^x)^i_{\,\, j}\, \mathbf{W}^{j\alpha}, \nonumber \\
&&[\mathbf{W}^{i\alpha},\mathbf{W}^{j\beta}] =
\epsilon^{ij}\, (\Lambda_a)^{\alpha\beta}\, T^a + \, \mathbb{C}^{\alpha\beta}\, \lambda_x^{ij}\, L^x
\label{genGD3}
\end{eqnarray}
where we have used the convention that symplectic indices are raised
and lowered with the symplectic metric, while adjoint representation
indices are raised and lowered with the Cartan-Killing metric.
\par
 For the reader's convenience the list of  Symmetric Special manifolds and of their Quaternionic K\"ahler  counterparts in the image of the c-map is recalled  in table \ref{homomodels} which reproduces the results of \cite{ToineCremmerOld}, according to which there is   a short list of Symmetric Homogeneous Special manifolds  comprising five discrete cases and two infinite series.
\begin{table}
\begin{center}
{\small
\begin{tabular}{||c|c||c||}
  \hline
   $\mathcal{SK}_n$  & $\mathcal{QM}_{4n+4}$ &    $\mbox{dim} \, \mathcal{SK}_n \, = \, $  \\
   Special K\"ahler manifold & Quaternionic K\"ahler manifold & $n$  \\
  \hline
\null & \null &\null \\
 $ \frac{\mathrm{SU(1,1)}}{\mathrm{U(1)}}$ & $ \frac{\mathrm{G_{2(2)}}}{\mathrm{SU(2)\times SU(2)}}$ & $n=1$\\
\null & \null &  \\
\hline
\null & \null &\null \\
  $ \frac{\mathrm{Sp(6,R)}}{\mathrm{SU(3)\times  U(1)}}$ & $ \frac{\mathrm{F_{4(4)}}}{\mathrm{USp(6)\times SU(2)}}$  &$n=6$\\
 \null & \null &  \\
 \null & \null &\null \\
\hline
\null & \null &\null \\
 $ \frac{\mathrm{SU(3,3)}}{\mathrm{SU(3)\times SU(3) \times U(1)}}$ & $ \frac{\mathrm{E_{6(2)}}}{\mathrm{SU(6)\times SU(2)}}$    &$n=9$\\
 \null & \null &  \\
\null & \null &\null \\
\hline
\null & \null &\null \\
 $ \frac{\mathrm{SO^\star(12)}}{\mathrm{SU(6)\times U(1)}}$ & $ \frac{\mathrm{E_{7(-5)}}}{\mathrm{SO(12)\times SU(2)}}$  & $n=15$ \\
\null & \null &  \\
\null & \null &\null \\
\hline
\null & \null &\null \\
$ \frac{\mathrm{E_{7(-25)}}}{\mathrm{E_{6(-78)} \times U(1)}}$ & $ \frac{\mathrm{E_{8(-24)}}}{\mathrm{E_{7(-133)}\times SU(2)}}$    &  $n=27$ \\
\null & \null &  \\
\hline
\null & \null &\null \\
 $ \frac{\mathrm{SL(2,\mathbb{R})}}{\mathrm{SO(2)}}\times\frac{\mathrm{SO(2,2+p)}}{\mathrm{SO(2)\times SO(2+p)}}$ & $ \frac{\mathrm{SO(4,4+p)}}{\mathrm{SO(4)\times SO(4+p)}}$   & $n=3+p$  \\
  \null & \null &  \\
\hline
\null & \null &\null \\
$ \frac{\mathrm{SU(p+1,1)}}{\mathrm{SU(p+1)\times U(1)}}$ & $ \frac{\mathrm{SU(p+2,2)}}{\mathrm{SU(p+2)\times SU(2)}}$    & $n=p+1$ \\
\null & \null &\null \\
\hline
\end{tabular}
}
\caption{List of special K\"ahler symmetric spaces with their Quaternionic K\"ahler  c-map images. The number $n$ denotes the complex dimension  of the Special K\"ahler preimage. On the other hand $4n+4$ is the real dimension of the Quaternionic K\"ahler c-map image. \label{homomodels}}
\end{center}
\end{table}
\par
Inspecting eq.s (\ref{genGD3}) we immediately realize that the Lie Algebra $\mathbb{U}_{Q}$ contains two universal Heisenberg subalgebras of dimension $(2n+3)$, namely:
\begin{eqnarray}
\mathbb{U}_{\mathcal{Q}} \, \supset \, \mathbb{H}\mathrm{eis}_1 \, &=& \mbox{span}_\mathbb{R}\, \left\{\mathbf{W}^{1\alpha} \, , \, \mathbb{Z}_1 \right\} \quad ; \quad  \mathbb{Z}_1  \, =\, L_+  \, \equiv \, L^1\, + \, L^2 \nonumber\\
  \null &\null & \left[\mathbf{W}^{1\alpha}\, , \, \mathbf{\mathbf{W}}^{1\beta} \right]\, = \, -\, \frac{1}{2} \, \mathbb{C}^{\alpha\beta} \, \mathbb{Z}_1  \quad ; \quad \left[\mathbb{Z}_1\, , \, \mathbf{W}^{1\beta} \right]\, = \,0 \label{Heinber1}\\
  \mathbb{U}_{\mathcal{Q}} \, \supset \, \mathbb{H}\mathrm{eis}_2 \, &=& \mbox{span}_\mathbb{R}\, \left\{\mathbf{W}^{2\alpha} \, , \, \mathbb{Z}_2 \right\} \quad ; \quad  \mathbb{Z}_2  \, =\, L_-  \, \equiv \, L^1\, - \, L^2 \nonumber\\
  \null &\null & \left[\mathbf{W}^{2\alpha}\, , \, \mathbf{W}^{2\beta} \right]\, = \, -\, \frac{1}{2} \, \mathbb{C}^{\alpha\beta} \, \mathbb{Z}_2  \quad ; \quad \left[\mathbb{Z}_2\, , \, \mathbf{W}^{2\beta} \right]\, = \,0 \label{Heinber2}
\end{eqnarray}
The first of these Heisenberg subalgebras of isometries is the universal one that exists for all Quaternionic K\"ahler manifolds $\mathcal{QM}_{4n+4}$ lying in the image of the $c$-map, irrespectively  whether the pre-image Special K\"ahler manifold $\mathcal{SK}_n$ is a symmetric space or not.  The tri-holomorphic moment map of these isometries was presented in eq.s(\ref{triholoHeis}) and (\ref{triholoHeisZ}). The second Heisenberg algebra exists only in the case when the Quaternionic K\"ahler manifold $\mathcal{QM}_{4n+4}$ is a symmetric space.
\par
From this discussion we also realize that the central charge $\mathbb{Z}_1$ is just the $L_+$ generator of a universal $\slal(2,R)_E$ Lie algebra that exists only in the symmetric space case and which was named the Ehlers algebra in the context of dimensional reduction analysis from $D=4$ to $D=3$ \cite{PietroSashaMarioBH1}.  When $\slal(2,R)_E$ does exist we can introduce the universal compact generator:
\begin{equation}\label{ruotogrande}
    \mathfrak{S}\, \equiv\, L_+ \, - \, L_- \,= \, 2 \, \lambda^2
\end{equation}
which rotates the two sets of Heisenberg translations one into the other:
\begin{equation}\label{tabarro}
    \left[ \mathfrak{S}\, , \, \mathbf{W}^{i\alpha}\right ] \, = \, \epsilon^{ij} \, \mathbf{W}^{j\alpha}
\end{equation}
As we shall see, the gauging of this generator is a rather essential ingredient in the inclusion of one-field cosmological models into gauged $\mathcal{N}=2$ supergravity.
\paragraph{The embedding tensor formulation of the gauging.} It is useful to encode the choice of the gauge algebra in an \emph{embedding tensor} $\theta_{\Lambda}{}^{\mathcal{A}}$ \cite{Cordaro:1998tx,Nicolai:2001sv,deWit:2002vt,deWit:2005ub} though which the gauge generators are expressed in terms of the global symmetry ones. If we denote by $\{t_{\mathcal{A}}\}\equiv \{T^a,\,L^x,\,{\bf W}^{i\,\alpha}\}$ the generators of $\mathbb{U}_{\mathcal{Q}}$ and by $X_{\Lambda}$
the gauge generators, since we are gauging only (abelian) isometries of the quaternionic manifold,  we can write:
\begin{equation}
X_{\Lambda}=\theta_{\Lambda}{}^{\mathcal{A}}\,t_{\mathcal{A}}\,,
\end{equation}
where the index $\Lambda$ runs over all the vector fields. Of these only a subset will actually gauge the chosen isometries. This subset will be labelled by boldface latin indices ${\bf I},\,{\bf J},\dots$.
The only condition on this tensor originates from the structure of the algebra:
\begin{equation}
[X_{\Lambda},\,X_{\Sigma}]=0\,\,\Rightarrow\,\,\,\,\,\theta_{\Lambda}{}^{\mathcal{A}}\theta_{\Lambda}{}^{\mathcal{B}}\,f_{\mathcal{A}\mathcal{B}}{}^\mathcal{C}=0\,,
\end{equation}
where $f_{\mathcal{A}\mathcal{B}}{}^\mathcal{C}$ are the structure constants of $\mathbb{U}_{\mathcal{Q}}$. The triholomorphic moment maps $\mathcal{P}^x_{\Lambda}$ and the Killing vectors $k_\Lambda^u$ can then be expressed in the following way:
\begin{equation}
\mathcal{P}^x_{\Lambda}=\theta_{\Lambda}{}^{\mathcal{A}}\,\mathcal{P}^x_{\mathcal{A}}\,\,;\,\,\,\,k_\Lambda^u=\theta_{\Lambda}{}^{\mathcal{A}}\,k_{\mathcal{A}}^u\,,
\end{equation}
where $\mathcal{P}^x_{\mathcal{A}},\,k_{\mathcal{A}}^u$ are the \emph{intrinsic} moment maps and Killing vectors associated with the quaternionic isometries.
In order to make the analysis independent of the initial symplectic frame of the vector multiplet sector, it is useful to describe  the gauge algebra generators by the (redundant) symplectic  notation $X_M=(X_\Lambda,\,X^\Lambda)=\theta_{M}{}^{\mathcal{A}}\,t_{\mathcal{A}}$, see \cite{deWit:2005ub}. The tensor $\theta_{M}{}^{\mathcal{A}}$ should then satisfy the locality constraint:
\begin{equation}
\theta_{\Lambda}{}^{\mathcal{A}}\theta^{\Lambda\,\mathcal{B}}-\theta_{\Lambda}{}^{\mathcal{B}}\theta^{\Lambda\,\mathcal{A}}=0\,,
\end{equation}
which guarantees that the tensor can be rotated, by means of a symplectic transformation, to an \emph{electric frame} in which $\theta^{\Lambda\,\mathcal{A}}=0$. For the restricted kind of gauging that we shall be dealing with, by extending the arguments given in \cite{thesearch}, we can work in the electric frame to start with, with no loss of generality.

\subsubsection{\sc The tri-holomorphic moment map in homogeneous symmetric Quaternionic K\"ahler manifolds}
\label{MezhduAlgGeom}
In the case the Quaternionic K\"ahler manifold $\mathcal{QM}_{4n+4}$  is a homogeneous symmetric space $\frac{\mathrm{U}_\mathcal{Q}}{\mathrm{H}_\mathcal{Q}}$, the tri-holomorphic moment map associated with any generator of $\mathfrak{t} \, \in \, \mathbb{U}_\mathcal{Q}$ of the isometry Lie algebra can be easily constructed by means of the formula:
\begin{equation}\label{generMapformula}
    \mathcal{P}_\mathfrak{t}^x \, = \, \mbox{Tr}_{[{\mathbf{fun}}]}\, \left( J^x \, \mathbb{L}_{Solv}^{-1} \,  \mathfrak{t} \, \mathbb{L}_{Solv} \right)
\end{equation}
where:
\begin{description}
  \item[a)] $J^x$ are the three generators of the $\su(2)$ factor in the isotropy subalgebra $\mathbb{H}\, = \, \su(2) \, \oplus \, \mathbb{H}^\prime$, satisfying the quaternionic algebra (\ref{quatKvera}). They should  be normalized in such a way  as to realize the following condition. Naming:
      \begin{equation}\label{MaurAmiCartan}
        \Xi \, = \, \mathbb{L}_{Solv}^{-1}(q) \, \mathrm{d} \mathbb{L}_{Solv}(q)
      \end{equation}
the Maurer Cartan differential one-form its projection on  $J^x$ should precisely yield the $\su(2)$ one-form defined in eq. (\ref{omegaSu2}):
\begin{equation}\label{omegacorrusco}
    \omega \, = \, - \, \frac{{\rm i}}{ \sqrt{2} N_f } \, \sum_{x=1}^3 \,\mbox{Tr}_{[\mathbf{fun}]}\, \left( J^x \, \Xi\right) \, \gamma_x\, = \, \left( \begin{array}{cc}
                            -\frac{\rm i}{2} \, \mathcal{Q}  \, - \, \frac{\rm i}{4} \,e^{-U} \, \Phi  & e^{-\frac{U}{2}} \, V^T \, \mathbb{C} \, \mathrm{d}\mathbf{Z}\\
                            - \, e^{-\frac{U}{2}} \, \overline{V}^T \, \mathbb{C} \, \mathrm{d}\mathbf{Z} & \frac{\rm i}{2} \, \mathcal{Q}  \, + \, \frac{\rm i}{4} \,e^{-U} \, \Phi
                          \end{array}
    \right)
\end{equation}
In the above equation, which provides the precise link between the $c$-map description and the coset manifold description of the same geometry, $N_f \, = \, \mbox{dim}\,\mathbf{fun}$ denotes the dimension of the fundamental representation of $\mathbb{U}_\mathcal{Q}$.
\item[b)] The solvable coset representative $\mathbb{L}_{Solv}(q)$ is obtained by exponentiation of the Solvable Lie algebra:
\begin{equation}\label{solvoexpo}
    \mathbb{L}_{Solv}(q) \, \simeq \, \exp \left[ q \, \cdot \, Solv\left( \frac{\mathrm{U}_\mathcal{Q}}{\mathcal{H}_\mathcal{Q}}\right) \right]
\end{equation}
but the detailed exponentiation rule has to be determined in such a way that projecting the same Maurer Cartan form (\ref{MaurAmiCartan}) along an appropriate basis of generators $T_{I|Solv}$ of the solvable Lie algebra $Solv\left( \frac{\mathrm{U}_\mathcal{Q}}{\mathcal{H}_\mathcal{Q}}\right)$ we precisely obtain the vielbein $E^I_{QM}$ defined in eq.(\ref{filibaine}). This summarized in the following general equations:
\begin{eqnarray}\label{sopore}
    E^I_{\mathcal{QM}} & = & \mbox{Tr}_{\mathbf{[fun]}} \left( T^I_{Solv} \, \Xi \right) \nonumber\\
    \delta^{I}_{J} & = & \mbox{Tr}_{\mathbf{[fun]}} \left( T^I_{Solv} \, T_{I|Solv} \right) \nonumber\\
    \Xi  & = & E^I_{\mathcal{QM}} \, T_{I|Solv}
\end{eqnarray}
\end{description}
In eq.(\ref{sopore}) by $T^I_{Solv}$ we have denoted the conjugate (with respect to the trace) of the solvable Lie algebra generators.
\par
A general comment is in order. The precise calibration of the basis of the solvable generators $T^I_{Solv}$ and of their exponentiation outlined in eq.(\ref{solvoexpo}) which allows the identification (\ref{sopore}) is a necessary and quite laborious task in order to establish the bridge between the general $c$-map description of the quaternionic geometry and its actual realization in each symmetric coset model. This is also an unavoidable step in order to give a precise meaning to the very handy formula (\ref{generMapformula}) for the tri-holomorphic map. It should also be noted that although (\ref{generMapformula}) covers all the cases, the result of such a purely algebraic  calculation is difficult to be guessed a priori. Hence educated guesses on the choice of generators whose gauging produces a priori determined features are difficult to be inferred from (\ref{generMapformula}). The analytic structure of the tri-holomorphic moment map instead is much clearer in the $c$-map framework of formulae (\ref{triholoHeis},\ref{triholoHeisZ},\ref{triholoSK}). The use of  both languages and the construction of the precise bridge between them in each model is therefore an essential ingredient to understand the nature and the properties of candidate gaugings in whatever physical application.
\newpage
\part{\sc Abelian Gaugings  and General Properties of  their Potentials in the $c$-map Framework}
\label{grandiscussia}
As we stressed in the introduction the inclusion into $\mathcal{N}=2$ supergravity obtained in \cite{thesearch} of inflaton potentials such as  the Starobinsky potential\footnote{Just as in \cite{thesearch} we mention scalar fields that typically have non canonical kinetic terms}
\begin{equation}\label{starotto}
    V_{Starobinsky}(\phi) \, \equiv \, \left(1 \, - \,  \exp\left[-\phi\right]\right)^2
\end{equation}
is not occasional and limited to the case of hypermultiplets lying in  $\frac{\mathrm{G_{(2,2)}}}{\mathrm{SU(2)} \times \mathrm{SU(2)}}$, rather it follows a general pattern that can be uncovered and relies on the properties of the $c$-map. In this way the mechanisms of the \cite{thesearch} can be generalized to
larger Quaternionic K\"ahler manifolds opening a quite interesting new playground for the search of inflaton potentials that can be classified and understood in their geometrical origin.
\par
Let us schematically summarize the main ingredients of the approach pioneered in \cite{thesearch} whose generalization we pursue in this paper:
\begin{description}
  \item[A)] The inflaton field $\phi$ is assumed to belong  to the hypermultiplet Quaternionic K\"ahler manifold $\mathcal{QM}$.
    \item[B)] In analogy with the construction in \cite{thesearch}, we require  the graviphoton not to be minimally coupled to any other field. This condition originally followed from the general argument that in  the  dual to the $R+R^2$ supergravity the central charge is  gauged. This will amount to a constraint on the form of the embedding tensor $\theta$ defining the gauge algebra.
  \item[C)] The inflaton potential is generated by the gauging of an abelian subalgebra $\mathcal{A} \subset \mbox{iso} \left[\mathcal{QM}\right]$ of the isometry algebra of the hypermultipet manifold.
  \item[D)] Since $\mathcal{A}$ is abelian it is not required to have any action on the vector multiplet scalars $\omega^i$ which are inert. Actually it is quite desirable that the potential $V_{gauging}$  generated by the gauging allows to fix all the $\omega^i$ to their values at some reference point, say $\omega^i \, = \, 0$:
      \begin{equation}\label{fixing}
       \left. \frac{\partial}{\partial \omega^i} \, V_{gauging} \right|_{\omega^i \, = \, 0} \, = \, 0
      \end{equation}
      As shown in \cite{thesearch}, one can generically guarantee the fixing conditions (\ref{fixing}) if the Special K\"ahler Geometry of the vector multiplets is chosen to be that of the so named Minimal Coupling, defined below in eq.s (\ref{minicup1}-\ref{minicup3}).
   \item[E)] With the above choice of the vector multiplet geometry, after fixing the scalars $\omega^i$ the effective potential reduces to a sum of squares of the tri-holomorphic moment maps $P_{\mathcal{A}}^x$ which still depend on the variables $\left\{Z,U,a,z^i,{\bar z}^{i^\star}\right\}$. In order to approach effective potentials recognizable also as $\mathcal{N}=1$ supergravity potentials one would like to be able to fix all the Heisenberg fields $\mathbf{Z}$ (and  possibly also the other fields $U$ and $a$) to zero, remaining only with the complex fields $\left ( z^i,{\bar z}^{i^\star}\right)$ of the inner Special K\"ahler manifold. Looking at the general form (\ref{triholoHeis}) of the tri-holomorphic moment map for the Heisenberg algebra generators and (\ref{triholoSK}) for the tri-holomorphic moment map of the inner Special K\"ahler isometries we immediately realize that, gauging these  isometries \emph{separately},  the condition:
       \begin{equation}\label{secondfixing}
       \left. \frac{\partial}{\partial \mathbf{Z}^\alpha} \sum_{\mathfrak{t} \,\in \, \mathcal{A}} \, \left(\mathcal{P}_{\mathfrak{t}}\right)^2 \right|_{\mathbf{Z}\, = \, 0} \, = \, 0
      \end{equation}
     is always satisfied. A gauge generator which is a combination of a translation in the Heisenberg algebra and a Special K\"ahler isometry, yields in general a scalar potential exhibiting linear terms in ${\bf Z}$, so that (\ref{secondfixing}) provides a non-trivial constraint.\par
      The definition of the locus $\mathcal{L}$ involves setting to zero a certain number of fields $\phi^r$ belonging to $\mathcal{SK}_n$ so that we should also realize the consistency condition:
      \begin{equation}\label{thirdfixing}
       \left. \frac{\partial}{\partial \phi^r} \sum_{\mathfrak{t} \,\in \, \mathcal{A}} \, \left(\mathcal{P}_{\mathfrak{t}}\right)^2 \right|_{\begin{array}{ccc}\mathbf{Z}& = & 0\\
       \phi^r & = & 0 \\
       \end{array}} \, = \, 0\,.
      \end{equation}
      As mentioned earlier, the gauging yielding Starobinsky-like potentials need also involve the  compact generator $\mathfrak{S}$. As we shall show in the following, if the gauged isometry is a combination of $\mathfrak{S}$ and an $\mathcal{SK}_n$ isometry, (\ref{secondfixing}) poses no  constraint on the gauging.
  \item[F)] A favorite, though  not mandatory, choice corresponds to looking for abelian generators of $\mbox{iso}\left[\mathcal{SK}_n\right]$ such that the locus which satisfies conditions (\ref{thirdfixing}) is defined by setting to zero all the axions $p_r$, namely all the fields associated with nilpotent generators of the solvable Lie algebra of $\mathcal{SK}_n$. The inclusion of the Starobinsky potential in supergravity was obtained in \cite{thesearch} precisely in this way. In section \ref{generalonuovo} we show a generalization of the same mechanism in the case of a bigger manifold $\mathcal{QM}_{4n+4}$, obtaining what can be denominated a multi Starobinsky model.
  \item[G)] \textbf{The $U$-problem}. If we use only the type of isometries yielding the tri-holomorphic moment maps (\ref{triholoHeis}), (\ref{triholoHeisZ}) and (\ref{triholoSK}) we face a serious problem with the fields $U$. It appears only through exponentials all of the same sign ($\exp[ - 2 \, U]$ or $\exp[ -  \, U]$ in front of perfect squares. Hence the field $U$ cannot be stabilized unless all such squares are zero which means no residual potential. To overcome such a problem one should have moment maps with the opposite sign of $U$ in the exponential and this can happen only by introducing in the gauging either $L^E_-$ or generators $\mathbf{W}^{2,\alpha}$ this means that such generators should exist, namely the manifold $\mathcal{QM}_{4n+4}$ should be a symmetric space. In \cite{thesearch} the $U$-problem was solved by adding to a parabolic generator of a $\mathcal{SK}_n$-isometry the universal compact generator (\ref{ruotogrande}). As we have emphasized the Ehlers subalgebra exists in all symmetric spaces and so does the compact generator (\ref{ruotogrande}). This implies that the mechanism leading to the inclusion of the Starobinsky model found in \cite{thesearch} is actually rather universal and can be generalized in several ways.
\end{description}
The above discussion provides a framework for the search of other inflaton potentials.
\section{\sc Minimal Coupling Special Geometry}
\label{minicoup}
In this section we shortly describe the structure of the Minimal Coupling Special K\"ahler manifold $\mathcal{MSK}_{p+1}$, mostly in order to fix our conventions and to establish our notations. As announced in the introduction, this kind of Special Geometry is our favorite choice for the vector multiplet sector of the $\mathcal{N}=2$ lagrangian which allows us to construct an entire class of theories where the vector multiplet scalars can be stabilized and the effective potential of an abelian gauging is reduced only to the hypermultiplet sector. In view of such a use of  $\mathcal{MSK}_{p+1}$, all items of its Special Geometry will be  denoted with a hat, and its complex coordinates will be named $\omega_i$ rather than $z^i$. However it is clear that  $\mathcal{MSK}_{p+1}$ might also be used as $c$-map preimage of a Quaternionic K\"ahler manifold describing hypermultiplets.
\par
As a manifold $\mathcal{MSK}_{p+1}$ is the following coset:
\begin{equation}\label{minicup1}
    \mathcal{MSK}_{p+1} \, = \, \frac{\mathrm{SU(1,p+1)}}{\mathrm{U(1)} \times \mathrm{SU(p+1)}}
\end{equation}
In terms of the complex coordinates $\omega^i$ a convenient choice of the $(2 \,p \,+\, 4)$-dimensional holomorphic symplect section is the following one:
\begin{equation}\label{minicup2}
    \widehat{\Omega} \, = \, \left( \begin{array}{c}
                                      \widehat{X}^\Lambda \\
                                      \hline
                                      \widehat{F}_\Sigma
                                    \end{array}
    \right) \, = \,\left( \begin{array}{c}
                                      1 \\
                                      \omega^i\\
                                      \hline
                                      -{\rm i}\\
                                      {\rm i}\, \omega^i
                                    \end{array}
    \right) \quad ; \quad (i\,= \, 1, \dots \, p+1 )
\end{equation}
which leads to the following K\"ahler potential:
\begin{equation}\label{minicup3}
    \widehat{\mathcal{K}} \, = \, - \, \log \,\left[ - \, {\rm i} \widehat{\Omega} \, \widehat{\mathbb{C}} \, \widehat{{\overline{\Omega}}}\right ] \, = \, - \, \log \, \left[\,2 \,\left( 1 \, - \,  \omega\, \cdot \,\bar{\omega}\right)\right]
\end{equation}
and to the following K\"ahler metric:
\begin{equation}\label{minicup4}
    \widehat{g}_{ij^\star} \, = \, \partial_i\,\partial_{j^\star} \,\widehat{\mathcal{K}}\, = \, \frac{1}{\left(1-\omega\cdot\ \bar{\omega} \right)^2}  \, \left( \delta^{ij}\, \left(1-\omega\cdot\ \bar{\omega} \right)\, + \, \bar{\omega}^i \, \omega^j\right)
\end{equation}
Defining the K\"ahler covariant derivatives of the covariantly holomorphic sections as in eq.s (\ref{uvector}) we obtain three results that are very important for the discussion of reduced scalar potentials in the present paper.
Firstly we get:
\begin{equation}\label{minicup4barra}
\nabla_i \, \widehat{U}_j \, \equiv \,     \nabla_i \, \nabla_j \widehat{V} \, = \,0
\end{equation}
which compared with eq.(\ref{defaltern}) implies the vanishing of the three-index symmetric tensor $\widehat{C}_{ijk}$. This unique property of the special K\"ahler manifold $\mathcal{MSK}_{p+1}$ defined by eq.(\ref{minicup1}) is the reason why it has been named the Minimal Coupling Special Geometry, the interpretation of the tensor $C_{ijk}$ in phenomenological applications being that of Yukawa couplings of the gauginos. In ref. \cite{thesearch} it was shown that the vanishing of $\widehat{C}_{ijk}$ guarantees the consistency (see eq. (3.10) of the quoted reference) of the truncation of the classical supergravity theory to the hypermultiplet quaternionic scalars by fixing the vector multiplet scalars to the origin of their manifold:
\begin{equation}\label{golubchika}
    \omega^i \, = \, 0
\end{equation}
Secondly we evaluate the the covariantly symplectic holomorphic section in the origin of the manifold and we obtain:
\begin{equation}\label{miniculpan1}
   \left. \widehat{V} \right|_{\omega\, =\,0} \, = \,\frac{1}{\sqrt{2}}\,\left\{\begin{array}{c|c||c|c}
                                                              1 & 0 & -\,{\rm i} & 0
                                                            \end{array}
   \right\}
\end{equation}
In the same point we have:
\begin{equation}\label{fantamini}
\left.\left(\widehat{g}^{ij^\star}\,\nabla_i \widehat{V}^\alpha \,\nabla_{j^\star} \widehat{\overline{V}}\right)\right|_{\omega\, =\,0} \, = \,  \frac{1}{2} \,   \left(\begin{array}{c|c||c|c}
            0 & 0 & 0 & 0 \\
            \hline
            0 & \mathbf{1}_{(p+1)\times(p+1)} & 0 & {\rm i} \,\mathbf{1}_{(p+1)\times(p+1)} \\
            \hline
            \hline
            0 & 0 & 0 & 0 \\
            \hline
            0 & - \,{\rm i} \,\mathbf{1}_{(p+1)\times(p+1)} & 0 & \mathbf{1}_{(p+1)\times(p+1)}
          \end{array}
    \right)
\end{equation}
\subsection{\sc Gauging abelian isometries of the hypermultiplets}
Relying on these results we see that if the hypermultiplet Quaternionic manifold $\mathcal{QM}_{4m}$ possesses a $p+1$-dimensional abelian Lie algebra of isometries, we can always gauge them by using, for the vector multiplets, the Special K\"ahler manifold
$\mathcal{MSK}_{p+1}$ introducing also the following embedding tensor:
\begin{equation}\label{gioisco}
    \theta_{M}^{I} \, \equiv \, \left\{\theta_{\Lambda}^I\, , \, \theta^{\Sigma|I }\right\} \, = \, \left \{\theta_0{}^{I} \, = \,0 \, , \, \theta_J{}^{I} \, = \, \delta^I_J \, , \, \theta^{\Sigma|I }\, = \, 0\right\}\,.
\end{equation}
Notice that the choice of setting $\theta_0{}^{I} \, = \,0$ follows from the requirement $B)$ that the graviphoton should not be gauged. This indeed amounts to
requiring:
\begin{equation}
 \left. \widehat{V} \right|_{\omega\, =\,0}^M\,\theta_M{}^{\mathcal{A}}=0\,\,\,\Rightarrow\,\,\,\,\,\,\theta_0{}^{I} \, = \,0\,.
\end{equation}
In such a theory the scalar potential has the following general form:
\begin{equation}\label{cornettoalgida}
    \mathcal{V}_{scalar}(\omega, \bar{\omega},q) \, = \, 4 \, k_I^u k_J^v \, h_{uv} \, \widehat{V}^I \, \widehat{\overline{V}}{}^J
    \, + \,
    \left(\widehat{g}^{ij^\star}\,\nabla_i \widehat{V}^I \,\nabla_{j^\star} \widehat{\overline{V}}{}^J \, - \, 3 \,
    \widehat{V}^I \,\widehat{\overline{V}}{}^J \right) \, \mathcal{P}^x_I  \, \mathcal{P}^x_J
\end{equation}
setting $\omega^i \, = \, 0$ is a consistent truncation and the reduced potential takes the following universal general form
which is positive definite by construction:
\begin{equation}\label{gartolini}
   \mathcal{V}_{scalar}(0, 0,q) \, = \, \sum_{I=1}^{p+1} \,\mathcal{P}^x_I(q)  \, \mathcal{P}^x_I(q)
\end{equation}
In the next section \ref{starobin1} we reconsider the derivation of the Starobinsky potential obtained in \cite{thesearch} from a parabolic gauging as a master example that can be generalized  to bigger manifolds.
\subsection{\sc The Starobinsky potential}
\label{starobin1}
Last year a great deal of activity was devoted to the inclusion of phenomenologically interesting inflaton potentials into $\mathcal{N}=1$ supergravity as we  recalled in the introduction. A first wave of investigations considered the possible generation of potentials by means of suitably chosen superpotentials, subsequently, after an important new viewpoint was introduced in \cite{minimalsergioKLP} and was  subsequently developed in
\cite{primosashapietro},\cite{piesashatwo},\cite{Ferrara:2013wka},\cite{Ferrara:2013kca},
\cite{pietrosergiosasha1},\cite{pietrosergiosasha2}, it became clear that positive definite inflaton potentials can be generated by the gauging of some isometry of the K\"ahler manifold of scalar multiplets. Such potentials have the form of squares of  K\"ahler moment maps. In \cite{pietrosergiosasha1} this mechanism was applied to the case of constant curvature one-dimensional K\"ahler manifolds and it was shown that  Starobinsky-like potentials \cite{Starobinsky:1980te} emerge from the moment map of a parabolic isometry in $\mathrm{SL(2,\mathbb{R})} \simeq \mathrm{SU(1,1)}$ with the addition of a Fayet Iliopoulos term. In particular the standard Starobinsky model that  is dual to an $R+R^2$ supergravity emerges from gauging the parabolic shift isometry of an $\frac{\mathrm{SU(1,1)}}{\mathrm{U(1)}}$ manifold with K\"ahler potential $\mathcal{K}\, = \, - 3\, \log (z-{\bar z})$ which is precisely the Special K\"ahler manifold $S^3$. Let us now consider eq.(\ref{triholoSK}) and we can learn an important lesson. If in the $c$-map image of some $\mathcal{SK}$ Special K\"ahler manifold, for instance the  $S^3$ model, we gauge, according to the scheme discussed in section \ref{grandiscussia}, some nilpotent Lie algebra element $\mathfrak{N}_+\, \in \, \mathbb{U}_{\mathcal{SK}} \, \subset \mathbb{U}_\mathcal{Q}$ identical with the parabolic shift generator that we would have gauged in  $\mathcal{N}=1$ supergravity, (for instance the generator $L_+\, \in \, \slal(2,\mathbb{R})$ in the case of the $S^3$ model), we obtain a moment map that contains precisely the $\mathcal{P}_\mathbf{I}$ of the $\mathcal{N}=1$ case, modified by $\mathbf{Z}$ dependent terms. In case the $\mathbf{Z}$ can be stabilized to zero the remaining effective potential is that of the corresponding $\mathcal{N}=1$ theory, apart from the Fayet Iliopoulos term. There are two remaining problems. The generation of a Fayet Iliopoulos term and the stabilization of the $U$ field. They are solved in one stroke by modifying the parabolic generator of the inner Special K\"ahler isometry with the addition of the universal Ehlers rotation (\ref{ruotogrande}).
\par
Let us see how this works.
\par
With reference to eq.s (\ref{generillini}) let us consider the following generator:
\begin{equation}\label{costianovo}
    \mathfrak{p}\, = \, \mathfrak{N}_+ \, + \, \kappa \, \mathfrak{S}
\end{equation}
where $\mathfrak{N}_+$ is the previously mentioned  nilpotent element of the Special K\"ahler subalgebra ($\mathfrak{N}_+^r \, = \, 0$, for some positive integer $r$) and $\kappa$ is a parameter.  Let us then calculate the tri-holomorphic moment map $\mathcal{P}^x_\mathfrak{p}$ according to formula (\ref{fisterone}).
\par
Because of the linearity of the momentum map in  Lie algebra elements we have:
\begin{eqnarray}
  \mathfrak{P}_\mathfrak{p} &=& \mathfrak{P}_{\mathfrak{N}_+} \, + \, \mathfrak{P}_{\mathfrak{S}}\nonumber \\
  \mathfrak{P}_{\mathfrak{N}_+} &=& \left( \begin{array}{c|c}
                                             \frac{\rm i}{4} \, \mathcal{P}_{\mathfrak{N}_+} \, + \, \mathcal{O} \, \left( \mathbf{Z}^2\right)& \mathcal{O} \, \left( \mathbf{Z}\right) \\
                                             \hline
                                             \mathcal{O} \, \left( \mathbf{Z}\right) & \, - \, \frac{\rm i}{4} \, \mathcal{P}_{\mathfrak{N}_+}\, - \, \mathcal{O} \, \left( \mathbf{Z}^2\right)
                                           \end{array}
  \right) \nonumber\\
 \mathfrak{P}_{\mathfrak{S}} &=& \left( \begin{array}{c|c}
                                             \frac{\rm i}{8} \, e^{-U} \,\left(1  +  a^2 \, + \, e^{2U}\right)\,  + \, \mathcal{O} \, \left( \mathbf{Z}^2\right)& \mathcal{O} \, \left( \mathbf{Z}\right) \\
                                             \hline
                                             \mathcal{O} \, \left( \mathbf{Z}\right) & \, - \, \frac{\rm i}{8} \, e^{-U} \,\left(1  + a^2 \, + \, e^{2U}\right)\,  - \, \mathcal{O} \, \left( \mathbf{Z}^2\right)
                                           \end{array}
  \right) \nonumber\\
  \label{gomorratano}
\end{eqnarray}
where $\mathcal{P}_{\mathfrak{N}_+}$ is the K\"ahlerian moment map of the Killing vector associated with the generator $\mathfrak{N}_+$ as defined in eq.(\ref{sisalvichipuo}).
It is evident by the above  completely universal formulae that the potential:
\begin{equation}\label{potentus}
    V_{gauging} \, = \, \mbox{const} \, \mbox{Tr} \left[ \, \mathfrak{P}_\mathfrak{p}\, \cdot \, \mathfrak{P}_\mathfrak{p} \right]
\end{equation}
possesses the following universal property:
\begin{equation}\label{Zetaseneva}
  \left.  \frac{\partial}{\partial \mathbf{Z}^\alpha} \, V_{gauging}\right|_{\mathbf{Z}=0} \, = \,0
\end{equation}
allowing for a consistent truncation of the Heisenberg fields. After such truncation we find:
\begin{equation}\label{Vzero}
    V_{eff}(U,a, z,\bar z) \, =\, \left.V_{gauging}\right|_{\mathbf{Z}=0} \, = \, \mbox{const} \, \times \, \left[ \mathcal{P}_{\mathfrak{N}_+} \, + \, \frac{\kappa}{2} \, e^{-U} \left(1+a^2+e^{2U}\right)\right]^2
\end{equation}
From equation (\ref{Vzero}) we further learn that we can consistently truncate the fields $a$ and $U$ setting them to zero since
\begin{equation}\label{ciurlonelmanicozzo}
    \left.  \frac{\partial}{\partial U} \, V_{eff}\right|_{U=a=0} \, = \,0 \quad ; \quad \left.  \frac{\partial}{\partial a} \, V_{eff}\right|_{U=a=0} \, = \,0
\end{equation}
We find:
\begin{equation}\label{curiosone}
  V_{infl}(z,\bar z) \, \equiv \,  V_{eff}(0,0, z,\bar z) \, = \, \left( \, \mathcal{P}_{\mathfrak{N}_+} \, + \, \kappa \right)^2
\end{equation}
which clearly shows how the universal generator $\mathfrak{S}$ provides, after stabilization of the $U$ field, the mechanism that generates the Fayet Iliopoulos term \cite{Fayet:1974jb} essential for inflation.

\newpage
\part{\sc Examples}
As an illustration of the general patterns and mechanisms described in the previous pages we consider two examples of Quaternionic K\"ahler manifolds $\mathcal{QM}_{4n+4}$ obtained from the $c$-map of two homogeneous symmetric Special K\"ahler manifolds $\mathcal{SK}_n$.
\begin{enumerate}
  \item The manifold $\frac{\mathrm{G_{(2,2)}}}{\mathrm{SU(2)} \times \mathrm{SU(2)}}$ which is the $c$-map image of the Special K\"ahler manifold $\frac{\mathrm{SU(1,1)}}{\mathrm{U(1)}}$ with cubic embedding of $\mathrm{SU(1,1)}$ in $\mathrm{Sp(4,\mathbb{R})}$. In this case $n=1$ and the corresponding coupling of one vector multiplet to supergravity is usually named the $S^3$ model in the literature.
  \item The manifold $\frac{\mathrm{F_{(4,4)}}}{\mathrm{SU(2)} \times \mathrm{USp(6)}}$ which is the $c$-map image of the Special K\"ahler manifold $\frac{\mathrm{Sp(6,\mathbb{R})}}{\mathrm{SU(3) \times U(1)}}$ . In this case $n=6$.
\end{enumerate}
For these two models we provide a full fledged construction of all the geometrical items and in particular we realize the bridge between the algebraic description and the analytic one advocated at the end of section \ref{MezhduAlgGeom}. This allows us to discuss a couple of examples of gaugings. In particular in the case of the of the first model which we utilized as a calibration device for our general formulae we retrieve the inclusion of the Starobinsky model first demonstrated in \cite{thesearch}.
\par
The detailed construction of the second model, which we plan to utilize  in future publications for an extensive and possibly exhaustive analysis  of gaugings in larger hypermultiplet spaces, is utilized in the present paper to provide an example of  generalization of the results of \cite{thesearch} by means of the inclusion of a multi Starobinsky model.
\section{\sc The $S^3$ model and its quaternionic image $\frac{\mathrm{G_{(2,2)}}}{\mathrm{SU(2)} \times \mathrm{SU(2)}}$}
 In this which is the simplest example $n=1$, namely the Special K\"ahler manifold has complex dimension $1$ and it can be identified with the time honored Poincar\'e Lobachevsky plane:
\begin{equation}\label{tripini}
    \mathcal{SK}_{1} \, = \, \frac{\mathrm{SU(1,1)}}{\mathrm{U(1)}}
\end{equation}
\subsection{\sc The special K\"ahler structure of $S^3$}
The corresponding K\"ahler potential is:
\begin{equation}\label{kelero1}
    \mathcal{K} \, = \, - \, \log \, \left[ (z \, - \, \bar{z})^3\right ]
\end{equation}
which leads to the K\"ahler metric:
\begin{equation}\label{kelero2}
    g_{z\bar{z}} \, = \, \frac{3}{4} \, \frac{1}{(z \, - \, \bar{z})^2}
\end{equation}
Setting:
\begin{equation}\label{realcordo}
    z \, = \, {\rm i} \, \exp[h] \, + \, y
\end{equation}
we get:
\begin{equation}\label{metrullone}
     g_{z\bar{z}} \, = \, \frac{3}{2} \, \left(dh^2 \, + \, \exp[-2h] \, dy^2\right)
\end{equation}
In the notations of \cite{PietroSashaMarioBH1} the holomorphic symplectic section governing this special geometry is given
by the following four component vector:
\begin{equation}\label{seziona}
    \Omega \, = \,\left\{-\sqrt{3}z^2,z^3,\sqrt{3} z,1\right\}
\end{equation}
In this case the $\mathbf{W}$-representation is the spin $j\, = \, \frac{3}{2}$ of the $\mathrm{SL(2,\mathbb{R})} \sim \mathrm{SU(1,1)}$ group that happens to be
four dimensional symplectic:
\begin{equation}
\label{frilli}
\mathrm{SL(2,\mathbb{R})} \, \ni \,\left(\begin{array}{ll}
 a & b \\
 c & d
\end{array} \right) \, \Longrightarrow \, \left(
\begin{array}{llll}
 d a^2+2 b c a & -\sqrt{3} a^2
   c & -c b^2-2 a d b &
   -\sqrt{3} b^2 d \\
 -\sqrt{3} a^2 b & a^3 &
   \sqrt{3} a b^2 & b^3 \\
 -b c^2-2 a d c & \sqrt{3} a
   c^2 & a d^2+2 b c d &
   \sqrt{3} b d^2 \\
 -\sqrt{3} c^2 d & c^3 &
   \sqrt{3} c d^2 & d^3
\end{array}
\right) \, \in \, \mathrm{Sp(4,\mathbb{R})}
\end{equation}
the preserved symplectic metric being the following one:
\begin{equation}\label{goriaci}
 \mathbb{C} \, = \,   \left(
\begin{array}{llll}
 0 & 0 & 1 & 0 \\
 0 & 0 & 0 & 1 \\
 -1 & 0 & 0 & 0 \\
 0 & -1 & 0 & 0
\end{array}
\right)
\end{equation}
According to the general rule the K\"ahler potential (\ref{kelero1}) is retrieved by setting:
\begin{equation}\label{fagano}
    \mathcal{K}(z,{\bar z}) \, = \, - \log \left[ - {\rm i} \Omega \, \mathbb{C} \, \
    \overline{\Omega} \right ]
\end{equation}
\subsection{\sc The matrix  $\mathcal{M}_4^{-1}$ and the $c$-map}
For the $S^3$ model  the matrix $\mathcal{M}_4$ and its inverse have the following  explicit appearance:
\begin{equation}\label{m4diretto}
 \mathcal{M}_4 \, = \,   \left(
\begin{array}{llll}
 \frac{4 i z {\bar z} \left(z^2+4 {\bar z}
   z+{\bar z}^2\right)}{(z-{\bar z})^3} & -\frac{4 i \sqrt{3} z^2
   {\bar z}^2 (z+{\bar z})}{(z-{\bar z})^3} & -\frac{i
   (z+{\bar z}) \left(z^2+10 {\bar z}
   z+{\bar z}^2\right)}{(z-{\bar z})^3} & -\frac{2 i \sqrt{3}
   (z+{\bar z})^2}{(z-{\bar z})^3} \\
 -\frac{4 i \sqrt{3} z^2 {\bar z}^2 (z+{\bar z})}{(z-{\bar z})^3} &
   \frac{8 i z^3 {\bar z}^3}{(z-{\bar z})^3} & \frac{2 i \sqrt{3} z
   {\bar z} (z+{\bar z})^2}{(z-{\bar z})^3} & \frac{i
   (z+{\bar z})^3}{(z-{\bar z})^3} \\
 -\frac{i (z+{\bar z}) \left(z^2+10 {\bar z}
   z+{\bar z}^2\right)}{(z-{\bar z})^3} & \frac{2 i \sqrt{3} z
   {\bar z} (z+{\bar z})^2}{(z-{\bar z})^3} & \frac{4 i \left(z^2+4
   {\bar z} z+{\bar z}^2\right)}{(z-{\bar z})^3} & \frac{4 i
   \sqrt{3} (z+{\bar z})}{(z-{\bar z})^3} \\
 -\frac{2 i \sqrt{3} (z+{\bar z})^2}{(z-{\bar z})^3} & \frac{i
   (z+{\bar z})^3}{(z-{\bar z})^3} & \frac{4 i \sqrt{3}
   (z+{\bar z})}{(z-{\bar z})^3} & \frac{8 i}{(z-{\bar z})^3}
\end{array}
\right)
\end{equation}
its inverse being:
\begin{equation}\label{m4inverso}
    \mathcal{M}_4^{-1} \, = \,\left(
\begin{array}{llll}
 \frac{4 i \left(z^2+4 {\bar z}
   z+{\bar z}^2\right)}{(z-{\bar z})^3} & \frac{4 i \sqrt{3}
   (z+{\bar z})}{(z-{\bar z})^3} & \frac{i \left(z^3+11 {\bar z}
   z^2+11 {\bar z}^2 z+{\bar z}^3\right)}{(z-{\bar z})^3} &
   -\frac{2 i \sqrt{3} z {\bar z} (z+{\bar z})^2}{(z-{\bar z})^3}
   \\
 \frac{4 i \sqrt{3} (z+{\bar z})}{(z-{\bar z})^3} & \frac{8
   i}{(z-{\bar z})^3} & \frac{2 i \sqrt{3}
   (z+{\bar z})^2}{(z-{\bar z})^3} & -\frac{i
   (z+{\bar z})^3}{(z-{\bar z})^3} \\
 \frac{i \left(z^3+11 {\bar z} z^2+11 {\bar z}^2
   z+{\bar z}^3\right)}{(z-{\bar z})^3} & \frac{2 i \sqrt{3}
   (z+{\bar z})^2}{(z-{\bar z})^3} & \frac{4 i z {\bar z}
   \left(z^2+4 {\bar z} z+{\bar z}^2\right)}{(z-{\bar z})^3} &
   -\frac{4 i \sqrt{3} z^2 {\bar z}^2 (z+{\bar z})}{(z-{\bar z})^3}
   \\
 -\frac{2 i \sqrt{3} z {\bar z} (z+{\bar z})^2}{(z-{\bar z})^3} &
   -\frac{i (z+{\bar z})^3}{(z-{\bar z})^3} & -\frac{4 i \sqrt{3}
   z^2 {\bar z}^2 (z+{\bar z})}{(z-{\bar z})^3} & \frac{8 i z^3
   {\bar z}^3}{(z-{\bar z})^3}
\end{array}
\right)
\end{equation}
Furthermore, in this case a convenient reference point is given by $z_0 \, = \, {\rm i}$ that can be mapped into any point of the upper complex plane by means of the element:
\begin{equation}\label{trasluco}
    \mathfrak{g}_z \, = \, \  \left(
\begin{array}{ll}
 e^{h/2} & e^{-h/2} y \\
 0 & e^{-h/2}
\end{array}
\right) \, \in \, \mathrm{SL(2,\mathbb{R})}
\end{equation}
acting by means of fractional linear transformations. The explicit form of the $\Lambda(\mathfrak{g})$ matrix in the $\mathbf{W}$-representation was given in eq.(\ref{frilli}). This provides us with all the necessary information in order to write down the explicit form of the $E^I_{\mathcal{QM}} $ vielbein for the $S^3$ case.
\subsection{\sc The vielbein and the  borellian Maurer Cartan equations}
They are the following ones:
\begin{equation}\label{EBHfilo}
    E^I_{\mathcal{QM}} \, = \, \frac{1}{2} \left(
\begin{array}{l}
 \mathrm{dU} \\
 \sqrt{3} \mathrm{dh} \\
 \sqrt{3} \mathrm{dy} e^{-h} \\
 e^{-U} \left(\mathrm{da}+\mathrm{dZ}_3 Z_1+\mathrm{dZ}_4 Z_2-\mathrm{dZ}_1 Z_3-\mathrm{dZ}_2
   Z_4\right) \\
 \sqrt{2} e^{-\frac{h}{2}-\frac{U}{2}}
   \left(\mathrm{dZ}_1+y \left(2
   \mathrm{dZ}_3-\sqrt{3} y
   \mathrm{dZ}_4\right)\right) \\
 \sqrt{2} e^{-\frac{3 h}{2}-\frac{U}{2}}
   \left(\left(\sqrt{3} \mathrm{dZ}_3-y
   \mathrm{dZ}_4\right) y^2+\sqrt{3} \mathrm{dZ}_1
   y+\mathrm{dZ}_2\right)\\
 \sqrt{2} e^{\frac{h-U}{2}}
   \left(\mathrm{dZ}_3-\sqrt{3} y
   \mathrm{dZ}_4\right) \\
 \sqrt{2} e^{\frac{3 h}{2}-\frac{U}{2}}
   \mathrm{dZ}_4
\end{array}
\right)
\end{equation}
Furthermore we find $\mathcal{M}_4^{-1}({\rm i},-{\rm i}) \, = \, - \, \mathbf{1}_{4\times4}$ so that the quadratic form (\ref{quadrotta}) is just:
\begin{equation}\label{doremito}
    \mathfrak{q}_{AB} \, = \, \mbox{diag} \,\left( 1,1,1,1,1,1,1,1\right)
\end{equation}
The next step consists of calculating the geometry of the space described by the above vielbein and flat metric (\ref{doremito}). To this effect we have first to calculate the contorsion, namely the exterior derivatives of the vielbein and then using such a result the spin connection $\omega^{IJ}$, finally the curvature two-form from which we extract the Riemann and the Ricci tensor.
\par
Addressing the first step, namely the contorsion, we have the first important surprise. The exterior derivatives of the vielbein are expressed in terms of  wedge-quadratic products of the same vielbein with constant numerical coefficients. This means that the above constructed vielbein satisfy a set of Maurer Cartan equations describing a Lie algebra, namely\footnote{Note that here, for simplicity we have dropped the suffix $\mathcal{SK}$. This is done for simplicity since there is no risk of confusion.}:
\begin{equation}\label{glorioso}
    dE^I \, - \, \frac{1}{2} \, f_{JK}^{\phantom{JK}I} \, E^I \, \wedge \, E^J \, = \, 0
\end{equation}
the tensor $f_{BC}^{\phantom{BC}A}$ being the structure constants of such a Lie algebra. Explicitly for the $S^3$ model we get:
\begin{equation}\label{primocartano}
\begin{array}{l}
 0 \, = \, dE^{1} \\
 0 \, = \, dE^{2} \\
 0 \, = \, dE^{3}\, + \, 2\,\frac{E^{2}\wedge
   E^{3}}{\sqrt{3}} \\
 0 \, = \, dE^{4}\, +\, 2 \, E^{1}\wedge
   E^{4}\, -\, 2 \, E^{5}\wedge
   E^{7}\, -\, 2 \, E^{6}\wedge E^{8}
   \\
 0 \, = \, dE^{5}+ E^{1}\wedge
   E^{5}\, +\,  \frac{E^{2}\wedge
   E^{5}}{ \sqrt{3}}-\frac{4 \,
   E^{3}\wedge E^{7}}{\sqrt{3}} \\
 0 \, = \, dE^{6}+ E^{1}\wedge
   E^{6}+ \sqrt{3}
   E^{2}\wedge
   E^{6}\, -\, 2\, E^{3}\wedge E^{5}
   \\
 0 \, = \, dE^{7}+ E^{1}\wedge
   E^{7}-\frac{E^{2}\wedge
   E^{7}}{\sqrt{3}}\, +\, 2 \, E^{3}\wedge
   E^{8} \\
 0 \, = \, dE^{8}+E^{1}\wedge
   E^{8}- \sqrt{3}
   E^{2}\wedge E^{8}
\end{array}
\end{equation}
Hence it arises the following  question: which Lie algebra is described by such Maurer Cartan equations?
Utilizing the standard method of diagonalizing  the adjoint action of the two commuting generators $H_{1,2}$ dual to $E^{1,2}$ we find that the eigenvalues are just the positive roots of $\mathfrak{g}_{2,2}$:
 \begin{equation}
\label{g2rootsystem}
\begin{array}{rclcrcl}
\alpha_1&=&(1,0)&;&\alpha_2&=&\frac{\sqrt{3}}{2}\,(-\sqrt{3},1)\\
\alpha_3 \, =\, \alpha_1+\alpha_2&=&\frac{1}{2}\,(-1,\sqrt{3}) &;&
\alpha_4 \, = \, 2\,\alpha_1+\alpha_2 &=&
\frac{1}{2}\,(1,\sqrt{3}) \\
\alpha_5 \, = \, 3\,\alpha_1+\alpha_2&=&\frac{\sqrt{3}}{2}\,(\sqrt{3},1)&;&\alpha_6 \, = \,  3\,\alpha_1+2\,\alpha_2 &=&
(0,\sqrt{3})\
\end{array}
\end{equation}
As it is well known the complex Lie algebra $\mathfrak{g}_2(\mathbb{C})$ has rank two and it is defined by the $2\times 2$ Cartan matrix encoded in the following Dynkin diagram:
\begin{center}
\begin{picture}(110,30)
\put (-60,20){$\mathfrak{g}_2$}
\put (10,23){\circle {10}}
\put (15,25){\line (1,0){20}}
\put (15,23){\line (1,0){20}}
\put (20.5,19){{\LARGE$>$}}
\put (15,20.5){\line (1,0){20}}
\put (40,23){\circle {10}}
\put (65,21){$=\quad\quad\left (\begin{array}{cc}
  2 & -3\\
  -1 & 2
\end{array} \right)$}
\end{picture}
\end{center}
The real form $\mathfrak{g}_{2,2}$ is the maximally split form of the above complex Lie algebra.
With a little bit of more work we can put eq.s(\ref{primocartano}) into the standard Cartan Weyl form for the Borel subalgebra of $\mathfrak{g}_{2,2}$, composed by the Cartan generators and by all the positive root step operators. Naming $T_J$ the generators dual to the vielbein $E^I$ such that $E^I(T_J) \, = \delta^I_J$, we find that the appropriate identifications are the following ones:
\begin{equation}\label{interpretaziag22}
    \begin{array}{l}
 T_2 \, = \, 2 \, \frac{\mathcal{H}_1}{\sqrt{3}} \\
 T_1 \, = \, 2 \, \frac{\mathcal{H}_2}{\sqrt{3}} \\
 T_3 \, = \, 2 \,  E^{\alpha  _1} \\
 T_4 \, = \, 2 \,  E^{\alpha  _6} \\
 T_8 \, = \, 2 \,   E^{\alpha  _2}  \\
 T_7 \, = \, 2 \,   E^{\alpha  _3 } \\
 T_5 \, = \, 2 \, E^{\alpha  _4} \\
 T_6 \, = \, 2 \,  E^{\alpha  _5}
\end{array}
\end{equation}
We conclude that the manifold on which the metric (\ref{geodaction}) is constructed is homeomorphic to the solvable group-manifold $\mbox{Bor}(\mathfrak{g}_{2,2})$.
\subsection{\sc The spin connection}
Next, calculating  the Levi-Civita spin connection  from its definition, namely the vanishing torsion condition:
\begin{equation}\label{surcallo}
    0 \, = \, dE^I \, + \, \omega^{IJ} \, \wedge \, E^J
\end{equation}
we find the following result:
\begin{equation}\label{ristolone}
  \omega^{IJ} \, = \,   \left(
\begin{array}{llllllll}
 0 & 0 & 0 & E^{4} &
   \frac{E^{5}}{2} &
   \frac{E^{6}}{2} &
   \frac{E^{7}}{2} &
   \frac{E^{8}}{2} \\
 0 & 0 & \frac{E^{3}}{\sqrt{3}} & 0 &
   \frac{E^{5}}{2 \sqrt{3}} & \frac{1}{2}
   \sqrt{3} E^{6} & -\frac{E^{7}}{2
   \sqrt{3}} & -\frac{1}{2} \sqrt{3} E^{8}
   \\
 0 & -\frac{E^{3}}{\sqrt{3}} & 0 & 0 &
   -\frac{E^{6}}{2}-\frac{E^{7}}{\sqrt{3}} & -\frac{E^{5}}{2} &
   \frac{E^{8}}{2}-\frac{E^{5}}{\sqrt{3}} & \frac{E^{7}}{2} \\
 -E^{4} & 0 & 0 & 0 &
   \frac{E^{7}}{2} &
   \frac{E^{8}}{2} &
   -\frac{E^{5}}{2} &
   -\frac{E^{6}}{2} \\
 -\frac{E^{5}}{2} & -\frac{E^{5}}{2
   \sqrt{3}} &
   \frac{E^{6}}{2}+\frac{E^{7}}{\sqrt{3}} & -\frac{E^{7}}{2} & 0 &
   \frac{E^{3}}{2} &
   -\frac{E^{3}}{\sqrt{3}}-\frac{E^4}{2} & 0 \\
 -\frac{E^{6}}{2} & -\frac{1}{2} \sqrt{3} \,
   E^{6} & \frac{E^{5}}{2} &
   -\frac{E^{8}}{2} &
   -\frac{E^{3}}{2} & 0 & 0 &
   -\frac{E^{4}}{2} \\
 -\frac{E^{7}}{2} & \frac{E^{7}}{2
   \sqrt{3}} &
   \frac{E^{5}}{\sqrt{3}}-\frac{E^{
   8}}{2} & \frac{E^{5}}{2} &
   \frac{E^{3}}{\sqrt{3}}+\frac{E^{
   4}}{2} & 0 & 0 & \frac{E^{3}}{2} \\
 -\frac{E^{8}}{2} & \frac{1}{2} \sqrt{3}
   E^{8} & -\frac{E^{7}}{2} &
   \frac{E^{6}}{2} & 0 &
   \frac{E^{4}}{2} &
   -\frac{E^{3}}{2} & 0
\end{array}
\right)
\end{equation}
which can be decomposed in the way we now describe.
\subsection{\sc Holonomy algebra and decompostion of the spin connection}
Let us introduce   two triplets  $J^x_{[I]}$ and $J^x_{[II]}$ of  $8 \times 8$ matrices that can be read off explicitly as the coefficients of $\alpha_x$ and $\beta_x$  in the following linear combinations:
\begin{eqnarray}\label{gringo1}
&\sum_{x=1}^3 \, \alpha_x \, J^x_{[I]}  \, = \, & \nonumber\\
  &  \left(
\begin{array}{llllllll}
 0 & 0 & 0 & -\frac{\alpha _1}{2} & -\frac{1}{4}
   \sqrt{3} \alpha _3 & -\frac{\alpha _2}{4} &
   \frac{\sqrt{3} \alpha _2}{4} & -\frac{\alpha
   _3}{4} \\
 0 & 0 & \frac{\alpha _1}{2} & 0 & -\frac{\alpha
   _3}{4} & -\frac{1}{4} \sqrt{3} \alpha _2 &
   -\frac{\alpha _2}{4} & \frac{\sqrt{3} \alpha
   _3}{4} \\
 0 & -\frac{\alpha _1}{2} & 0 & 0 & -\frac{\alpha
   _2}{4} & \frac{\sqrt{3} \alpha _3}{4} &
   \frac{\alpha _3}{4} & \frac{\sqrt{3} \alpha
   _2}{4} \\
 \frac{\alpha _1}{2} & 0 & 0 & 0 & \frac{\sqrt{3}
   \alpha _2}{4} & -\frac{\alpha _3}{4} &
   \frac{\sqrt{3} \alpha _3}{4} & \frac{\alpha
   _2}{4} \\
 \frac{\sqrt{3} \alpha _3}{4} & \frac{\alpha
   _3}{4} & \frac{\alpha _2}{4} & -\frac{1}{4}
   \sqrt{3} \alpha _2 & 0 & \frac{\sqrt{3} \alpha
   _1}{4} & -\frac{\alpha _1}{4} & 0 \\
 \frac{\alpha _2}{4} & \frac{\sqrt{3} \alpha
   _2}{4} & -\frac{1}{4} \sqrt{3} \alpha _3 &
   \frac{\alpha _3}{4} & -\frac{1}{4} \sqrt{3}
   \alpha _1 & 0 & 0 & \frac{\alpha _1}{4} \\
 -\frac{1}{4} \sqrt{3} \alpha _2 & \frac{\alpha
   _2}{4} & -\frac{\alpha _3}{4} & -\frac{1}{4}
   \sqrt{3} \alpha _3 & \frac{\alpha _1}{4} & 0 &
   0 & \frac{\sqrt{3} \alpha _1}{4} \\
 \frac{\alpha _3}{4} & -\frac{1}{4} \sqrt{3}
   \alpha _3 & -\frac{1}{4} \sqrt{3} \alpha _2 &
   -\frac{\alpha _2}{4} & 0 & -\frac{\alpha
   _1}{4} & -\frac{1}{4} \sqrt{3} \alpha _1 & 0
\end{array}
\right) & \nonumber\\
\end{eqnarray}
\begin{eqnarray}\label{gringo2}
&\sum_{x=1}^3 \, \beta_x \, J^x_{[II]}  \, = \, & \nonumber\\
  &  \left(
\begin{array}{llllllll}
 0 & 0 & 0 & \frac{3 \beta _1}{2} & -\frac{1}{4}
   \sqrt{3} \beta _3 & -\frac{3 \beta _2}{4} &
   -\frac{1}{4} \sqrt{3} \beta _2 & \frac{3 \beta
   _3}{4} \\
 0 & 0 & \frac{\beta _1}{2} & 0 & -\frac{\beta
   _3}{4} & -\frac{3}{4} \sqrt{3} \beta _2 &
   \frac{\beta _2}{4} & -\frac{3}{4} \sqrt{3}
   \beta _3 \\
 0 & -\frac{\beta _1}{2} & 0 & 0 & \frac{5 \beta
   _2}{4} & \frac{\sqrt{3} \beta _3}{4} & \frac{5
   \beta _3}{4} & -\frac{1}{4} \sqrt{3} \beta _2
   \\
 -\frac{3 \beta _1}{2} & 0 & 0 & 0 & -\frac{1}{4}
   \sqrt{3} \beta _2 & \frac{3 \beta _3}{4} &
   \frac{\sqrt{3} \beta _3}{4} & \frac{3 \beta
   _2}{4} \\
 \frac{\sqrt{3} \beta _3}{4} & \frac{\beta _3}{4}
   & -\frac{5 \beta _2}{4} & \frac{\sqrt{3} \beta
   _2}{4} & 0 & \frac{\sqrt{3} \beta _1}{4} &
   -\frac{5 \beta _1}{4} & 0 \\
 \frac{3 \beta _2}{4} & \frac{3 \sqrt{3} \beta
   _2}{4} & -\frac{1}{4} \sqrt{3} \beta _3 &
   -\frac{3 \beta _3}{4} & -\frac{1}{4} \sqrt{3}
   \beta _1 & 0 & 0 & -\frac{3 \beta _1}{4} \\
 \frac{\sqrt{3} \beta _2}{4} & -\frac{\beta
   _2}{4} & -\frac{5 \beta _3}{4} & -\frac{1}{4}
   \sqrt{3} \beta _3 & \frac{5 \beta _1}{4} & 0 &
   0 & \frac{\sqrt{3} \beta _1}{4} \\
 -\frac{3 \beta _3}{4} & \frac{3 \sqrt{3} \beta
   _3}{4} & \frac{\sqrt{3} \beta _2}{4} &
   -\frac{3 \beta _2}{4} & 0 & \frac{3 \beta
   _1}{4} & -\frac{1}{4} \sqrt{3} \beta _1 & 0
\end{array}
\right) & \nonumber\\
\end{eqnarray}
Both triplets form an 8-dimensional representation of the $\su(2)$ Lie algebra and the two triplets commute with each other:
\begin{eqnarray}
\left[ \,J^x_{[I]} \, , \, J^y_{[I]} \right ]  &=& \epsilon^{xyz} \, J^y_{[I]} \nonumber\\
  \left[ \,J^x_{[II]} \, , \, J^y_{[II]} \right ]  &=& \epsilon^{xyz} \, J^y_{[II]} \nonumber\\
  \left[ \,J^x_{[I]} \, , \, J^y_{[II]} \right ]  &=&0 \label{guglielmotell}
\end{eqnarray}
Furthermore all matrices are antisymmetric so that the two Lie algebras $\su_{\mathrm{I}}(2)$ and $\su_{\mathrm{II}}(2)$ are both subalgebras of  $\so(8)$.
The distinction between these two representations becomes clear when we calculate the Casimir operator for both of them. We obtain:
\begin{equation}\label{gaglioffo}
    \sum_{x=1}^3 \, J^x_{[I]}\cdot J^x_{[I]} \,  = \, - \, \frac{3}{4} \,  \mathbf{1} \quad ; \quad \sum_{x=1}^3 \, J^x_{[II]}\cdot J^x_{[II]} \, = \, - \, \frac{15}{4} \,  \mathbf{1}
\end{equation}
Hence the first $\su_{\mathrm{I}}(2)$ Lie algebra in realized on the considered eight--dimensional space in the $j=\frac{1}{2}$ representation, while the second
$\su_{\mathrm{II}}(2)$ Lie algebra in realized on the same space in the $j=\frac{3}{2}$. In other words, with respect to both subalgebras of $\so(8)$, the fundamental representation decomposes as follows:
\begin{equation}\label{cirimello}
    \mathbf{8} \, \stackrel{\su_{\mathrm{I}}(2) \oplus \su_{\mathrm{II}}(2) \subset \so(8)}{\Longrightarrow} \, (\mathbf{2},\mathbf{4})
\end{equation}
By direct calculation we verify that the spin connection displayed in equation (\ref{ristolone}) has the following structure:
\begin{equation}\label{decompospincon}
    \omega \, = \, \omega^{[I]}_x \, J^x_{[I]} \, \oplus \, \omega^{[II]}_x \, J^x_{[II]}
\end{equation}
where:
\begin{equation}\label{cucurucu}
    \omega^{[I]}_x \, = \, \left(
\begin{array}{l}
  \sqrt{3} \mathrm{E}^3- \mathrm{E}^4 \\
  \sqrt{3} \mathrm{E}^7- \mathrm{E}^6 \\
 - \mathrm{E}^8- \sqrt{3} \mathrm{E}^5
\end{array}
\right)  \quad ; \quad \omega^{[II]}_x \, = \, \left(
\begin{array}{l}
 \frac{\mathrm{E}^4}{2}+\frac{\mathrm{E}^3}{2
   \sqrt{3}} \\
 -\frac{\mathrm{E}^7}{2
   \sqrt{3}}-\frac{\mathrm{E}^6}{2} \\
 \frac{\mathrm{E}^8}{2}-\frac{\mathrm{E}^5}{2
   \sqrt{3}}
\end{array}
\right)
\end{equation}
This structure clearly demonstrates the reduced holonomy of the Quaternionic K\"ahler manifold. Indeed, according to eq.(\ref{cirimello}) the vielbein transforms in the doublet of  $\su_{\mathrm{I}}(2)$ tensored with the fundamental representation of $\sym(4,\mathbb{R})$. In the present case the symplectic $4 \times 4$ matrices are actually reduced to the subalgebra $\su_{\mathrm{II}}(2) \subset \sym(4,\mathbb{R})$ with respect to which the fundamental of $\sym(4,\mathbb{R})$ remains irreducible and coincides with the $j=\frac{3}{2}$ representation of $\su_{\mathrm{II}}(2)$. The above discussion can be summarized by the statement:
\begin{equation}\label{ollonomio}
\so(8) \, \subset \, \su(2) \, \oplus \, \usp(4) \, \subset \, \mathrm{Hol}  \, = \, \su_{\mathrm{I}}(2) \, \oplus \, \su_{\mathrm{II}}(2)
\end{equation}
by definition the holonomy algebra being the Lie algebra in which the Levi-Civita spin connection takes values.
\subsection{\sc Structure of the isotropy subalgebra $\mathbb{H}$}
It remains  to single out the structure of the denominator subalgebra ${\mathbb{H}} \, \subset \, \mathbb{U} \, \equiv \, {\mathfrak{g}}_{2,2}$ in the orthogonal decomposition:
 \begin{equation}\label{ortogonallodecompo}
    \mathbb{U} \, = \, \mathbb{H} \, \oplus \, \mathbb{K} \quad ; \quad \left \{ \begin{array}{ccc}
                                                                                   \left[ \mathbb{H} \, , \, \mathbb{H} \right] & \subset & \mathbb{H}  \\
                                                                                   \left[ \mathbb{H} \, , \, \mathbb{K} \right] & \subset & \mathbb{K} \\
                                                                                   \left[ \mathbb{K} \, , \, \mathbb{K} \right] & \subset & \mathbb{H} \\
                                                                                 \end{array}\right.
 \end{equation}
 Since our quaternionic K\"ahler manifold is a symmetric space it follows that Lie algebra $\mathbb{H}$ must be isomorphic with the holonomy algebra $\mathrm{Hol} \, = \,\su_{\mathrm{I}}(2) \oplus \su_{\mathrm{II}}(2)$ that we have calculated in the previous subsection.
 By definition the Lie algebra $\mathbb{H}$ is the maximal compact subalgebra which for maximal split algebras has a universal definition in terms of the step operators associated with the positive roots $E^\alpha$ and their conjugates $E^{-\alpha}$. In the case of $\mathfrak{g}_{2,2}$ which has six positive roots we can write:
 \begin{equation}\label{Hfavola}
\mathbb{H} \, \equiv \,\mbox{span}_{\mathbb{R}}  \left \{ E^{\alpha  _1} -  E^{\alpha  _1}\,  , \, E^{\alpha  _2} -  E^{\alpha  _2} \, , \, E^{\alpha  _3} -  E^{\alpha  _3} \, , \, E^{\alpha  _4} -  E^{\alpha  _4}\, , \, E^{\alpha  _5} -  E^{\alpha  _5} \, , \,  E^{\alpha  _6} -  E^{\alpha  _6} \right \}
 \end{equation}
The structure of (\ref{Hfavola}) is the following:
 \begin{equation}\label{fattucco}
    \mathbb{H} \, = \, \su_{\mathrm{I}}(2) \, \oplus \, \su_{\mathrm{II}}(2)
 \end{equation}
 where the generators of the two subalgebras are:
 \begin{equation}\label{friccouno}
    j^x_{[I]} \, = \,\left(
\begin{array}{l}
 \frac{-3 \mathrm{E}^{-\alpha_1}+3
   \mathrm{E}^{\alpha_1}+\sqrt{3}
   \left(\mathrm{E}^{\alpha_6}-\mathrm{E}^{-\alpha_6}\right)}{6 \sqrt{2}} \\
 \frac{3 \mathrm{E}^{-\alpha_3}-3
   \mathrm{E}^{\alpha_3}+\sqrt{3}
   \left(\mathrm{E}^{-\alpha_5}-\mathrm{E}^{\alpha_5}\right)}{6 \sqrt{2}} \\

 \frac{\sqrt{3} \left(\mathrm{E}^{\alpha_2}-\mathrm{E}^{-\alpha_2}\right)+3
   \left(\mathrm{E}^{-\alpha_4}-\mathrm{E}^{\alpha_4}\right)}{6 \sqrt{2}}
\end{array}
\right)
\end{equation}
and
\begin{equation}\label{friccodue}
      j^x_{[II]} \, = \,\left(
\begin{array}{l}
 \frac{-\mathrm{E}^{-\alpha_1}+\mathrm{E}^{\alpha_1}+\sqrt{3} \left(\mathrm{E}^{-\alpha_6}-\mathrm{E}^{\alpha_6}\right)}{2 \sqrt{2}}
   \\
 \frac{-\mathrm{E}^{-\alpha_3}+\mathrm{E}^{\alpha_3}+\sqrt{3} \left(\mathrm{E}^{-\alpha_5}-\mathrm{E}^{\alpha_5}\right)}{2 \sqrt{2}}
   \\
 \frac{\sqrt{3} \left(\mathrm{E}^{-\alpha_2}-\mathrm{E}^{\alpha_2}\right)+\mathrm{E}^{-\alpha_4}-\mathrm{E}^{\alpha_4}}{2 \sqrt{2}}
\end{array}
\right)
\end{equation}
and satisfy among themselves the same relations (\ref{guglielmotell}) as their homologous generators $ J^x_{[I]}$ and $ J^x_{[II]}$.
In eq.(\ref{fattucco}) we have used the same notation as in eq.(\ref{ollonomio}) using the obligatory homomorphism between the the holonomy algebra $\mathrm{Hol}$ and the isotropy subalgebra $\mathbb{H}$. The precise correspondence between generators of one algebra and generators of the other will be establishe in the next subsection by means of the use of the coset representative.
\subsection{\sc The coset representative}
The next step in the development of the coset approach is the construction of the solvable coset representative $\mathbb{L}_{Solv}(\phi)$, advocated in eq.s(\ref{solvoexpo}-\ref{sopore} ), namely a coordinate dependent element of the Borel group of $\mathfrak{g}_{(2,2)}$ such that the Maurer Cartan form
\begin{equation}\label{rupiaindu}
    \Xi \, = \, \mathbb{L}_{Solv}(\phi)^{-1} \, \mathrm{d}\mathbb{L}_{Solv}(\phi)
\end{equation}
projected along the Borel algebra generators, as given in eq.(\ref{interpretaziag22}), reproduces the vielbein of eq.(\ref{EBHfilo}).
The appropriate coset representative is obtained by exponentiating the Borel Lie algebra and the precise recipe is provided below.
First define:
\begin{eqnarray}\label{generillini}
    \mathbf{L}^E_0 & = & \frac{1}{\sqrt{3}} \,\mathcal{H}_2 \quad ; \quad \mathbf{L}^E_+ \, = \, -\sqrt{\frac{2}{3}}\, E^{\alpha_6} \nonumber\\
    \mathbf{L}_0 & = &\,\mathcal{H}_1 \quad ; \quad \mathbf{L}_+ \, = \, \sqrt{2} E^{\alpha_1} \nonumber\\
    \mathbf{W}^I & = & \sqrt{\frac{2}{3}} \, \left \{ E^{\alpha_4} \, , \, E^{\alpha_5} \, , \, - \, E^{\alpha_3}\, , \, - \, E^{\alpha_2} \, , \, \right \}
\end{eqnarray}
and then set:
\begin{eqnarray}\label{fornarina}
    \mathbb{L}& = & \exp\left [a \mathbf{L}^E_+\right]\,\cdot \, \exp\left [\sqrt{2} \left(Z_1 \,  \mathbf{W}^1 \, + \, Z_3 \,  \mathbf{W}^3\right )\right]\,\cdot \, \nonumber\\
     &&\,\cdot \,\exp\left [\sqrt{2} \left(Z_1 \,  \mathbf{W}^1 \, + \, Z_3 \,  \mathbf{W}^3\right )\right]\,\cdot \,
    \exp\left [y \mathbf{L}_+\right]\,\cdot \,\exp\left [h \mathbf{L}_0\right]\,\cdot \,\exp\left [U \mathbf{L}^E_0\right]
\end{eqnarray}
By explicit evaluation we obtain the result displayed in the appendix in formulae (\ref{fornoalegna}) and (\ref{cosettuspresento})
and we verify that, if we set:
\begin{equation}\label{sequenzus}
    \mathfrak{T}_I \, = \, \left\{ \mathbf{L}^E_0 \, , \, \mathbf{L}_0 \, , \, \mathbf{L}^E_+ \, , \, \mathbf{L}_+ \, , \,\mathbf{W}_1\, , \,\mathbf{W}_2\, , \,\mathbf{W}_3\, , \,\mathbf{W}_4\right\}
\end{equation}
upon substitution of  (\ref{cosettuspresento}) into the Maurer Cartan form (\ref{rupiaindu})
we obtain:
\begin{equation}\label{rupiamoresca}
    \mathbb{L}_{Solv}(\phi)^{-1} \, \mathrm{d}\mathbb{L}_{Solv}(\phi) \, = \, \sum_{I=1}^8 \, \mathfrak{T}_I \, E^I_{\mathcal{QM}}
\end{equation}
the forms $E^I_{\mathcal{QM}}$ being given in equation (\ref{EBHfilo}). Alternatively we can also write:
\begin{equation}\label{curlandico}
    \mathbb{L}_{Solv}(\phi)^{-1} \, \mathrm{d}\mathbb{L}_{Solv}(\phi) \, = \, \sum_{x=1}^3 \, \left (\omega^{[I]}_x \,  j^x_{[I]}\, \oplus \, \omega^{[II]}_x \,  j^x_{[II]}\right) \, \oplus \,  \sum_{I}^8 \, \mathbf{T}_I \, E^I_{\mathcal{SQ}}
\end{equation}
In the above equation $\omega^{[I]}_x$ and $\omega^{[II]}_x$ are the components of the spin connections given in eq. (\ref{cucurucu}),
$j^x_{[I]}$ and $j^x_{[II]}$ are the generators of $\mathbb{H}$ defined in eq. (\ref{friccouno},\ref{friccodue}) and $\mathbf{T}_I $ denotes a suitable base of generators in the $\mathbb{K}$ subspace of $\mathfrak{g}_{(2,2)}$ defined as:
\begin{eqnarray}\label{cofimus}
    \mathbb{K} & \equiv & ,\mbox{span}_{\mathbb{R}}  \left \{ \mathcal{H}_1 \, , \, \mathcal{H}_2 \, , \, E^{\alpha  _1} +  E^{\alpha  _1}\,  , \, E^{\alpha  _2} +  E^{\alpha  _2} \, , \, E^{\alpha  _3} +  E^{\alpha  _3} \right. \nonumber\\
    & &\left. E^{\alpha  _4} +  E^{\alpha  _4}\, , \, E^{\alpha  _5} +  E^{\alpha  _5} \, , \,  E^{\alpha  _6} +  E^{\alpha  _6} \right \}
\end{eqnarray}
The precise form of the generators $\mathbf{T}_I $ is not relevant to our purposes and we omit it. The key point is instead the identification of the generators $j^x_{[I]}$ of $\mathbb{H}$ with generators $J^x_{[I]}$ of the holonomy algebra. This provides us with the knowledge of the quaternionic complex structures within the algebra $\mathbb{U}_\mathcal{Q}$ and allows to calculate the tri-holomorphic moment map of any generator $\mathbf{t}\, \in \, \mathbb{U}_\mathcal{Q}$ by means of the formula (\ref{generMapformula}) which in our case reads:
\begin{equation}\label{fisterone}
    \mathcal{P}_{\mathbf{t}}^x \, = \,\frac{1}{2} \mbox{Tr}_{\mathbf{7}} \left ( j^x_{[I]} \, \mathbb{L}_{Solv}^{-1} \, \mathbf{t} \, \mathbb{L}_{Solv} \right )
\end{equation}
having denoted by $\mathbf{7}$ the $7$-dimensional fundamental representation of $\mathfrak{g}_{(2,2)}$.
\paragraph{The Starobinsky potential}
As an immediate application   of  eq.(\ref{fisterone}) one can retrieve  the results of \cite{thesearch} on the inclusion of the Starobinsky potential into supergravity. In section \ref{starobin1} we presented a general discussion of the gaugings of nilpotent generators in the Special K\"ahler subalgebra $\mathbb{U}_{\mathcal{SK}} \, \subset \, \mathbb{U}_\mathcal{Q}$. In the present case where $\mathbb{U}_{\mathcal{SK}} \, = \, \slal(2,\mathbb{R})$ the only available nilpotent operator is $L_+$ and from the general formula (\ref{sisalvichipuo}) applied to the case where the metric is given by (\ref{kelero2}) and the complex coordinate is parameterized as in eq.(\ref{realcordo}) we find:
\begin{equation}\label{cubitagorio}
    \mathcal{P}_{L_+} \, = \, \mbox{const} \, \times \, \exp[-h] \, = \, \mbox{const} \, \times \, \left(\mbox{Im} \,z\right)^{-1}
\end{equation}
This result inserted into the general formula (\ref{curiosone}) yields
\begin{equation}\label{starobbo}
  V(h) \, = \, \mbox{const} \, \times   \, \left(\exp[-h] \, + \, \kappa \right)^2
\end{equation}
which is indeed the Starobinsky potential, since, once expressed in terms of $h$, the K\"ahler potential is exactly $\mathcal{K}\, = \, 3 \, h$.
The same result is directly obtained with precise coefficients by inserting in eq.(\ref{fisterone}) the $7$-dimensional image of $L_+$ in the fundamental representation of $\mathfrak{g}_{(2,2)}$.
\newpage
\section{\sc The $\mathrm{Sp(6,\mathbb{R})}/\mathrm{SU(3)\times U(1)}$ - model and its c-map image.}
Next we consider  the Special K\"ahler manifold
\begin{equation}\label{gurto}
    \mathcal{M}_{\Sp 6} \, = \, \frac{\Sp(6,\mathbb{R})}{\mathrm{SU(3)\times U(1)}}
\end{equation}
and its c-map image which is the following quaternionic manifold:
\begin{equation}\label{quatergurto}
    \mbox{$c$-map} \quad : \quad \mathcal{M}_{\Sp 6} \, \mapsto \, \mathcal{QM}_{F4} \, \equiv \, \frac{\mathrm{F_{(4,4)}}}{\mathrm{SU(2)} \times \mathrm{USp(6)}}
\end{equation}
$\mathcal{M}_{\Sp 6}$ belongs to the magic square of exceptional special K\"ahler manifolds whose quaternionic $c$-map is a homogeneous symmetric space having, as it is evident from (\ref{quatergurto}), an exceptional Lie group as isometry group.
\par
We begin by illustrating some general properties of this remarkable manifold. First of all, in order to discuss them adequately we need to choose  a basis for the $\sym(6,\mathbb{R})$ Lie algebra. Since we are not interested in solving Lax equations we do not choose the basis where the matrices of the Borel subalgebra are upper triangular. We rather use the basis where the symplectic preserved metric is the standard one, namely:
\begin{equation}\label{Cmatra}
    \mathbb{C} \, = \, \left(
\begin{array}{llllll}
 0 & 0 & 0 & 1 & 0 & 0 \\
 0 & 0 & 0 & 0 & 1 & 0 \\
 0 & 0 & 0 & 0 & 0 & 1 \\
 -1 & 0 & 0 & 0 & 0 & 0 \\
 0 & -1 & 0 & 0 & 0 & 0 \\
 0 & 0 & -1 & 0 & 0 & 0
\end{array}
\right)
\end{equation}
This traditional choice allows to describe in a simple way other aspects of the manifold geometry that are more relevant to our present purposes.
\par
According to the above choice, an element of the $\Sp(6,\mathbb{R})$ group and an element of the $\sym(6,\mathbb{R})$ Lie-algebra are  matrices respectively  fulfilling the following two constraints:
\begin{equation}\label{giluro}
  \left( \begin{array}{c|c}
             A & B \\
             \hline
             C & D
           \end{array}
    \right)^T \, \mathbb{C} \, \left( \begin{array}{c|c}
             A & B \\
             \hline
             C & D
           \end{array}
    \right) \, = \, \mathbb{C} \quad ; \quad \left( \begin{array}{c|c}
             \mathbf{A} & \mathbf{B} \\
             \hline
             \mathbf{C} & \mathbf{D}
           \end{array}
    \right)^T \, \mathbb{C} + \mathbb{C} \left( \begin{array}{c|c}
             \mathbf{A} & \mathbf{B} \\
             \hline
             \mathbf{C} & \mathbf{D}
           \end{array}
    \right) \, = \, 0
\end{equation}
where $A,B,C,D$, $\mathbf{A,B,C,D}$ are $3\times 3$ blocks. By means of the so called Cayley transformation
\begin{equation}\label{caylus}
    \mathcal{C} \, = \, \left(
\begin{array}{llllll}
 \frac{1}{\sqrt{2}} & 0 & 0 & \frac{i}{\sqrt{2}} & 0 & 0 \\
 0 & \frac{1}{\sqrt{2}} & 0 & 0 & \frac{i}{\sqrt{2}} & 0 \\
 0 & 0 & \frac{1}{\sqrt{2}} & 0 & 0 & \frac{i}{\sqrt{2}} \\
 \frac{1}{\sqrt{2}} & 0 & 0 & -\frac{i}{\sqrt{2}} & 0 & 0 \\
 0 & \frac{1}{\sqrt{2}} & 0 & 0 & -\frac{i}{\sqrt{2}} & 0 \\
 0 & 0 & \frac{1}{\sqrt{2}} & 0 & 0 & -\frac{i}{\sqrt{2}}
\end{array}
\right)
\end{equation}
a real element of the symplectic group (or algebra) can be mapped into a  matrix that is simultaneously symplectic and pseudounitary:
\begin{equation}\label{grillo}
    \mathcal{S} \, = \, \mathcal{C}^\dagger \, \, \left( \begin{array}{c|c}
             {A} & {B} \\
            \hline
             {C} & {D}
           \end{array}
    \right) \,\mathcal{C} \, = \, \left(\begin{array}{cc}
                             U_0 & U_1^\star \\
                             U_1 & U_0 ^\star
                           \end{array}
    \right) \,\, \in \, \Sp(6,\mathbb{C}) \bigcap \mathrm{SU(3,3)}
\end{equation}
The diagonal blocks $U_0 \in \mathrm{U(3)}$ span the $\mathrm{H}$-subgroup of the coset (\ref{gurto}). This allows to introduce a set projective coordinates that parameterize the points of the manifold (\ref{gurto}) and have a nice fractional linear transformation under the action of the group $\Sp(6,\mathbb{R})$.  Given any coset parameterization
\begin{equation}\label{cosetpar}
     \left( \begin{array}{c|c}
             {A}(\phi) & {B}(\phi) \\
             \hline
             {C}(\phi) & {D}(\phi)
           \end{array}
    \right) \, \in \, \Sp(6,\mathbb{R})
\end{equation}
namely a family of   symplectic group elements depending  on 12 parameters $\phi^i$ such that  each different choice of the $\phi^i$ provides a representative of a different equivalence class in (\ref{gurto}), we can construct the following, \textit{symmetric complex matrix}:
\begin{equation}\label{Zmatra}
    Z(\phi) \, \equiv \, \left(A(\phi) \, - \, {\rm i} \,B(\phi) \right) \, \left( C(\phi) \, - \, {\rm i} D(\phi) \right)^{-1}
\end{equation}
which has a very simple transformation under the action of the symplectic group. Let us consider the action of any element of $\Sp(6,\mathbb{R})$ on the coset representative. We have:
\begin{equation}\label{canalito}
    \underbrace{\left( \begin{array}{c|c}
             {\hat{A}} & {\hat{B}} \\
             \hline
             {\hat{C}} & {\hat{D}}
           \end{array}
    \right)}_{= \mathfrak{g}\in  \,\Sp(6,\mathbb{R})}  \left( \begin{array}{c|c}
             {A}(\phi) & {B}(\phi) \\
             \hline
             {C}(\phi) & {D}(\phi)
           \end{array}
    \right)   \, = \, \left( \begin{array}{c|c}
             {A}(\phi^\prime) & {B}(\phi^\prime) \\
             \hline
             {C}(\phi^\prime) & {D}(\phi^\prime)
           \end{array}
    \right) \, \mathrm{H}(\phi, \mathfrak{g})
\end{equation}
where $\phi^\prime$ is the label of a new equivalence class and $\mathrm{H}(\phi, \mathfrak{g}) \, \in \, \mathrm{U(3)}$ is a suitable $\mathrm{H}$-compensator.  Calculating the matrix $Z(\phi^\prime)$ according to the definition (\ref{Zmatra}) we find that it is related to $Z(\phi)$ by a simple linear fractional transformation (generalized to matrices):
\begin{equation}\label{gourmet}
    Z(\phi^\prime) \, = \, \left( A Z(\phi) \, + \, B\right) \, \left( C Z(\phi) \, + \, D\right)^{-1}
\end{equation}
Formula (\ref{gourmet}) is of crucial relevance and requires several comments. From a mathematical point of view, (\ref{gourmet}) is the well known generalization of the action of the $\mathrm{SL(2,\mathbb{R})}\, \simeq \, \Sp(2,\mathbb{R})$ group on the upper complex plane of Poincar\'e-Lobachevsky. The complex numbers $z$ with positive imaginary parts (${\rm Im} z > 0$) are replaced by the complex symmetric matrices $Z_{ij}$ whose imaginary part is positive definite. Such matrices constitute the so named \textbf{upper Siegel plane}, which indeed is homeomorphic to the coset $\Sp(2n,\mathbb{R})/\mathrm{U(n)}$. From the physical point of view (\ref{gourmet}) is just identical to the Gaillard-Zumino formula for the construction of the kinetic matrix $\mathcal{N}_{\Lambda\Sigma}$ which appears in the lagrangian of the vector fields in $\mathcal{N}=2$ supergravity and  is rooted in the structure of special K\"ahler geometry. Indeed for any special K\"ahler manifold $\mathcal{M}_{n}$ of complex dimension $n$ that is also a symmetric space $\mathrm{G/H}$,  there exists a so named $\mathbf{W}$-representation of $\mathrm{G}$, which is symplectic, has dimension $2n+2$ and hosts the electric and magnetic field strengths of the model. Such a representation defines a symplectic embedding:
\begin{equation}\label{gongolini}
    \mathrm{G} \, \rightarrow \, \Sp\left( 2n+2,\mathbb{R}\right )
\end{equation}
which associates to any coset representative $\mathfrak{g}(\phi) \in \mathrm{G/H}$ its corresponding symplectic $(2n+2) \times (2n +2)$
representation $\left( \begin{array}{c|c}
             {A}(\phi) & {B}(\phi) \\
             \hline
             {C}(\phi) & {D}(\phi)
           \end{array}
    \right)$. From this latter, utilizing the recipe provided by formula (\ref{gourmet}) we obtain an $(n+1) \times (n+1)$ complex symmetric matrix to be identified with the appropriate $\mathcal{N}$ kinetic matrix largely discussed and utilized in section \ref{scrittaN}.
    \par
The peculiarity of the $\mathcal{N}=2$ model under investigation is that the original isometry  group $\mathrm{G}$ is already symplectic so that we can utilize the Gaillard-Zumino formula (\ref{gourmet}) in the fundamental $6$ dimensional representation in order to construct a Siegel parametrization of the coset in terms of a symmetric complex $ 3 \times 3$ matrix $Z$. The $\mathbf{W}$-representation is the $\mathbf{14}^\prime$ and this defines the embedding:
\begin{equation}\label{gallettoalladiavola}
    \Sp(6,\mathbb{R}) \mapsto \Sp(14,\mathbb{R})
\end{equation}
from which we can construct the $ 7\times 7 $ kinetic matrix $\mathcal{N}(Z)$.
\paragraph{\sc The transitive action of $\Sp(6,\mathbb{R})$ on the upper Siegel plane.} Before proceeding with the actual construction of the Lie algebra let us comment on the transitive action of the symplectic group on the Siegel plane. Focusing on the the formula (\ref{gourmet}), consider the $\Sp(6,\mathbb{R})$ parabolic subgroup composed by the following matrices:
\begin{equation}\label{caripollo}
    \mathfrak{g}(B) \, = \, \left(\begin{array}{c|c}
                               \mathbf{1}_{3\times 3} & B \\
                               \hline
                               \mathbf{0}_{3\times 3} & \mathbf{1}_{3\times 3}
                             \end{array}
    \right)
\end{equation}
where $B$ is symmetric and real. By means of such a subgroup we can always map a generic $Z$ matric into one that has vanishing real part $\mbox{Re} Z \, = \,0$.
Next consider the action on the residual imaginary part of $Z$ of the $\mathrm{GL(3,\mathbb{R})} \subset \Sp(6,\mathbb{R})$ subgroup
composed by the matrices:
\begin{equation}\label{caripollo2}
    \mathfrak{g}(B) \, = \, \left(\begin{array}{c|c}
                               \mathcal{A} & \mathbf{0}_{3\times 3} \\
                               \hline
                               \mathbf{0}_{3\times 3} & \left(\mathcal{A}^T\right) ^{-1}
                             \end{array}
    \right) \quad ; \quad \mathcal{A} \, \in \, \mathrm{GL(3,\mathbb{R})}
\end{equation}
We obtain:
\begin{equation}\label{struzzo}
    \mbox{Im} Z \, \mapsto \, \mathcal{A} \, \mbox{Im} Z \, \mathcal{A}^T
\end{equation}
Choosing $\mathcal{A} \, = \, (\mbox{Im} Z)^{\frac 12}$, which is always possible since $\mbox{Im} Z$ is positive definite we can reduce the imaginary part to the identity matrix. This shows the transitive action of the symplectic group on the Siegel plane and also provides a nice coset parameterization of the coset manifold. Indeed we can introduce the following matrix:
\begin{equation}\label{godereccio}
    \mathfrak{g}(Z) \, \equiv \, \left (\begin{array}{c|c}
                                          \left(\mbox{Im} Z\right)^{\frac 12} &\mbox{Re} Z \, \left(\mbox{Im} Z\right)^{-\frac 12} \\
                                          \hline
                                          \mathbf{0} & \left(\mbox{Im} Z\right)^{-\frac 12}
                                        \end{array}
     \right)
\end{equation}
which maps the origin of the manifold ${\rm i} \mathbf{1}_{3\times 3}$ in the complex symmetric matrix $Z$.
\subsection{\sc The $\sym(6,\mathbb{R})$ Lie algebra}
\begin{figure}
\centering
\begin{picture}(90,50)
      \put (-70,35){$C_3$}  \put
(10,35){\circle {10}} \put (7,20){$\alpha_1$} \put (15,35){\line
(1,0){20}} \put (40,35){\circle {10}} \put (37,20){$\alpha_2$}\put (45,38){\line (1,0){20}}
\put (55,35){\line (1,1){10}} \put (55,35){\line (1,-1){10}}\put (45,33){\line (1,0){20}} \put
(70,35){\circle {10}} \put (67,20){$\alpha_{3}$}
\end{picture}
 \caption{The Dynkin diagram of $C_{3}$.   \label{C3dynk}}
\end{figure}
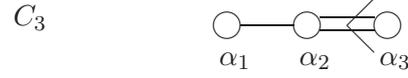

From the point of view of  the Dynkin classification the Lie algebra $\sym(6,\mathbb{R})$ is the maximally split real section of the complex Lie algebra $C_3$ whose Dynkin diagram is displayed in fig.\ref{C3dynk}.
The root system is composed of $18$-roots whose subset of $9$ positive ones is displayed here below:
\begin{equation}\label{cornette}
 \left[\begin{array}{c}
         \alpha_1 \\
         \alpha_2 \\
       \alpha_3\\
        \alpha_4 \\
         \alpha_5 \\
         \alpha_6 \\
         \alpha_7 \\
         \alpha_8 \\
         \alpha_9
       \end{array}
 \right]  \, = \, \left[
\begin{array}{ll}
 \alpha _1 & \{1,-1,0\} \\
 \alpha _2 & \{0,1,-1\} \\
 \alpha _3 & \{0,0,2\} \\
 \alpha _1+\alpha _2 &
   \{1,0,-1\} \\
 \alpha _2+\alpha _3 &
   \{0,1,1\} \\
 \alpha _1+\alpha _2+\alpha _3
   & \{1,0,1\} \\
 2 \alpha _2+\alpha _3 &
   \{0,2,0\} \\
 \alpha _1+2 \alpha _2+\alpha
   _3 & \{1,1,0\} \\
 2 \alpha _1+2 \alpha
   _2+\alpha _3 & \{2,0,0\}
\end{array}
\right]
\end{equation}
The simple roots are the first three. Of the remaining $6$ we have provided both their expression in terms of the simple roots and their realization as three-vectors in $\mathbb{R}^3$. Such a realization is spelled out also for the simple roots.
Next we present the basis of $6\times6$ matrices that fulfill the standard commutation relations of the Lie Algebra in the Cartan Weyl basis.
\paragraph{Cartan Generators.} The Cartan generators are named $\mathcal{H}^i$ and can be easily read-off from the following formula:
\begin{equation}\label{cartolini}
    \sum_{i=1}^{3} \, h_i \, \mathcal{H}^i \, = \,\left(
\begin{array}{llllll}
 h_1 & 0 & 0 & 0 & 0 & 0 \\
 0 & h_2 & 0 & 0 & 0 & 0 \\
 0 & 0 & h_3 & 0 & 0 & 0 \\
 0 & 0 & 0 & -h_1 & 0 & 0 \\
 0 & 0 & 0 & 0 & -h_2 & 0 \\
 0 & 0 & 0 & 0 & 0 & -h_3
\end{array}
\right)
\end{equation}
by collecting the coefficient of the parameter $h_i$.
\paragraph{\sc Positive Root Step Operators.} The step operator associated with the positive root $\alpha_i$ is named $\mathcal{E}^{\alpha_i}$ and can be easily read-off from the following formula:
\begin{equation}\label{steppinisu}
    \sum_{i=1}^{9} \, a_i \, \mathcal{E}^{\alpha_i} \, = \,\left(
\begin{array}{llllll}
 0 & a_1 & a_4 & \sqrt{2} a_9
   & a_8 & a_6 \\
 0 & 0 & a_2 & a_8 & \sqrt{2}
   a_7 & a_5 \\
 0 & 0 & 0 & a_6 & a_5 &
   \sqrt{2} a_3 \\
 0 & 0 & 0 & 0 & 0 & 0 \\
 0 & 0 & 0 & -a_1 & 0 & 0 \\
 0 & 0 & 0 & -a_4 & -a_2 & 0
\end{array}
\right)
\end{equation}
by collecting the coefficient of the parameter $a_i$.
\paragraph{\sc Negative Root Step Operators.} The step operator associated with the negative root $-\alpha_i$ is named $\mathcal{E}^{-\alpha_i}$ and can be easily read-off from the following formula:
\begin{equation}\label{steppinigiu}
    \sum_{i=1}^{9} \, b_i \, \mathcal{E}^{-\alpha_i} \, = \,\left(
\begin{array}{llllll}
 0 & 0 & 0 & 0 & 0 & 0 \\
 b_1 & 0 & 0 & 0 & 0 & 0 \\
 b_4 & b_2 & 0 & 0 & 0 & 0 \\
 \sqrt{2} b_9 & b_8 & b_6 & 0
   & -b_1 & -b_4 \\
 b_8 & \sqrt{2} b_7 & b_5 & 0
   & 0 & -b_2 \\
 b_6 & b_5 & \sqrt{2} b_3 & 0
   & 0 & 0
\end{array}
\right)
\end{equation}
by collecting the coefficient of the parameter $b_i$.
\subsection{\sc The representation $14^\prime$}
\label{seziona14}
The $\mathbf{14}^\prime$ representation of $\sym(6,\mathbb{R})$ which plays the role $\mathbf{W}$-representation for the special manifold under consideration is defined as the representation obeyed by the  three-times antisymmetric tensors with vanishing $\mathbb{C}$-traces, namely:
\begin{equation}\label{goliutico}
    \underbrace{t_{ABC}}_{\mbox{antisymmetric in} A,B,C} \times \quad \mathbb{C}^{BC} \, = \,0
\end{equation}
The generators are constructed in the appendix, subsection \ref{seziona14}, and  displayed in eq.s (\ref{cartolini14}), (\ref{steppinisu14}) and (\ref{steppinigiu14}).
\subsection{\sc The holomorphic symplectic section and its transformation in the $14^\prime$}
\label{holoseziona}
In order to construct the special geometry of the manifold (\ref{gurto}) we need to introduce the holomorphic symplectic section that, by definition, should transform in the $14^\prime$ representation of $\Sp(6,\mathbb{R})$. To this effect, we choose as special coordinates the components of the symmetric complex matrix defined by eq.(\ref{gourmet}) and we choose a lexicographic order to enumerate its independent components, namely we set:
\begin{equation}\label{specoord}
    Z \, = \, \left(
\begin{array}{lll}
 z_1 & z_2 & z_3 \\
 z_2 & z_4 & z_5 \\
 z_3 & z_5 & z_6
\end{array}
\right)
\end{equation}
Next we introduce the holomorphic prepotential defined by:
\begin{eqnarray}
  \mathcal{F} & \equiv & Z_{a,i} \, Z_{b,j} \, Z_{c,k} \, \epsilon^{abc} \,\epsilon^{ijk}\nonumber \\
  \null &=& -6 \left(z_6 z_2^2-2 z_3 z_5
   z_2+z_3^2 z_4+z_1
   \left(z_5^2-z_4
   z_6\right)\right)
\end{eqnarray}
and we can introduce a first ansatz for the symplectic section by writing:
\begin{eqnarray}
  \tilde{\Omega} &=& \left \{1, \, z^I , \, \mathcal{F} , \, \frac{\partial \mathcal{F} }{\partial z^J} \right\} \nonumber \\
  \null &=& \left\{1,z_1,z_2,z_3,z_4,z_5,z
   _6,-6 \left(z_6 z_2^2-2 z_3
   z_5 z_2+z_3^2 z_4+z_1
   \left(z_5^2-z_4
   z_6\right)\right), \right.\nonumber\\
   &&\left.6 z_4
   z_6-6 z_5^2,12 \left(z_3
   z_5-z_2 z_6\right),12
   \left(z_2 z_5-z_3
   z_4\right),6 z_1 z_6-6
   z_3^2,12 \left(z_2 z_3-z_1
   z_5\right),6 z_1 z_4-6
   z_2^2\right\}\nonumber\\
   \label{preomega}
\end{eqnarray}
In order to match the transformation of this holomorphic section with the transformations of the $\mathbf{14}^\prime$ representation as we defined it in  subsection \ref{holoseziona} we still need a change of basis. Consider the following matrix
{\scriptsize
 \begin{equation}\label{paola}
   \mathfrak{S}\, = \, \left(
\begin{array}{llllllllllllll}
 0 & 0 & 0 & 0 & 0 & 0 & \sqrt{2} & 0 & 0 & 0 & 0 & 0 & 0 & 0 \\
 0 & 0 & 0 & 0 & 0 & 0 & 0 & 0 & 0 & 0 & 0 & \frac{1}{3 \sqrt{2}} & 0 & 0 \\
 0 & 0 & 0 & 0 & 0 & 0 & 0 & 0 & -\frac{1}{3 \sqrt{2}} & 0 & 0 & 0 & 0 & 0 \\
 0 & 0 & 0 & 0 & 0 & 0 & 0 & \frac{1}{3 \sqrt{2}} & 0 & 0 & 0 & 0 & 0 & 0 \\
 0 & 0 & 0 & 0 & 0 & 0 & 0 & 0 & 0 & 0 & 0 & 0 & \frac{1}{6} & 0 \\
 0 & 0 & 0 & 0 & 0 & 0 & 0 & 0 & 0 & 0 & -\frac{1}{6} & 0 & 0 & 0 \\
 0 & 0 & 0 & 0 & 0 & 0 & 0 & 0 & 0 & \frac{1}{6} & 0 & 0 & 0 & 0 \\
 0 & 0 & 0 & 0 & 0 & 0 & 0 & 0 & 0 & 0 & 0 & 0 & 0 & \frac{1}{3 \sqrt{2}} \\
 0 & 0 & 0 & 0 & -\sqrt{2} & 0 & 0 & 0 & 0 & 0 & 0 & 0 & 0 & 0 \\
 0 & \sqrt{2} & 0 & 0 & 0 & 0 & 0 & 0 & 0 & 0 & 0 & 0 & 0 & 0 \\
 \sqrt{2} & 0 & 0 & 0 & 0 & 0 & 0 & 0 & 0 & 0 & 0 & 0 & 0 & 0 \\
 0 & 0 & 0 & 0 & 0 & -2 & 0 & 0 & 0 & 0 & 0 & 0 & 0 & 0 \\
 0 & 0 & 0 & 2 & 0 & 0 & 0 & 0 & 0 & 0 & 0 & 0 & 0 & 0 \\
 0 & 0 & -2 & 0 & 0 & 0 & 0 & 0 & 0 & 0 & 0 & 0 & 0 & 0
\end{array}
\right)
\end{equation}
}
and define:
\begin{equation}\label{gargamelle}
    \Omega\left( Z\right) \, = \, \mathfrak{S}\, \tilde{\Omega}(Z) \, = \, \left(
\begin{array}{l}
 \sqrt{2} z_6 \\
 \sqrt{2} \left(z_1 z_6-z_3^2\right) \\
 \sqrt{2} \left(z_5^2-z_4 z_6\right) \\
 -\sqrt{2} \left(z_6 z_2^2-2 z_3 z_5 z_2+z_3^2 z_4+z_1 \left(z_5^2-z_4 z_6\right)\right) \\
 2 z_2 z_3-2 z_1 z_5 \\
 2 z_3 z_4-2 z_2 z_5 \\
 2 z_3 z_5-2 z_2 z_6 \\
 \sqrt{2} \left(z_1 z_4-z_2^2\right) \\
 -\sqrt{2} z_4 \\
 \sqrt{2} z_1 \\
 \sqrt{2} \\
 -2 z_5 \\
 2 z_3 \\
 -2 z_2
\end{array}
\right)
\end{equation}
Naming $\mathcal{D}_{14}\left[g\right]$ the 14-dimensional representation of a finite element $g , \in \, \Sp(6,\mathbb{R})$ of the symplectic group that corresponds to the  representation of the algebra as we constructed it above,  the holomorphic symplectic section (\ref{gargamelle}) transforms in the following way:
\begin{equation}\label{buonotrasformo}
    \Omega \left[ (A\,Z \, + \, B) \, (C \, Z \, + \, D )^{-1} \right] \, = \,      \frac{1}{\mbox{Det}\left(C \, Z \, + \, D\right)} \, \mathcal{D}_{14}\left[\left(\begin{array}{c|c}
                               A & B \\
                               \hline
                               C & D
                             \end{array}\right)
    \right] \, \Omega[Z]
\end{equation}
The formula (\ref{buonotrasformo}) can be in particular applied to the case where the original $Z$ is the origin of the coset manifold:
$Z_0 \, = \, {\rm i} \, \mathbf{1}_{3\times 3}$. In that case, recalling eq. (\ref{godereccio}) we find:
\begin{equation}\label{frittellaprima}
 \Omega[Z_0] \, = \, \left\{i \sqrt{2},-\sqrt{2},\sqrt{2},-i
   \sqrt{2},0,0,0,-\sqrt{2},-i \sqrt{2},i
   \sqrt{2},\sqrt{2},0,0,0\right\}
  \end{equation}
  and
\begin{equation}\label{caviccio}
  \Omega[Z_0] \, = \, \sqrt{\mbox{Det}\left[\mbox{Im} \, Z\right] } \, \times \, \mathcal{D}_{14}\left[ \left (\begin{array}{c|c}
                                          \left(\mbox{Im} Z\right)^{\frac 12} &\mbox{Re} Z \, \left(\mbox{Im} Z\right)^{-\frac 12} \\
                                          \hline
                                          \mathbf{0} & \left(\mbox{Im} Z\right)^{-\frac 12}
                                        \end{array}
     \right)\right] \, \cdot \, \Omega[Z_0]
\end{equation}

\subsection{\sc The K\"ahler potential and the metric}
Provided with this information we can now write the explicit form of the K\"ahler potential and of the K\"ahler metric for the manifold (\ref{gurto}) according to the rules of special K\"ahler geometry. We have:
\begin{eqnarray}
  \mathcal{K} & \equiv & - \log \left( {\rm i} \Omega[Z] \,  \mathbb{C}_{14} \, \overline{\Omega}[\bar{Z}]\right) \nonumber\\
  \null &=& -\log \left(2 i \left(-z_6
   z_2^2+{\bar z}_6 z_2^2+2 z_6
   {\bar z}_2 z_2-2 z_5 {\bar z}_3
   z_2+2 {\bar z}_3 {\bar z}_5 z_2-2
   {\bar z}_2 {\bar z}_6 z_2-z_6
   {\bar z}_2^2-z_4
   {\bar z}_3^2+{\bar z}_1
   {\bar z}_5^2+\right.\right. \nonumber\\
   &&\left.\left. z_5^2 {\bar z}_1-z_4
   z_6 {\bar z}_1+2 z_5 {\bar z}_2
   {\bar z}_3+{\bar z}_3^2
   {\bar z}_4+z_6 {\bar z}_1
   {\bar z}_4 +z_3^2
   \left({\bar z}_4-z_4\right)-2 z_5
   {\bar z}_1 {\bar z}_5-2
   {\bar z}_2 {\bar z}_3
   {\bar z}_5 \right.\right.\nonumber\\
   &&\left.\left.+2 z_3 \left(-z_5
   {\bar z}_2+{\bar z}_5
   {\bar z}_2+z_4
   {\bar z}_3-{\bar z}_3
   {\bar z}_4+z_2
   \left(z_5-{\bar z}_5\right)\right)+
   {\bar z}_2^2 {\bar z}_6+z_4
   {\bar z}_1 {\bar z}_6-{\bar z}_1
   {\bar z}_4 {\bar z}_6 \right.\right.\nonumber\\
   &&\left.\left. -z_1
   \left(z_5^2-2 {\bar z}_5
   z_5+{\bar z}_5^2+z_6
   {\bar z}_4-{\bar z}_4
   {\bar z}_6+z_4
   \left({\bar z}_6-z_6\right)\right)\right)\right) \nonumber\\
\end{eqnarray}
and the line element on the manifold, in terms of the special coordinates $z_i$ takes the standard form:
\begin{equation}\label{gumilevo}
    ds_K^2 \, = \, \frac{\partial}{\partial z^i} \, \frac{\partial}{\partial \bar{z}^{j}} \, \mathcal{K} \, dz^i \otimes d\bar{z}^j
\end{equation}
The explicit form of $ds_K^2$ in terms of the special coordinate $z^i$ can be worked out by simple derivatives, yet its explicit form is quite lengthy and so much involved that we think it better not to display it. For the purposes that we pursue we rather prefer to  write the  form of the metric in terms of solvable real coordinates.
\subsubsection{\sc The solvable parametrization}
The transition to a solvable parametrization of the coset is rather simple. Let us define the solvable coset representative as the product of the exponentials of all the generators of the Borel subalgebra of $\sym(6,\mathbb{R})$:
\begin{eqnarray}
 & \mathbb{L}(h,p)\, = \, \prod_{i=1}^9 \, \exp\left [ p_{10-i} \, \mathcal{E}^{\alpha_{10-i}}\right ] \, \prod_{j=3}^3 \exp\left [ h_{j} \, \mathcal{H}^{j}\right ] \, = \, & \nonumber\\
   & \mbox{\scriptsize $\left(
\begin{array}{l|l|l|l|l|l}
 e^{h_1} & e^{h_2} p_1 & e^{h_3} p_4 &
   \begin{array}{l}
            e^{-h_1}  \left(\sqrt{2} p_1 p_2 p_3
   p_4 \right.\\
           \left.   +\left(p_1 p_2-p_4\right) p_6\right.\\
           \left.-p_1
   p_8+\sqrt{2} p_9 \right)
            \end{array}
    & \begin{array}{c}
                    e^{-h_2}
   \left(-\sqrt{2} p_2 p_3 p_4\right. \\
                \left.    -p_2
   p_6+p_8\right)
                  \end{array}
   & e^{-h_3}
   \left(\sqrt{2} p_3 p_4+p_6\right) \\
   \hline
 0 & e^{h_2} & e^{h_3} p_2 &
  \begin{array}{c}
                                      e^{-h_1}
   \left(\left(p_1 p_2-p_4\right)
   p_5\right. \\
                                      \left.-\sqrt{2} p_1 p_7+p_8\right)
                                    \end{array}
   &
   \begin{array}{c}
                e^{-h_2} \left(\sqrt{2} p_7 \right. \\
                \left. -p_2
   p_5\right)
              \end{array}
   & e^{-h_3} p_5
   \\
   \hline
 0 & 0 & e^{h_3} & e^{-h_1}
   \left(\sqrt{2} p_1 p_2 p_3-p_1
   p_5+p_6\right) & e^{-h_2}
   \left(p_5-\sqrt{2} p_2 p_3\right) &
   \sqrt{2} e^{-h_3} p_3 \\
   \hline
 0 & 0 & 0 & e^{-h_1} & 0 & 0 \\
 \hline
 0 & 0 & 0 & -e^{-h_1} p_1 & e^{-h_2} &
   0 \\
   \hline
 0 & 0 & 0 & e^{-h_1} \left(p_1
   p_2-p_4\right) & -e^{-h_2} p_2 &
   e^{-h_3}
\end{array}
\right) $}& \nonumber\\
\label{baltico}
\end{eqnarray}
The real coordinates of the manifold are now the $12$ parameters:
\begin{equation}\label{coordine}
 \mbox{coordinates} \, \equiv \,    \left\{ h_1, \dots , h_3, \, p_1 , \, \dots \, p_9 \right \}
\end{equation}
Extracting the complex matrix $Z$ from the symplectic matrix $\mathbb{L}(h,p)$ we find:
\begin{eqnarray}\label{zph}
   & Z(h,p) \, = \, & \nonumber\\
   & \mbox{\scriptsize $\left(
\begin{array}{lll}
 i e^{2 h_2} p_1^2+i e^{2
   h_1}+\left(\sqrt{2} p_3+i e^{2
   h_3}\right) p_4^2+\sqrt{2} p_9 & i
   e^{2 h_2} p_1+i e^{2 h_3} p_2
   p_4+p_8 & \left(\sqrt{2} p_3+i e^{2
   h_3}\right) p_4+p_6 \\
 i e^{2 h_2} p_1+i e^{2 h_3} p_2
   p_4+p_8 & i e^{2 h_3} p_2^2+i e^{2
   h_2}+\sqrt{2} p_7 & i e^{2 h_3}
   p_2+p_5 \\
 \left(\sqrt{2} p_3+i e^{2 h_3}\right)
   p_4+p_6 & i e^{2 h_3} p_2+p_5 &
   \sqrt{2} p_3+i e^{2 h_3}
\end{array}
\right)$} &
\end{eqnarray}
which defines the coordinate transformation from the special to the solvable coordinates:
\begin{equation}\label{carlinus}
\left(
\begin{array}{l}
 z_1 \\
 z_2 \\
 z_3 \\
 z_4 \\
 z_5 \\
 z_6
\end{array}
\right)\, = \,     \left(
\begin{array}{l}
 \sqrt{2} p_3 p_4^2+i \left(e^{2 h_2}
   p_1^2+e^{2 h_1}+e^{2 h_3}
   p_4^2\right)+\sqrt{2} p_9 \\
 i \left(e^{2 h_2} p_1+e^{2 h_3} p_2
   p_4\right)+p_8 \\
 i e^{2 h_3} p_4+\sqrt{2} p_3 p_4+p_6
   \\
 i \left(e^{2 h_3} p_2^2+e^{2
   h_2}\right)+\sqrt{2} p_7 \\
 i e^{2 h_3} p_2+p_5 \\
 \sqrt{2} p_3+i e^{2 h_3}
\end{array}
\right)
\end{equation}
Inserting such a coordinate transformation into the K\"ahler metric (\ref{gumilevo}) we obtain its form in terms of the real coordinates
(\ref{coordine}). For the explicit form of the metric, we refer the reader to the appendix, eq. (\ref{formidabile}).
The complete metric is quite formidable (\ref{formidabile}) since it contains a total of 100 terms. It has however quite simple properties when we sit in the neighborhood of the coset origin, in particular at $p_i \sim 0$. In this case it drastically simplifies and becomes diagonal:
\begin{eqnarray}\label{pippa}
  &&ds_K^2 \, \stackrel{p_i \rightarrow 0}{\Longrightarrow} \,   \mathrm{dh}_1^2+\mathrm{dh}_2^2+\mathrm{dh}_3
   ^2+\frac{1}{2} e^{2 h_2-2 h_1}
   \mathrm{dp}_1^2+\frac{1}{2} e^{2 h_3-2
   h_2} \mathrm{dp}_2^2+\frac{1}{2} e^{-4
   h_3} \mathrm{dp}_3^2+\frac{1}{2} e^{2
   h_3-2 h_1} \mathrm{dp}_4^2\nonumber\\
   &&+\frac{1}{2}
   e^{-2 h_2-2 h_3}
   \mathrm{dp}_5^2+\frac{1}{2} e^{-2
   h_1-2 h_3} \mathrm{dp}_6^2+\frac{1}{2}
   e^{-4 h_2} \mathrm{dp}_7^2+\frac{1}{2}
   e^{-2 h_1-2 h_2}
   \mathrm{dp}_8^2+\frac{1}{2} e^{-4 h_1}
   \mathrm{dp}_9^2
\end{eqnarray}
which shows that it is positive definite as it should be. It is also interesting to note that if the truncation to the Cartan is permitted by the potential, then we just have three dilatons with canonical kinetic terms.
\subsection{\sc The quartic invariant in the $14^\prime$}
Of crucial relevance for the analysis of Black Hole charges and in general for the classification of orbits in the $\mathbf{W}$-representation is the quartic symplectic invariant.  Given a $14$-vector
\begin{equation}\label{gongolando}
    \mathcal{Q} \, = \, \left\{ q_1,\, q_2 \, \dots ,\,q_{14}\right\}
\end{equation}
the standard form of this invariant can be expressed in the following manifestly ${\rm Sp}(6,\mathbb{R})$-invariant form (see for instance \cite{Ferrara:2013zga})
\begin{eqnarray}
  \mathfrak{J}_4(\mathcal{Q}) &=& -\frac{n_V(2n_V+1)}{6d}\,(\Lambda_a)_{\alpha\beta}\,(\Lambda^a)_{\gamma\delta}\,\mathcal{Q}^\alpha\,\mathcal{Q}^\beta\,\mathcal{Q}^\gamma\,\mathcal{Q}^\delta\,,
\end{eqnarray}
where in our case $n_V=7$ and $d={\rm dim}{\rm Sp}(6,\mathbb{R})=21$, the symplectic indices are raised and lowered by $\mathbb{C}_14^{\alpha\beta}$  and $\mathbb{C}_{14\,\alpha\beta}$ and  the index $a$ is raised by the inverse of $\eta_{ab}\equiv
{\rm Tr}(\Lambda_a\,\Lambda_b)$.
The explicit form of  $\mathfrak{J}_4(\mathcal{Q})$ reads:
\begin{eqnarray}
  \mathfrak{J}_4(\mathcal{Q}) &=& -2 q_1 q_9 q_5^2+2 q_3 q_{11} q_5^2-2 \sqrt{2} q_6 q_7 q_{11} q_5-2 q_1 q_8 q_{12}
   q_5+2 q_2 q_9 q_{12} q_5-2 q_3 q_{10} q_{12} q_5\nonumber\\
   &&+2 q_4 q_{11} q_{12} q_5-2
   \sqrt{2} q_7 q_9 q_{13} q_5  +2 \sqrt{2} q_1 q_6 q_{14} q_5+2 \sqrt{2} q_3 q_{13}
   q_{14} q_5+q_1^2 q_8^2\nonumber\\
   &&+q_2^2 q_9^2+q_3^2 q_{10}^2+q_4^2 q_{11}^2+2 q_2 q_8
   q_{12}^2-2 q_4 q_{10} q_{12}^2-2 q_3 q_8 q_{13}^2+2 q_4 q_9 q_{13}^2-2 q_2 q_3
   q_{14}^2\nonumber\\
   &&-2 q_1 q_4 q_{14}^2+2 q_1 q_2 q_8 q_9+2 q_1 q_6^2 q_{10}+2 q_1 q_3 q_8
   q_{10}-2 q_7^2 q_9 q_{10}-2 q_2 q_3 q_9 q_{10}\nonumber\\
   &&-4 q_1 q_4 q_9 q_{10}-2 q_2 q_6^2
   q_{11}-2 q_7^2 q_8 q_{11}-4 q_2 q_3 q_8 q_{11}\nonumber\\
   &&-2 q_1 q_4 q_8 q_{11}+2 q_2 q_4
   q_9 q_{11}+2 q_3 q_4 q_{10} q_{11}\nonumber\\
   &&+2 \sqrt{2} q_6 q_7 q_{10} q_{12}-2 q_1 q_6
   q_8 q_{13}-2 q_2 q_6 q_9 q_{13}+2 q_3 q_6 q_{10} q_{13}\nonumber\\
   &&+2 q_4 q_6 q_{11}
   q_{13}-2 \sqrt{2} q_7 q_8 q_{12} q_{13}+2 q_1 q_7 q_8 q_{14}\nonumber\\
   &&+2 q_2 q_7 q_9
   q_{14}+2 q_3 q_7 q_{10} q_{14}+2 q_4 q_7 q_{11} q_{14}-2 \sqrt{2} q_2 q_6 q_{12}
   q_{14}+2 \sqrt{2} q_4 q_{12} q_{13} q_{14}\nonumber\\
   \label{I4inv}
\end{eqnarray}
\subsection{\sc Truncation to the $STU$-model}
\label{truncazia}
Next we analyze how the $STU$-model is embedded into the $\Sp(6,\mathbb{R})$-model.
At the level of the special coordinates the truncation to the $STU$-model is very simply done. It suffices to
set to zero the complex coordinates $z_2,z_3,z_5$ keeping only $z_1,z_4,z_6$ that can be identified with the fields $S,T,U$. When we do so the symplectic section reduces as follows:
\begin{equation}\label{stuSymsec}
 \Omega\left[Z \left(
                 \begin{array}{ccc}
                   z_1 & 0 & 0 \\
                   0 & z_4 & 0 \\
                   0 & 0 & z_6 \\
                 \end{array}
               \right)
 \right] \, = \,  \left(
\begin{array}{l}
 \sqrt{2} z_6 \\
 \sqrt{2} z_1 z_6 \\
 -\sqrt{2} z_4 z_6 \\
 \sqrt{2} z_1 z_4 z_6 \\
 0 \\
 0 \\
 0 \\
 \sqrt{2} z_1 z_4 \\
 -\sqrt{2} z_4 \\
 \sqrt{2} z_1 \\
 \sqrt{2} \\
 0 \\
 0 \\
 0
\end{array}
\right)
\end{equation}
and the K\"ahler potential reduces to:
\begin{equation}\label{stuKalpot}
    \mathcal{K} \, \rightarrow\,  -\log \left[2 i \left(z_1-{\bar z}_1\right) \left(z_4-{\bar z}_4\right)
   \left(z_6-{\bar z}_6\right)\right]
\end{equation}
which yields three copies of the Poincar\'e metric, one for each of the three $\frac{\mathrm{SL(2,\mathbb{R})}}{\mathrm{SO(2)}}$
submanifolds.
\par
The result (\ref{stuSymsec}) is in agreement with the decomposition of the $\mathbf{14}^\prime$ of $\sym(6,\mathbb{R})$ with respect to the three subalgebras $\slal(2)$:
\begin{equation}\label{decumpo}
    \mathbf{14}^\prime \, \stackrel{\slal(2) \times \slal(2) \times \slal(2)}{\Longrightarrow} \, \left(\mathbf{2,2,2}\right)
    \oplus \left(\mathbf{2,1,1}\right) \oplus \left(\mathbf{1,2,1}\right) \oplus  \left(\mathbf{1,1,2}\right)
\end{equation}
From (\ref{stuSymsec}) we also learn that the directions $\{1,2,3,4,8,9,10,11\}$ of the $\mathbf{14}^\prime$ vector space span the representation $\left(\mathbf{2,2,2}\right)$, while the directions $\{5,6,7,12,13,14\}$ of the same space span the representations $\left(\mathbf{2,1,1}\right) \oplus \left(\mathbf{1,2,1}\right) \oplus  \left(\mathbf{1,1,2}\right)$.
The adjoint representation of $\sym(6,\mathbb{R})$ decomposes instead in the following way:
\begin{eqnarray}
    \mbox{adj} \,\left[\sym(6,\mathbb{R})\right] & \stackrel{\slal(2) \times \slal(2) \times \slal(2)}{\Longrightarrow} &
    \left(\mathbf{3,1,1}\right) \oplus \left(\mathbf{1,3,1}\right) \oplus \left(\mathbf{1,1,3}\right) \nonumber\\
    &&
    \oplus \left(\mathbf{2,2,1}\right) \oplus \left(\mathbf{2,1,2}\right) \oplus \left(\mathbf{1,2,2}\right) \label{adjointo}
\end{eqnarray}
as it is evident by a quick inspection of the roots (\ref{cornette}). In terms of the Cartan-Weyl basis the three
$\mathrm{\slal(2,\mathbb{R})}$ subalgebra contains the three Cartan generators $\mathcal{H}_i$ and the step operators
$\mathcal{E}^{\pm \alpha_3}$, $\mathcal{E}^{\pm \alpha_7}$ , $\mathcal{E}^{\pm \alpha_9}$. The remaining 12 step operators
span the representation $\left(\mathbf{2,2,1}\right) \oplus \left(\mathbf{2,1,2}\right) \oplus \left(\mathbf{1,2,2}\right)$, namely:
\begin{equation}\label{spannone}
   \left(\mathbf{2,2,1}\right) \oplus \left(\mathbf{2,1,2}\right) \oplus \left(\mathbf{1,2,2}\right) \, = \, \mbox{span} \left[
   \mathcal{E}^{\pm \alpha_1}, \mathcal{E}^{\pm \alpha_2} , \mathcal{E}^{\pm \alpha_4} ,\mathcal{E}^{\pm \alpha_5}, \mathcal{E}^{\pm \alpha_6} , \mathcal{E}^{\pm \alpha_8}
   \right]
\end{equation}
The explicit form of an $\sym(6,\mathbb{R})$ Lie algebra element reduced to the $\slal(2)^3$ subalgebra is the following one:
\begin{equation}\label{pagnocco}
    \left(
\begin{array}{llllll}
 h_1 & 0 & 0 & b_1 & 0 & 0 \\
 0 & h_2 & 0 & 0 & b_2 & 0 \\
 0 & 0 & h_3 & 0 & 0 & b_3 \\
 c_1 & 0 & 0 & -h_1 & 0 & 0 \\
 0 & c_2 & 0 & 0 & -h_2 & 0 \\
 0 & 0 & c_3 & 0 & 0 & -h_3
\end{array}
\right) \, \in \, \slal(2) \otimes \slal(2) \otimes \slal(2) \, \subset \, \sym(6,\mathbb{R})
\end{equation}
\subsection{\sc Reduction of the charge vector to the $(2,2,2)$}
In order to study the orbits of the charge vectors in the $\mathbf{14}^\prime$ our first step consists of reducing it to normal form, namely to the $(2,2,2)$ representation. We claim that for generic charge vectors this is always possible by means of $\Sp(6,\mathbb{R})$ rotations generated by elements of the $\left(\mathbf{2,2,1}\right) \oplus \left(\mathbf{2,1,2}\right) \oplus \left(\mathbf{1,2,2}\right)$ subspace. To show this let us consider the six dimensional compact Lie algebra element:
\begin{eqnarray}\label{cappotto}
    \mathbb{K}_\psi & = & \psi_1 \,\left(\mathcal{E}^{\alpha_1} -\mathcal{E}^{-\alpha_1}\right) +\psi_2 \, \left(\mathcal{E}^{\alpha_2} -\mathcal{E}^{-\alpha_2}\right) +\psi_3 \, \left(\mathcal{E}^{\alpha_4} -\mathcal{E}^{-\alpha_4}\right)\nonumber\\
    &&\psi_4 \,\left(\mathcal{E}^{\alpha_5} -\mathcal{E}^{-\alpha_5}\right) +\psi_5 \, \left(\mathcal{E}^{\alpha_6} -\mathcal{E}^{-\alpha_6}\right) +\psi_6 \, \left(\mathcal{E}^{\alpha_8} -\mathcal{E}^{-\alpha_8}\right)
\end{eqnarray}
and a generic charge vector that has components only in the $\mathbf{(2,2,2)}$ subspace.
\begin{equation}\label{toroidallo}
    \mathcal{Q}_{2,2,2} \, = \, \left\{\Theta _1,\Theta _2,\Theta _3,\Theta _4,0,0,0,\Theta _5,\Theta _6,\Theta
   _7,\Theta _8,0,0,0\right\}
\end{equation}
If we apply the $\mathbf{14}^\prime$ representation of $\mathbb{K}_\psi$ to the charge vector $\mathcal{Q}_N$ we obtain:
\begin{equation}\label{gorgonzola}
    \mathcal{D}_{14}\left( \mathbb{K}_\psi\right) \,\mathcal{Q}_{2,2,2} \, = \, \left(
\begin{array}{l}
 0 \\
 0 \\
 0 \\
 0 \\
 -\sqrt{2} \Theta _2 \psi _2+\sqrt{2} \Theta _5 \psi _2-\sqrt{2} \Theta _4 \psi
   _4-\sqrt{2} \Theta _7 \psi _4 \\
 -\sqrt{2} \Theta _3 \psi _3-\sqrt{2} \Theta _5 \psi _3+\sqrt{2} \Theta _4 \psi
   _5-\sqrt{2} \Theta _6 \psi _5 \\
 \sqrt{2} \Theta _2 \psi _1+\sqrt{2} \Theta _3 \psi _1-\sqrt{2} \Theta _1 \psi
   _6-\sqrt{2} \Theta _4 \psi _6 \\
 0 \\
 0 \\
 0 \\
 0 \\
 -\sqrt{2} \Theta _1 \psi _2-\sqrt{2} \Theta _6 \psi _2+\sqrt{2} \Theta _3 \psi
   _4-\sqrt{2} \Theta _8 \psi _4 \\
 \sqrt{2} \Theta _1 \psi _3-\sqrt{2} \Theta _7 \psi _3+\sqrt{2} \Theta _2 \psi
   _5+\sqrt{2} \Theta _8 \psi _5 \\
 \sqrt{2} \Theta _6 \psi _1+\sqrt{2} \Theta _7 \psi _1-\sqrt{2} \Theta _5 \psi
   _6-\sqrt{2} \Theta _8 \psi _6
\end{array}
\right)
\end{equation}
which clearly shows that the six parameters $\psi_{1,\dots,6}$ are sufficient to generate arbitrary components $\{5,6,7,12,13,14\}$ of the charge vector starting from vanishing ones. Reverting the path this means that by means of the same rotations, apart from singular orbits that deserve a separate study we can always fix the gauge where the six components $\{5,6,7,12,13,14\}$ vanish.
\subsubsection{\sc Further reduction to normal form of the charge vector}
Once the charge vector is reduced to $(2,2,2)$ representation, we can further act on it with the $\mathrm{SL(2,\mathbb{R})}^3$ group in order to further reduce its components. By using the three parameters of the abelian translation group $\mathbb{R}^3$ contained in $\mathrm{SL(2,\mathbb{R})}^3$ we can put to zero three of the eight charges and a possible normal form of the charge vector is the following one:
\begin{equation}\label{normaldata}
    \mathcal{Q}_{N} \, = \,  \left\{0,P_1,P_2,P_3,0,0,0,P_4,0,0,P_5,0,0,0\right\}
\end{equation}
The corresponding quartic invariant is:
\begin{equation}\label{normaldata2}
    \mathfrak{J}_4 \left(\mathcal{Q}_{N}\right)\, = \, P_3^2 P_5^2-4 P_1 P_2 P_4 P_5
\end{equation}
\section{\sc The $\frac{\mathrm{F_{(4,4)}}}{\mathrm{SU(2)} \times \mathrm{USp(6)}}$ quaternionic K\"ahler manifold}
Let us now come to the $c$-map image of the Special K\"ahler manifold (\ref{gurto}), namely to the quaternionic K\"ahler manifold (\ref{quatergurto}). The $F_{(4,4)}$ Lie algebra has rank four and its structure is codified in the Dynkin diagram presented in fig.\ref{F4dynk}.
\begin{figure}
\centering
\begin{picture}(90,50)
      \put (-70,35){$F_4$}  \put
(10,35){\circle {10}} \put (7,20){$\beta_4$} \put (15,35){\line
(1,0){20}} \put (40,35){\circle {10}} \put (37,20){$\beta_3$}\put (45,38){\line (1,0){20}}
\put (55,35){\line (1,1){10}} \put (55,35){\line (1,-1){10}}\put (45,33){\line (1,0){20}} \put
(70,35){\circle {10}} \put (67,20){$\beta_{2}$} \put
(75,35){\line (1,0){20}} \put (100,35){\circle {10}} \put (100,35){\circle {9}}\put (100,35){\circle {8}}\put (100,35){\circle {10}}\put (100,35){\circle {7}}
\put (100,35){\circle {6}}\put (100,35){\circle {5}}\put (100,35){\circle {4}}\put (100,35){\circle {3}}\put (100,35){\circle {2}}\put (100,35){\circle {1}}\put
(97,20){$\beta_{1}$}
\end{picture}
$$ \begin{array}{l}\psi \, = \, \beta_{24} \, = \,  2\beta_1 +3\beta_2 +4\beta_3+2\beta_4   \\(\psi \, , \,\beta_1) = 2
\quad; \quad  (\psi \, , \, \beta_i ) = 0 \quad i \ne 1 \
\end{array}     $$
\vskip 1cm \caption{The Dynkin diagram of $F_{4(4)}$. The only root which is not orthogonal
to the highest root is $\beta_V = \beta_1$. The root $\beta_V = \beta_1 $
is the highest weight of the $\mathbf{W}$-representation of $\sym(6,\mathbb{R})$  \label{F4dynk}}
\end{figure}
The complete set of positive roots contains $24$ elements that are listed below:
\begin{equation}\label{radicioneF4}
\begin{array}{lllll}
 \beta _1 & = & \beta _1 & = & \{-1,-1,-1,1\} \\
 \beta _2 & = & \beta _2 & = & \{0,0,2,0\} \\
 \beta _3 & = & \beta _3 & = & \{0,1,-1,0\} \\
 \beta _4 & = & \beta _4 & = & \{1,-1,0,0\} \\
 \beta _5 & = & \beta _1+\beta _2 & = & \{-1,-1,1,1\} \\
 \beta _6 & = & \beta _2+\beta _3 & = & \{0,1,1,0\} \\
 \beta _7 & = & \beta _3+\beta _4 & = & \{1,0,-1,0\} \\
 \beta _8 & = & \beta _1+\beta _2+\beta _3 & = &
   \{-1,0,0,1\} \\
 \beta _9 & = & \beta _2+2 \beta _3 & = & \{0,2,0,0\} \\
 \beta _{10} & = & \beta _2+\beta _3+\beta _4 & = &
   \{1,0,1,0\} \\
 \beta _{11} & = & \beta _1+\beta _2+2 \beta _3 & = &
   \{-1,1,-1,1\} \\
 \beta _{12} & = & \beta _1+\beta _2+\beta _3+\beta _4 & = &
   \{0,-1,0,1\} \\
 \beta _{13} & = & \beta _2+2 \beta _3+\beta _4 & = &
   \{1,1,0,0\} \\
 \beta _{14} & = & \beta _1+2 \beta _2+2 \beta _3 & = &
   \{-1,1,1,1\} \\
 \beta _{15} & = & \beta _1+\beta _2+2 \beta _3+\beta _4 & =
   & \{0,0,-1,1\} \\
 \beta _{16} & = & \beta _2+2 \beta _3+2 \beta _4 & = &
   \{2,0,0,0\} \\
 \beta _{17} & = & \beta _1+2 \beta _2+2 \beta _3+\beta _4 &
   = & \{0,0,1,1\} \\
 \beta _{18} & = & \beta _1+\beta _2+2 \beta _3+2 \beta _4 &
   = & \{1,-1,-1,1\} \\
 \beta _{19} & = & \beta _1+2 \beta _2+3 \beta _3+\beta _4 &
   = & \{0,1,0,1\} \\
 \beta _{20} & = & \beta _1+2 \beta _2+2 \beta _3+2 \beta _4
   & = & \{1,-1,1,1\} \\
 \beta _{21} & = & \beta _1+2 \beta _2+3 \beta _3+2 \beta _4
   & = & \{1,0,0,1\} \\
 \beta _{22} & = & \beta _1+2 \beta _2+4 \beta _3+2 \beta _4
   & = & \{1,1,-1,1\} \\
 \beta _{23} & = & \beta _1+3 \beta _2+4 \beta _3+2 \beta _4
   & = & \{1,1,1,1\} \\
 \beta _{24} & = & 2 \beta _1+3 \beta _2+4 \beta _3+2 \beta
   _4 & = & \{0,0,0,2\}
\end{array}
\end{equation}
In eq.(\ref{radicioneF4}) the first column is the name of the root, the second column gives its decomposition in terms of simple roots, while the last column provides the component of the root vector in $\mathbb{R}^4$.
\par
The standard Cartan-Weyl form of the Lie algebra is as follows:
\begin{eqnarray}
  \left[ \mathcal{H}_i  \, , \, E^{\pm\beta}\right] &=& \pm \, \beta^i \, E^{\pm\beta_I} \, \label{weylus1} \\
  \left[ E^{\beta} \, , \, E^{-\,\beta}\right] &=&  \, \beta \, \cdot \, \mathcal{H} \, \label{weylus2} \\
   \left[ E^{\beta}\, , \, E^{\gamma}\right] &=& \left\{ \begin{array}{lc}N_{\beta \gamma} \, E^{\beta \, + \, \gamma}& \mbox{if $\beta+\gamma$ is a root}\\
   0 &\mbox{if $\beta+\gamma$ is  not a root}\end{array} \right. \label{weylus3}
\end{eqnarray}
where $N_{\beta \gamma}$ are numbers that can be predicted from Lie algebra theory. They are irrelevant, since they follows from commutators, when one has an explicit matrix realization of all the Cartan generators $\mathcal{H}_i$ and of the step operators $E^{\pm\beta}$. The fundamental representation of $F_{(4,4)}$ is $26$-dimensional and real. We have constructed in a MATHEMATICA code all the $26 \times 26$ matrix representations of the $52$ generators of the Cartan Weyl basis. Obviously we can not present them here because they are too big. However all the rest of the construction can be easily presented in terms of these Weyl generators sand this is what we presently do.
\subsection{\sc The maximal compact subalgebra $\mathbb{H} \, = \, \su(2) \oplus \usp(6)$}
The maximal compact subalgebra $\mathbb{H}$ of a maximally split simple Lie algebra such as $F_{(4,4)}$, is just the real span of all the independent compact generators $E^{\beta_i} \, - \, E^{- \beta_i}$. In our case we have $24$ positive roots and we can write:
\begin{equation}\label{algebraH}
    \mathbb{H} \, = \, \mbox{span}_\mathbb{R} \, \left\{ H_1\, , \, H_2 \, , \, \dots \, , \, H_{24} \right\}
\end{equation}
where we have defined:
\begin{equation}\label{Higene}
    H_i \, = \, E^{\beta_i} \, - \, E^{- \beta_i}
\end{equation}
the positive roots being numbered as in eq.(\ref{radicioneF4}).  We know from theory that this maximal compact subalgebra has the structure:
\begin{equation}\label{furtivo}
    \mathbb{H} \, = \, \su(2) \oplus \usp(6)
\end{equation}
It is important to derive an explicit basis of generators satisfying the standard commutation relations of the two simple factors in eq.(\ref{furtivo}) for holonomy calculations of the coset manifold. Particularly important are the three generators $J^x$ of the $\su(2)$ subalgebra since they will act as quaternionic complex structures in the calculation of the tri-holomorphic moment map. By means of standard techniques of diagonalization of the adjoint action of generators we have retrieved the required basis rearrangement.
\subsubsection{\sc The $\su(2)$ Lie algebra} The three generators $J^x$ have tho following explicit form:
\begin{eqnarray}
  J^1 &=& \frac{H_1-H_{14}+H_{20}-H_{22}}{4 \sqrt{2}} \nonumber\\
  J^2 &=& \frac{H_5+H_{11}-H_{18}+H_{23}}{4 \sqrt{2}}\nonumber \\
  J^3 &=& -\frac{H_2-H_9+H_{16}+H_{24}}{4 \sqrt{2}}\label{su2generati}
\end{eqnarray}
and close the standard commutation relations:
\begin{equation}\label{stundasu2}
    \left[ J^x  \, , \, J^y \right] \, = \, \epsilon^{xyz} \, J^y
\end{equation}
\subsubsection{\sc The $\usp(6)$ Lie algebra} The $21$ generators of the $\usp(6)$ Lie algebra are given by the following combinations.
First we have three mutually commuting generators (the compact Cartan generators):
\begin{equation}\label{cartacantabene}
    \left[ \mathcal{L}^i  \, , \, \mathcal{L}^j  \right]\, = \, 0
\end{equation}
that are given by the following combinations:
\begin{equation}\label{compacarta}
   \begin{array}{lll}
 \mathcal{L}^1 & = &
   -\frac{H_2}{2}-\frac{H_9}{2}+\frac{H_{16
   }}{2}-\frac{H_{24}}{2} \\
 \mathcal{L}^2 & = &
   -\frac{H_2}{2}+\frac{H_9}{2}+\frac{H_{16
   }}{2}+\frac{H_{24}}{2} \\
 \mathcal{L}^3 & = &
   \frac{H_2}{2}+\frac{H_9}{2}+\frac{H_{16}
   }{2}-\frac{H_{24}}{2}
\end{array}
\end{equation}
Secondly we have $9$ pairs of generators $\left\{X_{i}\, , \, Y_{i}\right\}$ which are in correspondence with the $9$ positive roots of the $\sym(6,C)$ Lie algebra (see eq.(\ref{cornette}). Explicitly we have:
\begin{equation}\label{coppietteXY}
    \begin{array}{lllllll}
 X_1 & = & H_{10} & ; &
   Y_1 & = & H_7 \\
 X_2 & = & H_4 & ; & Y_2 &
   = & -H_{13} \\
 X_3 & = & H_6 & ; & Y_3 &
   = & -H_3 \\
 X_4 & = &
   -H_1+H_{14}+H_{20}-H_{22} & ; &
   Y_4 & = &
   -H_5-H_{11}-H_{18}+H_{23} \\
 X_5 & = & H_{21} & ; &
   Y_5 & = & -H_8 \\
 X_6 & = & H_1+H_{14}+H_{20}+H_{22}
   & ; & Y_6 & = &
   H_5-H_{11}-H_{18}-H_{23} \\
 X_7 & = &
   -H_1-H_{14}+H_{20}+H_{22} & ; &
   Y_7 & = &
   H_5-H_{11}+H_{18}+H_{23} \\
 X_8 & = & H_{17} & ; &
   Y_8 & = & H_{15} \\
 X_9 & = & H_{12} & ; &
   Y_9 & = & H_{19}
\end{array}
\end{equation}
The commutation relations with the compact Cartan generators are as follows:
\begin{equation}\label{cirocondo}
    \left[ \mathcal{L}^i \, , \, X_I  \right] \, = \, \alpha_I^i \, Y_I \quad ; \quad \left[ \mathcal{L}^i \, , \, Y_I  \right] \, = \, - \, \alpha_I^i \, X_I
\end{equation}
where $\alpha_I$ are the roots of eq.(\ref{cornette}). The remaining commutation relations mix the $Y$ and the $X$ among themselves and reproduce the Cartan generators.
\subsection{\sc The  subalgebra $\slal(2,\mathbb{R})_E \, \oplus \, \sym(6,\mathbb{R})$ and the $\mathbf{W}$-generators}
Of great relevance in all applications of the (pseudo)-quaternionic geometry either in the construction of Black-Hole solutions or in the quest of inflaton potentials by means of the gauging of hypermultiplet isometries is the identification of the subalgebra:
\begin{equation}\label{cannula}
    \slal(2,\mathbb{R})_E \, \oplus \, \sym(6,\mathbb{R}) \, \subset \, \mathfrak{f}_{(4,4)}
\end{equation}
and the recasting of $\mathfrak{f}_{(4,4)}$ in the general form \ref{genGD3} by means of the identification of the $\mathbf{W}$-generators.
\par
To this effect a very powerful tool is provided by the comparison of  the $\mathfrak{f}_{(4,4)}$ root system displayed in eq.(\ref{radicioneF4}) with the $\sym(6,\mathbb{R})$ root system displayed in eq.(\ref{cornette}). The step operators associated with the highest (lowest) root $\pm\beta_{24}$ are the only ones that have a grading $\pm 2$ with respect to the fourth Cartan generator $\mathcal{H}_4$. These three operators close among themselves the Lie algebra $\slal(2,\mathbb{R})_E $. There are $9$ roots that have grading zero with respect to $\mathcal{H}_4$. Projected onto the plane $\mathcal{H}_4\, = \,0$ these $9$ roots form, together with their negatives, a $\sym(6,\mathbb{R})$ root system. Correspondingly the $\sym(6,\mathbb{R})$ subalgebra is generated by the step operators associated with these $9$ roots (and with their negaives) plus the first $3$ Cartan generators. Finally there are $14$ positive roots $\beta$ that have have grading $1$ with respect to $\mathcal{H}_4$. The step operators associated with these $14$ roots form the $\mathbf{W}$-generators with index $1$ of $\mathrm{SL(2,\mathbb{R})}_E$. Their partners with index $2$ are provided by the corresponding negative root step operators.
\par
It is quite important to arrange the generators $\mathbf{W}$ in such a way that under any element $\mathfrak{g }\, \in \, \sym(6,\mathbb{R}) \, \subset \, \mathfrak{f}_{(4,4)} $ they transform exactly with the $\mathcal{D}_{14}(\mathfrak{g})$ matrices defined in eq.s(\ref{cartolini14}), (\ref{steppinisu14}) and (\ref{steppinigiu14}).
\par
The precise definition of all the generators that satisfy the specified requirements is given below.
\subsubsection{\sc The Ehlers subalgebra $\slal(2,\mathbb{R})_E$.} The standard commutation relations:
\begin{eqnarray}
  \left[ L_0^E \, , \, L^E_\pm\right] &=& \pm \, L^E_\pm \\
  \left[ L^E_+\, , \, L^E_- \right] &=& 2 \, L^E_0 \label{EhlersF4}
\end{eqnarray}
are satisfied by the following generators:
\begin{eqnarray}
  L_0^E&=& \frac{1}{2} \, \mathcal{H}_4 \nonumber \\
  L_+^E &=& \frac{1}{\sqrt{2}} \, \mathcal{E}^{\beta_{24}}\nonumber \\
  L_- ^E&=& \frac{1}{\sqrt{2}} \, \mathcal{E}^{- \,\beta_{24}} \label{EhlersF4BIS}
\end{eqnarray}
\subsubsection{\sc The subalgebra $\sym(6,\mathbb{R})$.} The Cartan generators are the following ones:
\begin{eqnarray}
\mathcal{H}_1 &=& \mathcal{H}_1 \nonumber\\
 \mathcal{H}_2 &=& \mathcal{H}_2 \nonumber\\
  \mathcal{H}_3 &=& \mathcal{H}_3 \label{samicarti}
\end{eqnarray}
while the step operators are identified as follows
\begin{equation}
    \begin{array}{lll}
 \mathcal{E}^{\pm \alpha _1} & = & \mathcal{E}^{\pm\beta_{4}} \\
 \mathcal{E}^{\pm \alpha _2} & = & \mathcal{E}^{\pm\beta_{3}} \\
 \mathcal{E}^{\pm \alpha _3} & = & \mathcal{E}^{\pm\beta_{2}} \\
 \mathcal{E}^{\pm \alpha _4} & = & \mathcal{E}^{\pm\beta_{7}} \\
 \mathcal{E}^{\pm \alpha _5} & = & -\mathcal{E}^{\pm\beta_{6}}\\
 \mathcal{E}^{\pm \alpha _6} & = & \mathcal{E}^{\pm\beta_{10}}\\
 \mathcal{E}^{\pm \alpha _7} & = & -\mathcal{E}^{\pm\beta_{9}}\\
 \mathcal{E}^{\pm \alpha _8} & = & \mathcal{E}^{\pm\beta_{13}}\\
 \mathcal{E}^{\pm \alpha _9} & = & \mathcal{E}^{\pm\beta_{16}}
\end{array}\label{fognato}
\end{equation}
We would like to attract the attention of the reader on the two minus signs introduced in the identifications (\ref{fognato}). Together with the other minus signs that appear below in the identification of the $W$-generators these signs are essential in order for the transformations of the $W$.s to be identical with those given by the  previously defined $\mathcal{D}_{14}(\mathfrak{g})$ matrices.
\subsubsection{\sc The $\mathbf{W}$-generators} Casting the $\mathfrak{f}_{(4,4)}$ Lie algebra in the general form (\ref{genGD3}) is completed by the identification of the $\mathbf{W}$-generators. We find:
\begin{equation}\label{widentifio}
   \begin{array}{lll}
 \mathbf{W}^{1,1} & = & \mathcal{E}^{ \beta_{5}} \\
 \mathbf{W}^{1,2} & = & \mathcal{E}^{ \beta_{20}} \\
 \mathbf{W}^{1,3} & = & \mathcal{E}^{ \beta_{14}} \\
 \mathbf{W}^{1,4} & = & -\mathcal{E}^{ \beta_{23}} \\
 \mathbf{W}^{1,5} & = & \mathcal{E}^{ \beta_{21}} \\
 \mathbf{W}^{1,6} & = & \mathcal{E}^{ \beta_{19}} \\
 \mathbf{W}^{1,7} & = & -\mathcal{E}^{ \beta_{17}} \\
 \mathbf{W}^{1,8} & = & -\mathcal{E}^{ \beta_{22}} \\
 \mathbf{W}^{1,9} & = & -\mathcal{E}^{ \beta_{11}} \\
 \mathbf{W}^{1,10} & = & -\mathcal{E}^{ \beta_{18}} \\
 \mathbf{W}^{1,11} & = & -\mathcal{E}^{ \beta_{1}} \\
 \mathbf{W}^{1,12} & = & -\mathcal{E}^{ \beta_{8}} \\
 \mathbf{W}^{1,13} & = & -\mathcal{E}^{ \beta_{12}} \\
 \mathbf{W}^{1,14} & = & -\mathcal{E}^{ \beta_{15}}
\end{array}
\end{equation}
and for all  $\mathfrak{g }\, \in \, \sym(6,\mathbb{R}) \, \subset \, \mathfrak{f}_{(4,4)} $ we have:
\begin{equation}\label{ciucciatiquesta}
    \left[ \mathfrak{g } \, , \, \mathbf{W}^{1,\alpha} \right] \, = \, \mathcal{D}_{14}(\mathfrak{g})^\alpha_{\phantom{\alpha}\gamma} \, \mathbf{W}^{1,\gamma}
\end{equation}
The generators $\mathbf{W}^{2,\alpha}$ are then easily obtained from by means of a rotation with the unique compact generator of the Ehlers subalgebra introduced in eq.(\ref{ruotogrande}):
\begin{equation}\label{fungatorepazzo}
    \left[ \mathfrak{S}  \, , \, \mathbf{W}^{1,\alpha} \right] \, = \,\mathbf{W}^{2,\alpha}
\end{equation}
\subsection{\sc The solvable coset representative}
The precise constructions of the previous sections enable us to introduce the solvable coset representative $\mathbb{L}_{Solv}\left(a,U,h,p,Z\right)$ of the manifold (\ref{quatergurto}) such that the Maurer Cartan form:
\begin{equation}\label{maurocartaneggia}
    \Xi \, \equiv \, \mathbb{L}_{Solv}^{-1} \, \mathrm{d} \mathbb{L}_{Solv}
\end{equation}
decomposed along the generators of the Borel Lie algebra:
\begin{eqnarray}\label{solubilino}
    \Xi &  = & E^I_{\mathcal{QM}} \, T_I\nonumber\\
    T_I & = & \left\{ \underbrace{L_0^E \, , \, L^E_+ }_{2 \,\hookrightarrow\, Solv\left[\slal(2)\right]} \, , \, \underbrace{\mathcal{H}^i \, , \, \mathcal{E}^{\alpha_i}}_{12 \, \hookrightarrow \, Solv\left[\sym(6)\right]} \, ,\, \underbrace{\mathbf{W}^{1\alpha}}_{14 \, \hookrightarrow \, \mathbb{H}\mathrm{eis}}\right\}
\end{eqnarray}
provides the vielbein $E^I_{\mathcal{QM}}$ mentioned in eq.(\ref{filibaine}) and by squaring the metric (\ref{cornish}).
\par
In full analogy with eq.s (\ref{fornarina}) and (\ref{baltico}) we write:
\begin{eqnarray}\label{SolvCosF4}
    \mathbb{L}_{Solv} & = & \exp \left[ a \, L^E_+\right] \, \cdot \, \exp\left[\sum_{j=1}^7 \, \mathbf{Z}_{2j-1} \, \mathbf{W}^{1,2j-1}\right] \, \cdot \, \exp\left[\sum_{j=1}^7 \, \mathbf{Z}_{2j} \, \mathbf{W}^{1,2j}\right] \, \times \nonumber \\
    && \times  \prod_{i=1}^9 \, \exp\left [ p_{10-i} \, \mathcal{E}^{\alpha_{10-i}}\right ] \, \cdot \, \prod_{j=3}^3 \exp\left [ h_{j} \, \mathcal{H}^{j}\right ] \, \cdot \, \exp\left[U \, L^E_0 \right]
\end{eqnarray}
The explicit expression of $\mathbb{L}_{Solv}$ in the fundamental $26$-dimensional representation is obviously very large but it can be dealt with by means of an appropriate MATHEMATICA code.
\par
We are finally in the position of calculating the tri-holomorphic moment map of any element $ \mathfrak{t}\, \in \, \mathfrak{f}_{(4,4)}$
 of the isometry Lie algebra of $\mathcal{QM}$ through the formula:
 \begin{equation}\label{cordiglierAndina}
    \mathcal{P}^x_\mathfrak{t} \, = \, \mbox{Tr}_{\mathbf{26}} \, \left( J^x \, \mathbb{L}_{Solv}^{-1} \, \mathfrak{t} \, \mathbb{L}_{Solv} \right)
 \end{equation}
\subsection{\sc The example of the inclusion of  multi Starobinsky models}
\label{generalonuovo}
In section \ref{truncazia} we studied the truncation of the $\sym(6,\mathbb{R})$ model to the STU model. There we showed that setting to zero the three complex coordinates $z_2,z_3,z_5$, the remaining ones $z_1,z_4,z_6$ span the STU model, namely they parameterize three copies of the Lobachevsky-Poincar\'e hyperbolic plane. Inspecting eq. (\ref{carlinus}) we also see that the three coordinates $z_1,z_4,z_6$ are the only ones that survive when all the axions $p_i$ are set to zero. We also recall from sect. \ref{truncazia} that the three parabolic generators of the three $\mathrm{SL(2,\mathbb{\mathbb{R}})}$ groups spanning the STU model are $\mathcal{E}^{\alpha_3},\, \mathcal{E}^{\alpha_7},\,\mathcal{E}^{\alpha_9}$ whose identification with $\mathfrak{f}_{(4,4)}$ generators is provided by eq.(\ref{fognato}). Correspondingly we introduce the following generator:
\begin{equation}\label{carnevaleSTU}
    \mathfrak{t}_{STU}\, = \, \beta_3 \, \mathcal{E}^{\alpha_3}\, + \, \beta_2 \,  \mathcal{E}^{\alpha_7}\, + \, \beta_1 \, \mathcal{E}^{\alpha_9}\, - \, \kappa \, \mathfrak{S}
\end{equation}
and we calculate its tri-holomorphic moment map, by means of eq.(\ref{cordiglierAndina}). Defining the potential:
\begin{equation}\label{gogamigoga}
    V_{STU} \, = \, \sum_{x=1}^3 \, \left(\mathcal{P}^x_{\mathfrak{t}_{STU}}\right)^2
\end{equation}
We can verify that:
\begin{eqnarray}
  \left.  \frac{\partial}{\partial \mathbf{Z}^\alpha} \, V_{STU}\right|_{\mathbf{Z}=U=a =0} & = &0 \nonumber\\
  \left.  \frac{\partial}{\partial U} \, V_{STU}\right|_{\mathbf{Z}=U=a =0} & = &0 \nonumber\\
  \left.  \frac{\partial}{\partial a} \, V_{STU}\right|_{\mathbf{Z}=U=a =0} & = &0 \label{UaZetaseneva}
\end{eqnarray}
Hence we can consistently truncate  $U$, $a$ and the Heisenberg fields $\mathbf{Z}$. We find:
\begin{equation}\label{multistarobbo}
   \left.  V_{STU}\right|_{\mathbf{Z}=U=a =0} \, = \, \frac{9}{4} \left( \, 2\kappa \,  - \, \sqrt{2} \sum_{i=1}^3
  \beta_i \,e^{-2 h_i} \right)^2
\end{equation}
The above potential can be named a multi-Starobinsky model with three independent dilatons.
\par
First of all let us note that in the above model the absolute value of $\beta_i$  is irrelevant since we can always reabsorb it by a constant shift $h_i \to h_i \, - \, \log|\beta_i|$. The only relevant thing are the signs of $\beta_i$ including in this notion also zero, namely $\beta_i $ can be  $\pm 1$ or $0$. Secondly we observe that when all the non vanishing $\beta_i$ have the same sign we can make a consistent one field truncation to
\begin{equation}\label{correlo}
  h_i \, = \, h \quad ; \quad \mbox{for all $i$ such that $\beta_i \ne 0$}
\end{equation}
After this truncation the potential (\ref{multistarobbo}) becomes the following:
\begin{equation}\label{riducione}
V_{eff} \, = \, \frac{9}{4} \left(\, 2\kappa \, - \, \sqrt{2} \, q\, e^{-2 h}\right)^2
\end{equation}
where $q$ is the number of equal sign non zero $\beta_i$, which obviously can take only three values $q=1,2,3$. In order to compare this result with the definition of $\alpha$-attractors introduced in \cite{alfatrattori}, we just have to compare the potential (\ref{riducione}) with the normalization of the scalar kinetic terms in the lagrangian:
\begin{equation}\label{kinescalo}
  \mathcal{L} \, = \, \dots \, + \, \frac{1}{4} \, (\partial U)^2 \, + \, (\partial h_1)^2\, + \, (\partial h_2)^2 \,  + \, (\partial h_3)^2 \, + \, \dots
\end{equation}
 which follows from eq.s(\ref{geodaction}, \ref{formidabile}). Renaming $h\, = \, \frac{1}{\sqrt{2 \, q}} \, \phi$, so that the new field $\phi$ has canonical kinetic term $\frac{1}{2} \, (\partial \phi)^2$, we obtain a potential:
 \begin{equation}\label{bambolone}
   V_{eff} \, = \, \mbox{const} \, \times \, \left( 2\kappa \, - \, \sqrt{2} \, q\, \exp\left [ - \, \sqrt{\frac{2}{q}} \, \phi\right]\right)^2
 \end{equation}
which, in the notation of  \cite{alfatrattori}, corresponds  to $\alpha = \frac{q}{3} $, namely to:
\begin{equation}\label{valoretti}
  \alpha \, = \, 1 \, , \, \frac{2}{3} \, , \, \frac{1}{3}
\end{equation}
The above result has been obtained by gauging only one generator, namely (\ref{carnevaleSTU}). Correspondingly we have generated Starobinsky-like models with only one massive vector that is the gauge vector associated with the gauged generator. There is another way of obtaining the same potential but with $q$-massive vectors (one for each constituent Starobinsky model with $q=\frac{1}{3}$). This is very simply understood remarking that the $\mathfrak{f}_{(4,4)}$ algebra contains an $\slal(2,\mathbb{R})^4$ subalgebra singled out as follows:
\begin{equation}\label{governolato}
  \mathfrak{f}_{(4,4)} \, \supset \, \slal(2,\mathbb{R})_E \, \oplus \, \underbrace{\slal(2,\mathbb{R})_S \, \oplus \, \slal(2,\mathbb{R})_T\, \oplus \, \slal(2,\mathbb{R})_U}_{\subset \, \sym(6,\mathbb{R})}
\end{equation}
where $\slal(2,\mathbb{R})_S \, \oplus \, \slal(2,\mathbb{R})_T\, \oplus \, \slal(2,\mathbb{R})_U$ describes the STU model embedded in the K\"ahler manifold (\ref{gurto}). These four $\slal(2,\mathbb{R})$ algebras are completely symmetric among themselves and the gauging of their generators produce identical results. So we can introduce the abelian gauge algebra spanned by the following  three commuting generators:
\begin{eqnarray}
  \mathfrak{t}_{S} &=&\beta_3 \, \mathcal{E}^{\alpha_3}\,  - \, \kappa_3 \, \mathfrak{S} \nonumber \\
  \mathfrak{t}_{T} &=&  \beta_2 \,  \mathcal{E}^{\alpha_7}\,  - \, \kappa_2 \, \mathfrak{S}\nonumber\\
    \mathfrak{t}_{U}  &=&  \, \beta_1 \, \mathcal{E}^{\alpha_9}\, - \, \kappa_1\, \mathfrak{S}
\end{eqnarray}
Gauging with three separate vectors each of the above generators we obtain a new  potential:
 \begin{equation}\label{STUnew}
   \widehat{V}_{STU} \, = \, \sum_{x=1}^3 \, \left(\mathcal{P}^x_{\mathfrak{t}_{S}}\right)^2 \, + \, \sum_{x=1}^3 \, \left(\mathcal{P}^x_{\mathfrak{t}_{T}}\right)^2 \, + \, \sum_{x=1}^3 \, \left(\mathcal{P}^x_{\mathfrak{t}_{U}}\right)^2 \,
 \end{equation}
 that has the same property as the potential (\ref{gogamigoga}), namely it allows us to   truncate consistently to zero all the axions $p_i$, all the Heisenberg fields $\mathbf{Z}^\alpha$ and the Taub NUT field $a$.  The reduced potential after such a truncation has the form:
 \begin{equation}\label{ciurlone}
  \widehat{V}_{red}\, = \,   \frac{9}{4}  \, \sum_{i=1}^3 \, \left( \, 2\kappa_i \, - \, \sqrt{2}
   e^{-2 h_i} \beta_i \right)^2
 \end{equation}
 As we already remarked before, the absolute value of the $\beta_i$ parameters is irrelevant: what matters is only the relative signs of the $\beta_i$ with respect to the sign of their corresponding $\kappa_i$. If for all non vanishing $\beta_i$ we have $\frac{\beta_i}{\kappa_i} \, = 1$, then we can consistently perform the same truncation (\ref{correlo}) as before and we reobtain the potentials (\ref{bambolone}) with the same spectrum of $\alpha$-values (\ref{valoretti}). The difference with the previous case is, as we emphasized at the beginning o this discussion, that now the number of massive fields is $q$, namely as many as the elementary non trivial constituent Starobinsky-like models.
 \subsection{\sc Nilpotent gaugings and truncations}
 \label{orbitando}
 Let us now put the above obtained results in the general framework discussed in sect.\ref{starobin1}. The issue is the classification of orbits of nilpotent operators and the question whether for each of these orbits we can find a consistent one-field reduction that produces a Starobinsky-like model with an appropriate value of $\alpha$.
 \par
 To answer this question we have followed the algorithm described in the second paper of \cite{noinilpotenti}. According to a general mathematical set, up to conjugation, every nilpotent orbit is associated with a standard triple $\left \{ x,y,h \right \}$ satisfying the standard commutation relations of the $\slal(2)$ Lie algebra, namely:
\begin{equation}\label{basictriple}
   \left [ h \, , \, x\right ] \, = \, x \quad ; \quad \left [ h \, , \, y\right ] \, = -\, y
   \quad ; \quad \left [ x \, , \, y \right ] \, = \, 2 \, h
\end{equation}
Interesting for us is the classification of nilpotent orbits in the K\"ahler subalgebra $\sym(6,\mathbb{R})$ and, according to the above mathematical theory, this is just the classification of embeddings of an $\slal(2)$ Lie algebra in the ambient one, modulo conjugation by the full group $\mathrm{Sp(6,\mathbb{R})}$.
The second relevant point emphasized in \cite{noinilpotenti} is that embeddings of subalgebras $\mathfrak{h} \subset \mathfrak{g}$ are characterized by the branching law of any representation of $\mathfrak{g}$ into irreducible representations of $\mathfrak{h}$. Clearly two embeddings might be conjugate only if their branching laws are identical. Embeddings with different branching laws necessarily belong to different orbits. In the case of the $\slal(2) \sim \so(1,2)$ Lie algebra, irreducible representations are uniquely identified by their spin $j$, so that the branching law is expressed by listing the angular momenta $\left\{ j_1 , j_2 , \dots j_n\right \}$ of the irreducible blocks into which any representation of the original algebra, for instance the fundamental, decomposes with respect to the embedded subalgebra. The dimensions of each irreducible module is $2j+1$ so that an a priori constraint on the labels $\left\{ j_1 , j_2 , \dots j_n\right \}$ characterizing an irreducible orbit of $\sym(6,\mathbb{R})$ is the summation rule:
\begin{equation}\label{summarulla}
    \sum_{i=1}^{n} (2 j_i +1) \, = \, 6\, = \, \mbox{dimension of the fundamental representation}
\end{equation}
Therefore we have considered all possible partitions of the number $6$ into integers and for each partition we have constructed a candidate $h$ element in the Cartan subalgebra of $\sym(6,\mathbb{R})$ containing as eigenvalues all the $J_3$ values of the corresponding $\left\{ j_1 , j_2 , \dots j_n\right \}$ representation. To clarify what we mean by this it suffices to consider the example of the first partition $6=6$. In this case the $6$ dimensional representation of $\slal(2)$ is the $j \, = \, \frac{5}{2}$ and the $6$ eigenvalues are $\pm \frac{5}{2}$, $\pm \frac{3}{2}$, $\pm \frac{1}{2}$. Having so fixed the so named central element $h$ of the candidate standard triplet we have tried to construct the corresponding $x$ and $y$. Imposing the standard commutation relations (\ref{basictriple}) one obtains quadratic equations on the coefficients of the linear combinations expressing the candidate $x$ and $y$ that may have or may not have solutions. If the solutions exist, then the corresponding standard triple is found, the orbit exists and we have constructed one representative $x$.
\par
Next, given the existing orbits and the corresponding standard triples, for each of them we have  constructed a Lobachevsky complex plane immersed in the Special K\"ahler manifold $\mathcal{M}_{Sp6}$ defined by eq.(\ref{gurto}). The construction is very simple. One calculates the group element $\mathfrak{g}(\lambda,\psi) \, \in \, \sym(6,\mathbb{R})$ defined below:
\begin{equation}\label{lambopsi}
    \mathfrak{g}(\lambda,\psi) \, = \, \exp\left[ \psi \, x\right] \, \cdot \, \, \exp\left[ \lambda \, h\right] \, = \, \left (\begin{array}{c|c}
                                                                                                                                  \mathbf{A}(\lambda,\psi) & \mathbf{B}(\lambda,\psi) \\
                                                                                                                                  \hline
                                                                                                                                  \mathbf{C}(\lambda,\psi) & \mathbf{D}(\lambda,\psi)
                                                                                                                                \end{array}
    \right)
\end{equation}
and using equation (\ref{Zmatra}), we write:
\begin{eqnarray}\label{embeddone}
    Z(\lambda\, , \, \psi) & = & \left( \mathbf{A}(\lambda,\psi) \, - \, {\rm i} \mathbf{B}(\lambda,\psi)\right) \, \cdot \, \left( \mathbf{C}(\lambda,\psi) \, - \, {\rm i} \mathbf{D}(\lambda,\psi)\right)^{-1} \, \nonumber\\
    & \equiv & \left( \begin{array}{ccc}
                                                                                                   z_1(\lambda,\psi) & z_2(\lambda,\psi) & z_3(\lambda,\psi) \\
                                                                                                   z_2(\lambda,\psi) & z_4(\lambda,\psi)  & z_5(\lambda,\psi)  \\
                                                                                                   z_3(\lambda,\psi) & z_5(\lambda,\psi)  & z_6(\lambda,\psi)
                                                                                                 \end{array}
    \right)
\end{eqnarray}
which defines the explicit embedding:
 \begin{equation}\label{embeddus}
    \phi \, : \, \frac{\mathrm{SL(2,\mathbb{R})}}{\mathrm{SO(2)}} \, \rightarrow \, \frac{\mathrm{Sp(6,\mathbb{R})}}{\mathrm{SU(3)} \times \mathrm{U(1)}} \, \equiv \, \mathcal{M}_{Sp6}
 \end{equation}
 of the Lobachevsky plane in $\mathcal{M}_{Sp6}$. Indeed from (\ref{embeddone}) we read off the parameterization of the complex coordinates $z_i$ ($i=1,\dots ,6$) as functions of $\lambda \, = \, \log \mbox{Im}\, w$ and $\psi \, = \, \mbox{Re}\,w$, the complex variable $w$ being the local variable over the embedded Poincar\'e-Lobachevsky plane.
\par
The question is whether the field equations of the scalar fields:
\begin{equation}\label{finocchione}
\partial_i \, \partial_{j^\star} \,\mathcal{K}\, \partial^\mu\partial_\mu \,\bar{z}^{j^\star}  \, + \,  \partial_i  \, \partial_{j^\star}\,\partial_{k^\star} \,  \mathcal{K} \, \partial_\mu z^{j^\star} \, \partial_\mu z^{k^\star}\, - \, \frac{1}{4}\,\partial_i \, V_{gauging}\left(z\, ,\, {\bar z}\right) \, = \, 0
\end{equation}
admit first a consistent reduction to the complex scalar field $w$ and then a consistent truncation to a vanishing axion $\psi \, = \, 0$. Consistency of the truncation can be verified or disproved in the following simple way. The pull-back on the immersed surface $\phi^\star \left(\frac{\mathrm{SL(2,\mathbb{R})}}{\mathrm{SO(2)}}\right) \, \subset \, \mathcal{M}_{Sp6}$ of the twelve field equations (\ref{finocchione}) (six complex equations) should be consistent among themselves and be identical with the two field equations obtained from the variation of the pull-back $\phi^\star(\mathcal{L})$ on the immersed surface of the Lagrangian $\mathcal{L}$ from which eq.s (\ref{finocchione}) derive, namely:
\begin{equation}\label{lagrandona}
 \mathcal{L} \, = \,   4 \, \partial_i \, \partial_{j^\star} \, \mathcal{K}\, \partial_\mu \,{z}^{i} \,\partial^\mu \,\bar{z}^{j^\star} \, - \, V_{gauging}\left(z\, ,\, {\bar z}\right)
\end{equation}
In other words, defining $ w\, = \, {\rm i} \, e^{\lambda} \, + \, \psi $, the truncation is consistent if the following diagram is commutative:
\begin{eqnarray}
    &&\begin{array}{ccc}
       \mathcal{L}(z,{\bar z}) & \stackrel{\phi^\star}{\Longrightarrow} & \phi^\star \mathcal{L}(w,{\bar w}) \\
       \downarrow  & \null & \downarrow \\
       \partial^\mu \, \frac{\partial \mathcal{L}}{\partial (\partial_\mu z)} \, - \, \frac{\partial \mathcal{L}}{\partial z} & \stackrel{\phi^\star}{\Longrightarrow} &  \partial^\mu \, \frac{\partial \phi^\star \mathcal{L}}{\partial (\partial_\mu w)} \, - \, \frac{\partial \phi^\star \mathcal{L}}{\partial w}
     \end{array} \label{coomodiag}
\end{eqnarray}
\begin{center}
\begin{tabular}{|c|c|c|c|}
  \hline
Partition&J.s&Orbit Name&One field reduction \\
\hline
  6=6 & $\left(\frac{5}{2}\right)$ & $\mathfrak{O}_1$ & NO \\
  6=5+1 & $\left(2,0\right)$ & Orbit does not exist & NO \\
  6=4+2 & $\left(\frac{3}{2},\frac{1}{2}\right)$ & $\mathfrak{O}_2$ & NO \\
  6=3+3 & $\left(1,1\right)$ & $\mathfrak{O}_3$ & NO \\
  6=3+2+1 & $\left(1,\frac{1}{2},0\right)$ &Orbit does not exist & NO \\
  6=3+1+1+1 & $\left(1,0,0,0\right)$ & Orbit does not exist & NO \\
  6=2+2+2 & $\left(\frac{1}{2},\frac{1}{2},\frac{1}{2}\right)$ & $\mathfrak{O}_4$ & YES \\
  6=2+2+1+1 & $\left(\frac{1}{2},\frac{1}{2},0,0\right)$ & $\mathfrak{O}_5$ & YES \\
6=2+1+1+1+1 & $\left(\frac{1}{2},0,0,0,0\right)$ & $\mathfrak{O}_6$ & YES \\
  \hline
\end{tabular}
\end{center}
In the above table we have summarized the results of this simple investigation. There is a total of six orbits (up to possible further splitting in Weyl group orbits which we have not analyzed) and for each of them the corresponding immersion formulae in the $\mathcal{M}_{\mathrm{Sp6}}$ manifolds are those described below.
\paragraph{Orbit $\mathfrak{O}_1$: ($j=\frac{5}{2}$).}
\begin{eqnarray}
 && \left( \begin{array}{ccc}
                                                                                                   z_1& z_2 & z_3\\
                                                                                                   z_2& z_4 & z_5 \\
                                                                                                   z_3 & z_5  & z_6
                                                                                                 \end{array}
    \right) = \, \nonumber\\
    && \left(
\begin{array}{lll}
 -6 \psi ^5+10 i e^{\lambda } \psi ^4+5 i e^{3 \lambda } \psi ^2+i
   e^{5 \lambda } & \sqrt{5} \psi  \left(3 \psi ^3-4 i e^{\lambda
   } \psi ^2-i e^{3 \lambda }\right) & i \sqrt{10} \left(i \psi
   +e^{\lambda }\right) \psi ^2 \\
 \sqrt{5} \psi  \left(3 \psi ^3-4 i e^{\lambda } \psi ^2-i e^{3
   \lambda }\right) & i \left(8 i \psi ^3+8 e^{\lambda } \psi
   ^2+e^{3 \lambda }\right) & \sqrt{2} \psi  \left(3 \psi -2 i
   e^{\lambda }\right) \\
 i \sqrt{10} \left(i \psi +e^{\lambda }\right) \psi ^2 & \sqrt{2}
   \psi  \left(3 \psi -2 i e^{\lambda }\right) & i e^{\lambda }-3
   \psi
\end{array}
\right) \nonumber\\
 && w \,=\, {\rm i} \, e^\lambda \, + \, \psi \label{immerO1}
\end{eqnarray}
The pull-back of the lagrangian is the following one:
\begin{equation}\label{orbittaprima}
    \phi^\star \mathcal{L} \, = \, 35 \, \left( \partial^\mu\psi\, \partial^\mu \psi  \, e^{-2\lambda} \, + \, \partial^\mu\lambda\, \partial^\mu \lambda\right) \, - \, \frac{1}{4}\,  g^2 \, \left(3\,e^{-\lambda} \, - \, \kappa\right)^2
\end{equation}
The pull-backs of the scalar field equations are inconsistent among themselves and  differ from the equations derived from the pull-back of the lagrangian (\ref{orbittaprima}), hence the truncation is not consistent. No Starobinsky--like model can be obtained from this orbit.
\par
One might wonder whether the inconsistency is due to the particularly chosen coset representative (\ref{lambopsi}) and to the explicit  form of the embedding (\ref{immerO1}) which turns out to be non-holomorphic. To clarify such a doubt and show that the inconsistency of the equations is  an intrinsic property of the orbit, we have addressed the problem from a different view point which leads  to a perfectly holomorphic embedding of the Lobachevsky plane associated with the considered orbit into the target Special K\"ahler manifold (\ref{gurto}).
\par
The argument is the following one. Having  fixed the embedding  $\slal(2,\mathbb{R})\mapsto \sym(6,\mathbb{R})$ at the level of the fundamental representation $\mathbf{6}$ it is fixed also in all other representations and we can wonder what is the branching rule of the $\mathbf{W}$-representation $\mathbf{14}^\prime$  such an embedding. By direct evaluation of the Casimir we obtain the following branching:
\begin{equation}\label{frikandello1}
  \mathbf{14}^\prime \, \stackrel{\slal(2,\mathbb{R})}{\longrightarrow} \, \left( j\, =\, \frac{9}{2}\right) \, \oplus \, \left( j\, =\, \frac{3}{2}\right)
\end{equation}
This means that the symplectic section (\ref{gargamelle}) splits into the sum of two  vectors, one lying in the 10-dimensional space of the first representation, the other in the $4$-dimensional space of the second representation. Imposing the vanishing of the lowest spin representation introduces a set of $4$ holomorphic constraints on the six coordinates $z_i$. By construction these constraints are $\slal(2,\mathbb{R})$ invariant: therefore  the sought for Lobachevsky plane certainly lies in the  complex two-folds defined by the vanishing of these constraints. With a little bit of work one can further eliminate one of the two remaining complex coordinates in such a way that the ten entries of the $ \left( j\, =\, \frac{9}{2}\right)$ representation correspond to all the powers $w^r$, with $r=0, 1, \dots , 9$  of a complex parameter $w$. Because of this very property $w$  can be interpreted as the  local coordinate of the sought for Lobachevsky plane embedded in the K\"ahler manifold (\ref{gurto}) according to the specified orbit. Indeed if $w$ transforms by fractional linear transformation under some algebra $\slal(2)$, then the $2j+1$ first powers of $w$ provide a basis for the $j$-representation of that $\slal(2)$. Viceversa, if a vector, which is known to transform in the $j$-represenation of a given $\slal(2)$ (up to an overall function of $w$),  is made by linear combinations of the first $2j+1$ powers of a coordinate $w$, then that $w$ is the local coordinate on a Lobachevsky plane transitive under the action of that very $\slal(2)$.
\par
In our case the four holomorphic constraints that express the vanishing of the $j=\frac{3}{2}$ representation inside the $14^\prime$ are the following ones:
\begin{eqnarray}
  \sqrt{\frac{2}{7}} \left(\sqrt{5} \left(z_4
   z_6-z_5^2\right)-2 z_2\right) &=& 0 \nonumber \\
  \frac{8 z_3-\sqrt{10} z_4}{\sqrt{21}}&=& 0 \nonumber\\
  \sqrt{\frac{2}{7}} \left(\sqrt{5} z_1+2 z_3 z_5-2 z_2
   z_6\right) &=& 0 \nonumber \\
  \frac{-\sqrt{10} z_3^2+8 z_4 z_3-8 z_2 z_5+\sqrt{10}
   z_1 z_6}{\sqrt{21}}&=& 0 \label{bamboccione}
\end{eqnarray}
The explicit form of (\ref{bamboccione}) obviously depends on the standard triple chosen as representative of the orbit, yet for whatever representative the four constraints are holomorphic. The next point consists in solving (\ref{bamboccione}) in terms of a parameter $w$ so that the complementary set of ten polynomials of the $z_i$ spanning the $j=\frac{9}{2}$ representation provide all the powers of $w$ from $0$ to $9$.
The requested solution is given by:
\begin{equation}\label{sicumerolo}
z_1\to \frac{3 w^5}{16}, \, \, z_2\to \frac{3\sqrt{5} w^4}{16},\, \, z_3\to \frac{1}{4}
   \sqrt{\frac{5}{2}} w^3,\, \, z_4\to w^3,\, \,z_5\to \frac{3w^2}{2 \sqrt{2}},\, \, z_6\to \frac{3 w}{2}
\end{equation}
Implementing the transformation (\ref{sicumerolo}) in the symplectic section (\ref{gargamelle}) one finds:
\begin{equation}\label{novemezzisezia}
  \Omega[Z]  \, \stackrel{\phi}{\Longrightarrow} \, \Omega_{\frac{9}{2}}[w] \, = \, \left(
\begin{array}{l}
 \frac{3 w}{\sqrt{2}} \\
 \frac{w^6}{4 \sqrt{2}} \\
 -\frac{3 w^4}{4 \sqrt{2}} \\
 \frac{w^9}{256 \sqrt{2}} \\
 -\frac{3 w^7}{32 \sqrt{2}} \\
 -\frac{1}{16} \sqrt{\frac{5}{2}} w^6 \\
 -\frac{3}{16} \sqrt{5} w^5 \\
 \frac{3 w^8}{128 \sqrt{2}} \\
 -\sqrt{2} w^3 \\
 \frac{3 w^5}{8 \sqrt{2}} \\
 \sqrt{2} \\
 -\frac{3 w^2}{\sqrt{2}} \\
 \frac{1}{2} \sqrt{\frac{5}{2}} w^3 \\
 -\frac{3}{8} \sqrt{5} w^4
\end{array}\right)
\end{equation}
which as requested contains all the powers of $w$ and has vanishing projection on the $j=\frac{3}{2}$ representation. Calculating the K\"ahler potential from such a section we obtain:
\begin{equation}\label{furutto}
    \mathcal{K}_{\frac{9}{2}} \, = \, - \, \log \,\left(\bar{\Omega}_{\frac{9}{2}}[\bar{w}]\,\mathbb{C}_{14} \, \Omega_{\frac{9}{2}}[w] \right) \, = \, \log\left ( -\frac{\rm i}{256} \,(w-\bar{w})^9\right)
\end{equation}
Now the question of consistency can be readdressed in the present  context. Implementing the substitution (\ref{sicumerolo}) in the six complex equations (\ref{finocchione}) (with for instance vanishing potential) do we obtain six consistent equations or not? The answer is no. The six equations (\ref{finocchione}) are inconsistent and this confirms in a holomorphic set up the same result we had previously obtained in the direct approach of eq.s (\ref{lambopsi}-\ref{embeddone}). Hence the $\slal(2)$ embedding of orbit $\mathfrak{O}_1$ leads to inconsistent truncations and has to be excluded.
\paragraph{Orbit $\mathfrak{O}_2$: ($j_1=\frac{3}{2}, \, j_2=\frac{1}{2}$).}
For the second orbit, the direct approach (\ref{lambopsi}-\ref{embeddone}) leads to:
\begin{eqnarray}
 && \left( \begin{array}{ccc}
                                                                                                   z_1& z_2 & z_3\\
                                                                                                   z_2& z_4 & z_5 \\
                                                                                                   z_3 & z_5  & z_6
                                                                                                 \end{array}
    \right) = \, \nonumber\\
    && \left(
\begin{array}{lll}
 \left(e^{\lambda }-i \psi \right)^2 \left(2 \psi -i e^{\lambda
   }\right) & 0 & \sqrt{3} \psi  \left(\psi +i e^{\lambda }\right)
   \\
 0 & -\psi -i e^{\lambda } & 0 \\
 \sqrt{3} \psi  \left(\psi +i e^{\lambda }\right) & 0 & -2 \psi -i
   e^{\lambda }
\end{array}
\right) \nonumber\\
 && w \,=\, {\rm i} \, e^\lambda \, + \, \psi \label{immerO2}
\end{eqnarray}
The pull-back of the lagrangian is the following one:
\begin{equation}\label{orbittaseconda}
    \phi^\star \mathcal{L} \, = \, 11 \, \left( \partial^\mu\psi\, \partial^\mu \psi  \, e^{-2\lambda} \, + \, \partial^\mu\lambda\, \partial^\mu \lambda\right) \, - \, \frac{1}{4}\,  g^2 \, \left(3\,e^{-\lambda} \, - \, \kappa\right)^2
\end{equation}
Also in this case the pull-back of the scalar field equations yields an  inconsistent set and there is no truncation. No Starobinsky--like model can be obtained from this orbit. In a similar way to the previous case we can discuss the same issue in a holomorphic set up. The branching rule of the $\mathbf{14}^\prime$ representation in the considered embedding  is the following one:
\begin{equation}\label{frikandello2}
  \mathbf{14}^\prime \, \stackrel{\slal(2,\mathbb{R})}{\longrightarrow} \, \left( j\, =\, \frac{5}{2}\right) \, \oplus \, \left( j\, =\, \frac{3}{2}\right) \, \oplus \, \left( j\, =\, \frac{3}{2}\right)
\end{equation}
and we can impose holomorphic constraints that suppress the two lowest spin representations $\left( j\, =\, \frac{3}{2}\right)$ leaving only the top one $\left( j\, =\, \frac{5}{2}\right)$  spanned by the powers of a parameter $w$ from $0$ to $5$.
Such a holomorphic embedding is given:
\begin{equation}\label{frikandellus2}
    z_1  \frac{2 w^3}{3^{3/4}}, \, \, z_2\to 0,\, \, z_3\to w^2,z_4\to
   \frac{w}{\sqrt[4]{3}},\, \, z_5\to 0,\, \, z_6\to \frac{2
   w}{\sqrt[4]{3}}
\end{equation}
Substitution of eq.s (\ref{frikandellus2}) into the field equations (\ref{finocchione}) confirms that their pull-back on this surface is inconsistent.
\paragraph{Orbit $\mathfrak{O}_3$: ($j_1=1, \, j_2=1$).}
For the third orbit, the direct approach (\ref{lambopsi}-\ref{embeddone}) leads to
\begin{eqnarray}
 && \left( \begin{array}{ccc}
                                                                                                   z_1& z_2 & z_3\\
                                                                                                   z_2& z_4 & z_5 \\
                                                                                                   z_3 & z_5  & z_6
                                                                                                 \end{array}
    \right) = \, \nonumber\\
    && \left(
\begin{array}{lll}
 -i e^{2 \lambda } & -\psi ^2 & -\sqrt{2} \psi  \\
 -\psi ^2 & -i \left(2 \psi ^2+e^{2 \lambda }\right) & -i \sqrt{2}
   \psi  \\
 -\sqrt{2} \psi  & -i \sqrt{2} \psi  & -i
\end{array}
\right) \nonumber\\
 && w \,=\, {\rm i} \, e^\lambda \, + \, \psi \label{immerO3}
\end{eqnarray}
The pull-back of the lagrangian is the following one:
\begin{equation}\label{orbittaterza}
    \phi^\star \mathcal{L} \, = \, 8 \, \left( \partial^\mu\psi\, \partial^\mu \psi  \, e^{-2\lambda} \, + \, \partial^\mu\lambda\, \partial^\mu \lambda\right) \, - \, \frac{1}{4}\,  g^2 \,  \kappa^2
\end{equation}
while the pull-back of the scalar field equations is an inconsistent  set.  Hence this truncation is not consistent and no Starobinsky--like model can be obtained from this orbit. As in the previous two cases we can confirm the same result in a holomorphic set up, yet we consider it useless to repeat once more the same type of calculations. What is relevant to mention in view of our subsequent considerations
is the branching rule of the $\mathbf{14}^\prime$ representation under this forbidden embedding leading to inconsistent field equations::
\begin{equation}\label{frikandello3}
  \mathbf{14}^\prime \, \stackrel{\slal(2,\mathbb{R})}{\longrightarrow} \, \left( j\, =\, 2\right) \, \oplus \, \left( j\, =\, 2\right)  \,
  \oplus \, 4 \, \times \, \left( j\, =\, 0\right)
\end{equation}
\paragraph{Orbit $\mathfrak{O}_4$: ($j_1=\frac{1}{2}, \, j_2=\frac{1}{2},j_3=\frac{1}{2}$).} For the fourth orbit, the direct approach (\ref{lambopsi}-\ref{embeddone}) leads to
\begin{eqnarray}
 && \left( \begin{array}{ccc}
                                                                                                   z_1& z_2 & z_3\\
                                                                                                   z_2& z_4 & z_5 \\
                                                                                                   z_3 & z_5  & z_6
                                                                                                 \end{array}
    \right) = \, \nonumber\\
    && \left(
\begin{array}{lll}
 i e^{\lambda }-\psi  & 0 & 0 \\
 0 & i e^{\lambda }-\psi  & 0 \\
 0 & 0 & i e^{\lambda }-\psi
\end{array}
\right)
\nonumber\\
 && w \,=\, {\rm i} \, e^\lambda \, - \, \psi \label{immerO4}
\end{eqnarray}
The pull-back of the lagrangian is the following one:
\begin{equation}\label{orbittaquarta}
    \phi^\star \mathcal{L} \, = \, 3 \, \left( \partial^\mu\psi\, \partial^\mu \psi  \, e^{-2\lambda} \, + \, \partial^\mu\lambda\, \partial^\mu \lambda\right) \, - \,   g^2 \, \frac{1}{4} \,\left( 3 \, e^{-\lambda}\, - \,  2\,  \kappa\right)^2
\end{equation}
The pull-back of the scalar field equations produces  equations consistent among themselves which coincide  with the equations derived from the pull-back of the lagrangian (\ref{orbittaquarta}), hence the truncation is consistent. We reobtain the Starobinsky model discussed in the previous section with $q=3$ and hence with $\alpha \, = \, 1$.  In this case the consistent truncation is  already produced form holomorphic constraints. Indeed equation (\ref{immerO4}) can be summarized as:
\begin{equation}\label{holomorph4}
    z_2 \, = \, z_3 \, = \, z_5 \, = \, 0 \quad ; \quad z_1 \, = \, z_4 \, = \, z_6 \, = \,  w
\end{equation}
It is interesting and important for our future consideration to mention the branching rule of the $\mathbf{14}^\prime$ representation under this $\slal(2)$ subalgebra:
\begin{equation}\label{frikandello4}
  \mathbf{14}^\prime \, \stackrel{\slal(2,\mathbb{R})}{\longrightarrow} \, \left( j\, =\, \frac{3}{2}\right) \,
  \oplus \, 5 \, \times \, \left( j\, =\, \frac{1}{2}\right)
\end{equation}
and the constraints (\ref{holomorph4}) precisely are the conditions under which the five representations $\left( j\, =\, \frac{1}{2}\right)$ vanish and we are left with the representation $\left( j\, =\, \frac{3}{2}\right)$ duely spanned by the powers $1,w,w^2,w^3$.
\paragraph{Orbit $\mathfrak{O}_5$: ($j_1=\frac{1}{2}, \, j_2=\frac{1}{2},j_3=0$).} For the fifth orbit, the direct approach (\ref{lambopsi}-\ref{embeddone}) leads to
\begin{eqnarray}
 && \left( \begin{array}{ccc}
                                                                                                   z_1& z_2 & z_3\\
                                                                                                   z_2& z_4 & z_5 \\
                                                                                                   z_3 & z_5  & z_6
                                                                                                 \end{array}
    \right) = \, \nonumber\\
    && \left(
\begin{array}{lll}
 i e^{\lambda }-\psi  & 0 & 0 \\
 0 & i e^{\lambda }-\psi  & 0 \\
 0 & 0 & i
\end{array}
\right) \nonumber\\
 && w \,=\, {\rm i} \, e^\lambda \, - \, \psi \label{immerO5}
\end{eqnarray}
The pull-back of the lagrangian is the following one:
\begin{equation}\label{orbittaquinta}
    \phi^\star \mathcal{L} \, = \, 2 \, \left( \partial^\mu\psi\, \partial^\mu \psi  \, e^{-2\lambda} \, + \, \partial^\mu\lambda\, \partial^\mu \lambda\right) \, - \, g^2 \,  \left(e^{-\lambda} \, - \, \kappa\right)^2
\end{equation}
The pull-back of the scalar field equations yields a consistent system identical with the field equations derived from the pull-back of the lagrangian (\ref{orbittasesta}), hence the truncation is consistent. We reobtain the Starobinsky--like model discussed in the previous section with $q=2$ and hence with $\alpha \, = \, \frac{2}{3}$.
\par
In this, as in the previous case, the consistent truncation is  produced from holomorphic constraints. Indeed equation (\ref{immerO4}) can be summarized as:
\begin{equation}
    z_2 \, = \, z_3 \, = \, z_5 \, = \, 0 \quad ; \quad z_1 \, = \, z_4 \, = \, w \quad ; \quad  z_6 \, = \,  {\rm i} \label{belgiuro}
\end{equation}
In this case the branching rule of the $\mathbf{14}^\prime$ representation under the considered $\slal(2)$ subalgebra is the following one:
\begin{equation}\label{frikandello5}
  \mathbf{14}^\prime \, \stackrel{\slal(2,\mathbb{R})}{\longrightarrow} \, \left( j\, =\, 1\right) \,
  \oplus \,\left( j\, =\, 1\right)\, \oplus \,  2 \, \times \, \left( j\, =\, \frac{1}{2}\right) \, + \, 4 \, \times \, \left( j\, =\, 0\right)
\end{equation}
and the constraint (\ref{belgiuro}) guarantees that the singlets and the $\left( j\, =\, \frac{1}{2}\right)$ representations are all set to zero.s
\paragraph{Orbit $\mathfrak{O}_6$: ($j_1=\frac{1}{2}, \, j_2=0,\, j_3=0$).} For the sixth orbit, the direct approach (\ref{lambopsi}-\ref{embeddone}) leads to
\begin{eqnarray}
 && \left( \begin{array}{ccc}
                                                                                                   z_1& z_2 & z_3\\
                                                                                                   z_2& z_4 & z_5 \\
                                                                                                   z_3 & z_5  & z_6
                                                                                                 \end{array}
    \right) = \, \nonumber\\
    && \left(
\begin{array}{lll}
 \psi +i e^{\lambda } & 0 & 0 \\
 0 & i & 0 \\
 0 & 0 & i
\end{array}
\right) \nonumber\\
 && w \,=\, {\rm i} \, e^\lambda \, + \, \psi \label{immerO6}
\end{eqnarray}
The pull-back of the lagrangian is the following one:
\begin{equation}\label{orbittasesta}
    \phi^\star \mathcal{L} \, = \,  \left( \partial^\mu\psi\, \partial^\mu \psi  \, e^{-2\lambda} \, + \, \partial^\mu\lambda\, \partial^\mu \lambda\right) \, - \, \frac{1}{4}\, g^2  \, \left( e^{\lambda}\, + \, \kappa\right)^2
\end{equation}
The pull-back of the scalar field equations yields a consistent system  coinciding  with the equations derived from the pull-back of the lagrangian (\ref{orbittasesta}). So we have a consistent truncation and we reobtain the Starobinsky--like model discussed in the previous section with $q=1$. It corresponds to  $\alpha \, = \, \frac{1}{3}$.   Equation (\ref{immerO6}) can be summarized as:
\begin{equation}\label{holomorph6}
    z_2 \, = \, z_3 \, = \, z_5 \, = \, 0 \quad ; \quad z_1 \, = \, w \quad ; \quad  z_4 \, = \,  z_6 \, = \,  {\rm i}
\end{equation}
The branching of the $\mathbf{14}^\prime$ dimensional representation under this $\slal(2)$ subalgebra is the following one:
\begin{equation}\label{frikandello6}
  \mathbf{14}^\prime \, \stackrel{\slal(2,\mathbb{R})}{\longrightarrow} \,  5 \, \times \,\left( j\, =\, \frac{1}{2}\right)  \,
  \oplus  \, 4 \, \times \, \left( j\, =\, 0\right)
\end{equation}
\subsubsection{\sc Conclusion of the above discussion}
This concludes our preliminary study of the orbits and shows that the embedded Starobinsky-like models described in section \ref{generalonuovo} exhaust the list of possible embeddings, the values of $\alpha\, = \, 1, \frac{2}{3} ,\frac{1}{3}$ being, apparently the only admissible ones. Next let us observe that the branching rules of the $\mathbf{14}^\prime$ dimensional representation which lead to consistent truncations, namely, (\ref{frikandello4},\ref{frikandello5},\ref{frikandello6}) are the only possible ones that we can obtain by embedding:
\begin{equation}\label{grullonevero}
    \slal(2) \, \mapsto \, \slal(2) \, \times \, \slal(2) \,\times \, \slal(2)
\end{equation}
if the considered $\mathbf{14}^\prime$ representation of  $\slal(2) \, \times \, \slal(2) \,\times \, \slal(2)$ is the following one:
\begin{equation}\label{cannoleccio}
    \mathbf{14}^\prime \, = \, \left(\frac{1}{2} \, , \, \frac{1}{2} \, , \, \frac{1}{2}\right)\, \oplus \, \left(\frac{1}{2},0,0\right) \, \oplus \,
    \left(0, \frac{1}{2},0\right) \, \oplus \, \left(0, 0, \frac{1}{2}\right)
\end{equation}
This has a profound meaning. It implies that the only consistent truncations occur when the $\slal(2)$ Lie algebra is embedded in the \textit{sub-Tits-Satake} Lie algebra, which as we discuss in the conclusive part is universal for all $\mathcal{N}=2$ models. This allows us to make the bold statement that the only values of $\alpha$ one can obtain form the gauging of hypermultiplet isometries in any supergravity theory based on symmetric manifolds is just $\alpha \, = \, 1, \frac{2}{3} , \,  \frac{1}{3}$.
\newpage
\part{\sc Conclusions and perspectives}
\par
Considering the structure of the $c$-map and the results relative to the inclusion of Starobinsky like potentials that we have concretely obtained in the case of the $\mathrm{Sp(6,\mathbb{R})}$-model, we interpret them within the framework provided by Tits-Satake subalgebras and Tits Satake universality classes. This allows us to advocate that the mechanism underlying the generation of such cosmological potentials is universal and the prediction on the possible values of $\alpha$ equally general. Actually the analysis we are going to present in the next section supports the following general conclusion: the cosmological potentials classified in \cite{pietrosergiosasha1}, that follow from the gauging of a generator respectively elliptic, hyperbolic or parabolic of a constant curvature K\"ahler surface admit a universal uplifting to all $\mathcal{N}=2$ models based on symmetric spaces for the hypermultiplets, provided the curvature of the K\"ahler surface is duely quantized. Indeed there exist consistent one-field truncation only if the gauging occurs inside the universal \textit{sub-Tits-Satake} Lie algebra.
\section{\sc The Tits Satake projection}
In most cases of lower supersymmetry, neither the algebra $\mathbb{U}_{\mathcal{SK}}$ nor the
algebra $\mathbb{U}_{\mathcal{QM}}$ are \textbf{maximally split}. In short
this means that the non-compact rank $r_{nc} < r $ is less than
the rank of $\mathbb{U}$, namely not all the Cartan generators are
non-compact. Rigorously $r_{nc}$ is defined as follows:
\begin{equation}
  r_{nc}\, = \, \mbox{rank} \left( \mathrm{U/H}\right)  \, \equiv \, \mbox{dim} \,
  \mathcal{H}^{n.c.} \quad ; \quad \mathcal{H}^{n.c.} \, \equiv \,
  \mbox{CSA}_{\mathbb{U}(\mathbb{C})} \, \bigcap \, \mathbb{K}
\label{rncdefi}
\end{equation}
When this happens it means that the structure of  both black-hole-like  and cosmological-like solutions of supergravity is effectively determined by a
\textit{maximally split subalgebra} $\mathbb{U}^{TS} \subset
\mathbb{U}$ named the \textit{Tits Satake} subalgebra of
$\mathbb{U}$, whose rank is equal to $r_{nc}$. Effectively determined
does not mean that  solutions of the big system
coincide with those
of the smaller system  rather it means that the former can be obtained from the
latter by means of rotations of a compact subgroup $\mathrm{G_{paint} \subset U}$ of the big group
which we name the \textit{paint group}, for whose precise definition we refer the reader to  \cite{titsusataku}.
Here we just emphasize few important facts, relevant for our goals. To this effect we recall that the Tits Satake algebra is obtained from the
original algebra via a projection of the root system of $\mathbb{U}$ onto the subspace orthogonal to the compact part of the Cartan subalgebra
of $\mathbb{U}^{TS}$:
\begin{equation}
  \Pi^{TS} \quad ; \quad \Delta_\mathbb{U} \,\mapsto \,
  \overline{\Delta}_{\mathbb{U}^{TS}}
\label{Tsproj}
\end{equation}
In euclidian geometry $\overline{\Delta}_{\mathbb{U}^{TS}}$ is just a collection of vectors in $r_{nc}$ dimensions; a priori there is no reason why it should be the root system of another Lie algebra. Yet in almost all cases, $\overline{\Delta}_{\mathbb{U}^{TS}}$ turns out to be a Lie algebra root system and the maximal split Lie algebra corresponding to it, $\mathbb{U}^{TS}$, is, by definition, the Tits Satake subalgebra of the original non maximally split Lie algebra: $\mathbb{U}^{TS} \subset \mathbb{U}$. Such algebras $\mathbb{U}$ are called \textit{non-exotic}. The \textit{exotic} non compact algebras are those for which the system $\overline{\Delta}_{\mathbb{U}^{TS}}$ is not an admissible root system. In such cases there is no Tits Satake subalgebra $\mathbb{U}^{TS}$. Exotic subalgebra are very few and in supergravity they appear only in three instances that display additional peculiarities relevant  for the black-hole and cosmological solutions. As for $\mathcal{N}=2$ theories the only exotic homogeneous symmetric Special K\"ahler manifolds are those of the Minimal Coupling series discussed in section \ref{minicoup}. Exotic are also the Quaternionic K\"ahler manifolds in the $c$-map image of the former.
\par
For the non exotic models  we have that the decomposition (\ref{gendecompo}) commutes with the projection,
namely:
\begin{equation}
\begin{array}{rcl}
\mbox{adj}(\mathbb{U}_{\mathcal{QM}}) &=&
\mbox{adj}(\mathbb{U}_{\mathcal{SK}})\oplus\mbox{adj}({\slal(2,\mathbb{R})_E})\oplus
\mathbf{W}_{(2,W)} \\
\null &\Downarrow & \null\\
\mbox{adj}(\mathbb{U}^{TS}_{\mathcal{QM}}) &=&
\mbox{adj}(\mathbb{U}^{TS}_{\mathcal{SK}})\oplus\mbox{adj}({\slal(2,\mathbb{R})_E})\oplus
\mathbf{W}_{(2,W^{TS})} \\
\end{array}
\label{gendecompo2}
\end{equation}
In other words the projection leaves the $A_1$ Ehlers subalgebra untouched and has a non trivial effect only
on the algebra $\mathbb{U}_{\mathcal{SK}}$. Furthermore the image under the projection of the highest root
of $\mathbb{U}$ is the highest root of $\mathbb{U}^{TS}$:
\begin{equation}
  \Pi^{TS} \quad : \quad \psi \, \rightarrow \, \psi^{TS}
\label{Pionpsi}
\end{equation}
The reason why the Tits Satake projection is relevant to us is that the classification of nilpotent orbits (standard triples)  and hence of abelian gaugings depends only on the Tits Satake subalgebra and therefore is universal for all members of the same Tits Satake universality class. By this name we mean all algebras that share the same Tits Satake projection. A similar property was extensively used in \cite{noinilpotenti} in the discussion of extremal black-hole solutions. Indeed the classification of extremal black-holes  also boils down to the classification of nilpotent orbits, so that the mathematical problem at stake is just the same.
\par
Having clarified these points we can proceed with the classification of homogeneous symmetric spaces relevant to $\mathcal{N}=2$ supergravity either in the vector multiplet sector (Special K\"ahler) or in the hypermultiplet sector (Quaternionic K\"ahler). These spaces are listed in table \ref{homomodelTS}.
\begin{table}
\begin{center}
{\tiny
\begin{tabular}{|l|c|c||c|c||c|c||c||}
  \hline
  \null & TS & TS & coset &coset &  Paint & subP &  susy\\
  $\#$ & $\mathcal{SK}_{TS}$ & $\mathcal{QM}_{TS}$ & $\mathcal{SK}$ & $\mathcal{QM}$ &  Group & Group &  \\
  \hline
  \hline
1 &\null & \null & \null & \null &\null & \null & \null \\
\null & $ \frac{\mathrm{SU(1,1)}}{\mathrm{U(1)}}$ & $ \frac{\mathrm{G_{2(2)}}}{\mathrm{SU(2)\times SU(2)}}$  & $ \frac{\mathrm{SU(1,1)}}{\mathrm{U(1)}}$ & $ \frac{\mathrm{G_{2(2)}}}{\mathrm{SU(2)\times SU(2)}}$  & $1$ & $1$ & $\mathcal{N}=2$ \\
\null &\null & \null & \null & \null &\null & \null & n=1 \\
\hline
\hline
2 &\null & \null & \null & \null &\null & \null & \null \\
\null & \null & \null & $ \frac{\mathrm{Sp(6,R)}}{\mathrm{SU(3)\times  U(1)}}$ & $ \frac{\mathrm{F_{4(4)}}}{\mathrm{USp(6)\times SU(2)}}$  & $1$ & $1$  & $\mathcal{N}=2$ \\
\null &\null & \null & \null & \null &\null & \null & $n=6$ \\
\cline{1-1} \cline{4-8}
3 &\null & \null & \null & \null &\null & \null & \null \\
\null & \null & \null & $ \frac{\mathrm{SU(3,3)}}{\mathrm{SU(3)\times SU(3) \times U(1)}}$ & $ \frac{\mathrm{E_{6(2)}}}{\mathrm{SU(6)\times SU(2)}}$  & $\mathrm{SO(2)\times SO(2)}$ & $1$  & $\mathcal{N}=2$ \\
\null &\null & \null & \null & \null &\null & \null & $n=9$ \\
\cline{1-1} \cline{4-8}
4 &\null & \null & \null & \null &\null & \null & $\mathcal{N}=6$ \\
\null & $ \frac{\mathrm{Sp(6,R)}}{\mathrm{SU(3)\times  U(1)}}$ & $ \frac{\mathrm{F_{4(4)}}}{\mathrm{Sp(6,R)\times SL(2,R)}}$ & $ \frac{\mathrm{SO^\star(12)}}{\mathrm{SU(6)\times U(1)}}$ & $ \frac{\mathrm{E_{7(-5)}}}{\mathrm{SO(12)\times SU(2)}}$  & $\mathrm{SO(3)\times SO(3)}$ & $\mathrm{SO(3)_{d}}$  & $\mathcal{N}=2$ \\
\null &\null & \null & \null & \null &$\mathrm{\times SO(3)}$ & \null & n=15 \\
\cline{1-1} \cline{4-8}
5 &\null & \null & \null & \null &\null & \null & \null \\
\null & \null & \null & $ \frac{\mathrm{E_{7(-25)}}}{\mathrm{E_{6(-78)} \times U(1)}}$ & $ \frac{\mathrm{E_{8(-24)}}}{\mathrm{E_{7(-133)}\times SU(2)}}$  & $\mathrm{SO(8)}$ & $G_{2(2)}$  & $\mathcal{N}=2$ \\
\null &\null & \null & \null & \null &\null & \null & $n=27$ \\
\hline
\hline
\hline
6 & \null \null & \null & \null & \null &\null & \null & \null \\
\null & $ \frac{\mathrm{SL(2,\mathbb{R})}}{\mathrm{SO(2)}}\times\frac{\mathrm{SO(2,1)}}{\mathrm{SO(2)}}$ & $ \frac{\mathrm{SO(4,3)}}{\mathrm{SO(4)\times SO(3)}}$  & $ \frac{\mathrm{SL(2,\mathbb{R})}}{\mathrm{SO(2)}}\times\frac{\mathrm{SO(2,1)}}{\mathrm{SO(2)}}$ & $ \frac{\mathrm{SO(4,3)}}{\mathrm{SO(4)\times SO(3)}}$ & $\mathrm{1}$ & $\mathrm{1}$ & $\mathcal{N}=2$ \\
\null &\null & \null & \null & \null &\null & \null & n=2 \\
\hline
7 & \null \null & \null & \null & \null &\null & \null & \null \\
\null & $ \frac{\mathrm{SL(2,\mathbb{R})}}{\mathrm{SO(2)}}\times\frac{\mathrm{SO(2,2)}}{\mathrm{SO(2)\times SO(2)}}$ & $ \frac{\mathrm{SO(4,4)}}{\mathrm{SO(4)\times SO(4)}}$  & $ \frac{\mathrm{SL(2,\mathbb{R})}}{\mathrm{SO(2)}}\times\frac{\mathrm{SO(2,2)}}{\mathrm{SO(2)\times SO(2)}}$ & $ \frac{\mathrm{SO(4,4)}}{\mathrm{SO(4)\times SO(4)}}$ & $\mathrm{1}$ & $\mathrm{1}$ & $\mathcal{N}=2$ \\
\null &\null & \null & \null & \null &\null & \null & n=3 \\
\hline
8 & \null \null & \null & \null & \null &\null & \null & \null \\
\null & $ \frac{\mathrm{SL(2,\mathbb{R})}}{\mathrm{SO(2)}}\times\frac{\mathrm{SO(2,3)}}{\mathrm{SO(2)\times SO(3)}}$ & $ \frac{\mathrm{SO(4,5)}}{\mathrm{SO(4)\times SO(5)}}$  & $ \frac{\mathrm{SL(2,\mathbb{R})}}{\mathrm{SO(2)}}\times\frac{\mathrm{SO(2,2+p)}}{\mathrm{SO(2)\times SO(2+p)}}$ & $ \frac{\mathrm{SO(4,4+p)}}{\mathrm{SO(4)\times SO(4+p)}}$ & $\mathrm{SO(p)}$ & $\mathrm{SO(p-1)}$ & $\mathcal{N}=2$ \\
\null &\null & \null & \null & \null &\null & \null & n=3+p \\
\hline
\hline
\hline
$\mathrm{exot}$&\null & \null & \null & \null &\null & \null & \null \\
\null \null & $ bc_1$ & $ bc_2$  & $ \frac{\mathrm{SU(p+1,1)}}{\mathrm{SU(p+1)\times U(1)}}$ & $ \frac{\mathrm{SU(p+2,2)}}{\mathrm{SU(p+1,1)\times SL(2,R)_{h^\star}}}$  & $\mathrm{U(1)\times U(1) \times U(p)}$ & $\mathrm{U(p-1)}$ & $\mathcal{N}=2$ \\
\null &\null & \null & \null & \null &\null& \null & n=p+1 \\
\hline
\end{tabular}
}
\caption{The first eight rows of this table list the  \textit{non-exotic} homogenous symmetric Special K\"ahler manifolds. They are displayed together with their Quaternionic K\"ahler $c$-map images and are organized in five Tits Satake universality classes.  Non exotic means that the Tits Satake projection of the root system is  a standard Lie Algebra root system. Of the five universality classes three contain only one maximally split element, one contains four elements filling the Tits magic square, while the last class contains an infinite number of elements. Within each class the models are distinguished by the different structure of the Paint Group and of its subPaint subgroup. The Paint group is a $c$-map invariant. It is the same in the Special K\"ahler  and in the Quaternionic K\"ahler case. The last line displays the unique family of exotic Special K\"ahler symmetric spaces for which the Tits Satake projection of the root system is not a root system. They correspond to the Minimal Coupling Models discussed in the main text. Their $c$-map images of the exotic models are also exotic. Notwithstanding this anomaly the concept of Paint Group, according to its definition as group of external automorphisms of the solvable Lie algebra generating the non compact coset manifold still exists.  The Paint group is the same for the K\"ahler  and the quaternionic K\"ahler case as in the non exotic cases.  \label{homomodelTS}}
\end{center}
\end{table}
In table \ref{homomodelTS} we have also listed the Paint groups and the subpaint groups. These latter are always compact and their different structures is what distinguishes the different elements belonging to the same class. As it was shown in \cite{titsusataku}, these groups are dimensional reduction invariant and therefore $c$-map invariant, namely they are the same in $\mathrm{U}_{\mathcal{SK}}$ and in $\mathrm{U}_{\mathcal{QM}}$. Hence the representation $\mathbf{W}$ which, as we have seen, hosts the symplectic section of Special Geometry and regulates, by means of its branching, the existence or non existence of consistent truncations, can be decomposed with respect to the Tits Satake subalgebra and the Paint group revealing a regularity structure inside each Tits Satake universality class which is what allows us to draw general conclusions and make universal predictions. In the case of black-holes the same Tits-Satake decomposition of the $\mathbf{W}$-representation is at the heart of the classification of \textit{charge orbits} of the hole, as we extensively discussed in \cite{noinilpotenti}.
\section{\sc Tits Satake Universality classes and the embedding of Starobinsky-like models}
In the present section we consider the decomposition of the $\mathbf{W}$-representations with respect to Tits-Satake subalgebras and Paint groups for all the non-exotic models.
\par
In \cite{titsusataku} the \textit{paint algebra} was defined as the algebra of external automorphisms of the solvable Lie algebra $\Solv_\mathcal{M}$ generating the non-compact symmetric space: $\mathcal{M}\, = \, \mathrm{U/H}$, namely
\begin{equation}
  \mathbb{G}_{\mathrm{paint}} \, = \, \mathrm{Aut}_{\mathrm{Ext}} \, \left[
\Solv_\mathcal{M}\right]. \label{pittureFuori}
\end{equation}
where:
\begin{equation}
  \mathrm{Aut}_{\mathrm{Ext}} \, \left[ \Solv_\mathcal{M}\right] \,
  \equiv \, \frac{\mathrm{Aut} \, \left[
  \Solv_\mathcal{M}\right]}{\Solv_\mathcal{M}},
\label{outerauto}
\end{equation}
Given the paint algebra $\mathbb{G}_{\mathrm{paint}} \, \subset \, \mathbb{U}$ and the Tits Satake subalgebra $\mathbb{G}_{\mathrm{TS}}\, \subset \, \mathbb{U}$, whose construction we have briefly recalled above, following \cite{titsusataku}
one introduces the \textit{sub Tits Satake} and \textit{sub paint} algebras as the centralizers of the paint algebra and of the Tits Satake algebra, respectively. In other words we have:
\begin{equation}\label{subTs}
    \mathfrak{s} \, \in \, \mathbb{G}_{\mathrm{subTS}} \, \subset \, \mathbb{G}_{\mathrm{TS}}\, \subset \, \mathbb{U} \quad \Leftrightarrow \, \quad \left[ \mathfrak{s} \, , \, \mathbb{G}_{\mathrm{paint}}\right] \, = \, 0
\end{equation}
and
\begin{equation}\label{subpaint}
    \mathfrak{t} \, \in \, \mathbb{G}_{\mathrm{subpaint}} \, \subset \, \mathbb{G}_{\mathrm{paint}}\, \subset \, \mathbb{U} \quad \Leftrightarrow \, \quad \left[ \mathfrak{t} \, , \, \mathbb{G}_{\mathrm{TS}}\right] \, = \, 0
\end{equation}
A very important property of the paint and subpaint algebras is that they are conserved int the $c$-map, namely they are the same for $\mathbb{U}_{\mathcal{SK}}$ and $\mathbb{U}_{\mathcal{QM}}$.
\par
In the next lines we analyze the decomposition of the $\mathbf{W}$-representations with respect to these subalgebras for each Tits Satake universality class of non maximally split models. In the case of maximally split models there is no paint algebra and there is nothing with respect to which to decompose.
\subsection{\sc Universality class $\sym(6,\mathbb{R})\Rightarrow \mathfrak{f}_{4(4)}$}
In this case the sub Tits Satake Lie algebra is
\begin{equation}\label{cuffio}
    \mathbb{G}_{\mathrm{subTS}} \, = \, \slal(2,\mathbb{R}) \oplus \slal(2,\mathbb{R}) \oplus \slal(2,\mathbb{R}) \subset \sym(6,\mathbb{R}) \, = \, \mathbb{G}_{\mathrm{TS}}
\end{equation}
and the $\mathbf{W}$-representation of the maximally split model decomposes as follows:
\begin{equation}\label{scompo14}
    \mathbf{14}^\prime  \, \stackrel{\mathbb{G}_{\mathrm{subTS}}}{\Longrightarrow} \, (\mathbf{2},\mathbf{1},\mathbf{1}) \oplus (\mathbf{1},\mathbf{2},\mathbf{1}) \oplus (\mathbf{1},\mathbf{1},\mathbf{2}) \oplus (\mathbf{2},\mathbf{2},\mathbf{2})
\end{equation}
This decomposition combines in the following way with the paint group representations in the various models belonging to the same universality class.
\subsubsection{\sc $\su(3,3)$ model}
For this case the paint algebra is
\begin{equation}\label{furadino}
     \mathbb{G}_{\mathrm{paint}} \, = \, \so(2) \oplus \so(2)
\end{equation}
and the $\mathbf{W}$-representation is the $\mathbf{20}$ dimensional of $\su(3,3)$ corresponding to an antisymmetric tensor with a reality condition of the form:
\begin{equation}\label{realata}
    t_{\alpha\beta\gamma}^\star = \frac{1}{3!} \, \epsilon_{\alpha\beta\gamma\delta\eta\theta} \, t_{\delta\eta\theta}
\end{equation}
The decomposition of this representation with respect to the Lie algebra $ \mathbb{G}_{\mathrm{paint}}\oplus {\mathbb{G}_{\mathrm{subTS}}}$ is the following one:
\begin{equation}\label{ruppatoA}
    \mathbf{20}  \, \stackrel{\mathbb{G}_{\mathrm{paint}} \oplus \mathbb{G}_{\mathrm{subTS}}}{\Longrightarrow} \, (2,q_1|\mathbf{2},\mathbf{1},\mathbf{1}) \oplus (2,q_2|\mathbf{1},\mathbf{2},\mathbf{1}) \oplus (2,q_3|\mathbf{1},\mathbf{1},\mathbf{2}) \oplus (1,0|\mathbf{2},\mathbf{2},\mathbf{2})
\end{equation}
where $(2,q)$ means a doublet of $\so(2)\oplus\so(2)$ with a
certain grading $q$ with respect to the generators, while $(1,0)$
means the singlet that has $0$ grading with respect to both
generators. The subpaint algebra in this case is $
\mathbb{G}_{\mathrm{subpaint}}\, = \,0$ and the decomposition of
the same $\mathbf{W}$-representation with respect to
$\mathbb{G}_{\mathrm{subpaint}} \oplus \mathbb{G}_{\mathrm{TS}}$
is:
\begin{equation}\label{ruppatoB}
    \mathbf{20}  \, \stackrel{\mathbb{G}_{\mathrm{subpaint}} \oplus \mathbb{G}_{\mathrm{TS}}}{\Longrightarrow} \, \mathbf{6} \, \oplus \, \mathbf{14}
\end{equation}
This follows from the decomposition of the $\mathbf{6}$ of $\sym(6,\mathbf{R})$ with respect to the sub Tits Satake algebra (\ref{cuffio}):
\begin{equation}\label{seifracco}
    \mathbf{6} \,  \stackrel{\mathbb{G}_{\mathrm{subTS}}}{\Longrightarrow}  \, (\mathbf{2},\mathbf{1},\mathbf{1}) \oplus (\mathbf{1},\mathbf{2},\mathbf{1}) \oplus (\mathbf{1},\mathbf{1},\mathbf{2})
\end{equation}
\subsubsection{\sc $\so^\star(12)$ model}
For this case the paint algebra is
\begin{equation}\label{furadinotwo}
     \mathbb{G}_{\mathrm{paint}} \, = \, \so(3) \oplus \so(3) \oplus \so(3)
\end{equation}
and the $\mathbf{W}$-representation is the $\mathbf{32}_s $
dimensional spinorial representation of $\so^\star(12)$. The
decomposition of this representation with respect to the Lie
algebra $ \mathbb{G}_{\mathrm{paint}}\oplus
{\mathbb{G}_{\mathrm{subTS}}}$ is the following one:
\begin{equation}\label{ruppatoAtwo}
    \mathbf{32}_s  \, \stackrel{\mathbb{G}_{\mathrm{paint}} \oplus \mathbb{G}_{\mathrm{subTS}}}{\Longrightarrow} \, (\underline{2},\underline{2},\underline{1}|\mathbf{2},\mathbf{1},\mathbf{1}) \oplus (\underline{2},\underline{1},\underline{2}|\mathbf{1},\mathbf{2},\mathbf{1}) \oplus (\underline{1},\underline{1},\underline{2}|\mathbf{1},\mathbf{1},\mathbf{2}) \oplus (\underline{1},\underline{1},\underline{1}|\mathbf{2},\mathbf{2},\mathbf{2})
\end{equation}
where $\underline{2}$ means the doublet spinor representation of
$\so(3)$. The subpaint algebra in this case is $
\mathbb{G}_{\mathrm{paint}}\, = \,\so(3)_{\mathrm{diag}}$ and the
decomposition of the same $\mathbf{W}$-representation with respect
to $\mathbb{G}_{\mathrm{subpaint}} \oplus
\mathbb{G}_{\mathrm{TS}}$ is:
\begin{equation}\label{ruppatoBtwo}
    \mathbf{32}_s  \, \stackrel{\mathbb{G}_{\mathrm{TS}}\oplus \mathbb{G}_{\mathrm{subpaint}}}{\Longrightarrow} \, (\mathbf{6}|\underline{3}) \, \oplus \, (\mathbf{14}^\prime|\underline{1})
\end{equation}
This follows from the decomposition of the product $\underline{2} \times \underline{2}$ of $\so(3)_{\mathrm{diag}}$ times the Tits Satake algebra (\ref{cuffio}):
\begin{equation}\label{seifraccotwo}
    \underline{2} \times \underline{2} \, =  \, \underline{3} \, \oplus \, \underline{1}
\end{equation}
\subsubsection{\sc $\mathfrak{e}_{7(-25)}$ model}
For this case the paint algebra is
\begin{equation}\label{furadinothree}
     \mathbb{G}_{\mathrm{paint}} \, = \, \so(8)
\end{equation}
and the $\mathbf{W}$-representation is the fundamental $\mathbf{56} $ dimensional  representation of $\mathfrak{e}_{7(-25)}$
The decomposition of this representation with respect to the Lie algebra $ \mathbb{G}_{\mathrm{paint}}\oplus {\mathbb{G}_{\mathrm{subTS}}}$ is the following one:
\begin{equation}\label{ruppatoAthree}
    \mathbf{56}  \, \stackrel{\mathbb{G}_{\mathrm{paint}} \oplus \mathbb{G}_{\mathrm{subTS}}}{\Longrightarrow} \, (\mathbf{8}_v|\mathbf{2},\mathbf{1},\mathbf{1}) \oplus (\mathbf{8}_s|\mathbf{1},\mathbf{2},\mathbf{1}) \oplus (\mathbf{8}_c|\mathbf{1},\mathbf{1},\mathbf{2}) \oplus (\mathbf{1}|\mathbf{2},\mathbf{2},\mathbf{2})
\end{equation}
where $\mathbf{8}_{v,s,c}$ are the three inequivalent eight-dimensional representations of $\so(8)$, the vector, the spinor and the conjugate spinor. The subpaint algebra in this case is $ \mathbb{G}_{\mathrm{paint}}\, = \,\mathfrak{g}_{2(-14)}$ with respect to which all three $8$-dimensional representations of $\so(8)$ branch as follows:
\begin{equation}\label{seifraccothree}
    \mathbf{8}_{v,s,c} \, \stackrel{\mathfrak{g}_{2(-14)}}{\Longrightarrow}  \, \mathbf{7 }\, \oplus \, \mathbf{1}
\end{equation}
In view of this the decomposition of the same $\mathbf{W}$-representation with respect to $\mathbb{G}_{\mathrm{subpaint}} \oplus \mathbb{G}_{\mathrm{TS}}$ is:
\begin{equation}\label{ruppatoBthree}
    \mathbf{56}  \, \stackrel{\mathbb{G}_{\mathrm{TS}} \oplus \mathbb{G}_{\mathrm{subpaint}}}{\Longrightarrow} \, (\mathbf{6}|\mathbf{7}) \, \oplus \, (\mathbf{14}^\prime|\mathbf{1})
\end{equation}
\subsection{\sc Universality class $\slal(2,\mathbb{R}) \oplus \so(2,3) \Rightarrow \so(4,5)$}
This case corresponds to one of the possible infinite families of $\mathcal{N}=2$ theories with a symmetric homogeneous special K\"ahler manifold and a number of vector multiplets larger than three ($n=3+p$).  The other infinite family corresponds instead to  one of the three exotic models.
\par
The generic element of this infinite class corresponds to the following algebras:
\begin{eqnarray}
  \mathbb{U}_{\mathcal{SK}} &=& \slal(2,\mathbb{R}) \oplus \so(2,2+p)\nonumber\\
  \mathbb{U}_{\mathcal{QM}} &=& \so(4,4+p) \label{feldane}
\end{eqnarray}
In this case the sub Tits Satake algebra is:
\begin{equation}\label{casilino1}
    \mathbb{G}_{\mathrm{subTS}} \, = \, \slal(2,\mathbb{R}) \oplus \slal(2,\mathbb{R}) \oplus \slal(2,\mathbb{R}) \, \simeq \, \slal(2,\mathbb{R}) \oplus \so(2,2) \, \, \subset \, \slal(2,\mathbb{R}) \oplus \so(2,3) \, = \, \mathbb{G}_{\mathrm{TS}}
\end{equation}
an the paint and subpaint algebras are as follows:
\begin{eqnarray}
  \mathbb{G}_{\mathrm{paint}} &=& \so(p) \nonumber\\
  \mathbb{G}_{\mathrm{subpaint}} &=& \so(p-1) \label{felanina1}
\end{eqnarray}
The symplectic $\mathbf{W}$ representation of $\mathbb{U}_{\mathcal{SK}}$ is the tensor product of the fundamental representation of $\slal(2)$ with the fundamental vector representation of $\so(2,2+p)$, namely
\begin{equation}\label{erast1}
    \mathbf{W} \, = \, \left( \mathbf{2 | 4}+p\right) \quad ; \quad \mbox{dim} \,\mathbf{W} \, = \, 8+2p
\end{equation}
The decomposition of this representation with respect to $\mathbb{G}_{\mathrm{subTS}} \oplus \mathbb{G}_{\mathrm{subpaint}}$ is the following one:
\begin{equation}\label{decompusOne}
    \mathbf{W} \, \stackrel{\mathbb{G}_{\mathrm{subTS}} \oplus \mathbb{G}_{\mathbf{subpaint}}}{\Longrightarrow} \,  \left(\mathbf{2,2,2|1}\right) \oplus \left(\mathbf{2,1,1|1}\right) \oplus \left(\mathbf{2,1,1}|p-1\right)
\end{equation}
where $\mathbf{2,2,2}$ denotes the tensor product of the three fundamental representations of $\slal(2,\mathbb{R})^3$. Similarly $\mathbf{2,1,1}$ denotes the doublet of the first $\slal(2,\mathbb{R})$ tensored with the singlets of the following two $\slal(2,\mathbb{R})$ algebras. The representations appearing in (\ref{decompusOne}) can be grouped in order to reconstruct full representations either of the complete Tits Satake or of the complete paint algebras. In this way one obtains:
\begin{eqnarray}
   \mathbf{W} & \stackrel{\mathbb{G}_{\mathrm{subTS}} \oplus \mathbb{G}_{\mathbf{paint}}}{\Longrightarrow} &  \left(\mathbf{2,2,2|1}\right) \oplus  \left(\mathbf{2,1,1}|p+1\right)\nonumber\\
  \mathbf{W} & \stackrel{\mathbb{G}_{\mathrm{TS}} \oplus \mathbb{G}_{\mathbf{subpaint}}}{\Longrightarrow} &  \left(\mathbf{2,5|1}\right) \oplus  \left(\mathbf{2,1}|p-1\right)
\end{eqnarray}
\subsection{\sc $\mathbf{W}$-representations of the maximally split non exotic models}
In the previous subsections we have analyzed the Tits-Satake decomposition of the $\mathbf{W}$-representation for all those models that are non maximally split. The remaining models are the maximally split ones for which there is no paint algebra and the Tits Satake projection is the identity map. There are essentially four type of models:
\begin{enumerate}
  \item The $\mathrm{SU(1,1)}$ non exotic model where the $\mathbf{W}$-representation is the $j=\ft 32$ of $\so(1,2)\sim \su(1,1)$
  \item The $\mathrm{Sp(6,\mathbb{R})}$ model where the $\mathbf{W}$-representation is the $\mathbf{14}^\prime$ (antisymmetric symplectic traceless three-tensor).
  \item The models $\slal(2,\mathbb{R}) \oplus \so(q,q)$ where the $\mathbf{W}$-representation is the $\left(\mathrm{2,2q}\right)$, namely the tensor product of the two fundamentals.
   \item The models $\slal(2,\mathbb{R}) \oplus \so(q,q+1)$ where the $\mathbf{W}$-representation is the $\left(\mathrm{2,2q+1}\right)$, namely the tensor product of the two fundamentals.
\end{enumerate}
Therefore, for the above maximally split models, we need the classification of  $\mathrm{U_{\mathcal{SK}}}$ orbits in the mentioned $\mathbf{W}$-representations. Actually  such orbits are sufficient also for the non maximally split models. Indeed each of the above $4$-models correspond to one Tits Satake universality class
and, within each universality class, the only relevant part of the $\mathbf{W}$-representation is the subpaint group singlet which is universal for all members of the class. This is precisely what we verified in the previous subsections.
\par
For instance for all members of the universality class of $\mathrm{Sp(6,\mathbb{R})}$, the $\mathbf{W}$-representation splits as follows with respect to the subalgebra $\sym(6,\mathbb{R})\oplus \mathbb{G}_{\mathrm{subpaint}}$:
\begin{equation}\label{gribochky}
    \mathbf{W} \, \stackrel{\sym(6,\mathbb{R})\oplus \mathbb{G}_{\mathrm{subpaint}}}{\Longrightarrow} \, \left( \mathbf{6}\, | \, \mathcal{D}_{\mathrm{subpaint}}\right) \, + \, \left( \mathbf{14}^\prime \, | \, \mathbf{1}_{\mathrm{subpaint}}\right)
\end{equation}
where the representation $\mathcal{D}_{\mathrm{subpaint}}$ is the following one for the three non-maximally split members of the class:
\begin{equation}\label{reppisubpitturi}
    \mathcal{D}_{\mathrm{subpaint}} \, = \, \left\{ \begin{array}{ccccc}
                                                      \mathbf{1} & \mbox{of} & \mathbf{1} & \mbox{for the} & \su(3,3)-\mbox{model} \\
                                                      \mathbf{3} & \mbox{of} & \so(3) & \mbox{for the} & \so^\star(12)-\mbox{model} \\
                                                      \mathbf{7} & \mbox{of} & \mathfrak{g}_{2(-14)} & \mbox{for the} & \mathfrak{e}_{7(-25)}-\mbox{model}\\
                                                    \end{array}\right.
\end{equation}
Clearly the condition:
\begin{equation}\label{condosputta}
    \left( \mathbf{6}\, | \, \mathcal{D}_{\mathrm{subpaint}}\right)\, = \, 0
\end{equation}
imposed on a vector in the $\mathbf{W}$-representation breaks the group $\mathrm{U}_{\mathcal{SK}}$  to its Tits Satake subgroup. The key point is that each $\mathbf{W}$-orbit of the big group $\mathrm{U}_{\mathcal{SK}}$ crosses the locus (\ref{condosputta}) so that the classification  of $\mathrm{Sp(6,\mathbb{R})}$ orbits in the $\mathbf{14}^\prime$-representation exhausts the classification of $\mathbf{W}$-orbits for all members of the universality class.
\par
In order to prove that the gauge (\ref{condosputta}) is always reachable it suffices to show that the representation $\left( \mathbf{6}\, | \,\mathcal{D}_{\mathrm{subpaint}}\right )$ always appears at least once in the decomposition of the Lie algebra $\mathbb{U}_{\mathcal{SK}}$ with respect to the subalgebra $\sym(6,\mathbb{R})\oplus \mathbb{G}_{\mathrm{subpaint}}$. The corresponding parameters of the big group can be used to set to zero the projection of the $\mathbf{W}$-vector onto $\left( \mathbf{6}\, | \,\mathcal{D}_{\mathrm{subpaint}}\right )$.
\par
The required condition is easily verified since we have:
\begin{eqnarray}
  \underbrace{\mbox{adj}\, \su(3,3)}_{\mathbf{35}} &\stackrel{\sym(6,\mathbb{R})}{\Longrightarrow} \,& \underbrace{\mbox{adj}\, \sym(6,\mathbb{R})}_{\mathbf{21}} \, \oplus \, \mathbf{6} \, \oplus \, \mathbf{6} \, \oplus \, \mathbf{1} \, \oplus \, \mathbf{1}\nonumber \\
 \underbrace{\mbox{adj}\, \so^\star(12)}_{\mathbf{66}} &\stackrel{\sym(6,\mathbb{R})\oplus \so(3)}{\Longrightarrow} \,& \underbrace{\mbox{adj}\, \sym(6,\mathbb{R})}_{\mathbf{21}} \, \oplus \, \underbrace{\mbox{adj}\, \so(3)}_{\mathbf{3}}\,
  \oplus \, \left(\mathbf{6} , \mathbf{3}\right)\, \oplus \, \left(\mathbf{6} , \mathbf{3}\right)\oplus \, \left(\mathbf{1},\mathbf{3}\right)\,\oplus \, \left(\mathbf{1},\mathbf{3}\right)\,\nonumber \\
 \underbrace{\mbox{adj}\, \mathfrak{e}_{7(-25)}}_{\mathbf{133}} &\stackrel{\sym(6,\mathbb{R})\oplus \mathfrak{g}_{2(-14)}}{\Longrightarrow} \,& \underbrace{\mbox{adj}\, \sym(6,\mathbb{R})}_{\mathbf{21}} \, \oplus \, \underbrace{\mbox{adj}\, \mathfrak{g}_{2(-14)}}_{\mathbf{14}}\,
  \oplus \, \left(\mathbf{6} , \mathbf{7}\right)\, \oplus \, \left(\mathbf{6} , \mathbf{7}\right)\oplus \, \left(\mathbf{1},\mathbf{7}\right)\,\oplus \, \left(\mathbf{1},\mathbf{7}\right)\,\nonumber \\
\end{eqnarray}
The reader cannot avoid being impressed by the striking similarity of the above decompositions which encode the very essence of  Tits Satake universality. Indeed the representations of the common Tits Satake subalgebra appearing in the decomposition of the adjoint are the same for all members of the class. They are simply uniformly assigned to the fundamental representation of the subpaint algebra which is different in the three cases. The representation $\left( \mathbf{6}\, | \,\mathcal{D}_{\mathrm{subpaint}}\right )$ appears twice in these decompositions and can be used to reach the gauge (\ref{condosputta}) as we claimed above.
\paragraph{\sc Holomorphic  consistent truncations} The next point to remark is that the condition (\ref{condosputta}) has another important interpretation if applied to the holomorphic section of special geometry. The key point is the following numerical identity valid for all members of the universality class:
\begin{equation}\label{curiosona}
    \mbox{dim} \, \frac{\mathrm{U}_{\mathcal{SK}}}{\mathrm{H}_{\mathcal{SK}}}\, = \,
     \mbox{dim}  \, \frac{\mathrm{U}^{TS}_{\mathcal{SK}}}{\mathrm{H}^{TS}_{\mathcal{SK}}}\, \oplus \, 6 \, \times \, \mbox{dim}\, \mathcal{D}_{\mathrm{subpaint}}
\end{equation}
This means that if we decompose the symplectic section of the big group according to the Tits-Satake subalgebra and we impose on it the condition (\ref{condosputta}) we just obtain the right number of holomorphic constraints to project onto the submanifold $\frac{\mathrm{U}_{\mathcal{SK}}}{\mathrm{H}_{\mathcal{SK}}}$. At the level of field equations this is certainly a consistent truncation, since we project onto the singlets of the subpaint group.
\par
On the other hand if we decompose the $\mathbf{W}$-representation   with respect to the sub-Tits-Satake subalgebra $\slal(2)\times \slal(2) \times \slal(2)$ we have the branching rule:
\begin{equation}\label{fillopona}
    \mathbf{W} \, \rightarrow \, \left(\mathcal{D}_1 |\mathbf{2},\mathbf{1},\mathbf{1}\right)\, \oplus \, \left(\mathcal{D}_2 |\mathbf{1},\mathbf{2},\mathbf{1}\right)\, \oplus \,\left(\mathcal{D}_3 |\mathbf{1},\mathbf{1},\mathbf{2}\right) \, \oplus \, \left(\mathbf{1} |\mathbf{2},\mathbf{2},\mathbf{2}\right)
\end{equation}
where $\mathcal{D}_{1,2,3}$ are three suitable representations of the Paint Group. Imposing on the symplectic section of the big model
the constraints:
\begin{eqnarray}
  \left(\mathcal{D}_1 |\mathbf{2},\mathbf{1},\mathbf{1}\right)&=& 0 \nonumber\\
  \left(\mathcal{D}_2 |\mathbf{1},\mathbf{2},\mathbf{1}\right) &=& 0 \nonumber\\
  \left(\mathcal{D}_3 |\mathbf{1},\mathbf{1},\mathbf{2}\right) &=& 0
\end{eqnarray}
yields precisely the correct number of holomorphic constraints that restrict the considered Special K\"ahler manifold to the Special K\"ahler manifold of the STU-model namely to $\left(\frac{\mathrm{SL(2,\mathbb{R})}}{\mathrm{SO(2)}}\right)^3$. This follows from the numerical identity true for all members of the universality class:
\begin{equation}\label{guttallaxa}
     \mbox{dim} \, \frac{\mathrm{U}_{\mathcal{SK}}}{\mathrm{H}_{\mathcal{SK}}}\, = \, \sum_{i=1}^3 \, 2 \, \times \, \mbox{dim} \, \mathcal{D}_i \, + \, 6
\end{equation}
The reason why the truncation to the STU-model is always a consistent truncation at the level of field equations is obvious in this set up. It corresponds to the truncation to the Paint Group singlets.
\paragraph{\sc $\mathbf{W}$-representations for the remaining models}
For the models of type $\slal(2,\mathbb{R})\oplus \so(q,q+p)$ having $\slal(2,\mathbb{R})\oplus \so(q,q+1)$ as Tits Satake subalgebra and $\so(p-1)$ as subpaint algebra the decomposition of the $\mathbf{W}$-representation is the following one:
\begin{equation}\label{funghifritti}
    \mathbf{W} \, = \, \left(\mathbf{2,2q+p}\right) \, \stackrel{\slal(2,\mathbb{R})\oplus \so(q,q+1)\oplus \so(p-1)}{\Longrightarrow} \, \left(\mathbf{2,2q+1}|\mathbf{1}\right) \, \oplus \, \left(\mathbf{2,1}|\mathbf{p-1}\right)
\end{equation}
and the question is whether each $\slal(2,\mathbb{R})\oplus \so(q,q+p)$ orbit in the $\left(\mathbf{2,2q+p}\right)$ representation intersects the $\slal(2,\mathbb{R})\oplus \so(q,q+1)\oplus \so(p-1)$-invariant locus:
\begin{equation}\label{fittone}
    \left(\mathbf{2,1}|\mathbf{p-1}\right) \, = \,0
\end{equation}
The answer is yes since we always have enough parameters in the coset
\begin{equation}\label{ciabatta}
    \frac{\mathrm{SL(2,\mathbb{R})}\times \mathrm{SO(q,q+p)}}{\mathrm{SL(2,\mathbb{R})\times SO(q,q+1)\times SO(p-1})}
\end{equation}
to reach the desired gauge (\ref{fittone}). Indeed let us observe the decomposition:
\begin{equation}\label{fruttodimare}
    \mbox{adj}\, \left[\slal(2,\mathbb{R})\oplus \so(q,q+p)\right] \, = \, \mbox{adj}\, \left[\slal(2,\mathbb{R})\right] \oplus \mbox{adj}\, \left[\so(q,q+1)\right] \, \oplus \mbox{adj}\, \left[\so(p-1)\right] \, \oplus \, \left(\mathbf{1,2q+1 | p-1}\right)
\end{equation}
The $2q+1$ vectors of $\so(p-1)$ appearing in (\ref{fruttodimare}) are certainly sufficient to set to zero the $2$ vectors of $\so(p-1)$ appearing in $\mathbf{W}$.
\par
Relevant for the case of $\mathcal{N}\,=2\,$ supersymmetry is the value $q=2$ and in this case the sub-Tits-Satake Lie algebra is :
\begin{equation}\label{grilloparlante}
    \mathbb{G}_{\mathrm{subTS}} \, = \, \slal(2,\mathbb{R})\oplus \so(2,2) \, = \, \slal(2,\mathbb{R})\oplus\slal(2,\mathbb{R})\oplus\slal(2,\mathbb{R})
\end{equation}
namely it is once again the Lie algebra of the STU-model. Reduction to the STU-model is consistent for the same reason as in the other universality classes: it corresponds to truncation to Paint Group singlets.
\subsection{\sc Gaugings with consistent one-field truncations}
On the basis of the analysis presented in the previous section we arrive at the following conclusion. By gauging a nilpotent element of the isometry subalgebra of $\mathcal{SK}$ inside $\mathcal{QM}$ we generate a potential. The structure of the theory depends on the nilpotent orbit, namely on the embedding of an $\slal(2)$ Lie algebra in $\mathbb{U}_{\mathcal{SK}}$ and there are many ways of doing this (the orbits), yet the gauged theory will admit a one-field truncation if and only if the $\slal(2)$ is embedded into the sub Tits Satake Lie algebra:
\begin{equation}\label{gongolato}
    \slal(2) \, \mapsto \, \mathbb{G}_{\mathrm{subTS}}\, \subset \, \mathbb{U}_{\mathcal{SK}}
\end{equation}
There are only three different embeddings of $\slal(2)$ into $\left( \slal(2) \right)^3$ and these correspond to the three admissible values $\alpha\, = \, 1 , \, \frac{2}{3}, \, \frac{1}{3}$ in the Starobinsky-like model.
\section{\sc Conclusions}
In this paper we have analyzed in detail the structure of the $c$-map from Special K\"ahler manifolds to Quaternionic K\"ahler manifolds in connection with the  abelian gaugings of hypermultiplet isometries in $\mathcal{N}=2$ supergravity and the generation of effective one field potentials that might describe inflaton dynamics.
\par
The motivations of such a study have been put forward in the introduction. Here we try to summarize the main results  we have obtained and the perspectives for future investigations that have emerged.
\paragraph{\sc Results}
\begin{description}
  \item[I] The $c$-map description of Quaternionic K\"ahler manifolds $\mathcal{QM}_{4n+4}$ allows to distribute the isometries of
  $\mathcal{QM}_{4n+4}$ into three classes:
  \begin{description}
    \item[Ia)] The isometries associated with the Heisenberg algebra which exists in all cases, even when the internal Special K\"ahler manifold $\mathcal{SK}_{n}$ has no continuous symmetries as it might happen when we deal the moduli space of a Calabi-Yau three-fold and the Heisenberg fields are the Ramond--Ramond scalars of a superstring compactification. The gauging of such perturbative isometries produces one field potentials that are of the pure exponential type.
    \item[Ib)] The isometries of the inner Special K\"ahler manifold $\mathcal{SK}_{n}$ that can always be promoted to isometries of the full Quaternionic manifold $\mathcal{QM}_{4n+4}$. Gauging these isometries one obtains a potential that at $\mathbf{Z}^\alpha \, =\,  a \, = \, U \, = \, 0$ is identical with the potential obtained from the gauging of the same K\"ahler isometries in $\mathcal{N}=1$ supersymmetry.  Hence one would like to be able to stabilize these fields. As for $\mathbf{Z}^\alpha \, =\,  a \, = \, 0$ there is no problem. We can always truncate them. The field $U$ instead, that in superstring interpretations of the hypermultiplets can be identified with the coupling constant dilaton field, appears through exponentials all of the same sign in these perturbative gaugings and cannot be stabilized.
   \item[Ic)] The non perturbative isometries that mix the Heisenberg symmetries with the K\"ahler symmetries. These exist only when $\mathcal{QM}_{4n+4}$ is a symmetric homogeneous space and the Lie algebra of the full isometry group takes the universal form (\ref{genGD3}). Gauging a compact non perturbative generator appears to be the only way of introducing exponentials with opposite sign of the field $U$ allowing for its stabilization.
  \end{description}
  \item[II] The Starobinsky-like  potential:
  \begin{equation}\label{bambolone2}
   V_{Starobinsky} \, = \, \mbox{const} \, \times \, \left( 1\, - \, \exp\left [ - \, \sqrt{\frac{2}{3 \, \alpha}} \, \phi\right]\right)^2
 \end{equation}
  can be obtained  universally  from all homogeneous symmetric Quaternionic K\"ahler manifolds by means of an admissible embedding  of a $\slal(2,\mathbb{R})$ Lie algebra into the Special K\"ahler subalgebra $\mathbb{U}_\mathcal{SK}\,\subset \, \mathbb{U}_\mathcal{Q}$. The problem of embedding of $\slal(2,\mathbb{R})$ algebras is the same as the problem of classifying standard triples and nilpotent algebras, yet the embedding must also be admissible, in the sense that it should allow for consistent one-field truncations. In the case of $\mathcal{SK} \,  = \, \sym(6,\mathbb{R})$ we exhausted the analysis of orbits and showed that admissible subalgebras correspond to embeddings of $\slal(2)$ into the maximal subalgebra $\left(\slal(2)\right)^3$ yielding $\alpha \, = \, 1, \, \frac{2}{3}, \, \frac{1}{3}$. By means of arguments based on Tits Satake universality classes, we have advocated that this result is general for all supergravity theories where the hypermultiplets are described by a homogeneous symmetric space.
  \item[III] The above results rely on the use of the minimal coupling Special Geometry for the description of the vector multiplets. Staying within such a framework one can gauge by different vector fields the generators of a maximal set of mutually commuting $\slal(2,\mathbb{R})$ Lie algebras and obtain as many massive fields. The number of massive vector fields appears therefore to be equal to the rank of the Tits Satake subalgebra $\mathbb{G}_{\mathrm{subTS}} \, \subset \, \mathbb{U}_\mathcal{SK}\,\subset \, \mathbb{U}_\mathcal{Q}$. The result of \cite{thesearch} where it was found that the maximal number of massive vector fields in this sort of gauging is one is confirmed by this general rule. In \cite{thesearch} we have $\mathbb{G}_{\mathrm{subTS}}\, =\, \slal(2,\mathbb{R}) \, = \, \mathbb{U}_\mathcal{SK}\,\subset \, \mathbb{U}_\mathcal{Q} \, \equiv \, \mathfrak{g}_{(2,2)}$ and $\mbox{rank}\,\slal(2,\mathbb{R})\, = \, 1$. In all other cases the rank of the sub Tits Satake Lie algebra is three.
\end{description}
\paragraph{\sc Perspectives and Generalizations}
In order to improve our understanding of the possible embedding  of inflaton dynamics within larger unified theories, we consider necessary and feasible to explore the following directions.
\begin{description}
  \item[A)] Enlarge the scope of gaugings of hypermultiplet isometries to non abelian algebras and consider the attractive situation where the Special K\"ahler manifold describing the vector multiplets and that, whose $c$-map provides the Quaternionic K\"ahler manifold description of the hypermultiplets are just the same or at least belong to the same class of homogeneous Special K\"ahler spaces.
  \item[B] Utilize the above $\mathcal{N}=2$ set up to promote the embedding of inflaton dynamics from $\mathcal{N}=2$ to higher $\mathcal{N}$, in particular to $\mathcal{N}=3$ and $\mathcal{N}=4$.
  \item[C] Consider such scenarios as superstring compactification on $T^2 \times K3$ and try to interpret the gaugings that produce inflaton dynamics in terms of fluxes.
\end{description}
We plan to address such questions in forthcoming publications.
\par
\section*{Acknowledgements}
We are grateful to our friends Sergio Ferrara and Antoine Van Proyen for useful discussions during the completion of this work.
The work of A.S. was supported in part by the RFBR Grants
No. 13-02-91330-NNIO-a, No. 13-02-90602-Arm-a and by the
Heisenberg-Landau program.
\newpage
\appendix
\begin{landscape}
\section{\sc Coset representatives and other large formulae not displayed in the main text}
In this appendix we collect some large formulae that need the landscape format and are not presented in the main text.
\subsection{\sc Concerning the $\mathrm{G_{(2,2)}}$ model}
\paragraph{\sc The solvable coset representative for $\frac{\mathrm{G_{(2,2)}}}{\mathrm{SU(2)} \times \mathrm{SU(2)}}$}
\begin{eqnarray}\label{fornoalegna}
    & \mathbb{L}  \, = \, & \nonumber\\
\end{eqnarray}
{\tiny
\begin{eqnarray}\label{cosettuspresento}
  &\left(
\begin{array}{lllllll}
 e^{\frac{h+U}{2}} & e^{\frac{U-h}{2}} y &
   -\sqrt{\frac{2}{3}} e^h Z_3 & -\frac{2
   Z_1}{\sqrt{3}}-\frac{2 y Z_3}{\sqrt{3}} &
   -\sqrt{\frac{2}{3}} e^{-h} Z_3 y^2-2
   \sqrt{\frac{2}{3}} e^{-h} Z_1 y-\sqrt{2}
   e^{-h} Z_2 & e^{\frac{h-U}{2}}
   \left(a-\frac{Z_1 Z_3}{3}-Z_2 Z_4\right) &
   e^{-\frac{h}{2}-\frac{U}{2}} \left(\frac{2 Z_2
   Z_3}{\sqrt{3}}-\frac{2
   Z_1^2}{3}\right)+e^{-\frac{h}{2}-\frac{U}{2}}
   y \left(a-\frac{Z_1 Z_3}{3}-Z_2 Z_4\right) \\
 0 & e^{\frac{U-h}{2}} & -\sqrt{2} e^h Z_4 &
   \frac{2 Z_3}{\sqrt{3}}-2 y Z_4 & -\sqrt{2}
   e^{-h} Z_4 y^2+2 \sqrt{\frac{2}{3}} e^{-h} Z_3
   y+\sqrt{\frac{2}{3}} e^{-h} Z_1 &
   e^{\frac{h-U}{2}} \left(\frac{2
   Z_3^2}{3}+\frac{2 Z_1 Z_4}{\sqrt{3}}\right) &
   e^{-\frac{h}{2}-\frac{U}{2}} y \left(\frac{2
   Z_3^2}{3}+\frac{2 Z_1
   Z_4}{\sqrt{3}}\right)+e^{-\frac{h}{2}-\frac{U}
   {2}} \left(a+\frac{Z_1 Z_3}{3}+Z_2 Z_4\right)
   \\
 0 & 0 & e^h & \sqrt{2} y & e^{-h} y^2 &
   -\sqrt{\frac{2}{3}} e^{\frac{h-U}{2}} Z_1 &
   -\sqrt{\frac{2}{3}}
   e^{-\frac{h}{2}-\frac{U}{2}} y Z_1-\sqrt{2}
   e^{-\frac{h}{2}-\frac{U}{2}} Z_2 \\
 0 & 0 & 0 & 1 & \sqrt{2} e^{-h} y & \frac{2
   e^{\frac{h-U}{2}} Z_3}{\sqrt{3}} & \frac{2
   e^{-\frac{h}{2}-\frac{U}{2}}
   Z_1}{\sqrt{3}}+\frac{2
   e^{-\frac{h}{2}-\frac{U}{2}} y Z_3}{\sqrt{3}}
   \\
 0 & 0 & 0 & 0 & e^{-h} & \sqrt{2}
   e^{\frac{h-U}{2}} Z_4 & \sqrt{2}
   e^{-\frac{h}{2}-\frac{U}{2}} y
   Z_4-\sqrt{\frac{2}{3}}
   e^{-\frac{h}{2}-\frac{U}{2}} Z_3 \\
 0 & 0 & 0 & 0 & 0 & e^{\frac{h-U}{2}} &
   e^{-\frac{h}{2}-\frac{U}{2}} y \\
 0 & 0 & 0 & 0 & 0 & 0 &
   e^{-\frac{h}{2}-\frac{U}{2}}
\end{array}
\right)&\nonumber\\
\end{eqnarray}
}
\subsection{\sc Explicit form of generators for the Lie algebra $\sym(6,\mathbb{R})$ in the $\mathbf{14}^\prime$ representation}
\label{seziona14}
The $\mathbf{14}^\prime$ representation of $\sym(6,\mathbb{R})$ which plays the role $\mathbf{W}$-representation for the special manifold under consideration is defined as the representation obeyed by the  three-times antisymmetric tensors with vanishing $\mathbb{C}$-traces, namely:
\begin{equation}\label{goliutico}
    \underbrace{t_{ABC}}_{\mbox{antisymmetric in} A,B,C} \times \quad \mathbb{C}^{BC} \, = \,0
\end{equation}
Let us then consider a lexicographic ordered basis for the $20$-dimensional reducible representation provided by the three times antisymmetric tensor:
{\small \begin{equation}\label{cirillo}
\left(
\begin{array}{l}
 V_1 \\
 V_2 \\
 V_3 \\
 V_4 \\
 V_5 \\
 V_6 \\
 V_7 \\
 V_8 \\
 V_9 \\
 V_{10} \\
 V_{11} \\
 V_{12} \\
 V_{13} \\
 V_{14} \\
 V_{15} \\
 V_{16} \\
 V_{17} \\
 V_{18} \\
 V_{19} \\
 V_{20}
\end{array}
\right) \, \equiv \,    \left(
\begin{array}{l}
 t_{ 1     2    3} \\
t_{ 1     2    4 }\\
 t_{ 1     3    4 }\\
t_{ 2     3    4} \\
t_{ 1    2    5 }\\
 t_{ 1    3    5 }\\
t_{ 2     3    5 }\\
t_{ 1      4    5 }\\
t_{ 2     4    5 }\\
t_{ 3     4    5 }\\
t_{ 1     2    6 }\\
 t_{ 1     3    6 }\\
t_{ 2     3    6 }\\
t_{ 1      4    6 }\\
t_{ 2     4    6 }\\
t_{ 3     4    6 }\\
 t_{ 1     5    6 }\\
 t_{ 2     5    6 }\\
t_{ 3    5    6 }\\
t_{ 4    5    6}
\end{array}
\right)
\end{equation}
}
The splitting into the two irreducible subspaces of dimension $\mathbf{14}$ and $\mathbf{6}$ respectively can be performed by defining the following new basis vectors:
\begin{equation}\label{splittu}
    \left(
\begin{array}{l}
 \Phi _1 \\
 \Phi _2 \\
 \Phi _3 \\
 \Phi _4 \\
 \Phi _5 \\
 \Phi _6 \\
 \Phi _7 \\
 \Phi _8 \\
 \Phi _9 \\
 \Phi _{10} \\
 \Phi _{11} \\
 \Phi _{12} \\
 \Phi _{13} \\
 \Phi _{14}
\end{array}
\right) \, = \, \left(
\begin{array}{l}
 V_1 \\
 V_4 \\
 V_6 \\
 V_{10} \\
 V_{11} \\
 V_{15} \\
 V_{17} \\
 V_{20} \\
 V_5-V_{12} \\
 -V_2-V_{13} \\
 V_7-V_3 \\
 V_{16}-V_9 \\
 V_8+V_{19} \\
 V_{14}-V_{18}
\end{array}
\right) \quad ; \quad \left(
\begin{array}{l}
 C_1 \\
 C_2 \\
 C_3 \\
 C_4 \\
 C_5 \\
 C_6
\end{array}
\right) \, = \, \left(
\begin{array}{l}
 V_5+V_{12} \\
 V_{13}-V_2 \\
 -V_3-V_7 \\
 -V_9-V_{16} \\
 V_8-V_{19} \\
 V_{14}+V_{18}
\end{array}
\right)
\end{equation}
Taking the antisymmetric cubic tensor product and using the splitting (\ref{splittu}), the matrices $\widehat{\mathcal{D}}_{14}(\mathfrak{g})$ representing any element  $\mathfrak{g} \in \sym(6,\mathbb{R})$ of the Lie algebra in the $\mathbf{14}^\prime$ representation can  be easily extracted. The so obtained $\widehat{\mathcal{D}}_{14}(\mathfrak{g})$ matrices are symplectic, since, by direct calculation one can determine a unique  antisymmetric matrix:
\begin{equation}\label{c14matra}
    \widehat{\mathbb{C}}_{14} \, = \,\left(
\begin{array}{llllllllllllll}
 0 & 0 & 0 & 0 & 0 & 0 & 0 & 2 & 0 & 0
   & 0 & 0 & 0 & 0 \\
 0 & 0 & 0 & 0 & 0 & 0 & -2 & 0 & 0 & 0
   & 0 & 0 & 0 & 0 \\
 0 & 0 & 0 & 0 & 0 & -2 & 0 & 0 & 0 & 0
   & 0 & 0 & 0 & 0 \\
 0 & 0 & 0 & 0 & 2 & 0 & 0 & 0 & 0 & 0
   & 0 & 0 & 0 & 0 \\
 0 & 0 & 0 & -2 & 0 & 0 & 0 & 0 & 0 & 0
   & 0 & 0 & 0 & 0 \\
 0 & 0 & 2 & 0 & 0 & 0 & 0 & 0 & 0 & 0
   & 0 & 0 & 0 & 0 \\
 0 & 2 & 0 & 0 & 0 & 0 & 0 & 0 & 0 & 0
   & 0 & 0 & 0 & 0 \\
 -2 & 0 & 0 & 0 & 0 & 0 & 0 & 0 & 0 & 0
   & 0 & 0 & 0 & 0 \\
 0 & 0 & 0 & 0 & 0 & 0 & 0 & 0 & 0 & 0
   & 0 & 1 & 0 & 0 \\
 0 & 0 & 0 & 0 & 0 & 0 & 0 & 0 & 0 & 0
   & 0 & 0 & 1 & 0 \\
 0 & 0 & 0 & 0 & 0 & 0 & 0 & 0 & 0 & 0
   & 0 & 0 & 0 & 1 \\
 0 & 0 & 0 & 0 & 0 & 0 & 0 & 0 & -1 & 0
   & 0 & 0 & 0 & 0 \\
 0 & 0 & 0 & 0 & 0 & 0 & 0 & 0 & 0 & -1
   & 0 & 0 & 0 & 0 \\
 0 & 0 & 0 & 0 & 0 & 0 & 0 & 0 & 0 & 0
   & -1 & 0 & 0 & 0
\end{array}
\right)
\end{equation}
which  verifies the relation:
\begin{equation}\label{nostandard}
 \forall \, \mathfrak{g} \, \in \, \sym(6,\mathbb{R}) \quad : \quad   \widehat{\mathcal{D}}_{14}(\mathfrak{g})^T \, \widehat{\mathbb{C}}_{14} \, + \, \widehat{\mathbb{C}}_{14} \, \widehat{\mathcal{D}}_{14}(\mathfrak{g}) \, = \, 0
\end{equation}
Unfortunately $\widehat{\mathbb{C}}_{14}$ is not yet the standard symplectic matrix for the Lie algebra $\sym(14,\mathbb{R})$. Hence we still need to perform a change of basis that brings $\widehat{\mathbb{C}}_{14}$ to its standard form:
\begin{equation}\label{c14normal}
    {\mathbb{C}}_{14} \, \equiv \,\left(
\begin{array}{llllllllllllll}
 0 & 0 & 0 & 0 & 0 & 0 & 0 & 1 & 0 & 0 & 0 & 0 & 0 & 0 \\
 0 & 0 & 0 & 0 & 0 & 0 & 0 & 0 & 1 & 0 & 0 & 0 & 0 & 0 \\
 0 & 0 & 0 & 0 & 0 & 0 & 0 & 0 & 0 & 1 & 0 & 0 & 0 & 0 \\
 0 & 0 & 0 & 0 & 0 & 0 & 0 & 0 & 0 & 0 & 1 & 0 & 0 & 0 \\
 0 & 0 & 0 & 0 & 0 & 0 & 0 & 0 & 0 & 0 & 0 & 1 & 0 & 0 \\
 0 & 0 & 0 & 0 & 0 & 0 & 0 & 0 & 0 & 0 & 0 & 0 & 1 & 0 \\
 0 & 0 & 0 & 0 & 0 & 0 & 0 & 0 & 0 & 0 & 0 & 0 & 0 & 1 \\
 -1 & 0 & 0 & 0 & 0 & 0 & 0 & 0 & 0 & 0 & 0 & 0 & 0 & 0 \\
 0 & -1 & 0 & 0 & 0 & 0 & 0 & 0 & 0 & 0 & 0 & 0 & 0 & 0 \\
 0 & 0 & -1 & 0 & 0 & 0 & 0 & 0 & 0 & 0 & 0 & 0 & 0 & 0 \\
 0 & 0 & 0 & -1 & 0 & 0 & 0 & 0 & 0 & 0 & 0 & 0 & 0 & 0 \\
 0 & 0 & 0 & 0 & -1 & 0 & 0 & 0 & 0 & 0 & 0 & 0 & 0 & 0 \\
 0 & 0 & 0 & 0 & 0 & -1 & 0 & 0 & 0 & 0 & 0 & 0 & 0 & 0 \\
 0 & 0 & 0 & 0 & 0 & 0 & -1 & 0 & 0 & 0 & 0 & 0 & 0 & 0
\end{array}
\right)
\end{equation}
Such a change of basis is provided by the matrix:
\begin{equation}\label{Ortdil}
    \Lambda \, = \, \left(
\begin{array}{llllllllllllll}
 0 & 0 & 0 & \frac{1}{\sqrt{2}} & 0 & 0 & 0 & 0 & 0 & 0 & 0 & 0 & 0 & 0 \\
 0 & 0 & -\frac{1}{\sqrt{2}} & 0 & 0 & 0 & 0 & 0 & 0 & 0 & 0 & 0 & 0 & 0 \\
 0 & -\frac{1}{\sqrt{2}} & 0 & 0 & 0 & 0 & 0 & 0 & 0 & 0 & 0 & 0 & 0 & 0 \\
 \frac{1}{\sqrt{2}} & 0 & 0 & 0 & 0 & 0 & 0 & 0 & 0 & 0 & 0 & 0 & 0 & 0 \\
 0 & 0 & 0 & 0 & 0 & 0 & 0 & \frac{1}{\sqrt{2}} & 0 & 0 & 0 & 0 & 0 & 0 \\
 0 & 0 & 0 & 0 & 0 & 0 & 0 & 0 & \frac{1}{\sqrt{2}} & 0 & 0 & 0 & 0 & 0 \\
 0 & 0 & 0 & 0 & 0 & 0 & 0 & 0 & 0 & \frac{1}{\sqrt{2}} & 0 & 0 & 0 & 0 \\
 0 & 0 & 0 & 0 & 0 & 0 & 0 & 0 & 0 & 0 & \frac{1}{\sqrt{2}} & 0 & 0 & 0 \\
 0 & 0 & 0 & 0 & 1 & 0 & 0 & 0 & 0 & 0 & 0 & 0 & 0 & 0 \\
 0 & 0 & 0 & 0 & 0 & 1 & 0 & 0 & 0 & 0 & 0 & 0 & 0 & 0 \\
 0 & 0 & 0 & 0 & 0 & 0 & 1 & 0 & 0 & 0 & 0 & 0 & 0 & 0 \\
 0 & 0 & 0 & 0 & 0 & 0 & 0 & 0 & 0 & 0 & 0 & 1 & 0 & 0 \\
 0 & 0 & 0 & 0 & 0 & 0 & 0 & 0 & 0 & 0 & 0 & 0 & 1 & 0 \\
 0 & 0 & 0 & 0 & 0 & 0 & 0 & 0 & 0 & 0 & 0 & 0 & 0 & 1
\end{array}
\right)
\end{equation}
Indeed we have:
\begin{eqnarray}
  \Lambda^T \, {\widehat{\mathbb{C}}}_{14} \Lambda^T &=& {{\mathbb{C}}}_{14} \nonumber\\
  {\mathcal{D}}_{14}(\mathfrak{g})& \equiv & \Lambda^{-1} \, \widehat{\mathcal{D}}_{14}(\mathfrak{g}) \, \Lambda \nonumber\\
  0 & = & {\mathcal{D}}_{14}(\mathfrak{g})^T {\mathbb{C}}_{14} \, + \, {\mathbb{C}}_{14} \, {\mathcal{D}}_{14}(\mathfrak{g}) \label{normalno14}
\end{eqnarray}
We do not present the intermediate matrices $\widehat{\mathcal{D}}_{14}(\mathfrak{g})$. We go directly to the presentation of the final ones ${\mathcal{D}}_{14}(\mathfrak{g})$ that are standard symplectic matrices of the $\sym(14,\mathbb{R})$
Lie algebra.
\paragraph{\sc Cartan Generators in the $\mathbf{14}^\prime$.} The Cartan generators are named $\mathcal{H}^i_{14} \equiv \mathcal{D}_{14}\left(\mathcal{H}^i\right)$ and can be easily read-off from the following formula:
\begin{eqnarray}\label{cartolini14}
    &\sum_{i=1}^{3} \, h_i \, \mathcal{H}^i_{14} \, = \,& \nonumber\\
    &
 \mbox{\scriptsize $ \left(
\begin{array}{llllllllllllll}
 -h_1-h_2+h_3 & 0 & 0 & 0 & 0 & 0 &
   0 & 0 & 0 & 0 & 0 & 0 & 0 & 0 \\
 0 & h_1-h_2+h_3 & 0 & 0 & 0 & 0 &
   0 & 0 & 0 & 0 & 0 & 0 & 0 & 0 \\
 0 & 0 & -h_1+h_2+h_3 & 0 & 0 & 0 &
   0 & 0 & 0 & 0 & 0 & 0 & 0 & 0 \\
 0 & 0 & 0 & h_1+h_2+h_3 & 0 & 0 &
   0 & 0 & 0 & 0 & 0 & 0 & 0 & 0 \\
 0 & 0 & 0 & 0 & h_1 & 0 & 0 & 0 &
   0 & 0 & 0 & 0 & 0 & 0 \\
 0 & 0 & 0 & 0 & 0 & h_2 & 0 & 0 &
   0 & 0 & 0 & 0 & 0 & 0 \\
 0 & 0 & 0 & 0 & 0 & 0 & h_3 & 0 &
   0 & 0 & 0 & 0 & 0 & 0 \\
 0 & 0 & 0 & 0 & 0 & 0 & 0 &
   h_1+h_2-h_3 & 0 & 0 & 0 & 0 & 0
   & 0 \\
 0 & 0 & 0 & 0 & 0 & 0 & 0 & 0 &
   -h_1+h_2-h_3 & 0 & 0 & 0 & 0 & 0
   \\
 0 & 0 & 0 & 0 & 0 & 0 & 0 & 0 & 0
   & h_1-h_2-h_3 & 0 & 0 & 0 & 0 \\
 0 & 0 & 0 & 0 & 0 & 0 & 0 & 0 & 0
   & 0 & -h_1-h_2-h_3 & 0 & 0 & 0
   \\
 0 & 0 & 0 & 0 & 0 & 0 & 0 & 0 & 0
   & 0 & 0 & -h_1 & 0 & 0 \\
 0 & 0 & 0 & 0 & 0 & 0 & 0 & 0 & 0
   & 0 & 0 & 0 & -h_2 & 0 \\
 0 & 0 & 0 & 0 & 0 & 0 & 0 & 0 & 0
   & 0 & 0 & 0 & 0 & -h_3
\end{array}
\right) $ }&\nonumber\\
\end{eqnarray}
by collecting the coefficient of the parameter $h_i$.
\paragraph{\sc Positive Root Step Operators in the $\mathbf{14}^\prime$} The step operator associated with the positive root $\alpha_i$ is named $\mathcal{E}^{\alpha_i}_{14} \, \equiv \, \mathcal{D}_{14}\left(\mathcal{E}^{\alpha_i} \right)$ and can be easily read-off from the following formula:
 \begin{eqnarray}\label{steppinisu14}
   & \sum_{i=1}^{9} \, a_i \, \mathcal{E}^{\alpha_i}_{14} \, = \, & \nonumber\\
   &\mbox{\scriptsize $\left(
\begin{array}{llllllllllllll}
 0 & 0 & 0 & 0 & 0 & 0 & 0 & 0 & 0
   & 0 & \sqrt{2} a _3 & 0 & 0
   & 0 \\
 \sqrt{2} a _9 & 0 & 0 & 0 & 0
   & 0 & -\sqrt{2} a _1 & 0 &
   0 & \sqrt{2} a _3 & 0 & 0 &
   -\sqrt{2} a _6 & 0 \\
 -\sqrt{2} a _7 & 0 & 0 & 0 &
   0 & 0 & 0 & 0 & \sqrt{2} a
   _3 & 0 & 0 & -\sqrt{2} a _5
   & 0 & 0 \\
 0 & \sqrt{2} a _7 & -\sqrt{2}
   a _9 & 0 & \sqrt{2} a
   _5 & -\sqrt{2} a _6 &
   \sqrt{2} a _8 & \sqrt{2}
   a _3 & 0 & 0 & 0 & 0 & 0 &
   0 \\
 0 & -\sqrt{2} a _2 & 0 & 0 &
   0 & a _1 & -a _4 & 0 &
   0 & -\sqrt{2} a _5 & 0 &
   \sqrt{2} a _9 & a _8 &
   -a _6 \\
 0 & 0 & -\sqrt{2} a _4 & 0 &
   0 & 0 & a _2 & 0 &
   -\sqrt{2} a _6 & 0 & 0 &
   a _8 & \sqrt{2} a _7 &
   a _5 \\
 -\sqrt{2} a _8 & 0 & \sqrt{2}
   a _1 & 0 & 0 & 0 & 0 & 0 &
   0 & 0 & 0 & -a _6 & a
   _5 & \sqrt{2} a _3 \\
 0 & 0 & 0 & 0 & -\sqrt{2} a
   _2 & \sqrt{2} a _4 & 0 & 0
   & -\sqrt{2} a _9 & \sqrt{2}
   a _7 & 0 & 0 & 0 & \sqrt{2}
   a _8 \\
 0 & 0 & 0 & 0 & 0 & 0 & 0 & 0 & 0
   & 0 & -\sqrt{2} a _7 &
   \sqrt{2} a _2 & 0 & 0 \\
 0 & 0 & 0 & 0 & 0 & 0 & 0 & 0 & 0
   & 0 & \sqrt{2} a _9 & 0 &
   \sqrt{2} a _4 & -\sqrt{2}
   a _1 \\
 0 & 0 & 0 & 0 & 0 & 0 & 0 & 0 & 0
   & 0 & 0 & 0 & 0 & 0 \\
 -\sqrt{2} a _2 & 0 & 0 & 0 &
   0 & 0 & 0 & 0 & 0 & 0 &
   -\sqrt{2} a _5 & 0 & 0 & 0
   \\
 \sqrt{2} a _4 & 0 & 0 & 0 & 0
   & 0 & 0 & 0 & 0 & 0 & \sqrt{2}
   a _6 & -a _1 & 0 & 0
   \\
 0 & 0 & 0 & 0 & 0 & 0 & 0 & 0 &
   \sqrt{2} a _1 & 0 &
   -\sqrt{2} a _8 & a _4
   & -a _2 & 0
\end{array}
\right) $  }& \nonumber\\
\end{eqnarray}
by collecting the coefficient of the parameter $a_i$.
\paragraph{\sc Negative Root Step Operators in the $14^\prime$} The step operator associated with the negative root $-\alpha_i$ is named $\mathcal{E}^{-\alpha_i}_{14} \, \equiv \, \mathcal{D}_{14}\left(-\mathcal{E}^{\alpha_i} \right)$ and can be easily read-off from the following formula:
\begin{eqnarray}\label{steppinigiu14}
  &  \sum_{i=1}^{9} \, b_i \, \mathcal{E}^{-\alpha_i}_{14} \, = \, &
  \mbox{\scriptsize $ \left(
\begin{array}{llllllllllllll}
 0 & \sqrt{2} b_9 & -\sqrt{2} b_7 &
   0 & 0 & 0 & -\sqrt{2} b_8 & 0 &
   0 & 0 & 0 & -\sqrt{2} b_2 &
   \sqrt{2} b_4 & 0 \\
 0 & 0 & 0 & \sqrt{2} b_7 &
   -\sqrt{2} b_2 & 0 & 0 & 0 & 0 &
   0 & 0 & 0 & 0 & 0 \\
 0 & 0 & 0 & -\sqrt{2} b_9 & 0 &
   -\sqrt{2} b_4 & \sqrt{2} b_1 & 0
   & 0 & 0 & 0 & 0 & 0 & 0 \\
 0 & 0 & 0 & 0 & 0 & 0 & 0 & 0 & 0
   & 0 & 0 & 0 & 0 & 0 \\
 0 & 0 & 0 & \sqrt{2} b_5 & 0 & 0 &
   0 & -\sqrt{2} b_2 & 0 & 0 & 0 &
   0 & 0 & 0 \\
 0 & 0 & 0 & -\sqrt{2} b_6 & b_1 &
   0 & 0 & \sqrt{2} b_4 & 0 & 0 & 0
   & 0 & 0 & 0 \\
 0 & -\sqrt{2} b_1 & 0 & \sqrt{2}
   b_8 & -b_4 & b_2 & 0 & 0 & 0 & 0
   & 0 & 0 & 0 & 0 \\
 0 & 0 & 0 & \sqrt{2} b_3 & 0 & 0 &
   0 & 0 & 0 & 0 & 0 & 0 & 0 & 0 \\
 0 & 0 & \sqrt{2} b_3 & 0 & 0 &
   -\sqrt{2} b_6 & 0 & -\sqrt{2}
   b_9 & 0 & 0 & 0 & 0 & 0 &
   \sqrt{2} b_1 \\
 0 & \sqrt{2} b_3 & 0 & 0 &
   -\sqrt{2} b_5 & 0 & 0 & \sqrt{2}
   b_7 & 0 & 0 & 0 & 0 & 0 & 0 \\
 \sqrt{2} b_3 & 0 & 0 & 0 & 0 & 0 &
   0 & 0 & -\sqrt{2} b_7 & \sqrt{2}
   b_9 & 0 & -\sqrt{2} b_5 &
   \sqrt{2} b_6 & -\sqrt{2} b_8 \\
 0 & 0 & -\sqrt{2} b_5 & 0 &
   \sqrt{2} b_9 & b_8 & -b_6 & 0 &
   \sqrt{2} b_2 & 0 & 0 & 0 & -b_1
   & b_4 \\
 0 & -\sqrt{2} b_6 & 0 & 0 & b_8 &
   \sqrt{2} b_7 & b_5 & 0 & 0 &
   \sqrt{2} b_4 & 0 & 0 & 0 & -b_2
   \\
 0 & 0 & 0 & 0 & -b_6 & b_5 &
   \sqrt{2} b_3 & \sqrt{2} b_8 & 0
   & -\sqrt{2} b_1 & 0 & 0 & 0 & 0
\end{array}
\right)$} \nonumber \\
\end{eqnarray}
by collecting the coefficient of the parameter $b_i$.
\subsubsection
The explicit form of the metric in solvable coordinates reads:
{\scriptsize \begin{eqnarray}
\label{formidabile}
  &&ds^2_K \,=\,  \mathrm{dh}_1^2+\mathrm{dh}_2^2+\mathrm{dh}_3
   ^2+\frac{1}{2} e^{2 h_2-2 h_1}
   \mathrm{dp}_1^2+\frac{1}{2} e^{2 h_3-2
   h_2} \mathrm{dp}_2^2 \nonumber\\
  &&+\frac{1}{2} e^{-4 h_3}
   \mathrm{dp}_3^2+\frac{1}{2} e^{2 h_3-2
   h_1} \mathrm{dp}_4^2+\frac{1}{2} e^{-2
   h_2-2 h_3} \mathrm{dp}_5^2+\frac{1}{2}
   e^{-2 h_1-2 h_3}
   \mathrm{dp}_6^2+\frac{1}{2} e^{-4 h_2}
   \mathrm{dp}_7^2 \nonumber\\
  &&+\frac{1}{2} e^{-2 h_1-2 h_2}
   \mathrm{dp}_8^2-\sqrt{2} e^{-2 h_1-2
   h_2} \mathrm{dp}_7 p_1
   \mathrm{dp}_8+\frac{1}{2} e^{-4 h_1}
   \mathrm{dp}_9^2-e^{2 h_3-2 h_1}
   \mathrm{dp}_2 \mathrm{dp}_4 p_1-e^{-2
   h_1-2 h_3} \mathrm{dp}_5 \mathrm{dp}_6
   p_1 \nonumber\\
  &&+\frac{1}{2} e^{2 h_3-2 h_1}
   \mathrm{dp}_2^2 p_1^2+\frac{1}{2}
   e^{-2 h_1-2 h_3} \mathrm{dp}_5^2
   p_1^2+e^{-2 h_1-2 h_2} \mathrm{dp}_7^2
   p_1^2+e^{-4 h_1} \mathrm{dp}_8^2
   p_1^2-\sqrt{2} e^{-4 h_1}
   \mathrm{dp}_8 \mathrm{dp}_9 p_1 \nonumber\\
  &&+\frac{1}{2} e^{-4 h_1} \mathrm{dp}_7^2
   p_1^4-\sqrt{2} e^{-4 h_1}
   \mathrm{dp}_7 \mathrm{dp}_8 p_1^3+e^{-4
   h_1} \mathrm{dp}_7 \mathrm{dp}_9
   p_1^2-\sqrt{2} e^{-2 h_2-2 h_3}
   \mathrm{dp}_3 \mathrm{dp}_5 p_2-\sqrt{2}
   e^{-4 h_2} \mathrm{dp}_5 \mathrm{dp}_7
   p_2\nonumber\\
  &&-e^{-2 h_1-2 h_2} \mathrm{dp}_6
   \mathrm{dp}_8 p_2+\sqrt{2} e^{-2 h_1-2
   h_3} \mathrm{dp}_3 \mathrm{dp}_6 p_1
   p_2+\sqrt{2} e^{-2 h_1-2 h_2}
   \mathrm{dp}_6 \mathrm{dp}_7 p_1 p_2+2
   e^{-2 h_1-2 h_2} \mathrm{dp}_5
   \mathrm{dp}_8 p_1 p_2\nonumber\\
   &&+\sqrt{2} e^{-4
   h_1} \mathrm{dp}_6 \mathrm{dp}_9 p_1 p_2 \nonumber\\
  &&+\sqrt{2} e^{-4 h_1} \mathrm{dp}_6
   \mathrm{dp}_7 p_2 p_1^3-\sqrt{2} e^{-2
   h_1-2 h_3} \mathrm{dp}_3 \mathrm{dp}_5
   p_2 p_1^2-2 \sqrt{2} e^{-2 h_1-2
   h_2} \mathrm{dp}_5 \mathrm{dp}_7 p_2
   p_1^2\nonumber\\
   &&-2 e^{-4 h_1} \mathrm{dp}_6
   \mathrm{dp}_8 p_2 p_1^2-\sqrt{2} e^{-4
   h_1} \mathrm{dp}_5 \mathrm{dp}_9 p_2
   p_1^2 \nonumber\\
  &&-\sqrt{2} e^{-4 h_1} \mathrm{dp}_5
   \mathrm{dp}_7 p_2 p_1^4+\sqrt{2} e^{-4
   h_1} \mathrm{dp}_6 \mathrm{dp}_7 p_2
   p_1^3+2 e^{-4 h_1} \mathrm{dp}_5
   \mathrm{dp}_8 p_2 p_1^3+e^{-2 h_2-2
   h_3} \mathrm{dp}_3^2 p_2^2\nonumber\\
   &&+e^{-4 h_2}
   \mathrm{dp}_5^2 p_2^2+\frac{1}{2}
   e^{-2 h_1-2 h_2} \mathrm{dp}_6^2 p_2^2\nonumber\\
  &&+ e^{-2 h_1-2 h_3} \mathrm{dp}_3^2 p_1^2
   p_2^2+2 e^{-2 h_1-2 h_2}
   \mathrm{dp}_5^2 p_1^2 p_2^2+e^{-4 h_2}
   \mathrm{dp}_3 \mathrm{dp}_7 p_2^2-2
   e^{-2 h_1-2 h_2} \mathrm{dp}_5
   \mathrm{dp}_6 p_1 p_2^2-\sqrt{2} e^{-2
   h_1-2 h_2} \mathrm{dp}_3 \mathrm{dp}_8
   p_1 p_2^2 \nonumber\\
  &&-2 e^{-4 h_1} \mathrm{dp}_5 \mathrm{dp}_6
   p_2^2 p_1^3-\sqrt{2} e^{-4 h_1}
   \mathrm{dp}_3 \mathrm{dp}_8 p_2^2
   p_1^3+2 e^{-2 h_1-2 h_2}
   \mathrm{dp}_5^2 p_2^2 p_1^2+e^{-4 h_1}
   \mathrm{dp}_6^2 p_2^2 p_1^2\nonumber\\
   &&+2 e^{-2
   h_1-2 h_2} \mathrm{dp}_3 \mathrm{dp}_7
   p_2^2 p_1^2+e^{-4 h_1} \mathrm{dp}_3
   \mathrm{dp}_9 p_2^2 p_1^2 \nonumber\\
  &&+e^{-4 h_1} \mathrm{dp}_5^2 p_2^2
   p_1^4+e^{-4 h_1} \mathrm{dp}_3
   \mathrm{dp}_7 p_2^2 p_1^4-2 \sqrt{2}
   e^{-2 h_1-2 h_2} \mathrm{dp}_3
   \mathrm{dp}_5 p_2^3 p_1^2+\sqrt{2}
   e^{-2 h_1-2 h_2} \mathrm{dp}_3
   \mathrm{dp}_6 p_2^3 p_1-\sqrt{2} e^{-4
   h_2} \mathrm{dp}_3 \mathrm{dp}_5 p_2^3\nonumber\\
  &&+\frac{1}{2} e^{-4 h_1} \mathrm{dp}_3^2
   p_2^4 p_1^4-\sqrt{2} e^{-4 h_1}
   \mathrm{dp}_3 \mathrm{dp}_5 p_2^3
   p_1^4+\sqrt{2} e^{-4 h_1}
   \mathrm{dp}_3 \mathrm{dp}_6 p_2^3
   p_1^3+e^{-2 h_1-2 h_2} \mathrm{dp}_3^2
   p_2^4 p_1^2\nonumber\\
   &&-2 \sqrt{2} e^{-2 h_1-2
   h_2} \mathrm{dp}_3 \mathrm{dp}_5 p_2^3
   p_1^2+\frac{1}{2} e^{-4 h_2}
   \mathrm{dp}_3^2 p_2^4\nonumber\\
  &&+\sqrt{2} e^{-2 h_1-2 h_3} \mathrm{dp}_4
   \mathrm{dp}_6 p_3-\sqrt{2} e^{-2 h_1-2
   h_3} \mathrm{dp}_4 \mathrm{dp}_5 p_1
   p_3\nonumber\\
   &&-\sqrt{2} e^{-2 h_1-2 h_2}
   \mathrm{dp}_4 \mathrm{dp}_8 p_2 p_3+2
   e^{-2 h_1-2 h_3} \mathrm{dp}_3
   \mathrm{dp}_4 p_1 p_2 p_3+2 e^{-2
   h_1-2 h_2} \mathrm{dp}_4 \mathrm{dp}_7
   p_1 p_2 p_3 \nonumber\\
  &&+2 e^{-4 h_1} \mathrm{dp}_4 \mathrm{dp}_7
   p_2 p_3 p_1^3-2 \sqrt{2} e^{-4 h_1}
   \mathrm{dp}_4 \mathrm{dp}_8 p_2 p_3
   p_1^2\nonumber\\
   &&-2 \sqrt{2} e^{-2 h_1-2 h_2}
   \mathrm{dp}_4 \mathrm{dp}_5 p_2^2 p_3
   p_1+2 e^{-4 h_1} \mathrm{dp}_4
   \mathrm{dp}_9 p_2 p_3 p_1+\sqrt{2}
   e^{-2 h_1-2 h_2} \mathrm{dp}_4
   \mathrm{dp}_6 p_2^2 p_3 \nonumber\\
  &&+2 e^{-4 h_1} \mathrm{dp}_3 \mathrm{dp}_4
   p_2^3 p_3 p_1^3-2 \sqrt{2} e^{-4
   h_1} \mathrm{dp}_4 \mathrm{dp}_5 p_2^2
   p_3 p_1^3+2 \sqrt{2} e^{-4 h_1}
   \mathrm{dp}_4 \mathrm{dp}_6 p_2^2 p_3
   p_1^2\nonumber\\
   &&+2 e^{-2 h_1-2 h_2} \mathrm{dp}_3
   \mathrm{dp}_4 p_2^3 p_3 p_1+e^{-2
   h_1-2 h_3} \mathrm{dp}_4^2 p_3^2 \nonumber\\
  &&+ e^{-2 h_1-2 h_2} \mathrm{dp}_4^2 p_2^2
   p_3^2+2 e^{-4 h_1} \mathrm{dp}_4^2
   p_1^2 p_2^2 p_3^2-e^{-2 h_1-2 h_2}
   \mathrm{dp}_5 \mathrm{dp}_8 p_4-\sqrt{2}
   e^{-4 h_1} \mathrm{dp}_6 \mathrm{dp}_9
   p_4+\sqrt{2} e^{-2 h_1-2 h_2}
   \mathrm{dp}_5 \mathrm{dp}_7 p_1 p_4\nonumber\\
  &&+ \sqrt{2} e^{-4 h_1} \mathrm{dp}_5
   \mathrm{dp}_7 p_4 p_1^3-\sqrt{2} e^{-4
   h_1} \mathrm{dp}_6 \mathrm{dp}_7 p_4
   p_1^2-2 e^{-4 h_1} \mathrm{dp}_5
   \mathrm{dp}_8 p_4 p_1^2+2 e^{-4 h_1}
   \mathrm{dp}_6 \mathrm{dp}_8 p_4
   p_1+\sqrt{2} e^{-4 h_1} \mathrm{dp}_5
   \mathrm{dp}_9 p_4 p_1 \nonumber\\
  &&-2 e^{-4 h_1} \mathrm{dp}_5^2 p_2 p_4
   p_1^3+4 e^{-4 h_1} \mathrm{dp}_5
   \mathrm{dp}_6 p_2 p_4 p_1^2-2 e^{-2
   h_1-2 h_2} \mathrm{dp}_5^2 p_2 p_4
   p_1-2 e^{-4 h_1} \mathrm{dp}_6^2 p_2
   p_4 p_1+e^{-2 h_1-2 h_2} \mathrm{dp}_5
   \mathrm{dp}_6 p_2 p_4\nonumber\\
  &&\sqrt{2} e^{-4 h_1} \mathrm{dp}_3
   \mathrm{dp}_5 p_2^2 p_4 p_1^3-\sqrt{2}
   e^{-4 h_1} \mathrm{dp}_3 \mathrm{dp}_6
   p_2^2 p_4 p_1^2+\sqrt{2} e^{-2 h_1-2
   h_2} \mathrm{dp}_3 \mathrm{dp}_5 p_2^2
   p_4 p_1\nonumber\\
   &&-2 \sqrt{2} e^{-4 h_1}
   \mathrm{dp}_4 \mathrm{dp}_6 p_2 p_3 p_4
   p_1+\sqrt{2} e^{-2 h_1-2 h_2}
   \mathrm{dp}_4 \mathrm{dp}_5 p_2 p_3 p_4 \nonumber\\
  &&+e^{-4 h_1} \mathrm{dp}_5^2 p_4^2 p_1^2+2
   \sqrt{2} e^{-4 h_1} \mathrm{dp}_4
   \mathrm{dp}_5 p_2 p_3 p_4 p_1^2-2
   e^{-4 h_1} \mathrm{dp}_5 \mathrm{dp}_6
   p_4^2 p_1+\frac{1}{2} e^{-2 h_1-2
   h_2} \mathrm{dp}_5^2 p_4^2+e^{-4 h_1}
   \mathrm{dp}_6^2 p_4^2 \nonumber\\
\end{eqnarray}}
\end{landscape}

\end{document}